\documentclass[12pt,letterpaper]{article}%
\usepackage{amsfonts}
\usepackage{amssymb}
\usepackage{amsmath}
\usepackage{amsthm}
\usepackage[round,authoryear]{natbib}
\usepackage{booktabs}
\usepackage{dcolumn}
\usepackage{graphicx,epstopdf}
\usepackage[T1]{fontenc}
\usepackage{lmodern}
\usepackage{enumitem}
\usepackage{subfig}
\usepackage{bm}
\usepackage[colorlinks,linkcolor=red,citecolor=blue,pdftex,breaklinks]%
{hyperref}
\usepackage[nameinlink]{cleveref}
\usepackage[flushleft]{threeparttable}
\usepackage{graphicx}%
\usepackage{comment}
\usepackage{anyfontsize}
\providecommand{\U}[1]{\protect\rule{.1in}{.1in}}
\hypersetup{bookmarksopen=true}
\Crefname{assumption}{Assumption}{Assumptions}

\makeatletter
\renewcommand*{\eqref}[1]{\hyperref[{#1}]{\textup{\tagform@{\ref*{#1}}}}}
\makeatother
\renewcommand{\baselinestretch}{1}
 \theoremstyle{plain}
\newtheorem{theorem}{Theorem}

\newtheorem{lemma}{Lemma}[section]
\theoremstyle{definition}
\newtheorem{assumption}{Assumption}
\newtheorem*{algorithm}{Algorithm}
\expandafter
\def \expandafter \normalsize \expandafter{\normalsize \setlength \abovedisplayskip{10pt plus 2pt minus 7pt}}
\expandafter
\def \expandafter \normalsize \expandafter{\normalsize \setlength \abovedisplayshortskip{0pt plus 2pt}}
\expandafter
\def \expandafter \normalsize \expandafter{\normalsize \setlength \belowdisplayskip{10pt plus 2pt minus 7pt}}
\expandafter
\def \expandafter \normalsize \expandafter{\normalsize \setlength \belowdisplayshortskip{5pt plus 2pt minus 3pt}}

\newcommand{\blind}{0}

\begin{document}

\def\spacingset#1{\renewcommand{\baselinestretch}%
{#1}\small\normalsize} \spacingset{1}

\if0\blind
{
  \title{\bf{Rolling Window Selection in FAR Models with Structural  Instabilities}\thanks{I am grateful for comments from  Silvia Gon\c calves, Jiaying Gu, Peter Hansen, Benoit Perron, Yuanyuan Yan, and seminar participants at the 2026 Africa Meeting of the Econometric Society, University of Toronto,   95th Annual Meeting of the Southern Economic Association, Conference Celebrating James MacKinnon at Queen's University, 2025 RESA Conference, 2025 CIREQ Econometrics Conference,
  7th International Conference on Econometrics and Statistics, and 17th International Conference on Computational and Financial Econometrics. The author gratefully acknowledges
financial support from the Social Sciences and Humanities Research
Council (SSHRC) of Canada.}}
  \author{Antoine A. Djogbenou \hspace{.2cm}\\
    Department of Economics, York University\\
    }
  \maketitle
} \fi

\if1\blind
{
  \bigskip
  \bigskip
  \bigskip
  \begin{center}
    {\LARGE\bf Window Selection in FAR Models with Structural  Instabilities}
\end{center}
  \medskip
} \fi

\begin{abstract}
The paper develops a theory for selecting the rolling-window size when generating out-of-sample forecasts with factor-augmented regression (FAR) models in the presence of structural instabilities. It shows how to select a rolling window by minimizing the conditional mean squared forecast error (MSFE) when the factors are latent and need to be estimated. Because the conditional MSFE is unobserved and the factors are latent, this paper proposes a feasible version of the criterion and derives conditions under which the new method is optimal. A simulation experiment documents the procedure's performance.
\end{abstract}

\noindent%

\vfill

\newpage
\spacingset{1.45} 

\section{Introduction}
 Large datasets of financial and macroeconomic variables are currently easily accessible. In many policy-relevant studies, empirical researchers have found it useful to extract common factors from a large set of economic variables. Originally used in economics by \citet*{sargent1977business} and \citet*{geweke1977dynamic}, factor models have been employed to model the dynamics of these common factors, which involve only a few time-series indexes that summarize information in a high-dimensional set of economic variables. 
 \citet*{stock2002forecasting} popularized the use of factor-augmented regression (FAR) models or diffusion indices to model a large set of macroeconomic variables and produce forecasts. In practice, a common approach is to rely on the principal component method (PCM) to extract factors and use the estimated factors as predictors in a forecasting equation. 
 For further discussion of the usefulness or applications in applied macro-econometrics of factor-augmented models for forecasting, see, for example, \citet*{BB2005}, \citet*{DD2013}, \citet*{ABT2015}, \citet*{Djogbenou2020}, \citet*{Djogbenou2024}, and \citet*{DS2024}. 
 PCM's popularity as a factor extraction procedure stems from its computational simplicity and well-known properties established in \citet*{BN2002}, \citet*{Bai2003}, and \citet*{BN2006}. 
 \citet*{BN2002}  propose criteria for selecting the optimal number of factors.  \citet*{Bai2003}  provides a theory for inference of the estimated factors and their loadings.  \citet*{BN2006} derive the conditions for the validity of the inference and prediction intervals for a regression model with estimated factors as predictors. \citet*{BN2013} discuss the conditions for identifying the factors. \citet*{GP2014} derive a formula for the asymptotic bias in the distribution of regression coefficients when using estimated rather than observed factors, and propose bootstrap approaches that account for uncertainty in the estimated factors.  \citet*{DGP2015}  extend their results to account for time dependence in 
 factor-augmented regression models.

 As is well known, very good in-sample predictive ability does not by itself ensure good out-of-sample predictive ability.  \citet*{Rossi2005} points out that although common econometric methods of identifying potentially useful predictors rely on in-sample significance tests, these approaches provide little assurance that the identified predictors are reliable and stable over time. The author discusses optimal model selection tests for regression models with observed regressors under structural instability. \citet*{CM2003} also discuss the harmful effect of structural change on forecast comparisons. 
 \citet*{GMP2017} show that the encompassing test proposed by \citet*{CM2001} and the Granger causality test by \citet*{Mccracken2007} are asymptotically valid when factors are estimated recursively. However, when a rolling-window version of these tests is employed, the effect of structural instability in the factor loadings and regression coefficients is unclear. It is also unclear how much past information to use for forecasting. 

Moreover, in the presence of structural instability, it is natural to rely on observations near the end of the sample.
\citet*{HI2015} argue that factor dynamics change when structural breaks occur because factors' sensitivities change in large data sets. Thus, neglecting these structural changes may reduce efficiency and worsen forecast performance.  They propose the Lagrange multiplier and Wald-type tests for structural breaks. 
 \citet*{BE2011} also suggest an alternative Lagrange multiplier test based on the idea that the number of factors may increase when a structural break occurs in the large panel. 
 
 In this paper, we focus on determining the optimal forecasting window size in the presence of estimated factors and structural instability. Several papers have developed procedures for selecting the window size $R$ in linear regression models with observed regressors. For example, \citet{IJR2017} propose minimizing an approximation to the conditional MSFE, since it is unknown in practice. \citet{CH2012} develop a test for
 smooth structural changes as well as abrupt structural breaks in linear regression models.
 However, none of the existing procedures discuss selecting $R$ in the presence of estimated factors. In FAR, the latent factors are unknown and can be identified only up to rotation.  \citet{BN2009} and \citet{Djogbenou2021} show that the MSFE is inflated in finite samples by a factor-estimation uncertainty term. The implications of factor estimation for selecting the window $R$ remain unstudied. This paper fills this gap.

This paper develops a method for selecting the window size when forecasting in the presence of time-varying parameters and latent factors. It proposes a procedure that minimizes an estimator of the forecaster's
loss function, that is, the conditional MSFE
up to a constant. Because the forecaster's loss function depends on rotated parameters and rotated latent factors, we propose a feasible version that relies on local linear estimators of the regression coefficients and factor loadings.  The paper studies the asymptotic loss efficiency of the new approach and derives conditions for its validity. Simulation experiments document the approach's performance and show a substantial reduction in mean squared forecast error under the new procedure.

The remainder of the paper is organized as follows.  In \Cref{Settings}, we discuss the framework. In \Cref{Main}, we present the assumptions and the paper's main theoretical results. \Cref{Simulation} contains the simulation results, while 
\Cref{Conclusion} concludes. We move the proofs to Appendix A, and Appendix B contains additional simulation results. We denote 
$\left\Vert \bm M\right\Vert =(
\mathrm{Trace}( \bm M^{\top} \bm M) )  ^{1/2}$, the Frobenius (Euclidean)
norm, $\bm Q$ $>0$, the positive definiteness for any square matrix $\bm Q$, and $c$ or
$C$, a generic finite constant.

\section{Settings and Framework}
\label{Settings}
Consider the following time-varying coefficient factor-augmented regression model 
\begin{equation}
y_{t+h}=(\bm\alpha(t/T))^\top \bm f_t  +(\bm\beta(t/T))^\top\bm w_t +\varepsilon_{t+h}, t=1,\ldots, T-h,
\label{FAR}
\end{equation}
where $t$ indicates the time period, $T$ is the time dimension,  $h$ is the horizon of forecast, and $\varepsilon_{t+h}$ is the regression error. 
The model assumes the dependent variable $y_{t+h}$ is explained by $q$ observed regressors, $\bm w_t$, and $r$ latent factors, $\bm f_t$, that summarize the comovements in a large set $N$ of variables, $x_{it},i=1,\ldots,N$, through the approximate factor model
\begin{equation}
x_{it}=(\bm\lambda_i(t/T))^\top \bm f_t +e_{it}, i=1,\ldots,N, t=1,\ldots, T,
\label{panel}
\end{equation}
where $e_{it}$ is the idiosyncratic error and the factor loading $\lambda_i(t/T)$ is an $r$-vector of unknown deterministic functions of $t$. The considered FAR model assumes the regression coefficients and the factor loadings smoothly change over time. 

Our interest is to predict $y_{T+h}$ using the past information and the most recent relevant information that is available, i.e., $\left(y_t, w_t, x_{1t}, \ldots, x_{Nt}\right), t=T-R+1,\ldots, T,$ where $R$ needs to be determined. We contribute to the literature by developing a theory for selecting the window $R$ that is robust to factor estimation and structural changes in factor loadings and regression coefficients. 

For any window $R$, the rolling window approach estimates parameters (factor loadings and regression coefficients) at the terminal date, where $t/T=1$,
and uses $R$ recent observations to generate a forecast.
Because the factors are latent, they must be estimated. 
The rolling window-based sample analog of the latent factor for the last $R$ dates and the associated factor loadings are therefore obtained by minimizing the sum of squared idiosyncratic error, $\sum_{i=1}^{N}\sum_{t=T-R+1}^{T}\left(x_{it}-\bm\lambda_i(1)^\top\bm f_t\right)^2$, under the restriction $\frac{1}{R}\sum_{t=T-R+1}^{T}\bm f_t\bm f_t^\top=\bm I_r$. The solution of the optimization problem, known as the principal component estimator is given by $\tilde{\bm F}_R=[\tilde{\bm f}_{T-R+1}, \ldots,\tilde{\bm f}_T]^\top: R\times r$, which is $\sqrt R$ times the eigenvectors corresponding to the $r$ largest eigenvalues of $\frac{1}{NR}\bm X_R^\top \bm X_R,$ where $\bm X_R=(x_{it})_{i=1,\ldots, N, t=T-R+1,\ldots, T}$. The estimated factor loadings are obtained by regressing the vector corresponding to each row of $\bm X_R$ on the estimated factors. Given the restriction on the estimated factors, the estimated $N\times r$-matrix of factor loadings is given by $\tilde{\bm \Lambda}_R(1)=[\tilde{\bm \lambda}_{1, R}(1), \ldots,\tilde{\bm \lambda}_{N, R}(1)]^\top=\frac{1}{R}\bm X_R \tilde{\bm F}_R$.\footnote{The estimation approach could be extended to allow for kernel smoothing. In that case, we would need to choose a kernel and address boundary issues. See, for example, \citet*{SW2017}, who developed local least-squares estimation for factors and factor loadings over time, for more details. We focus on our objective of estimating the factor at the end of the sample using a constant weight over a window that must be selected appropriately.}
 \citet{Bai2003} shows that estimated factors converge to a rotation of the latent factor, where the rotation matrix is given by
 $\tilde{\bm V}_T^{-1}\frac{1}{T}\tilde{\bm F}_T^\top\bm F_T\frac{1}{N}\bm \Lambda(1)^\top\bm \Lambda(1)$, when the factor loadings in the underlying factor model are time-invariant, where $\tilde{\bm V}_{R}$ is an $r\times r$-diagonal matrix, with the largest eigenvalues on its diagonal in decreasing order using a window of size $R$, and $\bm\Lambda(t/T)$ is the $N\times r$-matrix of factor loadings at time $t$.
 We extend that result to the time-varying setting. As we will see later, $\tilde{\bm f}_{T,R}=\bm H_R\bm f_T+o_P(1)$, 
where $\bm H_R=\tilde{\bm V}_R^{-1}\frac{1}{R}\sum_{t=T-R+1}^T\tilde{\bm f}_t\bm f_t^\top\frac{1}{N}(\bm \Lambda(t/T))^\top\bm \Lambda(1)$ 
is the new rotation matrix, which is unknown to the econometrician. 
We also show in the appendix that $\frac{1}{R}\sum_{t=T-R+1}^T\Vert \tilde{\bm f}_{t,R}-\bm H_{R}\bm f_t\Vert^2
=O_P(\delta_{NRT}^{-2})$,  which extends the finding in \citet{BN2002} that the order is
 $
O_P(\delta_{NT}^{-2})$ to the time-varying setting, where $\delta_{NRT}^2=\min[N,R,\frac{T^2}{R^2}]$ and $\delta_{NT}^2=\min[N,T]$.

Let $\bm\delta_R(1)=(\bm\alpha^\top(1)\bm H_R^{-1}, \bm\beta^\top(1))^\top$, $\hat{\bm z}_{t,R}=(\tilde{\bm f}_{t,R}^\top, \ \bm w_t^\top)^\top$. 
The rolling window based ordinary least-square estimator of $\bm\delta_R(1)$, i.e.,  $\bm\delta(t/T)$ when $t=T$, is obtained by regressing $y_{t+h}$ on $\hat{\bm z}_{t,R}$, for $t=T-R+1,\ldots,T-h,$ and given by
\begin{equation}
	\hat{\bm\delta}_R(1)=\bigg(\sum_{t=T-R+1}^{T-h}\hat{\bm z}_{t,R}\hat{\bm z}_{t,R}^\top\bigg)^{-1}\sum_{t=T-R+1}^{T-h}\hat{\bm z}_{t,R} y_{t+h}.
	\label{OLS}
\end{equation}
The forecast $\hat{y}_{T+h,R}$ of $y_{T+h}$ depends on the choice of the window $R$ and is given by
$	\hat{y}_{T+h,R}=\hat{\bm\delta}_R(1)^\top\hat{\bm z}_{T,R}.$ 
We can write the forecast error $y_{T+h}-\hat y_{T+h,R}$ as
\begin{equation}
	\varepsilon_{T+h}-(\hat{\bm\delta}^\top_R(1)\hat{\bm z}_{T,R}-\bm\delta^\top(1)\bm z_T)=\varepsilon_{T+h}-(\hat{\bm\delta}_R(1)-\bm\delta_R(1))^\top \hat{\bm z}_{T,R}  -\bm\alpha^\top(1)\bm H_R^{-1}(\tilde{\bm f}_{T,R}-\bm H_R \bm f_T),
	\label{FAR_V2}
\end{equation}
where $\bm\delta(1)=(\bm\alpha(1)^\top, \bm\beta(1)^\top)^\top$ and $\bm z_T=(\bm f_T^\top, \bm w_T^\top)^\top$.
The first term is the $ h$-step-ahead error term. The second component captures parameter-estimation uncertainty. Its order in probability is driven by the order of $\hat{\bm\delta}_R(1)-\bm\delta_R(1)=O_P\big(\frac{1}{\sqrt R}+\frac{R}{T}\big)$ as shown in the next section. 
The third term is specific to the factor-augmented regression model and arises because the factors are not observed but estimated. It is the factor estimation uncertainty component in the forecast error. 
As we show below $\tilde{\bm f}_{T,R}-\bm H_R\bm f_T=O_P\big(\frac{1}{\sqrt N}+\frac{1}{\sqrt R\delta_{NRT}}\big)$. This third quantity highlights the need to consider the impact of factor estimation when selecting the optimal rolling window. 

To find an estimate $\hat R$ for $R$, we aim to use a procedure which minimizes the conditional MSFE, $\mathrm{E}((y_{T+h}-\hat y_{T+h,R})^2|\mathcal{F}_T)$, where $\mathcal{F}_T$ is the information set up to time $T$. 
We write 
\begin{align}
    \mathrm{E}((y_{T+h}-\hat y_{T+h,R})^2|\mathcal{F}_T)=&\mathrm{E}\left(\varepsilon_{T+h}^2|\mathcal{F}_T\right)+
    \big(\hat{\bm\delta}^\top_R(1)\hat{\bm z}_{T,R}-\bm\delta^\top(1)\bm z_T\big)^2.
	\label{MSFE}
\end{align}
Because the variance of the $h$-period ahead forecast error $\mathrm{E}(\varepsilon_{T+h}^2|\mathcal{F}_T)=\sigma_T^2$ does not depend on $R$, it can be ignored when searching for $R$. Therefore, we focus on the infeasible loss function that changes with $R$, 
\begin{align}
	C_{NT}^0(R)=& \big(\hat{\bm\delta}^\top_R(1)\hat{\bm z}_{T,R}-\bm\delta^\top(1)\bm z_T\big)^2.
	\label{MSFE}
\end{align}
Because this conditional loss at time $T$ depends on the unobserved latent factor at time $T$ and the unknown parameter $\bm{\delta}(1)$, it cannot be directly minimized by the researcher. 
Therefore, we propose a feasible version $C_{NT}(R)$ of $C_{NT}^0(R)$. 

To obtain the sample counterpart for this unobserved quantity, we extend the approach of \citet*{IJR2017}, who only consider structural change in regression coefficients and observed regressors, to the estimated factor setting.\footnote{It is possible that when a large break occurs, the number of selected factors, using existing factor selection procedures, changes before the terminal date $T$. Our framework could be extended to accommodate this by selecting the number of factors for each candidate window. However, we assume the number of factors to be fixed to keep the theoretical development in Appendix A tractable.}

Let $R_0$ denote a given pilot window size such that $R_0 \gg R$. We rewrite $C_{NT}^0(R)=(\hat{\bm z}_{T,R}^\top\hat{\bm\delta}_R(1)-\bm z_{T}^\top\bm\delta(1))^2$ as 
\begin{align}
&\big(\hat{\bm z}_{T,R}^\top\hat{\bm\delta}_R(1)-\big[(\bm H^{0}\bm f_{T})^\top(((\bm H^{0})^{-1})^\top\bm\alpha(1))+\bm w_{T}^\top\bm\beta(1)\big]\big)^2,\nonumber%
\end{align}
where $\bm H^{0}$ is a rotation-associated matrix for pilot-window-based estimation.
To get a feasible version $C_{NT}(R)$ of $C_{NT}^0(R)$, local linear estimators  
$\ddot{\bm \alpha}(1), \ddot{\bm \beta}(1)\text{   and    }\ddot{\bm f}_{T}$
are used to replace 
$((\bm H^{0})^{-1})^\top\bm\alpha(1), \bm \beta(1)\text{   and   }\bm H^{0}\bm f_T.$  Denote $\tilde{\bm f}_{t}^0, t=T-R_0+1,\ldots,T$, the PCM estimated factors based on the last $R_0$ observations. 
 The proposed procedure minimizes a feasible version of $C_{NT}^0$, where sample estimators $\ddot{\bm \delta}(1)$ and  $\ddot{\bm f}_{T}$ replace a rotated version 
of the unknown parameters
and the latent factors 
at time $T$.
The criterion 
 is therefore given by 
\begin{align}
	C_{NT}(R)=
    &\big(\hat{\bm z}_{T,R}^\top\hat{\bm\delta}_R(1)-\ddot{\bm z}_{T}^\top\ddot{\bm\delta}(1)\big)^2,
	\nonumber
\end{align}
where $\ddot{\bm z}_{T}=(\ddot{\bm f}_{T}^\top,\bm w_T^\top)^\top $.

To find $\ddot{\bm f}_{T}$, we rely on the following local expansion for $\lambda_i(t/T)$ around one,
\begin{equation}
\bm\lambda_i\big(\frac{t}{T}\big)=\bm\lambda_i(1)+\bm\lambda_i^\prime(1)\big(\frac{t}{T}-1\big)+\frac{1}{2}\bm\lambda_i^{\prime\prime}(c_i)\big(\frac{t}{T}-1\big)^2,
\label{loading_expansion}
\end{equation}
where $c_i\in (t/T,1)$ for $t\leq T$, and $\bm\lambda_i^\prime(\cdot)$ and $\bm\lambda_i^{\prime\prime}(\cdot)$ are first and second derivatives of $\bm\lambda_i(\cdot)$. Replacing $\bm\lambda_i\big(\frac{t}{T}\big)$ in the panel factor model and using the identity $\bm H^0\bm f_t=\tilde{\bm f}_{t}^0-(\tilde{\bm f}_{t}^0-\bm H^0\bm f_t)$, we get for $t=T-R_0+1,\ldots,T,$ that
\begin{equation}
x_{it}=(((\bm H^0)^{\top})^{-1}\bm\lambda_i(1))^\top\tilde{\bm f}_{t}^0+(((\bm H^0)^{\top})^{ -1}\bm\lambda_i^\prime(1))^\top\big(\frac{t}{T}-1\big)\tilde{\bm f}_{t}^0+u_{it},
\label{panel_expansion}
\end{equation}
where $u_{it}$ is the new error term.

Now, we can obtain an alternative estimator $\bm \lambda_{i}^0(1)=\ddot{\bm \lambda}_{i}(1)$ for $((\bm H^0)^{\top})^{-1}\bm \lambda_{i}(1)$, by regressing $x_{it}$ 
on $\tilde{\bm f}_{t}^0 \text{ and } \big(t/T-1\big)\tilde{\bm f}_{t}^0, t=T-R_0+1,\ldots,T.$ Thus, the computed $\ddot{\bm \lambda}_{i}$ corresponds to the estimator of the factor loadings associated with $\ddot{\bm f}_T$. To find the alternative estimator $\ddot{\bm f}_T$ of $\bm H^{0}\bm f_T$, 
we note that $\bm H^{0}\bm f_T$
equals 
$$(\tilde{\bm V}^{0})^{-1}R_0^{-1}\sum_{t=T-R_0+1}^T\big((\tilde{\bm f}_{t}^0(\bm H^{0}\bm f_t)^\top )(N^{-1}\sum_{i=1}^N((\bm H^{0})^{-1 \top}\bm \lambda_i(t/T)) ((\bm H^{0})^{-1 \top}\bm \lambda_i)^\top)(\bm H^{0}\bm f_T)\big),$$
where $\tilde{\bm V}^{0}$ is the matrix of eigenvalues from the PCM in decreasing order.
 To get $\ddot{\bm f}_{T}$, the alternative estimator of $\bm H^0\bm f_{T}$, we use the formula
$$
(\tilde{\bm V}^{0})^{-1}\frac{1}{R_0}\sum_{t=T-R_0+1}^T\big((\tilde{\bm f}_{t}^0) (\tilde{\bm f}_{t}^0)^{\top}\big(
\frac{1}{N}\sum_{i=1}^N[\ddot{\bm\lambda}_i(1)+\frac{(t-T)}{T}\ddot{\bm\lambda}_i^\prime(1)]\ddot{\bm\lambda}_i^\top(1)\big)
\tilde{\bm f}_{T}^0\big),$$
replacing $\bm H^{0}\bm f_t$ with $\tilde{\bm f}_{t}^0$ and $\bm\lambda_i^0(t/T)=((\bm H^{0})^{-1})^{\top}\bm \lambda_i(t/T)$ with $\ddot{\bm \lambda}_{i}(1)+\frac{t-T}{T}\ddot{\bm \lambda}_{i}^\prime(1)$.

To obtain $\ddot{\bm \delta}(1)$, we follow \citet*{IJR2017} for the case without latent factors, and use a local linear expansion of $\bm\delta(\cdot)$, which is given by 
\begin{equation}
\bm\delta\bigg(\frac{t}{T}\bigg)=\bm\delta(1)+\bm\delta^\prime(1)\bigg(\frac{t}{T}-1\bigg)+\frac{1}{2}\bm\delta^{\prime\prime}(d)\bigg(\frac{t}{T}-1\bigg)^2,
\label{expansion}
\end{equation}
where $d\in (t/T,1)$ for $t\leq T$, and $\bm\delta^\prime(\cdot)$ and $\bm\delta^{\prime\prime}(\cdot)$ are first and second derivatives of $\bm\delta(\cdot)$.

Define $\hat{\bm z}_{t}^0=\begin{pmatrix} (\tilde{\bm f}_{t}^0)^\top, \bm w_t^\top \end{pmatrix}^\top, $ as well as the rotated parameters
$$  \bm\delta^0\big(\frac{t}{T}\big)=
 \begin{pmatrix}
  ((\bm H^0)^{ -1})^\top\bm\alpha\big(\frac{t}{T}\big) \\
  \bm\beta\big(\frac{t}{T}\big) \end{pmatrix}, \ \
  (\bm\delta^0)^\prime\big(\frac{t}{T}\big)=
 \begin{pmatrix}
  ((\bm H^0)^{ -1})^\top\bm\alpha^\prime\big(\frac{t}{T}\big) \\
  \bm\beta^\prime\big(\frac{t}{T}\big) \end{pmatrix}.$$  
Using the expansion \Cref{expansion}, 
we rewrite \Cref{FAR} when $t$ is close to $T$ as
\begin{align}
y_{t+h}=(\bm\delta^0(1))^\top \hat{\bm z}_{t}^0+(\bm\delta^{0\prime}(1))^\top \big(\frac{t}{T}-1\big) \hat{\bm z}_{t}^0+v_{t+h},
\label{expansion_y}
\end{align}
where $v_{t+h}$ is the corresponding error term.
By regressing $y_{t+h}$ on $\hat{\bm z}_t^0$ and $\frac{t-T}{T}\hat{\bm z}_t^0, t=T-R_0+1,\ldots, T-h$, we get $\ddot{\bm\delta}(1)$, which consists of the estimated coefficients for $\hat{\bm z}_t^0$. Hence, the first $r$ elements of $\ddot{\bm\delta}(1)$ provide $\ddot{\bm\alpha}(1)$. 
Next, we present the convergence rate of the estimators in our new framework as well as the optimality results.

\section{Main Results} 
\label{Main}

Before presenting the main results, we state the assumptions for any rolling window $R$.

\begin{assumption}
\label{Ass1}(Factor model and idiosyncratic errors)
\end{assumption}

\begin{description}
\item[(a)] $\mathrm{E}\left\Vert \bm{f}_{t}\right\Vert ^{8}\leq C$ and
$\bm{\Sigma}_{\bm{F}}=\mathrm{E}(\bm{f}_{t}\bm{f}_{t}^\top)=\bm I_r.$

\item[(b)] The factor loadings $\left\{  \bm{\lambda}_{i}(t/T)\right\}
_{i=1,\ldots,N, \ t=1,\ldots,T}$ are non-random and bounded. In addition, 
$\frac{1}{N}
\sum_{i=1}^N\bm{\lambda}_{i}(1)\bm{\lambda}_{i}^{\top}(1)  =\bm{\Sigma
}_{\bm{\Lambda}}+o(1)$, where $\bm{\Sigma }_{\bm{\Lambda}}$ has eigenvalues that are finite and bounded away from zero.

\item[(c)] The eigenvalues of the $r\times r$ matrix  
$\bm{\Sigma }_{\bm{\Lambda}}  $ 
are distinct.  

\item[(d)] $\mathrm{E}\left(  e_{it}\right)  =0,$ $\mathrm{E}\left\vert
e_{it}\right\vert ^{8}\leq C.$

\item[(e)] $\mathrm{E}\left(  e_{it}e_{js}\right)  =\sigma_{ij,ts}%
$,\ $\left\vert \sigma_{ij,ts}\right\vert \leq\overline{\sigma}_{ij}$ for all
$\left(  t,s\right)  $ and $\left\vert \sigma_{ij,ts}\right\vert \leq\tau
_{st}$ for all $\left(  i,j\right)  $, with $\frac{1}{N}\sum_{i,j=1}%
^{N}\overline{\sigma}_{ij}\leq C,$ $\frac{1}{R}\sum_{t,s=T-R+1}^{T}\tau_{st}\leq
C$ and $\frac{1}{NR}\sum_{i,j=1}^N\sum_{t,s=T-R+1}^T\left\vert \sigma_{ij,ts}\right\vert \leq
C.$

\item[(f)] $\mathrm{E}\left\vert \frac{1}{\sqrt{N}}\sum_{i=1}^{N}\left(
e_{it}e_{is}-E\left(  e_{it}e_{is}\right)  \right)  \right\vert ^{4}\leq C$
for all $\left(  t,s\right)  .$

\item[(g)] $\sum_{s=T-R+1}^T \big\vert \frac{1}{N}\sum_{i=1}^N \sigma_{ii,st} \big\vert\leq C$, for any $t$.
\end{description}

\begin{assumption}
\label{Ass2}(Moment conditions and weak dependence for $\{\bm{z}_{t}\}$ 
and 
$\{e_{it}\}$)
\end{assumption}

\begin{description}
\item[(a)] $
\mathrm{E}\left(  \frac{1}{N}\sum_{i=1}^{N}\left\Vert \frac{T^k}{R^k}\frac
{1}{\sqrt{R}}\sum_{t=T-R+1}^{T}\bm{f}_{t}e_{it}\frac{(T-t)^k}{T^k}\right\Vert
^{2}\right)  \leq C$,  $k=0,1,2$, where $\mathrm{E}\left(  \bm{f}_{t}%
e_{it}\right)  =\bm 0$.

\item[(b)] For each $t,$ $
\mathrm{E}\left\Vert \frac{1}{\sqrt{RN}}\sum
_{s=T-R+1}^{T-h}\sum_{i=1}^{N}\bm{z}_{s}\left(  e_{it}e_{is}-\mathrm{E}\left(
e_{it}e_{is}\right)  \right)  \right\Vert ^{2}\leq C$, where $\bm z_s=(\bm f_s^\top \ \bm w_s^\top)^\top$.

\item[(c)] $
\frac{1}{R^2}\sum_{t=T-R+1}^{T}\sum
_{s=T-R+1}^{T-h} \mathrm E\big(\big\Vert\frac{1}{\sqrt N}\sum_{i=1}^{N}\bm f_s^\top\bm \lambda_i\big(\frac{s}{T}\big)\bm e_{it}\big\Vert^2\big)\leq C$. 

\item[(d)] $
\mathrm{E}\left\Vert \frac{T^k}{R^k} \frac{1}{\sqrt{RN}}\sum_{t=T-R+1}^{T-h}\sum
_{i=1}^{N}\bm{z}_{t}\bm{\lambda}_{i}^{\top}(\frac{s}{T})e_{it} \frac{(T-t)^k}{T^k}
\right\Vert
^{2}\leq C$, $k=0,1,2$, for all $s$, where $\mathrm{E}(\bm{z}_{t}e_{it})  =\bm0$.

\item[(e)] $
\mathrm{E}\left(  
\left\Vert \frac
{1}{\sqrt{N}}\sum_{i=1}^{N}\bm{\lambda}_{i}(s/T)e_{it}\right\Vert
^{4}\right)  \leq C$, for all $(t,s)$.

\item[(f)] $
\mathrm{E}\left(  
\left\Vert \frac
{1}{\sqrt{N}}\sum_{i=1}^{N}\bm{\lambda}_{i}^\prime(1)e_{it}\right\Vert
^{2}\right)  \leq C$
and
 $
\mathrm{E}\left(  
\left\Vert \frac
{1}{\sqrt{N}}\sum_{i=1}^{N}\bm{\lambda}_{i}^{\prime\prime}(c_i)e_{it}\right\Vert
^{2}\right)  \leq C$
for all $(t,s)$.
\end{description}

\begin{assumption}
	\label{Ass3}(Moment conditions and weak dependence for $\{\varepsilon_{t+h}\}$ 
    and 
	$\{e_{it}\}$)
\end{assumption}

\begin{description}
	
	\item[(a)] For each $t,$ $
    \mathrm{E}\left\Vert \frac{1}{\sqrt{RN}}\sum
	_{s=T-R+1}^{T-h}\sum_{i=1}^{N}\varepsilon_{s+h}\left(  e_{it}e_{is}-\mathrm{E}\left(
	e_{it}e_{is}\right)  \right)  \right\Vert ^{2}\leq C$.
	
	\item[(b)] $
    \mathrm{E}\left\Vert \frac{T^k}{R^k}\frac{1}{\sqrt{RN}}\sum_{t=T-R+1}^{T-h}\sum
	_{i=1}^{N}\varepsilon_{t+h}\bm{\lambda}_{i}^{\top}(s/T)e_{it}\frac{(T-t)^k}{T^k}\right\Vert
	^{2}\leq C$, $k=0,1,2$, for all $s$, where $\mathrm{E}\left(  \varepsilon_{t+h}%
    e_{it}\right)  =\bm0$.
\end{description}

\begin{assumption}
	\label{Ass4}(Moments and dependence conditions for $\{v_t\}=\{(\bm z_{t-h}^\top, \varepsilon_t)^\top\}$) 
\end{assumption}

\begin{description}
\item[(a)] $\{\varepsilon_{t+h} \}_{t=1}^T$ is a sequence such that $\mathrm{E}\left(\varepsilon_{t+h} |\mathcal{F}_t\right)=\bm 0,$ $\mathrm{E}\left(\varepsilon_{t+h}^2 |\mathcal{F}_t\right)=\sigma_t^2,$ is well-defined almost surely, and  $\mathcal{F}_t=\sigma(\ldots,x_{1(t-1)},x_{1t} ,\ldots, x_{N(t-1)},x_{Nt},\ldots, \bm z_{t-1}^\top, \bm z_{t}^\top, \ldots, y_{t-1}, y_{t}),$ an information set up to time $t$.

\item[(b)] For $m>2$, the sequence $\{v_t\}$, 
$t=h+1,\ldots,T+h$, is $L_{4m/(m-1)}$-NED of size $-2$, with positive constants $d_t=O(\Vert \bm v_t -\mathrm{E}(v_t)\Vert_{4m/(m-1)})$, on the $\alpha$-mixing sequence  $\{g_t \}_{-\infty}^\infty$ of size $-2m/(m-2)$, and  $\Vert \mathrm{Vec}(\bm v_t \bm v_t^\top) \Vert_m\leq C$.

\item[(c)]  $
\frac{1}{R}\sum_{t=T-R+1}^{T-h}%
\mathrm(\bm{z}_{t}\bm{z}_{t}^{\top})\overset{P}{\longrightarrow}%
\bm{\Sigma }_{\bm{Z}}>0$, where $\bm{\Sigma }_{\bm{Z}}=\mathrm{E}( \bm z_t  \bm z_t^\top)$ is non-random, $\mathrm{E}(\Vert \bm z_t\Vert^8 )\leq C$ and $\mathrm{E}(\vert \varepsilon_{t+h}\vert^8 )\leq C$.

\item[(d)]  All eigenvalues of $\mathrm{E}(\bm z_t\bm z_t^\top)$ and $\mathrm{E}(\bm z_t\bm z_t^\top\varepsilon_{t+h}^2)$ are finite and bounded away from zero uniformly in $t$.
\end{description}

\begin{assumption}
	\label{Ass5}(Smoothness conditions)
\end{assumption}

\begin{description}
\item[(a)]  
$\bm \delta(\cdot)
$ is twice continuously differentiable, 
and $\Vert \bm \delta^{\prime}(t/T) \Vert$ and  $\Vert \bm \delta^{\prime\prime}(t/T) \Vert$  are bounded uniformly in $t$.
			
\item[(b)]  $\bm \lambda(\cdot)$ is twice continuously differentiable, and $\Vert\bm \lambda^{\prime}(t/T)\Vert$ and  $\Vert\bm \lambda^{\prime\prime}(t/T)\Vert$  are bounded uniformly in $t$.

\end{description}

\Cref{Ass1} (a)-(f) adapts Assumptions 1 and 2 on the factor model and the idiosyncratic error of \citet{DGP2015} to the time-varying setting. The key difference in \Cref{Ass1} (a)-(c) comes from the normalization of the latent factor moment to an identity matrix similar to Assumption 1 (a) in \citet{GMP2017}. \Cref{Ass1} (g) is the same as Assumption E (1) on weak dependence in \citet{Bai2003}.  \Cref{Ass2,Ass3} require weak dependence among $\bm z_t, e_{it}$, and $\varepsilon_{t+h}$. 
They modify the standard assumptions in \citet{GP2014} to accommodate local expansion around the terminal date in our derivation of the asymptotic results. \Cref{Ass4} contains conditions that resemble Assumptions 1-3 used by \citet{IJR2017} for the error term and the regressors in the time-varying forecasting equation.
\Cref{Ass5} states that the factor loadings and regression coefficients are twice differentiable and that their derivatives are bounded. The next theorem presents our first key result.

\begin{theorem}[Order for $C_{NT}^0(R)$]
If \Cref{Ass1,Ass2,Ass3,Ass4,Ass5}  hold, $T,$ $%
N\rightarrow \infty $, $\frac{\sqrt{N}}{R}\rightarrow 0$, and $\frac{\sqrt{R}}{N}\rightarrow c<\infty,$ then%
\begin{align}
&\hat{\bm{\delta}}_R(1)-\bm{\delta}_R(1)=O_P\bigg(\frac{1}{\sqrt{R}}+\frac{1}{N}+\frac{R}{T}\bigg)=O_P\bigg(\frac{1}{\sqrt{R}}+\frac{R}{T}\bigg),
	\label{Thm11}\\
&\tilde{\bm f}_{T,R}-\bm H_R\bm f_T=O_P\bigg(\frac{1}{\sqrt{N}}+\frac{1}{\sqrt{R}\delta_{NRT}}\bigg),
	\label{Thm12}\\
& \hat{\bm z}_T^\top\hat{\bm{\delta}}_R(1)-\bm z_T^\top\delta(1)=O_P\bigg(\frac{1}{\sqrt R}+\frac{1}{\sqrt{N}}+\frac{R}{T}\bigg). \label{Thm13}
\end{align}%
In  addition, the infeasible loss 
\begin{equation}
C_{NT}^0(R)=O_P\bigg(\frac{1}{R}+\frac{1}{N}+\frac{R^2}{T^2}\bigg).
\label{Thm14}
\end{equation}
	\label{Thm1}
\end{theorem}

\Cref{Thm1} \eqref{Thm11} and \eqref{Thm12} extend the order in probabilities for the asymptotic results from \citet{BN2006} and \citet{GP2014} to the time-varying context. In \Cref{Thm1} \eqref{Thm11}, we show that the convergence rate of the rolling window-based estimator of the regression coefficients in the presence of estimated factors matches that in \citet{IJR2017} when the factors are observed. When the window size relative to the sample size is small enough, the convergence rate becomes $\sqrt R$. \Cref{Thm1} \eqref{Thm12} is new and extends the $\sqrt N$ convergence rate from \citet{Bai2003} for the estimated factor at the end of the sample. 
Combining these findings, we show in \Cref{Thm1} \eqref{Thm14} that the number of variables $N$ can also affect the accuracy of the conditional mean estimate given information up to time $T$. Thus, factor estimation enters the order of the infeasible loss function in \Cref{Thm1} \eqref{Thm13} when $ R$ and $ N$ are comparable. The next theorem presents the results related to the local expansion.

\begin{theorem}[Order for local linear estimations]
If \Cref{Ass1,Ass2,Ass3,Ass4,Ass5} hold, $T,$ $%
N\rightarrow \infty $, $\frac{\sqrt{N}}{R_0}\rightarrow 0$, and $\frac{\sqrt{R_0}}{N}\rightarrow c<\infty,$ then%
\begin{align}
&\ddot{\bm{\delta}}(1)-\bm{\delta}^0(1)=O_P\bigg(\frac{1}{\sqrt{R_0}}+\frac{1}{N}+\frac{R_0}{T}
\bigg)=O_P\bigg(\frac{1}{\sqrt{R_0}}+\frac{R_0}{T}
\bigg),\label{Thm21}\\
&\ddot{\bm f}_{T}-\bm H^0\bm f_T=O_P\bigg(\frac{1}{\sqrt{N}}+\frac{1}{R_0}+\frac{R_0}{T}
\bigg)=O_P\bigg(\frac{1}{\sqrt{N}}+\frac{R_0}{T}
\bigg),\label{Thm23}\\
& 
\ddot{\bm z}_T^\top\ddot{\bm{\delta}}(1)-\bm z_T^\top\bm{\delta}(1)=O_P\bigg(\frac{1}{\sqrt{R_0}}+\frac{1}{\sqrt{N}}+\frac{R_0}{T}
\bigg),\label{Thm24}
\\
&  C_{NT}^0(\hat R)=O_P\big(\inf_{R\in \Theta_R}C_{NT}^0(R)+\frac{1}{R_0}+\frac{1}{N}+\frac{R_0^2}{T^2}\big),
    \label{Thm25}
\end{align}%
where $\hat{R}=\arg \min_{R\in \Theta_R}C_{NT}\left( R\right)$  and $\Theta_R$ contains all possible values of $R$.
\label{Thm2}
\end{theorem}

In \Cref{Thm2} \eqref{Thm21}, we find that the regression coefficient estimator based on local linear estimation is $O_P\big(\frac{1}{\sqrt{R_0}}+\frac{R_0}{T}
\big)$.   
\Cref{Thm2} \eqref{Thm23} shows that the factor estimator's order in probability is inflated by the pilot window relative to the full sample, compared to the standard asymptotic case. \Cref{Thm2} \eqref{Thm24} provides the resulting order for the conditional mean estimation. 
\Cref{Thm2} \eqref{Thm25} shows that the infeasible loss at the selected window is at most of the order of the smallest infeasible loss over $\Theta_R$ plus the squared order of the pilot-window conditional mean estimation error.  When $\frac{1}{R_0}+\frac{1}{N}+\frac{R_0^2}{T^2}$ is of the same order as, or of smaller order than, $\inf_{R \in\Theta_R}C_{NT}^0(R)$, the selected window attains the best convergence rate over $\Theta_R$. We make the following additional assumptions.

\begin{assumption}
	\label{Ass6}(Uniform order in probability) 
\end{assumption}

\begin{description}
\item[(a)] $\sup_{t,s}
\left( 
\left\Vert \frac
{1}{\sqrt{N}}\sum_{i=1}^{N}\bm{\lambda}_{i}(s/T)e_{it}\right\Vert
^{4}\right) =O_P(1)$.
\item[(b)] $
   \sup_{t,s}\left\vert \frac{1}{\sqrt{N}}\sum_{i=1}^{N}\left(
e_{it}e_{is}-E\left(  e_{it}e_{is}\right)  \right)  \right\vert 
=O_P(1)$.
\item[(c)] $\sup_R\frac{1}{R}\sum_{t=T-R+1}^{T}\Vert\frac{1}{\sqrt{RN}}\sum_{s=T-R+1}^{T-h}(\bm e_s^\top\bm e_t-\mathrm{E}(\bm e_s^\top\bm e_t))\bm z_s\Vert^2=O_P(1)$.
\item[(d)] $\sup_R\frac{1}{R}\sum_{t=T-R+1}^{T}\Vert\frac{1}{\sqrt{RN}}\sum_{s=T-R+1}^{T-h}(\bm e_s^\top\bm e_t-\mathrm{E}(\bm e_s^\top\bm e_t))\varepsilon_{s+h}\Vert^2=O_P(1)$.
\end{description}

\begin{assumption}
	\label{Ass7}(Conditions on $R$ and $R_0$)
\end{assumption}

\begin{description}
	\item[(a)] When $T\rightarrow \infty$, $R_0\rightarrow \infty$, $R\rightarrow \infty$, $R/\min(R_0,N) \rightarrow 0$, and $\frac{NR_0^2}{T^2}\rightarrow 0$
	
	\item[(b)] The cardinality of $\Theta_R \subseteq [\underline{R}, \  \overline{R}]\subset \mathbb{Z}^+$ satisfies $\underline{R}^\rho,$ for some $\rho\in (0,1)$. 
	The conditions on $R$ in \Cref{Ass7} (a) are satisfied by $\underline{R}$ and $\overline{R}$.
\end{description}

In \Cref{Ass6}, we strengthen \Cref{Ass2} and \Cref{Ass3} to establish the new procedure's optimality. In \Cref{Ass7}, we outline the conditions on the rolling window and the pilot window needed for implementation. \Cref{Ass7}
(b) is the same as in \citet{IJR2017}, while  \Cref{Ass7} (a)  is different and assumes the window to be smaller than the number of variables. \Cref{Lemma4} (1), (5) and (6) in Appendix A give bounds depending on some $\theta\in (0,1/2)$, $\nu \in (0,1/4)$, and $\mu\in (0, 1/8)$ chosen such that $\rho\leq \min(2\theta,4\nu,8\mu) $.

Let $m_{NRT}^2=\min(R^{1-2\theta},\,T^2/R^2,\,N)$ and some $\theta\in(0,1/2)$. 
We can deduce from the result in Appendix A, \Cref{Lemma4} (15), that $\sup_{R\in \Theta_R}\big(m_{NRT}^2C_{NT}^0(R)\big)=O_P(1)$. Hence, $m_{NRT}^2$ uniformly bounds the rate at which the infeasible loss converges to zero. It depends on the number of variables in the large panel, the time dimension $T$, and the window size $R$.

\begin{theorem}[Optimality]
If \Cref{Ass1,Ass2,Ass3,Ass4,Ass5,Ass6,Ass7} hold, $T,$ $%
N\rightarrow \infty $, $\frac{\sqrt{N}}{ R}\rightarrow 0$, 
and $\frac{\sqrt{R_0}}{N}\rightarrow c<\infty,$ then%
\begin{equation}
C_{NT}^0(\hat R)-\inf_{R\in \Theta_R} C_{NT}^0(R)=o_P\big(m_{N\hat RT}^{-2}+m_{N\hat R^0T}^{-2}\big),
    \label{Thm31}
\end{equation}
where $\hat{R}=\arg \min_{R\in \Theta_R}C_{NT}\left( R\right)$ and $\hat{R}^0=\arg \min_{R\in \Theta_R}C_{NT}^0\left( R\right)$.

If in addition, $\inf_{R\in\Theta_R}\, \big(m_{NRT}^2C^0_{NT}(R)\big)$ is bounded away from zero in probability, then
\begin{equation}
\frac{C_{NT}^0\left( \hat{R}\right) }{\inf_{R\in \Theta_R}C_{NT}^0\left( R\right)}\overset{P}{\longrightarrow} 1,
\label{Thm32}
\end{equation}%
and
\begin{equation}
\frac{C_{NT}\left( \hat{R}\right) }{\inf_{R\in \Theta_R}C_{NT}^0\left( R\right)}\overset{P}{\longrightarrow} 1.
\label{Thm33}
\end{equation}%
\label{Thm3}
\end{theorem}
Since $m^{-2}_{NRT}$ is a uniform upper bound on the order of $C^0_{NT}(R)$, 
\Cref{Thm3} \eqref{Thm31} shows that the excess loss from using the feasible criterion is of smaller order than this bound, both at the selected window $\hat R$ and at the infeasible optimum $\hat R^0$.
\Cref{Thm3} \eqref{Thm32} proves that the new procedure is asymptotically loss efficient in the sense that $\frac{C_{NT}^0(\hat R)}{\inf_{R\in \Theta_R}C_{NT}^0(R)}\overset{P}{\longrightarrow}1$.  
Therefore, the infeasible loss incurred at the selected window is asymptotically
as small as the smallest infeasible loss attainable over $\Theta_R$. Note that the asymptotic loss efficiency result does not imply consistent selection of the rolling window size, because we do not assume that there is a true window size that needs to be determined. Rather, we establish that the window selected by the feasible criterion attains asymptotically the same infeasible loss as the best window. Consequently, the infeasible loss of the forecaster who uses the selected window $\hat R$ is asymptotically equivalent to the smallest infeasible loss over $\Theta_R$. \Cref{Thm3} \eqref{Thm33} shows that, in addition, the minimized feasible criterion $C_{NT}\left( R\right)$ consistently estimates this smallest loss in ratio. For the ratio in \Cref{Thm32,Thm33} to be well-defined, we require $\inf_{R\in\Theta_R}\, \big(m_{NRT}^2\,C^0_{NT}(R)\big)$ to be bounded away from zero in probability. This condition may fail when there is no structural instability.\footnote{Our simulation results show that the proposed procedure yields a root mean square forecast error close to that of full-sample estimation when there are no structural instabilities.} Note that the optimality result in  \Cref{Thm3} \eqref{Thm31} does not require this lower bound condition. 
We summarize the procedure in the following algorithm.

\begin{algorithm} 
\label{algo1}
\   \

\begin{enumerate}
\item Choose a pilot window size, $R_0$.\footnote{We follow \citet{IJR2017} and select it using the cross-validation approach in \citet{PT2007}.} 
Extract principal components from the last $R_0$ columns of $\bm X$ to get
$\tilde{\bm f}^{0}_t$, $t=T-R_0+1,\dots,T$, with the number of factors $r$, 
and $\tilde{\bm V}^{0}$ is the diagonal matrix of the $r$ largest eigenvalues.
Set $\hat{\bm z}^{0}_t=(\tilde{\bm f}^{0\top}_t, \bm w_t^\top)^\top$.

\item For each $i=1,\dots,N$, regress $x_{it}$ on
$\tilde{\bm f}^{0}_t$ and $\big(\frac{t}{T}-1\big)\tilde{\bm f}^{0}_t$ over the pilot window to obtain
$\ddot{\bm \lambda}_i(1)$ and $\ddot{\bm\lambda}_i'(1)$.

\item Compute
$
\ddot{\bm f}_T=(\tilde{\bm V}^{0})^{-1}\,\frac{1}{R_0}\sum_{t=T-R_0+1}^{T}\big(\tilde{\bm f}^{0}_t\,(\tilde{\bm f}^{0}_t)^\top\,
\frac{1}{N}\sum_{i=1}^{N}\Big[\ddot{\bm\lambda}_i(1)+\frac{t-T}{T}\,\ddot{\bm\lambda}_i'(1)\Big]\ddot{\bm\lambda}_i(1)^\top\,\tilde{\bm f}^{0}_T\big).
$

\item Regress $y_{t+h}$ on $\hat{\bm z}^{0}_t$ and
$\frac{t-T}{T}\hat{\bm z}^{0}_t$ for $t=T-R_0+1,\dots,T-h$. The coefficient on $\hat{\bm z}^{0}_t$ is
$\ddot{\bm\delta}(1)$. Set $\ddot{\bm z}_T=(\ddot{\bm f}_T^\top,\bm w_T^\top)^\top$.

\item For each candidate rolling window $R$:
\begin{enumerate} 
  \item Extract principal components from $X_R=[x_{it}]_{t=T-R+1}^{T}$ to get $\tilde{\bm f}_{t,R}$;
  set $\hat{\bm z}_{t,R}=(\tilde{\bm f}_{t,R}^\top,\bm{w}_t^\top)^\top$.
  \item Estimate $\hat{\bm\delta}_R(1)$ by OLS of $y_{t+h}$ on $\hat{\bm z}_{t,R}$ over the last $R$ observations.
  \item Compute the feasible loss
  \[
  C_{NT}(R)=\Big(\hat{\bm z}_{T,R}^\top\hat{\bm\delta}_R(1)-\ddot{\bm z}_T^\top\ddot{\bm\delta}(1)\Big)^2 .
  \]
\end{enumerate}

\item Select the optimal window $\hat R=\arg\min_{R\in\Theta_R} C_{NT}(R)$, and generate the forecast $\hat y_{T+h}=\hat{\bm z}_{T,\hat R}^\top\,\hat{\bm\delta}_{\hat R}(1)$.
\end{enumerate}
\end{algorithm}
We provide simulation evidence of the performance of the proposed procedure in the next section.

\section{Simulation Experiments}
\label{Simulation}
These simulation experiments compare procedures for selecting the window size to generate forecasts when a latent factor must be estimated. We show how the proposed criterion improves window-size estimation when structural instabilities arise in the latent-factor model loadings and in the slope parameters of the forecasting equation.

To perform the simulation exercise, we consider a factor-augmented regression 
that allows time-varying regression coefficients and factor loadings,
and consider several data generating processes (DGPs). The factor-augmented regression model is given by 
$y_{t+1}=\alpha(t/T)f_t+\beta(t/T)+\varepsilon_{t+1}, t=0,\ldots,T,$
where $f_{t+1}=\rho_ff_t+u_{t+1}$, 
the initial value is $0$, $\rho_f=0.9$, 
 and the error terms  
 $\varepsilon_{t+1}$ and $u_{t+1}$ are $\mathrm{IIDN}(0, 1)$.
 The latent factors $f_t, t=1,\ldots,T$ summarize the comovement in the factor panel model
 $x_{it}=\lambda_i(t/T)f_t+e_{it}, t=1,\ldots, T, i=1,\ldots,N,$
where $\lambda_i(t/T)$ and $e_{it}$ are defined as follows. 

The first DGP, called DGP 1, assumes $\alpha(t/T)=1, \beta(t/T)=1,$ and the factor loadings $\lambda_i(t/T)\sim \mathrm{IIDU}[0,1]$,
where the mean is positive to avoid zero exposure of the variables $x_{it}$  to the latent factor, and we let 
the idiosyncratic errors  $e_{it}=\sigma_{i}v_{it}$, with $\sigma_{i}\sim \mathrm{IIDU}[0,1]$ and $v_{it}\sim \mathrm{IIDN}(0,1)$, to allow for heteroskedasticity in the idiosyncratic errors.
In the second specification called DGP 2, we keep $\alpha(t/T)=1, \beta(t/T)=1,$  $\lambda_i\sim\mathrm{IIDU}[0,1]$, but let $\bm e_t=(e_{1t},\ldots,e_{Nt})^\top \sim \mathrm{IIDN}(\bm 0_{N\times 1},\bm \Sigma_{\bm e})$, where $\bm \Sigma_{\bm e}$ is an $N\times N$ Toeplitz matrix with $(i,j)$-element given by $0.5^{|i-j|}$, similar to \citet{GP2014}, to introduce correlation among the cross-sectional units at each date. Although our theory focuses on smooth structural instabilities, we investigate discrete structural breaks in our experiment to see how the procedures perform. In particular, in DGPs 3 and 4, we modify DGPs 1 and 2 to allow for structural change in the regression coefficients and the factor loadings. In particular, we set $\lambda_{i}(t/T)=\lambda_{i0}+2\mathbb{I}(t\geq 0.5 T)$, where $\lambda_{i0}\sim\mathrm{IIDU}[0,1]$, $\alpha(t/T)=1+2\mathbb{I}(t\geq 0.5T)$, and $\beta(t/T)=1+4\mathbb{I}(t\geq 0.5T)$. 
To allow for smooth structural changes in the loadings, we let $\lambda_i(t/T)=\lambda_{i0}+2t/T$, with $\lambda_{i0}\sim\mathrm{IIDU}[0,1]$, 
 $\alpha(t/T)=1+2t/T$, and $\beta(t/T)=1+4t/T$,
in DGPs 1 and 2, to have DGPs 5 and 6, respectively. For the analysis, we set $T=100,200,300,400,500,$ and $600$, $N=100,200,$ and $300$, 
to discuss finite sample behavior when the sample sizes are not large enough or fall outside the theory. We simulate the data $M=1000$ times. In each iteration of the simulation experiment, we construct the large dataset $\bm X=[x_{it}]: N\times T$. 

Next, we select a pilot rolling window using the cross-validation approach described in \citet{PT2007} and considered by \citet{IJR2017}. We set the minimum rolling window, $R_{\min}$, to the closest integer to $\max(0.10 T, T^{4/5})$, and employ $0.25T$ periods for the evaluation step in the cross-validation.  
To select the number of factors, we apply the $IC_{p2}$ criterion proposed by \citet{BN2002} on recent data based on the pilot window size. Moreover, we compute the root mean square forecast error (RMSFE) under different rolling-window choices, relative to the RMSFE of the forecast based on the full sample. 

To do so, we first use $\bm X$ to estimate the latent factors 
and factor loadings using PCM based on the full sample. We then obtain the least-squares estimator and forecast $y_{T+1}$ using the estimated factors at time $T$. 
In addition, we construct the matrix $\bm X_R=[x_{it}]_{i=1,\ldots,N, t=T-R+1,\ldots,T}$, where $R_{min}\leq R\leq T$.  Therefore, we estimate the factors and factor loadings from the matrix $\bm X_R$. For each criterion discussed previously, we select the window $R$ and generate the forecasts. Furthermore, we select the rolling window using the approach for observed regressors (Method $j=1$) in \citet{IJR2017}, where we treat the estimated factors as observed, the new selection procedure (Method $j=2$), and the selection based on the minimization of the infeasible loss function (Method $ j=3$).\footnote{Note that one can also test for structural instabilities before applying the rolling window selection procedures as in \citet{IJR2017} to exclude situations where the rolling window selection is irrelevant.} We call the forecast based on the full sample $j=0$.

For each simulated data, $m=1,\ldots,M,$ let $\hat y_{T+1}^{(m,j)}$ denote the forecasted values by the method $j=0,1,2,3$,
and the simulated values of $y_{T+1}^{(m)}$. The RMSFE for method $j$ is given by $\sqrt{\frac{1}{M}\sum_{m=1}^{M}(\hat y_{T+1}^{(m,j)}-y_{T+1}^{(m)})^2}$. The results below report the relative RMSFE for methods $j=1,2$ and $3$, 
which is the RMSFE for each of these methods divided by the RMSFE based on the full-sample forecast.

\begin{table}[h!]
\begin{center}
\caption{Relative RMSFE when $N=100$}
\label{tab1}
\begin{tabular}{lcccccc}
\hline\hline
$T=100$ & DGP 1 & DGP 2 & DGP 3 & DGP 4 & DGP 5 & DGP 6 \\ \hline
Method 1 & 1.01 & 1.02 & 0.55 & 0.61 & 0.69 & 0.68 \\ 
Method 2 & 1.02 & 1.04 & 0.46 & 0.47 & 0.57 & 0.58 \\ 
Method 3 & 0.98 & 0.97 & 0.44 & 0.43 & 0.56 & 0.57 \\

\hline
$T=200$ & DGP 1 & DGP 2 & DGP 3 & DGP 4 & DGP 5 & DGP 6 \\ \hline
Method 1 & 1.00 & 1.01 & 0.51 & 0.52 & 0.65 & 0.64 \\ 
Method 2 & 1.00 & 1.01 & 0.45 & 0.47 & 0.55 & 0.55 \\ 
Method 3 & 0.99 & 0.97 & 0.44 & 0.45 & 0.55 & 0.55 \\ 

\hline
$T=300$ & DGP 1 & DGP 2 & DGP 3 & DGP 4 & DGP 5 & DGP 6 \\ \hline
Method 1 & 1.01 & 1.01 & 0.53 & 0.50 & 0.65 & 0.67 \\ 
Method 2 & 1.01 & 1.01 & 0.45 & 0.46 & 0.53 & 0.55 \\ 
Method 3 & 0.99 & 0.98 & 0.44 & 0.44 & 0.53 & 0.55 \\

\hline
$T=400$ & DGP 1 & DGP 2 & DGP 3 & DGP 4 & DGP 5 & DGP 6 \\ \hline
Method 1 & 1.00 & 1.00 & 0.50 & 0.53 & 0.65 & 0.64 \\ 
Method 2 & 1.01 & 1.01 & 0.45 & 0.47 & 0.53 & 0.53 \\ 
Method 3 & 0.99 & 0.98 & 0.45 & 0.45 & 0.53 & 0.53 \\ 

\hline
$T=500$ & DGP 1 & DGP 2 & DGP 3 & DGP 4 & DGP 5 & DGP 6 \\ \hline
Method 1 & 1.00 & 1.00 & 0.49 & 0.48 & 0.63 & 0.65 \\ 
Method 2 & 1.00 & 1.00 & 0.44 & 0.45 & 0.51 & 0.54 \\ 
Method 3 & 0.98 & 0.98 & 0.43 & 0.43 & 0.51 & 0.54 \\

\hline
$T=600$ & DGP 1 & DGP 2 & DGP 3 & DGP 4 & DGP 5 & DGP 6 \\ \hline
Method 1 & 1.00 & 1.00 & 0.50 & 0.50 & 0.64 & 0.64 \\ 
Method 2 & 1.00 & 1.00 & 0.45 & 0.45 & 0.52 & 0.53 \\ 
Method 3 & 0.99 & 0.99 & 0.44 & 0.43 & 0.52 & 0.52 \\

\hline\hline
\end{tabular}
\end{center}
\textit{Note:} This table reports the RMSFEs, which are the square root of the average mean square forecast errors based on minimizing $C_{NT}(R)$ (called Method 2), $C_{NT}^0(R)$ (called Method 3), and the feasible version, where we treat estimated factors as observed (called Method 1),  
divided by the square root of the average MSFE for the full sample forecast.
\end{table}

\begin{table}[h!]
\begin{center}
\caption{Relative RMSFE when $N=200$}
\label{tab2}
\begin{tabular}{lcccccc}
\hline\hline
$T=100$ & DGP 1 & DGP 2 & DGP 3 & DGP 4 & DGP 5 & DGP 6 \\ \hline
Method 1 & 1.01 & 1.01 & 0.62 & 0.60 & 0.72 & 0.70 \\ 
Method 2 & 1.02 & 1.02 & 0.47 & 0.48 & 0.60 & 0.59 \\ 
Method 3 & 0.99 & 0.98 & 0.44 & 0.43 & 0.59 & 0.58 \\
\hline
$T=200$ & DGP 1 & DGP 2 & DGP 3 & DGP 4 & DGP 5 & DGP 6 \\ \hline
Method 1 & 1.01 & 1.02 & 0.50 & 0.50 & 0.63 & 0.65 \\ 
Method 2 & 1.02 & 1.02 & 0.47 & 0.46 & 0.54 & 0.56 \\ 
Method 3 & 0.99 & 0.99 & 0.44 & 0.44 & 0.53 & 0.56 \\ 

\hline
$T=300$ & DGP 1 & DGP 2 & DGP 3 & DGP 4 & DGP 5 & DGP 6 \\ \hline
Method 1 & 1.00 & 1.01 & 0.53 & 0.50 & 0.67 & 0.65 \\ 
Method 2 & 1.00 & 1.01 & 0.47 & 0.46 & 0.55 & 0.54 \\ 
Method 3 & 0.99 & 0.98 & 0.46 & 0.44 & 0.55 & 0.54 \\ 
\hline
$T=400$ & DGP 1 & DGP 2 & DGP 3 & DGP 4 & DGP 5 & DGP 6 \\ \hline
Method 1 & 1.00 & 1.01 & 0.51 & 0.51 & 0.65 & 0.65 \\ 
Method 2 & 1.01 & 1.02 & 0.46 & 0.46 & 0.54 & 0.53 \\ 
Method 3 & 0.99 & 0.99 & 0.45 & 0.45 & 0.54 & 0.53 \\ 
\hline
$T=500$ & DGP 1 & DGP 2 & DGP 3 & DGP 4 & DGP 5 & DGP 6 \\ \hline
Method 1 & 1.01 & 1.01 & 0.51 & 0.48 & 0.64 & 0.65 \\ 
Method 2 & 1.01 & 1.01 & 0.45 & 0.45 & 0.52 & 0.53 \\ 
Method 3 & 0.99 & 0.99 & 0.44 & 0.44 & 0.52 & 0.53 \\ 

\hline
$T=600$ & DGP 1 & DGP 2 & DGP 3 & DGP 4 & DGP 5 & DGP 6 \\ \hline
Method 1 & 1.00 & 1.00 & 0.49 & 0.51 & 0.65 & 0.64 \\ 
Method 2 & 1.00 & 1.01 & 0.44 & 0.46 & 0.51 & 0.53 \\ 
Method 3 & 0.99 & 0.99 & 0.44 & 0.45 & 0.51 & 0.52 \\ 
\hline\hline
\end{tabular}
\newline
\textit{Note:} See note for \Cref{tab1}.
\end{center}
\end{table}

\begin{table}[h!]
\begin{center}
\caption{Relative RMSFE when $N=300$}
\label{tab3}
\begin{tabular}{lcccccc}
\hline\hline
$T=100$ & DGP 1 & DGP 2 & DGP 3 & DGP 4 & DGP 5 & DGP 6 \\ \hline
Method 1 & 1.02 & 1.02 & 0.63 & 0.61 & 0.69 & 0.72 \\ 
Method 2 & 1.02 & 1.02 & 0.50 & 0.48 & 0.59 & 0.60 \\ 
Method 3 & 0.99 & 0.99 & 0.47 & 0.44 & 0.58 & 0.59 \\

\hline
$T=200$ & DGP 1 & DGP 2 & DGP 3 & DGP 4 & DGP 5 & DGP 6 \\ \hline
Method 1 & 1.01 & 1.01 & 0.53 & 0.51 & 0.66 & 0.66 \\ 
Method 2 & 1.01 & 1.01 & 0.49 & 0.46 & 0.56 & 0.57 \\ 
Method 3 & 0.99 & 0.98 & 0.47 & 0.44 & 0.56 & 0.57 \\ 

\hline
$T=300$ & DGP 1 & DGP 2 & DGP 3 & DGP 4 & DGP 5 & DGP 6 \\ \hline
Method 1 & 1.00 & 1.00 & 0.52 & 0.51 & 0.68 & 0.66 \\ 
Method 2 & 1.01 & 1.00 & 0.47 & 0.46 & 0.55 & 0.55 \\ 
Method 3 & 1.00 & 0.99 & 0.46 & 0.45 & 0.55 & 0.55 \\

\hline
$T=400$ & DGP 1 & DGP 2 & DGP 3 & DGP 4 & DGP 5 & DGP 6 \\ \hline
Method 1 & 1.00 & 1.00 & 0.50 & 0.52 & 0.66 & 0.66 \\ 
Method 2 & 1.00 & 1.01 & 0.44 & 0.47 & 0.54 & 0.55 \\ 
Method 3 & 0.99 & 0.99 & 0.43 & 0.45 & 0.54 & 0.54 \\ 

\hline
$T=500$ & DGP 1 & DGP 2 & DGP 3 & DGP 4 & DGP 5 & DGP 6 \\ \hline
Method 1 & 1.00 & 1.01 & 0.50 & 0.50 & 0.66 & 0.65 \\ 
Method 2 & 1.00 & 1.01 & 0.46 & 0.45 & 0.53 & 0.52 \\ 
Method 3 & 0.99 & 0.99 & 0.45 & 0.45 & 0.53 & 0.52 \\ 

\hline
$T=600$ & DGP 1 & DGP 2 & DGP 3 & DGP 4 & DGP 5 & DGP 6 \\ \hline
Method 1 & 1.00 & 1.00 & 0.51 & 0.48 & 0.63 & 0.64 \\ 
Method 2 & 1.00 & 1.00 & 0.45 & 0.44 & 0.53 & 0.52 \\ 
Method 3 & 0.99 & 0.99 & 0.44 & 0.42 & 0.52 & 0.52 \\ 
\hline\hline
\end{tabular}
\newline
\textit{Note:} See note for \Cref{tab1}.
\end{center}
\end{table}

\clearpage
The results for DGPs 1-6 are presented in 
\Cref{tab1,tab2,tab3} for each value of $N$.  \Cref{tab1} shows the results for the six DGPs when $N=100$. The table contains six panels for the different values of $T$. Regardless of the method and sample size, DGPs 1 and 2 show that the RMSFEs are close to 1 when there is no structural instability. These findings suggest that selecting 
the rolling window yields mean squared forecast errors that are close to those from forecasting based on the full sample. When there is a structural break (DGPs 3 and 4) or smooth structural instability (DGPs 5 and 6), the relative RMSFE can be reduced by at least $0.28$ compared to full-sample estimation, across \Cref{tab1,tab2,tab3}. The most important improvement occurs when the gain in accuracy comes from minimizing the infeasible loss function, $C_{NT}^0(R)$. However, the econometrician does not know it, and it cannot be used in practice. The closest in performance is the proposed rolling-window selection procedure, which extends \citet{IJR2017} to account for factor estimation.  The \citet{IJR2017} approach performs reasonably well. Nevertheless, accounting for the fact that the factors are latent and must be estimated yields reduced mean-square forecast errors closer to those from the infeasible criterion. These patterns do not change with more observations and $T=200,300,400,500,600$. The conclusions are similar for $N=200$ in \Cref{tab2} and $N=300$ in \Cref{tab3}. 

We conduct further experiments to assess the selection procedure's performance. We change DGPs 1-6 to DGPs 7-12 by reducing the persistence of the common latent factor. More precisely, we set the autoregressive parameter $\rho_f$ to $0.5$. Moreover, we consider DGPs 13-18, where we augment the factor-augmented regression in DGPs 1-6 with an observed regressor, $y_{t}$. In particular, we use the model $y_{t+1}=\alpha(t/T)f_t+\beta_1(t/T)+\beta_2(t/T)y_t+\varepsilon_{t+1}, t=0,\ldots, T$, while keeping all the other features of DGPs 1-6 the same. Note that $\beta_1(t/T)$ remains the same as $\beta(t/T)$ in DGPs 1-6. For $\beta_2(t/T)$, we let it equal to $0.9$ for DGPs 13 and 14. In DGPs 15 and 16, $\beta_2(t/T)=0.9-0.4\mathbb{I}(t\geq 0.5T)$, while  $\beta_2(t/T)=0.9-0.4(t/T)$ for DGPs 17 and 18.

In \Cref{tab4,tab5,tab6} in Appendix B, we report results for DGPs 7-12, where we modify DGPs 1-6 by setting the autoregressive parameter $\rho_f$ to $0.5$ instead of $0.9$. The results are qualitatively similar, and RMSFE is often smaller than for DGPs with higher autoregressive persistence. In \Cref{tab7,tab8,tab9}, for DGPs 13-18, where we add an observed regressor to DGPs 1-6, the results show that in the presence of instabilities, the proposed approach yields a larger reduction in RMSFE than the procedure that treats the estimated factors as observed.

We also present in Appendix B the selected window sizes that yield the reported RMSFE. See \Cref{tab1b,tab2b,tab3b,tab4b,tab5b,tab6b,tab7b,tab8b,tab9b}. 
We observe that when the latent factor is less persistent and the autoregressive parameter $\rho_f$ changes from $0.9$ to $0.5$, Method 2 tends to select a shorter window size under structural instabilities. 
Also, under structural instabilities, the window sizes are generally closer between the new approach and the infeasible-loss minimization approach.

In summary, accounting for factor estimation matters when selecting the rolling window. We obtain smaller RMSFEs in the presence of structural instabilities. In addition, we achieve a larger reduction in RMSFE when structural changes are smooth than with the approach that ignores factor estimation.

\section{Conclusion}
\label{Conclusion}
This paper explores rolling-window selection for forecasting with linear regression models with unobserved factors. It focuses on the situation where the latent factors are estimated using the principal component method, an unsupervised machine learning technique. The proposed approach minimizes an approximation of a loss function that represents the mean-squared forecast error conditional on past information, up to a constant. The loss function is the squared distance between the fitted value and the unknown conditional mean at the end of the sample.
The paper has two key theoretical results. First, it shows that the infeasible loss function depends on factor-estimation uncertainty and on the number of variables used to estimate the factors.
Second, it derives the conditions under which the estimated rolling window based on the proposed criterion is asymptotically loss-efficient with respect to the infeasible loss function.
The new approach is theoretically valid, easy to implement, and documented with numerical results. Further research could include proving the validity of the proposed method when the number of selected factors changes over the sample.

\clearpage

\section*{Appendix A:  Proofs of Results}
\label{AppendixA}
\renewcommand{\theequation}{A.\arabic{equation}} \renewcommand{\thesection}{A} \setcounter{equation}{0}

\renewcommand{\thelemma}{A.\arabic{lemma}} \renewcommand{\thesection}{A} \setcounter{lemma}{0}

Let $\tilde{\bm V}_R$ be the diagonal matrix consisting of the eigenvalues of 
$\frac{1}{NR}\bm X_R^\top\bm X_R$, in decreasing order,
$\delta_{NRT}^2=\min[N,R,\frac{T^2}{R^2}]$, and $\phi_{NRT}^2=\min[N,R,\frac{T}{R}]$. Recall that $\bm z_t=(\bm f_t^\top,\bm w_t^\top)^\top$. By definition of eigenvectors, $\frac{1}{NR}\bm X_R^\top\bm X_R \tilde{\bm F}_R=\tilde{\bm F}_R\tilde{\bm V}_{R}.$ Therefore, we can replace $\bm X_s=\bm \Lambda(s/T)\bm f_s+\bm e_s$, with $\bm X_s=[x_{1s} \cdots x_{Ns}]^\top$ and $\bm \Lambda(s/T)=[\bm \lambda_1(s/T)\cdots \bm\lambda_N(s/T)]^\top$, in  the expression 
$$\tilde{\bm F}_R=\frac{1}{NR}\bm X_R^\top\sum_{s=T-R+1}^{T}\bm X_{s} \tilde{\bm f}_{s,R}^\top\tilde{\bm V}_{R}^{-1}$$ to have
\begin{equation}
\tilde{\bm F}_R=\frac{1}{NR}\bm X_R^\top\sum_{s=T-R+1}^{T}\bm \Lambda(s/T)\bm f_s\tilde{\bm f}_{s,R}^\top\tilde{\bm V}_{R}^{-1}+\frac{1}{NR}\bm X_R^\top\sum_{s=T-R+1}^{T}\bm e_s \tilde{\bm f}_{s,R}^\top\tilde{\bm V}_{R}^{-1}
\end{equation}
Consequently, we have
\begin{equation}
	\tilde{\bm f}_{t,R}=\tilde{\bm V}_{R}^{-1}\frac{1}{NR}\sum_{s=T-R+1}^{T}\tilde{\bm f}_{s,R}\bm f_s^\top\bm \Lambda^\top(s/T)\bm X_{t}+\tilde{\bm V}_{R}^{-1}\frac{1}{NR}\sum_{s=T-R+1}^{T}\tilde{\bm f}_{s,R}\bm e_s^\top\bm X_t
\end{equation}
and plugging in $\bm X_t$ leads to 
\begin{align}
	\tilde{\bm f}_{t,R}&=\tilde{\bm V}_{R}^{-1}\frac{1}{NR}\sum_{s=T-R+1}^{T}\tilde{\bm f}_{s,R}\bm f_s^\top\bm \Lambda^\top(s/T)\bm \Lambda(t/T)\bm f_t+\tilde{\bm V}_{R}^{-1}\frac{1}{NR}\sum_{s=T-R+1}^{T}\tilde{\bm f}_{s,R}\bm f_s^\top\bm \Lambda^\top(s/T)\bm e_{t}\nonumber\\
	&+\tilde{\bm V}_{R}^{-1}\frac{1}{NR}\sum_{s=T-R+1}^{T}\tilde{\bm f}_{s,R}\bm e_s^\top\bm \Lambda(t/T)\bm f_t+\tilde{\bm V}_{R}^{-1}\frac{1}{NR}\sum_{s=T-R+1}^{T}\tilde{\bm f}_{s,R}\bm e_s^\top\bm e_t.
	\nonumber
\end{align}
In the proofs, we therefore use the identity
\begin{equation}
	\tilde{\bm f}_{t,R}-\bm H_{R}\bm f_t 
	=\tilde{\bm V}_{R}^{-1}\bigg(\bm A_{1t}+\bm A_{2t}+\bm A_{3t}+\bm A_{4t}+\bm A_{5t}\bigg),
	\label{identity}
\end{equation}
with 
$\bm A_{1t}=\frac{1}{R}\sum_{s=T-R+1}^{T} \tilde{\bm f}_s\nu_{st},$ 
$\bm A_{2t}=\frac{1}{R}\sum_{s=T-R+1}^{T} \tilde{\bm f}_s\eta_{st},$ 
$\bm A_{3t}=\frac{1}{R}\sum_{s=T-R+1}^{T} \tilde{\bm f}_s\xi_{st},$ 
$\bm A_{4t}=\frac{1}{R}\sum_{s=T-R+1}^{T} \tilde{\bm f}_s\gamma_{st},$
$\bm A_{5t}=\frac{1}{R}\sum_{s=T-R+1}^{T} \tilde{\bm f}_s\zeta_{st},$
and the time-varying rotation  matrix
$\bm H_{R}=\tilde{\bm V}_{R}^{-1}\frac{1}{R}\sum_{s=T-R+1}^{T} \tilde{\bm f}_s\bm f_s^\top\bm \Lambda^\top(s/T)\bm \Lambda(1)/N,$ 
where $\nu_{st}=\bm f_s^\top\bm \Lambda^\top(s/T) [\Lambda(t/T)-\Lambda(1)]\bm f_{t}/N$,
$\eta_{st}=\bm f_s^\top\bm \Lambda^\top(s/T)\bm e_{t}/N,  \   \  
\xi_{st}=\bm e_s^\top\bm \Lambda(t/T)\bm f_t/N,    \   \  
\gamma_{st}=\mathrm E(\bm e_s^\top\bm e_t/N), $   
and
$\zeta_{st}=\bm e_s^\top\bm e_t/N-\gamma_{st}.$

To prove the main results in  \Cref{Thm1,Thm2,Thm3}, we use the auxiliary results in \Cref{Lemma1,Lemma2,Lemma3,Lemma4}.
\Cref{Lemma1} includes results that follow immediately from 
\citet{IJR2017}. 
 \Cref{Lemma2} extends the results in Lemmas A.1-A.3 of \citet{Bai2003} and Lemma A.1 of  \citet{BN2006} to the time-varying framework. 
\Cref{Lemma3,Lemma4} are needed because the two-step factor model is time-varying and to justify the validity of the selection procedure.

\begin{lemma} \label{Lemma1}
Suppose that \Cref{Ass4} (b)-(d) and \Cref{Ass5} (a) hold. It follows that: 
\begin{enumerate}
\item $\left\Vert  R^{-1}\sum_{t=T-R+1}^{T-h} 
\big[\mathrm{E}(\bm z_t\bm z_t^\top)\big(\frac{t-T}{T}\big)^k\big] \right\Vert\leq C\frac{R^k}{T^k}, \  \  k=0,1,2,3.$

\item $\left\Vert R^{-1}\sum_{t=T-R+1}^{T-h}\bm z_t\varepsilon_{t+h}\big(\frac{t-T}{T}\big)^k \right\Vert=O_P\big(\big(\frac{R}{T}\big)^k R^{-1/2}\big), \  \  k=0,1,2,3.$

\item $\left\Vert  R^{-1}\sum_{t=T-R+1}^{T-h} 
\big[\bm z_t\bm z_t^\top-\mathrm{E}(\bm z_t\bm z_t^\top)\big]\big(\frac{t-T}{T}\big)^k \right\Vert=O_P\big(\big(\frac{R}{T}\big)^k R^{-\frac{1}{2}}\big), \  \  k=0,1,2,3.$

\item $\sup_{R \in \Theta_R} R^{\frac{1}{2}-\theta}\left\Vert  R^{-1}\sum_{t=T-R+1}^{T-h} 
\big[\bm z_t\bm z_t^\top-\mathrm{E}(\bm z_t\bm z_t^\top)\big]\right\Vert=O_P(1),$ for some constant, $\theta \in (0, 1/2)$, and   $\left\Vert  R^{-1}\sum_{t=T-R+1}^{T-h} 
\big[\bm z_t\bm z_t^\top-\mathrm{E}(\bm z_t\bm z_t^\top)\big]\right\Vert=O_P(R^{-\frac{1}{2}}).$

\item $R^{-1}\sum_{t=T-R+1}^{T-h}\Vert\mathrm{E}(\bm z_t \bm z_t^\top) (\bm \delta
\big(\frac{t}{T}\big)-\bm\delta
(1))\Vert\leq C\frac{R}{T}.$

\item $\sup_{R \in \Theta_R} \frac{T}{ R^{\frac{1}{2}+\theta}}\Vert \frac{1}{R}\sum_{t=T-R+1}^{T-h}(\bm z_t \bm z_t^\top-\mathrm{E}(\bm z_t \bm z_t^\top)) (\bm \delta
\big(\frac{t}{T}\big)-\bm\delta
(1))\Vert=O_P(1)$,  for some constant,
 $\theta \in (0, 1/2)$,
and  
$\Vert \frac{1}{R}\sum_{t=T-R+1}^{T-h}(\bm z_t \bm z_t^\top-\mathrm{E}(\bm z_t \bm z_t^\top)) (\bm \delta
\big(\frac{t}{T}\big)-\bm\delta
(1))\Vert=O_P\big(\frac{R}{T}R^{-\frac{1}{2}}\big)$.

\item $\sup_{R \in \Theta_R} \Vert R^{-1}\sum_{t=T-R+1}^{T-h}
(\mathrm{E}(\bm z_t \bm z_t^\top)) \Vert \leq C$.

\item $\sup_{R \in \Theta_R} \Vert R^{-\frac{1}{2}-\theta}\sum_{t=T-R+1}^{T-h} \bm z_t\varepsilon_{t+h}\Vert =O_P(1)$, for some constant,
 $\theta \in (0, 1/2)$.

\item $\sup_{R \in \Theta_R} \Vert TR^{-2} \sum_{t=T-R+1}^{T-h}  \mathrm{E}(\bm z_t \bm z_t^\top)(\bm \delta
\big(\frac{t}{T}\big)-\bm\delta
(1))\Vert \leq C$.

\end{enumerate}
\end{lemma}

\begin{lemma}
\label{Lemma2}
Suppose that \Cref{Ass1,Ass2,Ass3,Ass4,Ass5} hold. It follows that
\begin{enumerate}
\item $\frac{1}{R^2}\sum_{s=T-R+1}^{T}\sum_{t=T-R+1}^{T} \Vert \frac{1}{N} \bm \Lambda^\top(s/T)\bm \Lambda(t/T)-\frac{1}{N}\bm \Lambda^\top(s/T)\bm \Lambda(1)
\Vert^k=O\big(\frac{R^{k}}{T^{k}}\big),$ for $k=1,2,3,4.$
\item $\frac{1}{R}\sum_{t=T-R+1}^T\Vert \tilde{\bm f}_{t,R}-\bm H_{R}\bm f_t\Vert^2
=O_P(\delta_{NRT}^{-2}).$
\item $\frac{1}{R}\sum_{t=T-R+1}^T\Vert \tilde{\bm f}_{t,R}-\bm H_{R}\bm f_t\Vert^2\frac{T-t}{T}= O_P\big(\delta_{NRT}^{-2}\frac{R}{T}\big).$
\item $\frac{1}{R}\sum_{t=T-R+1}^T\Vert \tilde{\bm f}_{t,R}-\bm H_{R}\bm f_t\Vert^2\frac{(t-T)^2}{T^2}=O_P\big(\delta_{NRT}^{-2}\frac{R^2}{T^2}\big).$
\item $\frac{1}{R}\sum_{t=T-R+1}^{T-h}( \tilde{\bm f}_{t,R}-\bm H_{R}\bm f_t)\bm z_{t}^\top=O_P(\phi_{NRT}^{-2}).$
\item $\frac{1}{R}\sum_{t=T-R+1}^{T-h}( \tilde{\bm f}_{t,R}-\bm H_{R}\bm f_t)\hat{\bm z}_{t,R}^\top=O_P(\phi_{NRT}^{-2}).$
\item $\frac{1}{R}\sum_{t=T-R+1}^{T-h}( \tilde{\bm f}_{t,R}-\bm H_{R}\bm f_t)\varepsilon_{t+h}=O_P(\phi_{NRT}^{-2}).$
\item $\frac{1}{R}\sum_{t=T-R+1}^{T-h}( \tilde{\bm f}_{t,R}-\bm H_{R}\bm f_t)\big(\frac{t-T}{T}\big)\bm z_{t}^\top=O_P(\phi_{NRT}^{-2}\frac{R}{T}).$
\item $\frac{1}{R}\sum_{t=T-R+1}^{T-h}( \tilde{\bm f}_{t,R}-\bm H_{R}\bm f_t)\big(\frac{t-T}{T}\big)\hat{\bm z}_{t,R}^\top=O_P(\phi_{NRT}^{-2}\frac{R}{T}).$
\item $\frac{1}{R}\sum_{t=T-R+1}^{T-h}( \tilde{\bm f}_{t,R}-\bm H_{R}\bm f_t)\big(\frac{t-T}{T}\big)\varepsilon_{t+h}=O_P(\phi_{NRT}^{-2}\frac{R}{T}).$
\item $\frac{1}{R}\sum_{t=T-R+1}^{T-h}( \tilde{\bm f}_{t,R}-\bm H_{R}\bm f_t)\big(\frac{t-T}{T}\big)^2\bm z_{t}^\top=O_P(\phi_{NRT}^{-2}\frac{R^2}{T^2}).$
\item $\frac{1}{R}\sum_{t=T-R+1}^{T-h}( \tilde{\bm f}_{t,R}-\bm H_{R}\bm f_t)\big(\frac{t-T}{T}\big)^2\hat{\bm z}_{t}^\top=O_P(\phi_{NRT}^{-2}\frac{R^2}{T^2}).$
\item $\frac{1}{R}\sum_{t=T-R+1}^{T-h}( \tilde{\bm f}_{t,R}-\bm H_{R}\bm f_t)\big(\frac{t-T}{T}\big)^2\varepsilon_{t+h}=O_P(\phi_{NRT}^{-2}\frac{R^2}{T^2}).$
\item $\frac{1}{R}\sum_{t=T-R+1}^{T-h}( \tilde{\bm f}_{t,R}-\bm H_{R}\bm f_t)\big(\frac{t-T}{T}\big)^3\bm z_{t}^\top=O_P(\phi_{NRT}^{-2}\frac{R^3}{T^3}).$
\item $\frac{1}{R}\sum_{t=T-R+1}^{T-h}( \tilde{\bm f}_{t,R}-\bm H_{R}\bm f_t)\big(\frac{t-T}{T}\big)^3\hat{\bm z}_{t}^\top=O_P(\phi_{NRT}^{-2}\frac{R^3}{T^3}).$
\item $\frac{1}{R}\sum_{t=T-R+1}^{T-h}( \tilde{\bm f}_{t,R}-\bm H_{R}\bm f_t)\big(\frac{t-T}{T}\big)^3\varepsilon_{t+h}=O_P(\phi_{NRT}^{-2}\frac{R^3}{T^3}).$
 \item $\frac{1}{R}\sum_{t=T-R+1}^{T-h}\hat{\bm z}_{t,R}\hat{\bm z}_{t,R}^\top-(\bm \Phi_R )^{\top}\big(\frac{1}{R}\sum_{t=T-R+1}^{T-h}\bm z_{t}\bm z_{t}^\top\big)\bm \Phi_R=o_P(1)$, where we denote $\bm \Phi_R=\mathrm{diag}(\bm H_R, \bm I_q) $.
\item $(\bm \Phi_R )^{\top}\big(\frac{1}{R}\sum_{t=T-R+1}^{T-h}\bm z_{t}\bm z_{t}^\top\big)\bm \Phi_R -(\bm \Phi )^{\top}\bm \Sigma_{\bm Z} \bm \Phi=o_P(1),$
where $\bm H=\mathrm{plim}\ \bm H_R$ and $\bm \Phi=\mathrm{diag}(\bm H, \bm I_q)$.

\item $\frac{1}{R}\sum_{t=T-R+1}^{T-h}(\hat{\bm z}_{t,R}-\bm \Phi_R\bm z_t)\bm z_{t}^\top(\bm\delta
(t/T)-\bm\delta
(1))=O_P\big(\phi_{NRT}^{-2}\frac{R}{T}\big)$ and $\frac{1}{R}\sum_{t=T-R+1}^{T-h}\bm z_{t}(\hat{\bm z}_{t,R}-\bm \Phi_R\bm z_t)^\top(\bm\delta_R(t/T)-\bm\delta_R(1))=O_P\big(\phi_{NRT}^{-2}\frac{R}{T}\big)$.

\item $\frac{1}{R}\sum_{t=T-R+1}^{T-h}(\hat{\bm z}_{t,R}-\bm \Phi_R\bm z_t)\hat{\bm z}_{t,R}^\top(\bm\delta_R(t/T)-\bm\delta_R(1))=O_P\big(\phi_{NRT}^{-2}\frac{R}{T}\big)$
and
$\frac{1}{R}\sum_{t=T-R+1}^{T-h}\hat{\bm z}_{t,R}\times(\hat{\bm z}_{t,R}-\bm \Phi_R\bm z_t)^\top(\bm\delta_R(t/T)-\bm\delta_R(1))=O_P\big(\phi_{NRT}^{-2}\frac{R}{T}\big)$.
\end{enumerate}
\end{lemma}

\begin{lemma}
\label{Lemma3}
Suppose that \Cref{Ass1}, \Cref{Ass2} and \Cref{Ass5} (b) hold. 
\begin{enumerate}
\item[1.] We have
\begin{align*}
\bm W_{1}&=(\bm H^{0\top})^{-1}\frac{1}{N}\sum_{i=1}^N\bm\lambda_i(1)\bm\lambda_i^\top(1)(\bm H^{0})^{-1}=O_P(1),\\
\bm W_{2}&=(\bm H^{0\top})^{-1}\frac{1}{N}\sum_{i=1}^N\bm\lambda_i^\prime(1)\bm\lambda_i^{\prime\top}(1)(\bm H^{0})^{-1}=O_P(1),\\
\bm W_{3}&=
\frac{1}{N}\sum_{i=1}^N\bm\lambda_i^{\prime\prime}(c_i)\bm\lambda_i^{\prime\prime\top}(c_i)=O(1).
\\
\bm W_{10}&=(\bm H^{0\top})^{-1}\frac{1}{N}\sum_{i=1}^N\bm\lambda_i(1)\bm\lambda_i^{\prime\top}(1)(\bm H^{0})^{-1}=O_P(1),\\
\bm W_{11}&=(\bm H^{0\top})^{-1}\frac{1}{N}\sum_{i=1}^N\bm\lambda_i(1)\bm\lambda_i^{\prime\prime\top}(c_i)
=O_P(1),\\
\bm W_{20}&=(\bm H^{0\top})^{-1}\frac{1}{N}\sum_{i=1}^N\bm\lambda_i^{\prime}(1)\bm\lambda_i^{\prime\prime\top}(c_i)=O_P(1).
\end{align*}
\end{enumerate}

\begin{enumerate}
\item[2.] We have
\end{enumerate}
\begin{align*}
\bm W_{4}&=\frac{1}{R_0^2}\sum\sum(\tilde{\bm f}^0_{t}-\bm H^0\bm f_t)(\tilde{\bm f}^0_{s}-\bm H^0\bm f_s)^\top \frac{1}{N}\sum_{i=1}^Ne_{it}e_{is}=O_P\big(\delta_{NR_0T}^{-2}\big),
\\
\bm W_{5}&=\frac{1}{R_0^2}\sum\sum(\frac{t}{T}-1)(\tilde{\bm f}^0_{t}-\bm H^0\bm f_t)(\tilde{\bm f}^0_{s}-\bm H^0\bm f_s)^\top \frac{1}{N}\sum_{i=1}^Ne_{it}e_{is}=O_P\big(\frac{R_0}{T}\delta_{NR_0T}^{-2}\big),
\\
\bm W_{6}&=\frac{1}{R_0^2}\sum\sum(\frac{t}{T}-1)(\frac{s}{T}-1)(\tilde{\bm f}^0_{t}-\bm H^0\bm f_t)(\tilde{\bm f}^0_{s}-\bm H^0\bm f_s)^\top \frac{1}{N}\sum_{i=1}^Ne_{it}e_{is}=O_P\big(\frac{R_0^2}{T^2}\delta_{NR_0T}^{-2}\big).
\end{align*}

\begin{enumerate}
\item[3.] We have
\end{enumerate}
\begin{align*}
\bm W_{7}&=\bm H^0\frac{1}{R_0^2}\sum\sum\bm f_{t}\bm f_{s}^\top  \frac{1}{N}\sum_{i=1}^N e_{it}e_{is}  \bm H^{0\top}=O_P\big(\frac{1}{R_0}\big),\\
\bm W_{8}&=\bm H^0\frac{1}{R_0^2}\sum\sum(\frac{t}{T}-1)\bm f_{t}\bm f_{s}^\top  \frac{1}{N}\sum_{i=1}^N e_{it}e_{is}\bm H^{0\top} =O_P\big(\frac{R_0}{T}\frac{1}{R_0}\big),\\
\bm W_{9}&=\bm H^0\frac{1}{R_0^2}\sum\sum(\frac{t}{T}-1)(\frac{s}{T}-1)\bm f_{t}\bm f_{s}^\top  \frac{1}{N}\sum_{i=1}^N e_{it}e_{is} \bm H^{0\top}=O_P\big(\frac{R_0^2}{T^2}\frac{1}{R_0}).
\end{align*}

\begin{enumerate}
\item[4.] We have
\end{enumerate}
\begin{align*}
\bm W_{12}&=\frac{1}{R_0^2}\sum\sum \tilde{\bm f}^0_{t}(\tilde{\bm f}^0_{t}-\bm H^0 \bm f_t)^\top (\bm H^{0\top})^{-1}\frac{1}{N}\sum_{i=1}^N\bm\lambda_i(1) e_{is}
(\tilde{\bm f}^0_{s}-\bm H^0\bm f_s)^\top\\
&=O_P\big(\frac{\phi_{NR_0T}^{-2}\delta_{NR_0T}^{-1}}{\sqrt N}\big),\\
\bm W_{13}&=\frac{1}{R_0^2}\sum\sum \tilde{\bm f}^0_{t}(\tilde{\bm f}^0_{t}-\bm H^0 \bm f_t)^\top (\bm H^{0\top})^{-1}\frac{1}{N}\sum_{i=1}^N\bm\lambda_i(1) e_{is}
(\frac{s}{T}-1)(\tilde{\bm f}^0_{s}-\bm H^0\bm f_s)^\top\\
&=O_P\big(\frac{R_0}{T}\frac{\phi_{NR_0T}^{-2}\delta_{NR_0T}^{-1}}{\sqrt N}\big),\\
\bm W_{14}&=\frac{1}{R_0^2}\sum\sum (\frac{t}{T}-1)\tilde{\bm f}^0_{t}(\tilde{\bm f}^0_{t}-\bm H^0 \bm f_t)^\top (\bm H^{0\top})^{-1}\frac{1}{N}\sum_{i=1}^N\bm\lambda_i(1) e_{is}
(\tilde{\bm f}^0_{s}-\bm H^0\bm f_s)^\top\\
&=O_P\big(\frac{R_0}{T}\frac{\phi_{NR_0T}^{-2}\delta_{NR_0T}^{-1}}{\sqrt N}\big),
\\
\bm W_{15}&=\frac{1}{R_0^2}\sum\sum (\frac{t}{T}-1)\tilde{\bm f}^0_{t}(\tilde{\bm f}^0_{t}-\bm H^0 \bm f_t)^\top (\bm H^{0\top})^{-1}\frac{1}{N}\sum_{i=1}^N\bm\lambda_i(1) e_{is}
(\frac{s}{T}-1)(\tilde{\bm f}^0_{s}-\bm H^0\bm f_s)^\top\\
&=O_P\big(\frac{R_0^2}{T^2}\frac{\phi_{NR_0T}^{-2}\delta_{NR_0T}^{-1}}{\sqrt N}\big).
\end{align*}

\begin{enumerate}
\item[5.] We have
\end{enumerate}
\begin{align*}
\bm W_{16}&=\frac{1}{R_0^2}\sum\sum \tilde{\bm f}^0_{t}(\tilde{\bm f}^0_{t}-\bm H^0 \bm f_t)^\top (\bm H^{0\top})^{-1}\frac{1}{N}\sum_{i=1}^N\bm\lambda_i(1) e_{is}
(\bm H^0\bm f_s)^\top\\
&=O_P\big(\frac{\phi_{NR_0T}^{-2}}{\sqrt{N}}\big), \\
\bm W_{17}&=\frac{1}{R_0^2}\sum\sum \tilde{\bm f}^0_{t}(\tilde{\bm f}^0_{t}-\bm H^0 \bm f_t)^\top (\bm H^{0\top})^{-1}\frac{1}{N}\sum_{i=1}^N\bm\lambda_i(1) e_{is}
(\frac{s}{T}-1)(\bm H^0\bm f_s)^\top \\
&=O_P\big(\frac{R_0}{T}\frac{\phi_{NR_0T}^{-2}}{\sqrt{N}}\big),\\
\bm W_{18}&=\frac{1}{R_0^2}\sum\sum (\frac{t}{T}-1)\tilde{\bm f}^0_{t}(\tilde{\bm f}^0_{t}-\bm H^0 \bm f_t)^\top (\bm H^{0\top})^{-1}\frac{1}{N}\sum_{i=1}^N\bm\lambda_i(1) e_{is}
(\bm H^0\bm f_s)^\top \\
&=O_P\big(\frac{R_0}{T}\frac{\phi_{NR_0T}^{-2}}{\sqrt{N}}\big),
\\
\bm W_{19}&=\frac{1}{R_0^2}\sum\sum (\frac{t}{T}-1)\tilde{\bm f}^0_{t}(\tilde{\bm f}^0_{t}-\bm H^0 \bm f_t)^\top (\bm H^{0\top})^{-1}\frac{1}{N}\sum_{i=1}^N\bm\lambda_i(1) e_{is}
(\frac{s}{T}-1)(\bm H^0\bm f_s)^\top \\
&=O_P\big(\frac{R_0^2}{T^2}\frac{\phi_{NR_0T}^{-2}}{\sqrt{N}}\big).
\end{align*}

\begin{enumerate}
\item[6.] We have
\end{enumerate}
\begin{align*}
\bm W_{21}&=\frac{1}{R_0^2}\sum\sum (\frac{t}{T}-1) \tilde{\bm f}^0_{t}(\tilde{\bm f}^0_{t}-\bm H^0 \bm f_t)^\top (\bm H^{0\top})^{-1}\frac{1}{N}\sum_{i=1}^N\bm\lambda_i^\prime(1) e_{is}
(\tilde{\bm f}^0_{s}-\bm H^0\bm f_s)^\top \\
&=O_P\big(\frac{R_0}{T}\frac{\phi_{NR_0T}^{-2}\delta_{NR_0T}^{-1}}{\sqrt N}\big),\\
\bm W_{22}&=\frac{1}{R_0^2}\sum\sum (\frac{t}{T}-1) \tilde{\bm f}^0_{t}(\tilde{\bm f}^0_{t}-\bm H^0 \bm f_t)^\top (\bm H^{0\top})^{-1}\frac{1}{N}\sum_{i=1}^N\bm\lambda_i^\prime(1) e_{is}
(\frac{s}{T}-1)(\tilde{\bm f}^0_{s}-\bm H^0\bm f_s)^\top \\
&=O_P\big(\frac{R_0^2}{T^2}\frac{\phi_{NR_0T}^{-2}\delta_{NR_0T}^{-1}}{\sqrt N}\big),\\
\bm W_{23}&=\frac{1}{R_0^2}\sum\sum (\frac{t}{T}-1)^2 \tilde{\bm f}^0_{t}(\tilde{\bm f}^0_{t}-\bm H^0 \bm f_t)^\top (\bm H^{0\top})^{-1}\frac{1}{N}\sum_{i=1}^N\bm\lambda_i^\prime(1) e_{is}
(\tilde{\bm f}^0_{s}-\bm H^0\bm f_s)^\top\\
&=O_P\big(\frac{R_0^2}{T^2}\frac{\phi_{NR_0T}^{-2}\delta_{NR_0T}^{-1}}{\sqrt N}\big).
\\
\bm W_{24}&=\frac{1}{R_0^2}\sum\sum (\frac{t}{T}-1)^2 \tilde{\bm f}^0_{t}(\tilde{\bm f}^0_{t}-\bm H^0 \bm f_t)^\top (\bm H^{0\top})^{-1}\frac{1}{N}\sum_{i=1}^N\bm\lambda_i^\prime(1) e_{is}
(\frac{s}{T}-1)(\tilde{\bm f}^0_{s}-\bm H^0\bm f_s)^\top\\
&=O_P\big(\frac{R_0^3}{T^3}\frac{\phi_{NR_0T}^{-2}\delta_{NR_0T}^{-1}}{\sqrt N}\big).
\end{align*}

\begin{enumerate}
\item[7.] We have
\end{enumerate}
\begin{align*}
\bm W_{25}&=\frac{1}{R_0^2}\sum\sum (\frac{t}{T}-1) \tilde{\bm f}^0_{t}(\tilde{\bm f}^0_{t}-\bm H^0 \bm f_t)^\top (\bm H^{0\top})^{-1}\frac{1}{N}\sum_{i=1}^N\bm\lambda_i^\prime(1) e_{is}
(\bm H^0\bm f_s)^\top \\
&=O_P\big(\frac{R_0}{T}\frac{\phi_{NR_0T}^{-2}}{\sqrt N}\big),\\
\bm W_{26}&=\frac{1}{R_0^2}\sum\sum (\frac{t}{T}-1) \tilde{\bm f}^0_{t}(\tilde{\bm f}^0_{t}-\bm H^0 \bm f_t)^\top (\bm H^{0\top})^{-1}\frac{1}{N}\sum_{i=1}^N\bm\lambda_i^\prime(1) e_{is}
(\frac{s}{T}-1)(\bm H^0\bm f_s)^\top\\
&=O_P\big(\frac{R_0^2}{T^2}\frac{\phi_{NR_0T}^{-2}}{\sqrt N}\big),\\
\bm W_{27}&=\frac{1}{R_0^2}\sum\sum (\frac{t}{T}-1)^2 \tilde{\bm f}^0_{t}(\tilde{\bm f}^0_{t}-\bm H^0 \bm f_t)^\top (\bm H^{0\top})^{-1}\frac{1}{N}\sum_{i=1}^N\bm\lambda_i^\prime(1) e_{is}
(\bm H^0\bm f_s)^\top \\
&=O_P\big(\frac{R_0^2}{T^2}\frac{\phi_{NR_0T}^{-2}}{\sqrt N}\big),\\
\bm W_{28}&=\frac{1}{R_0^2}\sum\sum (\frac{t}{T}-1)^2 \tilde{\bm f}^0_{t}(\tilde{\bm f}^0_{t}-\bm H^0 \bm f_t)^\top (\bm H^{0\top})^{-1}\frac{1}{N}\sum_{i=1}^N\bm\lambda_i^\prime(1) e_{is}
(\frac{s}{T}-1)(\bm H^0\bm f_s)^\top\\
&=O_P\big(\frac{R_0^3}{T^3}\frac{\phi_{NR_0T}^{-2}}{\sqrt N}\big). 
\end{align*}

\begin{enumerate}
\item[8.] We have
\end{enumerate}
\begin{align*}
\bm W_{29}&=\frac{1}{R_0^2}\sum\sum (\frac{t}{T}-1)^2 (\tilde{\bm f}^0_{t}-\bm H^0 \bm f_t)\bm f_t^\top\frac{1}{N}\sum_{i=1}^N\bm\lambda_i^{\prime\prime}(c_i) e_{is}
(\tilde{\bm f}^0_{s}-\bm H^0\bm f_s)^\top\\
&=O_P\big(\frac{R_0^2}{T^2}\frac{\phi_{NR_0T}^{-2}\delta_{NR_0T}^{-1}}{\sqrt N}\big),\\
\bm W_{30}&=\frac{1}{R_0^2}\sum\sum (\frac{t}{T}-1)^2 (\tilde{\bm f}^0_{t}-\bm H^0 \bm f_t)\bm f_t^\top\frac{1}{N}\sum_{i=1}^N\bm\lambda_i^{\prime\prime}(c_i) e_{is}
(\frac{s}{T}-1)(\tilde{\bm f}^0_{s}-\bm H^0\bm f_s)^\top\\
&=O_P\big(\frac{R_0^3}{T^3}\frac{\phi_{NR_0T}^{-2}\delta_{NR_0T}^{-1}}{\sqrt N}\big),\\
\bm W_{31}&=\frac{1}{R_0^2}\sum\sum (\frac{t}{T}-1)^3 (\tilde{\bm f}^0_{t}-\bm H^0 \bm f_t) \bm f_t^\top \frac{1}{N}\sum_{i=1}^N\bm\lambda_i^{\prime\prime}(c_i) e_{is}
(\tilde{\bm f}^0_{s}-\bm H^0\bm f_s)^\top\\
&=O_P\big(\frac{R_0^3}{T^3}\frac{\phi_{NR_0T}^{-2}\delta_{NR_0T}^{-1}}{\sqrt N}\big),\\
\bm W_{32}&=\frac{1}{R_0^2}\sum\sum (\frac{t}{T}-1)^3 (\tilde{\bm f}^0_{t}-\bm H^0 \bm f_t) \bm f_t^\top\frac{1}{N}\sum_{i=1}^N\bm\lambda_i^{\prime\prime}(c_i) e_{is}
(\frac{s}{T}-1)(\tilde{\bm f}^0_{s}-\bm H^0\bm f_s)^\top\\ &=O_P\big(\frac{R_0^4}{T^4}\frac{\phi_{NR_0T}^{-2}\delta_{NR_0T}^{-1}}{\sqrt N}\big).
\end{align*}

\begin{enumerate}
\item[9.] We have
\end{enumerate}
\begin{align*}
\bm W_{33}&=\frac{1}{R_0^2}\sum\sum (\frac{t}{T}-1)^2 (\tilde{\bm f}^0_{t}-\bm H^0 \bm f_t)\bm f_t^\top \frac{1}{N}\sum_{i=1}^N\bm\lambda_i^{\prime\prime}(c_i) e_{is}
(\bm H^0\bm f_s)^\top\\
&=O_P\big(\frac{R_0^2}{T^2}\frac{\phi_{NR_0T}^{-2}}{\sqrt N}\big),\\
\bm W_{34}&=\frac{1}{R_0^2}\sum\sum (\frac{t}{T}-1)^2 (\tilde{\bm f}^0_{t}-\bm H^0 \bm f_t) \bm f_t^\top \frac{1}{N}\sum_{i=1}^N\bm\lambda_i^{\prime\prime}(c_i) e_{is}
(\frac{s}{T}-1)(\bm H^0\bm f_s)^\top \\
&=O_P\big(\frac{R_0^3}{T^3}\frac{\phi_{NR_0T}^{-2}}{\sqrt N}\big),\\
\bm W_{35}&=\frac{1}{R_0^2}\sum\sum (\frac{t}{T}-1)^3 (\tilde{\bm f}^0_{t}-\bm H^0 \bm f_t)\bm f_t^\top \frac{1}{N}\sum_{i=1}^N\bm\lambda_i^{\prime\prime}(c_i) e_{is}
(\bm H^0\bm f_s)^\top \\
&=O_P\big(\frac{R_0^3}{T^3}\frac{\phi_{NR_0T}^{-2}}{\sqrt N}\big),
\\
\bm W_{36}&=\frac{1}{R_0^2}\sum\sum (\frac{t}{T}-1)^3 (\tilde{\bm f}^0_{t}-\bm H^0 \bm f_t)\bm f_t^\top \frac{1}{N}\sum_{i=1}^N\bm\lambda_i^{\prime\prime}(c_i) e_{is}
(\frac{s}{T}-1)(\bm H^0\bm f_s)^\top\\
&=O_P\big(\frac{R_0^4}{T^4}\frac{\phi_{NR_0T}^{-2}}{\sqrt N}\big).
\end{align*}

\begin{enumerate}
\item[10.] We have
\end{enumerate}
\begin{align*}
\bm W_{37}&=\bm H^0\frac{1}{R_0^2}\sum\sum (\frac{t}{T}-1)^2  \bm f_t\bm f_t^\top\frac{1}{N}\sum_{i=1}^N\bm\lambda_i^{\prime\prime}(c_i) e_{is}
(\tilde{\bm f}^0_{s}-\bm H^0\bm f_s)^\top\\
&=O_P\big(\frac{R_0^2}{T^2}\frac{\delta_{NR_0T}^{-1}}{\sqrt N}\big), \\
\bm W_{38}&=\bm H^0\frac{1}{R_0^2}\sum\sum (\frac{t}{T}-1)^2  \bm f_t\bm f_t^\top\frac{1}{N}\sum_{i=1}^N\bm\lambda_i^{\prime\prime}(c_i) e_{is}
(\frac{s}{T}-1)(\tilde{\bm f}^0_{s}-\bm H^0\bm f_s)^\top \\
&=O_P\big(\frac{R_0^3}{T^3}\frac{\delta_{NR_0T}^{-1}}{\sqrt N}\big), \\
\bm W_{39}&=\bm H^0\frac{1}{R_0^2}\sum\sum (\frac{t}{T}-1)^3 \bm f_t \bm f_t^\top \frac{1}{N}\sum_{i=1}^N\bm\lambda_i^{\prime\prime}(c_i) e_{is}
(\tilde{\bm f}^0_{s}-\bm H^0\bm f_s)^\top \\
&=O_P\big(\frac{R_0^3}{T^3}\frac{\delta_{NR_0T}^{-1}}{\sqrt N}\big),
\\
\bm W_{40}&=\bm H^0\frac{1}{R_0^2}\sum\sum (\frac{t}{T}-1)^3  \bm f_t \bm f_t^\top\frac{1}{N}\sum_{i=1}^N\bm\lambda_i^{\prime\prime}(c_i) e_{is}
(\frac{s}{T}-1)(\tilde{\bm f}^0_{s}-\bm H^0\bm f_s)^\top\\
&=O_P\big(\frac{R_0^4}{T^4}\frac{\delta_{NR_0T}^{-1}}{\sqrt N}\big). 
\end{align*}

\begin{enumerate}
\item[11.] We have
\begin{align*}
\bm W_{41}&=\bm H^0\frac{1}{R_0^2}\sum\sum (\frac{t}{T}-1)^2  \bm f_t\bm f_t^\top \frac{1}{N}\sum_{i=1}^N\bm\lambda_i^{\prime\prime}(c_i) e_{is}
(\bm H^0\bm f_s)^\top \\
&=O_P\big(\frac{R_0^2}{T^2}\frac{1}{\sqrt N}\big),\\
\bm W_{42}&=\bm H^0\frac{1}{R_0^2}\sum\sum (\frac{t}{T}-1)^2  \bm f_t \bm f_t^\top \frac{1}{N}\sum_{i=1}^N\bm\lambda_i^{\prime\prime}(c_i) e_{is}
(\frac{s}{T}-1)(\bm H^0\bm f_s)^\top\\
& =O_P\big(\frac{R_0^3}{T^3}\frac{1}{\sqrt N}\big),\\
\bm W_{43}&=\bm H^0\frac{1}{R_0^2}\sum\sum (\frac{t}{T}-1)^3 \bm f_t\bm f_t^\top \frac{1}{N}\sum_{i=1}^N\bm\lambda_i^{\prime\prime}(c_i) e_{is}
(\bm H^0\bm f_s)^\top\\
& =O_P\big(\frac{R_0^3}{T^3}\frac{1}{\sqrt N}\big),\\
\bm W_{44}&=\bm H^0\frac{1}{R_0^2}\sum\sum (\frac{t}{T}-1)^3 \bm f_t\bm f_t^\top \frac{1}{N}\sum_{i=1}^N\bm\lambda_i^{\prime\prime}(c_i) e_{is}
(\frac{s}{T}-1)(\bm H^0\bm f_s)^\top \\
&=O_P\big(\frac{R_0^4}{T^4}\frac{1}{\sqrt N}\big).
\end{align*}
\end{enumerate}

\begin{enumerate}
\item[12.] We have
\begin{align*}
\bm W_{45}&=\bm H^0\frac{1}{R_0^2}\sum\sum (\tilde{\bm f}^0_{t}-\bm H^0 \bm f_t)  \frac{1}{N}\sum_{i=1}^N e_{it}e_{is}
(\bm H^0\bm f_s)^\top\\
&=O_P\big(\frac{\delta_{NR_0T}^{-1}}{\sqrt{R_0}}\big) \\
\bm W_{46}&=\bm H^0\frac{1}{R_0^2}\sum\sum (\tilde{\bm f}^0_{t}-\bm H^0 \bm f_t) \frac{1}{N}\sum_{i=1}^N e_{it}e_{is}
(\frac{s}{T}-1)(\bm H^0\bm f_s)^\top\\
&=O_P\big(\frac{R_0}{T}\frac{\delta_{NR_0T}^{-1}}{\sqrt{R_0}}\big) \\
\bm W_{47}&=\bm H^0\frac{1}{R_0^2}\sum\sum (\frac{t}{T}-1) (\tilde{\bm f}^0_{t}-\bm H^0 \bm f_t) \frac{1}{N}\sum_{i=1}^N e_{it}e_{is}
(\bm H^0\bm f_s)^\top\\
&=O_P\big(\frac{R_0}{T}\frac{\delta_{NR_0T}^{-1}}{\sqrt{R_0}}\big)
\\
\bm W_{48}&=\bm H^0\frac{1}{R_0^2}\sum\sum (\frac{t}{T}-1) (\tilde{\bm f}^0_{t}-\bm H^0 \bm f_t) \frac{1}{N}\sum_{i=1}^N  e_{it}e_{is}
(\frac{s}{T}-1)(\bm H^0\bm f_s)^\top\\
&=O_P\big(\frac{R_0^2}{T^2}\frac{\delta_{NR_0T}^{-1}}{\sqrt{R_0}}\big).
\end{align*}
\end{enumerate}

\begin{enumerate}
\item[13.] We have
$$\frac{1}{N}\sum_{i=1}^N(\ddot{\bm\lambda}_i(1)-\bm\lambda_i^0(1))(\ddot{\bm\lambda}_i(1)-\bm\lambda_i^0(1))^\top=O_P\big(\phi_{NR_0T}^{-2}\big),$$
$$\frac{1}{N}\sum_{i=1}^N(\ddot{\bm\lambda}_i^\prime(1)-\bm\lambda_i^{0\prime}(1))(\ddot{\bm\lambda}_i(1)-\bm\lambda_i^0(1))^\top=O_P\big(\frac{T}{R_0}\phi_{NR_0T}^{-2}\big),$$
$$\frac{1}{N}\sum_{i=1}^N(\ddot{\bm\lambda}_i(1)-\bm\lambda_i^0(1))(\bm\lambda_i^0(1))^\top=O_P\big(\phi_{NR_0T}^{-2}\big),$$
$$\frac{1}{N}\sum_{i=1}^N(\ddot{\bm\lambda}_i^\prime(1)-\bm\lambda_i^{0\prime}(1))(\bm\lambda_i^0(1))^\top=O_P\big(\frac{T}{R_0}\phi_{NR_0T}^{-2}\big),$$
$$\frac{1}{N}\sum_{i=1}^N(\ddot{\bm\lambda}_i^\prime(1)-\bm\lambda_i^{0\prime}(1))(\bm\lambda_i^{0\prime}(1))^\top=O_P\big(\frac{T}{R_0}\phi_{NR_0T}^{-2}\big),$$
and
$$\frac{1}{N}\sum_{i=1}^N(\ddot{\bm\lambda}_i(1)-\bm\lambda_i^0(1))(\bm\lambda_i^{0\prime}(1))^\top=O_P\big(\phi_{NR_0T}^{-2}\big).$$
\end{enumerate}
\end{lemma}

\begin{lemma}
\label{Lemma4}
Suppose that \Cref{Ass1,Ass2,Ass3,Ass4,Ass5,Ass6,Ass7} hold. It follows that
\begin{enumerate}
\item $\sup_{R \in \Theta_R} \big(\frac{1}{R}\sum_{s=T-R+1}^{T}  \Vert \bm f_s\Vert^k\big)=O_P(1)$, $k=1,2,3,4$.

\item $\sup_{R \in \Theta_R}\frac{T^4}{R^6}\sum_{s=T-R+1}^{T}\sum_{t=T-R+1}^{T} \Vert  \bm \Lambda^\top(s/T)\bm \Lambda(t/T)/N-\bm \Lambda^\top(s/T)\bm \Lambda(1)/N
\Vert^4=O(1).$

\item $\sup_{R\in \Theta_R}\big(\delta_{NRT}^{2}\frac{1}{R}\sum_{t=T-R+1}^T \Vert \tilde{\bm f}_{t,R}-\bm H_{R}\bm f_t \Vert^2\big)=O_P(1).$

\item If $\sup_{R \in \Theta_R}\big( q_{NRT}\Vert \mathrm{Vec}(\bm Q_R)-\mathrm{Vec}(\bm Q_{R,0})\Vert\big)=O_P(1),$
then $\sup_R\big( q_{NRT}\Vert \mathrm{Vec}(\bm Q_R^{-1})-\mathrm{Vec}(\bm Q_{R,0}^{-1})\Vert\big)=O_P(1),$ as $N,T\rightarrow\infty$, with $q_{NRT}\rightarrow \infty $.

\item $\sup_{R \in \Theta_R}\big(\min(\sqrt{N},\sqrt R\delta_{NRT},R^{7/8-\mu})((\bm H_{R}^{-1})^{\top}\bm\alpha(1))^\top(\tilde{\bm f}_{T,R}-\bm H_R\bm f_T)\big)=O_P(1)$, for some $\mu \in (0,1/8)$.

\item $\sup_{R\in \Theta_R} \big\Vert\frac{\min(\phi_{NRT}^{2},R^{\frac{3}{4}-\nu})}{R}\sum_{t=T-R+1}^{T-h}( \tilde{\bm f}_{t,R}-\bm H_{R}\bm f_t)\bm z_{t}^\prime\big\Vert=O_P(1),$ for $\nu \in (0,1/4)$.

\item $\sup_{R\in \Theta_R} \big\Vert\frac{\min(\phi_{NRT}^{2},R^{\frac{3}{4}-\nu})}{R}\sum_{t=T-R+1}^{T-h}( \tilde{\bm f}_{t,R}-\bm H_{R}\bm f_t)\hat{\bm z}_{t}^\prime\big\Vert=O_P(1),$ for $\nu \in (0,1/4)$.

\item $\sup_{R\in \Theta_R} \big\Vert\frac{\min(\phi_{NRT}^{2},R^{\frac{3}{4}-\nu})}{R}\sum_{t=T-R+1}^{T-h}( \tilde{\bm f}_{t,R}-\bm H_{R}\bm f_t)\varepsilon_{t+h}\big\Vert=O_P(1),$ for $\nu \in (0,1/4)$.

\item $\sup_{R \in \Theta_R}\big\Vert \min(R^{\frac{1}{2}-\theta},\phi_{NRT}^2) \frac{1}{R}\sum_{t=T-R+1}^{T-h}\hat{\bm z}_{t,R}\varepsilon_{t+h}\big\Vert=O_P(1)$ for some $\theta \in (0,1/2)$.

 \item $\sup_{R \in \Theta_R}\big\Vert \frac{1}{R}\sum_{t=T-R+1}^{T-h}\hat{\bm z}_{t,R}\hat{\bm z}_{t,R}^\top\big\Vert=O_P(1)$.

 \item $\sup_{R\in \Theta_R} \big\Vert \big(\frac{1}{R}\sum_{t=T-R+1}^{T-h}\hat{\bm z}_{t,R}\hat{\bm z}_{t,R}^\top\big)^{-1}\big\Vert=O_P(1)$ 
\item $\sup_{R \in \Theta_R}\bigg\Vert \frac{T}{R^2}\sum_{t=T-R+1}^{T-h}\hat{\bm z}_{t,R}\hat{\bm z}_{t,R}^\top(\bm\delta_R(t/T)-\bm\delta_R(1))\bigg\Vert =O_P(1)$.

\item  $\sup_{R \in \Theta_R}\bigg\Vert
\frac{\min(\phi_{NRT}^{2},R^{\frac{3}{4}-\nu})}{R}\sum_{t=T-R+1}^{T-h}\hat{\bm z}_{t,R}(\tilde{\bm f}_{t,R}-\bm H_R  \bm f_t)^\top(\bm H_R^{\top} )^{-1}\bm\alpha(t/T)
\bigg\Vert=O_P(1)$ for some $\nu\in (0,1/4)$.

\item  $\sup_{R \in \Theta_R}\big(\min (R^{1/2-\theta},\frac{T}{R},N)\hat{\bm z}_{T,R}^\top(\hat{\bm\delta}_R(1)-\bm\delta_R(1))\big)=O_P(1)$,
for some $\theta\in (0,1/2)$.

\item  $\sup_R[\min(R^{\frac{1}{2}-\theta},\frac{T}{R},\sqrt{N})(\hat{\bm z}_{T,R}^\top\hat{\bm{\delta}}_R(1)-\bm z_T^\top\bm\delta(1))]=O_P(1)$, for some $\theta\in (0,1/2)$.

\item $\sup_{R \in \Theta_R} \big(\min(R^{1-2\theta},\frac{T^2}{R^2},N)\left\vert \left(C_{NT}( R)-C_{NT}^0( R)\right)\right\vert\big) =o_P(1).$
\end{enumerate}

\end{lemma}

For convenience, we use the notations $\sup_{R}, \sup_{t},$ and $ \sup_{s}$ for
$\sup_{R \in \Theta_R}, \sup_{R \in [T-R+1,R]},$ and $\sup_{s \in  [T-R+1,R]}$, respectively. Next, we prove
 \Cref{Thm1,Thm2,Thm3} followed by the auxiliary results in the lemmas.

\subsection*{Proof of \Cref{Thm1}}
\subsubsection*{Proof of \Cref{Thm11} in \Cref{Thm1}}

For any $R\in [\underline{R},\overline{R}]$, we recall that 
$\hat{\bm{\delta}}_R(1)=\bigg(\frac{1}{R}\sum_{t=T-R+1}^{T-h}\hat{\bm z}_{t,R}\hat{\bm z}_{t,R}^\top\bigg)^{-1}\frac{1}{R}\sum_{t=T-R+1}^{T-h}\hat{\bm z}_{t,R}y_{t+h},$ where we can write $y_{t+h}= \varepsilon_{t+h}+\hat{\bm z}_{t,R}^\top\bm\delta_R(t/T)  -(\tilde{\bm f}_{t,R}-\bm H_R \bm f_t)^\top\bm H_R^{\top -1}\bm\alpha(t/T).$ Therefore, we have
$\hat{\bm{\delta}}_R(1)-\bm{\delta}_R(1)=\bm C_1+\bm C_2-\bm C_3$, where 
\begin{align}
\bm C_1&=\bigg(\frac{1}{R}\sum_{t=T-R+1}^{T-h}\hat{\bm z}_{t,R}\hat{\bm z}_{t,R}^\top\bigg)^{-1}\frac{1}{R}\sum_{t=T-R+1}^{T-h}\hat{\bm z}_{t,R}\varepsilon_{t+h},\nonumber\\
\bm C_2&=\bigg(\frac{1}{R}\sum_{t=T-R+1}^{T-h}\hat{\bm z}_{t,R}\hat{\bm z}_{t,R}^\top\bigg)^{-1}\frac{1}{R}\sum_{t=T-R+1}^{T-h}\hat{\bm z}_{t,R}\hat{\bm z}_{t,R}^\top(\bm\delta_R(t/T)-\bm\delta_R(1)),\nonumber \\
\bm C_3 &=\bigg(\frac{1}{R}\sum_{t=T-R+1}^{T-h}\hat{\bm z}_{t,R}\hat{\bm z}_{t,R}^\top\bigg)^{-1}\frac{1}{R}\sum_{t=T-R+1}^{T-h}\hat{\bm z}_{t,R}(\tilde{\bm f}_{t,R}-\bm H_R  \bm f_t)^\top(\bm H_R^{\top} )^{-1}\bm\alpha(t/T).
\nonumber
\end{align}
We use $\bm C_1=O_P(\frac{1}{\sqrt{R}}+\frac{R}{T})$, $\bm C_2=O_P(\frac{R}{T}),$ $ \bm C_3=O_P(\phi_{NRT}^{-2})$, and $\frac{\sqrt{R}}{N}\rightarrow c,$ to have
$\hat{\bm{\delta}}_R(1)-\bm{\delta}_R(1)=O_P\big(\frac{R}{T}+\frac{1}{\min\{N,R\}}+\frac{1}{\sqrt{R}}\big)=O_P\big(\frac{R}{T}+\frac{1}{\sqrt{R}}\big)$.  Therefore, to complete the proof, we need to show these results hold.

For $\bm C_1$, we note that $\bm C_1=\bm C_{11}+\bm C_{12}+\bm C_{13}+\bm C_{14}$, where 
\begin{align*}
\bm C_{11}&=\bigg( \big(\frac{1}{R}\sum_{t=T-R+1}^{T-h}\hat{\bm z}_{t,R}\hat{\bm z}_{t,R}^\top\big)^{-1}-\big((\bm \Phi^{-1} )^{\top}\bm \Sigma_{\bm Z}^{-1} \bm \Phi^{-1}\big)\bigg)\frac{1}{R}\sum_{t=T-R+1}^{T-h}(\hat{\bm z}_{t,R}-\bm \Phi_R\bm z_t)\varepsilon_{t+h},\\
\bm C_{12}&=\bigg( \big(\frac{1}{R}\sum_{t=T-R+1}^{T-h}\hat{\bm z}_{t,R}\hat{\bm z}_{t,R}^\top\big)^{-1}-\big((\bm \Phi^{-1} )^{\top}\bm \Sigma_{\bm Z}^{-1} \bm \Phi^{-1}\big)\bigg)\bm \Phi_R\frac{1}{R}\sum_{t=T-R+1}^{T-h}\bm z_{t}\varepsilon_{t+h},\\
\bm C_{13}&=\big((\bm \Phi^{-1} )^{\top}\bm \Sigma_{\bm Z}^{-1} \bm \Phi^{-1}\big) \frac{1}{R}\sum_{t=T-R+1}^{T-h}(\hat{\bm z}_{t,R}-\bm \Phi_R\bm z_t)\varepsilon_{t+h},\\
\bm C_{14}&=\big((\bm \Phi^{-1} )^{\top}\bm \Sigma_{\bm Z}^{-1}\bm \Phi^{-1}\big)\bm \Phi_R \frac{1}{R}\sum_{t=T-R+1}^{T-h} \bm z_{t}\varepsilon_{t+h},
\end{align*}
where $\bm \Phi=\mathrm{diag}(\bm H, \bm I_q)$ and $\bm H$ is the limit in probability of $\bm H_R$.

Because 
$\big(\frac{1}{R}\sum_{t=T-R+1}^{T-h}\hat{\bm z}_{t,R}\hat{\bm z}_{t,R}^\top\big)^{-1}-(\bm \Phi^{-1} )^{\top}\bm \Sigma_{\bm Z}^{-1}\bm \Phi^{-1}=o_P(1)$ from \Cref{Lemma2} (17)-(18),
$\frac{1}{R}\sum_{t=T-R+1}^{T-h}(\hat{\bm z}_{t,R}-\bm \Phi_R\bm z_t)\varepsilon_{t+h}=O_P(\phi_{NRT}^{-2})$
from \Cref{Lemma2} (7),
$\frac{1}{R}\sum_{t=T-R+1}^{T-h} \bm z_{t}\varepsilon_{t+h}=O_P\big(\frac{1}{\sqrt R}\big)$
from \Cref{Lemma1} (2), $(\bm \Phi^{-1} )^{\top}\bm \Sigma_{\bm Z}^{-1}\bm \Phi^{-1}=O(1)$, $\bm \Phi_R=O_P(1)$, we deduce that
$\bm C_{11}=o_P(\phi_{NRT}^{-2})$,
$\bm C_{12}=o_P\big(\frac{1}{\sqrt R}\big)$,
$\bm C_{13}=O_P(\phi_{NRT}^{-2})$,
and
$\bm C_{14}=O_P\big(\frac{1}{\sqrt R}\big)$. 
Thus, $\bm C_1=O_P\big(\phi_{NRT}^{-2}\big)+O_P\big(\frac{1}{\sqrt R}\big)=O_P\big(\frac{R}{T}+\frac{1}{\sqrt R}\big)$.

For $\bm C_2$, we note that $\bm C_2=\bm C_{21}+\bm C_{22}+\bm C_{23}+\bm C_{24}+\bm C_{25}+\bm C_{26}$, where 
\begin{align*}
\bm C_{21}&=\big( \big(\frac{1}{R}\sum_{t=T-R+1}^{T-h}\hat{\bm z}_{t,R}\hat{\bm z}_{t,R}^\top\big)^{-1}-(\bm \Phi^{-1} )^{\top}\bm \Sigma_{\bm Z}^{-1} \bm \Phi^{-1}\big)\frac{1}{R}\sum_{t=T-R+1}^{T-h}(\hat{\bm z}_{t,R}-\bm \Phi_R\bm z_t)\hat{\bm z}_{t,R}^\top\\
&\times(\bm\delta_R(t/T)-\bm\delta_R(1)),\\
\bm C_{22}&=\big( \big(\frac{1}{R}\sum_{t=T-R+1}^{T-h}\hat{\bm z}_{t,R}\hat{\bm z}_{t,R}^\top\big)^{-1}-(\bm \Phi^{-1})^{\top}\bm \Sigma_{\bm Z}^{-1} \bm \Phi^{-1}\big)\bm \Phi_R\frac{1}{R}\sum_{t=T-R+1}^{T-h}\bm z_{t}(\hat{\bm z}_{t,R}-\bm \Phi_R\bm z_t)^\top\\
&\times(\bm\delta_R(t/T)-\bm\delta_R(1)),\\
\bm C_{23}&=\big( \big(\frac{1}{R}\sum_{t=T-R+1}^{T-h}\hat{\bm z}_{t,R}\hat{\bm z}_{t,R}^\top\big)^{-1}-(\bm \Phi^{-1})^{\top}\bm \Sigma_{\bm Z}^{-1} \bm \Phi^{-1}\big)\bm \Phi_R\frac{1}{R}\sum_{t=T-R+1}^{T-h}\bm z_{t}\bm z_{t}^\top\bm\Phi_R^\top\\
&\times(\bm\delta_R(t/T)-\bm\delta_R(1)),\\
\bm C_{24}&=\big((\bm \Phi^{-1} )^{\top}\bm \Sigma_{\bm Z}^{-1} \bm \Phi^{-1}\big) \frac{1}{R}\sum_{t=T-R+1}^{T-h}(\hat{\bm z}_{t,R}-\bm \Phi_R\bm z_t)
\hat{\bm z}_{t,R}
^\top(\bm\delta_R(t/T)-\bm\delta_R(1)),\\
\bm C_{25}&=\big((\bm \Phi^{-1} )^{\top}\bm \Sigma_{\bm Z}^{-1}\bm \Phi^{-1}\big)\bm \Phi_R \frac{1}{R}\sum_{t=T-R+1}^{T-h} \bm z_{t}(\hat{\bm z}_{t,R}-\bm \Phi_R\bm z_t)^\top(\bm\delta_R(t/T)-\bm\delta_R(1)),\\
\bm C_{26}&=\big((\bm \Phi^{-1} )^{\top}\bm \Sigma_{\bm Z}^{-1}\bm \Phi^{-1}\big)\bm \Phi_R \frac{1}{R}\sum_{t=T-R+1}^{T-h} \bm z_{t}\bm z_{t}^\top\bm\Phi_R^\top(\bm\delta_R(t/T)-\bm\delta_R(1)),
\end{align*}
where we note that
$ \big(\frac{1}{R}\sum_{t=T-R+1}^{T-h}\hat{\bm z}_{t,R}\hat{\bm z}_{t,R}^\top\big)^{-1}-(\bm \Phi^{-1} )^{\top}\bm \Sigma_{\bm Z}^{-1} \bm \Phi^{-1}=o_P(1)$ from \Cref{Lemma2} (17)-(18), 
$\frac{1}{R}\sum_{t=T-R+1}^{T-h}\bm z_{t}(\hat{\bm z}_{t,R}-\bm \Phi_R\bm z_t)^\top(\bm\delta_R(t/T)-\bm\delta_R(1))=O_P\big(\phi_{NRT}^{-2}\frac{R}{T}\big)$ as in \Cref{Lemma2} (19), and
$\frac{1}{R}\sum_{t=T-R+1}^{T-h}(\hat{\bm z}_{t,R}-\bm \Phi_R\bm z_t)\hat{\bm z}_{t}^\top(\bm\delta_R(t/T)-\bm\delta_R(1))=O_P\big(\phi_{NRT}^{-2}\frac{R}{T}\big),$ from \Cref{Lemma2} (20). In addition, 
$ \frac{1}{R}\sum_{t=T-R+1}^{T-h} \bm z_{t}\bm z_{t}^\top\bm\Phi_R^\top(\bm\delta_R(t/T)-\bm\delta_R(1))
=\frac{1}{R}\sum_{t=T-R+1}^{T-h} \mathrm E(\bm z_{t}\bm z_{t}^\top)(\bm\delta(t/T)-\bm\delta(1))+\frac{1}{R}\sum_{t=T-R+1}^{T-h} (\bm z_{t}\bm z_{t}^\top-\mathrm E(\bm z_{t}\bm z_{t}^\top))(\bm\delta(t/T)-\bm\delta(1))=O_P\big(\frac{R}{T}\big)+O_P\big(\frac{R}{T}R^{-1/2}\big)=O_P\big(\frac{R}{T}\big)$, from \Cref{Lemma1} (5) and (6). Therefore,
$\bm C_{21}=o_P\big(\phi_{NRT}^{-2}\frac{R}{T}\big), 
\bm C_{22}=o_P\big(\phi_{NRT}^{-2}\frac{R}{T}\big), 
\bm C_{23}=o_P\big(\frac{R}{T}\big), 
\bm C_{24}=O_P\big(\phi_{NRT}^{-2}\frac{R}{T}\big), 
\bm C_{25}=O_P\big(\phi_{NRT}^{-2}\frac{R}{T}\big),$ 
and 
$\bm C_{26}=O_P\big(\frac{R}{T}\big)$. Thus, $\bm C_{2}=O_P\big(\frac{R}{T}\big)$.

For $\bm C_3$, we note that $\bm C_3=\bm C_{31}+\bm C_{32}$, where 
\begin{align*}
\bm C_{31}&=\big( \big(\frac{1}{R}\sum_{t=T-R+1}^{T-h}\hat{\bm z}_{t,R}\hat{\bm z}_{t,R}^\top\big)^{-1}-(\bm \Phi^{-1})^{\top}\bm \Sigma_{\bm Z}^{-1} \bm \Phi^{-1}\big)\frac{1}{R}\sum_{t=T-R+1}^{T-h}\hat{\bm z}_{t}(\tilde{\bm f}_{t,R}-\bm H_R  \bm f_t)^\top(\bm H_R^{\top} )^{-1}\\
&\times\bm\alpha(t/T),\\
\bm C_{32}&=\big((\bm \Phi^{-1} )^{\top}\bm \Sigma_{\bm Z}^{-1}\bm \Phi^{-1}\big) \frac{1}{R}\sum_{t=T-R+1}^{T-h} \hat{\bm z}_{t}(\tilde{\bm f}_{t,R}-\bm H_R  \bm f_t)^\top(\bm H_R^{\top} )^{-1}\bm\alpha(t/T).
\end{align*}
We note that $\frac{1}{R}\sum_{t=T-R+1}^{T-h}\hat{\bm z}_{t}(\tilde{\bm f}_{t,R}-\bm H_R  \bm f_t)^\top(\bm H_R^{\top} )^{-1}\bm\alpha(t/T)$ equals
$\frac{1}{R}\sum_{t=T-R+1}^{T-h}\hat{\bm z}_{t}(\tilde{\bm f}_{t,R}-\bm H_R  \bm f_t)^\top(\bm H_R^{\top} )^{-1}\bm\alpha(1)
+\frac{1}{R}\sum_{t=T-R+1}^{T-h}\hat{\bm z}_{t}(\tilde{\bm f}_{t,R}-\bm H_R  \bm f_t)^\top\frac{t-T}{T}(\bm H_R^{\top} )^{-1}\bm\alpha^{\prime}(1)+\frac{1}{2R}\sum_{t=T-R+1}^{T-h}\hat{\bm z}_{t}(\tilde{\bm f}_{t,R}-\bm H_R  \bm f_t)^\top\frac{(t-T)^2}{T^2}(\bm H_R^{\top} )^{-1}\bm\alpha^{\prime\prime}(d)=O_P(\phi_{NRT}^{-2})$ using \Cref{Lemma2} (6), (9) and (12). In addition, $\big(\frac{1}{R}\sum_{t=T-R+1}^{T-h}\hat{\bm z}_{t,R}\hat{\bm z}_{t,R}^\top\big)^{-1}-(\bm \Phi^{-1})^{\top}\bm \Sigma_{\bm Z}^{-1} \bm \Phi^{-1}=o_P(1)$. Therefore, $\bm C_{31}=o_P(\phi_{NRT}^{-2})$,
$\bm C_{32}=O_P(\phi_{NRT}^{-2})$ and 
$\bm C_{3}=O_P(\phi_{NRT}^{-2})$.

Combining the order in probability for $\bm C_1$, $\bm C_2$ and $\bm C_3$,
we conclude that $\hat{\bm{\delta}}_R(1)-\bm{\delta}_R(1)=O_P\big(\frac{1}{\sqrt R}\big)+O_P\big(\frac{R}{T}\big)+O_P(\phi_{NRT}^{-2})=O_P\big(\frac{1}{\sqrt R}\big)+O_P\big(\frac{R}{T}\big)+O_P\big(\frac{1}{ N}\big)=O_P\big(\frac{1}{\sqrt R}\big)+O_P\big(\frac{R}{T}\big)$.

\subsubsection*{Proof of  \Cref{Thm12} in \Cref{Thm1}}

From \Cref{identity}, we note that
$\tilde{\bm f}_{T,R}-\bm H_R\bm f_T=\tilde{\bm V}_{R}^{-1}(\bm A_{1T}+\bm A_{2T}+\bm A_{3T}+\bm A_{4T}+\bm A_{5T}).$ Because the limit in probability of $\tilde{\bm V}_{R}$ is positive definite, $\tilde{\bm V}_{R}^{-1}=O_P(1)$, as in \citet{GMP2017}. Thus, we focus on determining the order in probability of $\bm A_{jT}, j=1,\ldots,5.$

For $\bm A_{1T},$ we note that $\bm A_{1T}=\frac{1}{R}\sum_{s=T-R+1}^{T} \tilde{\bm f}_s\nu_{sT}$, where $\nu_{sT}= 0.$ Therefore, $\bm A_{1T}=\bm 0$. 
For $\bm A_{2T},$ we start with the decomposition
$$\bm A_{2T}=\frac{1}{R}\sum_{s=T-R+1}^{T} (\tilde{\bm f}_s-\bm H_R \bm f_s)\eta_{sT}+\bm H_R\frac{1}{R}\sum_{s=T-R+1}^{T}  \bm f_s\eta_{sT}.$$
To bound the first term, we use the Cauchy-Schwarz inequality to have 
\begin{align*}
	\big\Vert\frac{1}{R}\sum_{s=T-R+1}^{T} (\tilde{\bm f}_s-\bm H_R\bm f_s)\eta_{sT}\big\Vert &\leq \big( \frac{1}{R}\sum_{s=T-R+1}^{T} \Vert\tilde{\bm f}_s-\bm H_R \bm f_s\Vert^2\big)^{1/2} \big( \frac{1}{R}\sum_{s=T-R+1}^{T} \Vert\eta_{sT}\Vert^2\big)^{1/2} \\
	&=O_P(\delta_{NRT}^{-1})O_P(N^{-1/2})=O_P(N^{-1/2}\delta_{NRT}^{-1}),
\end{align*}
because $\frac{1}{R}\sum_{s=T-R+1}^{T} \Vert\tilde{\bm f}_s-\bm H_R \bm f_s\Vert^2=O_P(\delta_{NRT}^{-2})$ from \Cref{Lemma2} (2) and 
\begin{align*}
	\frac{1}{R}\sum_{s=T-R+1}^{T} \Vert\eta_{sT}\Vert^2 
	& =\frac{1}{N}\frac{1}{RN}\sum_{s=T-R+1}^{T} \big\Vert \bm f_s^\top \bm \Lambda^\top(s/T)\bm e_{T}\big\Vert^2 =O_P(1/N),
\end{align*}
where we use  
$$\frac{1}{RN}\sum_{s=T-R+1}^{T} \big\Vert \bm f_s^\top \bm \Lambda^\top\big(\frac{s}{T}\big)\bm e_{T}\big\Vert^2 \leq 
\big(\frac{1}{R}\sum_{s=T-R+1}^T \Vert \bm f_s\Vert^4\big)^{\frac{1}{2}} \big(\frac{1}{R}\sum_{s=T-R+1}^{T} \big\Vert \frac{1}{\sqrt N} \bm \Lambda^\top\big(\frac{s}{T}\big)\bm e_{T}\big\Vert^4\big)^{\frac{1}{2}}
=O_P(1)$$
from \Cref{Ass1} (a) and \Cref{Ass2} (e).
For the second term, we have 
\begin{align*}
	\big\Vert\bm H_R\frac{1}{R}\sum_{s=T-R+1}^{T}  \bm f_s\eta_{sT}\big\Vert
	\leq \Vert \bm H_R\Vert
	\big( \frac{1}{R}\sum_{s=T-R+1}^{T} \Vert\bm f_s\Vert^2\big)^{1/2} \big( \frac{1}{T}\sum_{s=T-R+1}^{T} \Vert\eta_{sT}\Vert^2\big)^{1/2}=O_P(N^{-1/2}),
\end{align*}
where 
$\Vert \bm H_R\Vert=O_P(1)$, $ \frac{1}{R}\sum_{s=T-R+1}^{T} \Vert\bm f_s\Vert^2=O_P(1)$, $ \frac{1}{R}\sum_{s=T-R+1}^{T} \Vert\eta_{sT}\Vert^2=O_P(N^{-1})$. Consequently,  $\bm A_{2T}=O_P(N^{-1/2})$.

To show that $\bm A_{3T}=O_P\big(\frac{1}{\sqrt{N}\delta_{NRT}}\big),$ we consider with the decomposition
$$\bm A_{3T}=\frac{1}{R}\sum_{s=T-R+1}^{T}\tilde{\bm f}_s\xi_{sT}=\frac{1}{RN}\sum_{s=T-R+1}^{T} (\tilde{\bm f}_s-\bm H_R \bm f_s)\bm e_s^\top\bm \Lambda(1)\bm f_T+\bm H_R\frac{1}{RN}\sum_{s=T-R+1}^{T}  \bm f_s \bm e_s^\top\bm \Lambda(1)\bm f_T,$$
and we observe the proof becomes the same as \citet[Lemma A.2 (d)]{Bai2003}. 
We note that the proof of $ \bm A_{4T}=O_P\big(\frac{1}{\sqrt{R}\delta_{NRT}}\big)$ and $\bm A_{5T}=O_P\big(\frac{1}{\sqrt{N}\delta_{NRT}}\big)$ follow identical steps as the proofs of \citet[Lemma A.2 (a)-(b)]{Bai2003}. Thus, they are omitted.
Because 
$\bm A_{2T}=O_P\big(\frac{1}{\sqrt{N}}\big)$, $\bm A_{3T}=O_P\big(\frac{1}{\sqrt{N}\delta_{NRT}}\big)$,   $\bm A_{4T}=O_P\big(\frac{1}{\sqrt{R}\delta_{NRT}}\big)$, $\bm A_{5T}=O_P\big(\frac{1}{\sqrt{N}\delta_{NRT}}\big)$,  and $\tilde{\bm V}_{R}^{-1}=O_P(1)$, we conclude that $\tilde{\bm f}_{T,R}-\bm H_R\bm f_T=O_P\big(\frac{1}{\sqrt{N}}+\frac{1}{\sqrt{R}\delta_{NRT}}\big).$

\subsubsection*{Proof of \Cref{Thm13} in \Cref{Thm1}}

We consider the decomposition
$\hat{\bm z}_{T,R}^\top\hat{\bm{\delta}}_R(1)-\bm z_T^\top\bm\delta_R(1)=
\hat{\bm z}_{T,R}^\top(\hat{\bm{\delta}}_R(1)-\bm\delta_R(1))+(\hat{\bm z}_{T,R}-\bm \Phi_R\bm z_T)^\top\bm\delta_R(1),$
where  $\tilde{\bm f}_{T,R}-\bm H_R\bm f_T=O_P\big(\frac{1}{\sqrt{N}}+\frac{1}{\sqrt{R}\delta_{NRT}}\big)$ and 
$\hat{\bm{\delta}}_R(1)-\bm{\delta}(1)=O_P\big(\frac{1}{\sqrt{R}}+\frac{R}{T}\big)$. As a result,  
$\hat{\bm z}_{T,R}^\top\hat{\bm{\delta}}_R(1)-\bm z_T^\top\bm\delta(1)=O_P\big(\frac{1}{\sqrt{R}}+\frac{1}{\sqrt{N}}+\frac{R}{T}\big)=O_P(\delta_{NRT}^{-1})$.

\subsubsection*{Proof of \Cref{Thm14} in \Cref{Thm1}}

Recall that $C_{NT}^0=(\hat{\bm z}_{T,R}^\top\hat{\bm{\delta}}_R(1)-\bm z_T^\top\bm\delta(1))^2$.
Therefore, we have $C_{NT}^0=O_P(\frac{1}{R}+\frac{R^2}{T^2}+\frac{1}{N}),$
since $\hat{\bm z}_{T,R}^\top\hat{\bm{\delta}}_R(1)-\bm z_T^\top\bm\delta(1)=O_P\big(\frac{1}{\sqrt{R}}+\frac{1}{\sqrt{N}}+\frac{R}{T}\big)$.

\subsection*{Proof of \Cref{Thm2}}

\subsubsection*{Proof of \Cref{Thm21} 
in \Cref{Thm2}}

Let $\sum$ denote $\sum_{t=T-R_0+1}^{T-h}$.  We use an argument similar to the proof of \citet[Lemma 13]{IJR2017}, with the difference that we have to deal with estimated factors. Given the OLS estimation for the forecasting equation based on the pilot window,
we observe that 
\begin{equation}
\begin{bmatrix}
 \ddot{\bm\delta}(1) \\
  \ddot{\bm\delta}^{\prime}(1) 
\end{bmatrix}
=
{\underbrace{\begin{bmatrix}
\frac{1}{R_0}\sum\hat{\bm z}_{t}^0\hat{\bm z}_{t}^{0\top}  &  \frac{1}{R_0}\sum(\frac{t}{T}-1)\hat{\bm z}^0_{t}\hat{\bm z}_{t}^{0\top}
 \\
\frac{1}{R_0}\sum(\frac{t}{T}-1)\hat{\bm z}^0_{t}\hat{\bm z}_{t}^{0\top}  & \frac{1}{R_0}\sum(\frac{t}{T}-1)^2\hat{\bm z}^0_{t}\hat{\bm z}_{t}^{0\top}
\end{bmatrix}}_{\hat{\bm S}}
}^{-1}
\begin{bmatrix}
\frac{1}{R_0}\sum\hat{\bm z}^0_{t}y_{t+h}     \\
\frac{1}{R_0}\sum(\frac{t}{T}-1)\hat{\bm z}^0_{t}y_{t+h}  
\end{bmatrix}
.
\label{vector}
\end{equation}

In addition, we have from \Cref{expansion_y} that
\begin{align*}
y_{t+h}&=\bm\delta^{0\top}(1) \hat{\bm z}^0_{t}+\bm\delta^{0\prime\top}(1) \big(\frac{t}{T}-1\big) \hat{\bm z}^0_{t} -\bm\alpha^\top(1)(\bm H^0)^{-1}(\tilde{\bm f}^0_{t}-\bm H^0 \bm f_t)\\
&-\bm\alpha^{\prime\top}(1)(\bm H^{0})^{-1}(\tilde{\bm f}^0_{t}-\bm H^0 \bm f_t)(\frac{t}{T}-1)+\frac{1}{2}\bm\delta^{\prime\prime\top}(d)\big(\frac{t}{T}-1\big)^2\bm z_t+\varepsilon_{t+h}
\end{align*}
Consequently,
\begin{align*}
\begin{bmatrix}
\frac{1}{R_0}\sum\hat{\bm z}^0_{t}y_{t+h}     \\
\frac{1}{R_0}\sum(\frac{t}{T}-1)\hat{\bm z}^0_{t}y_{t+h}  
\end{bmatrix}
&=\hat{\bm S}
\begin{bmatrix}
\bm\delta^0(1)   \\
\bm\delta^{0\prime}(1)
\end{bmatrix}
-\begin{bmatrix}
\frac{1}{R_0}\sum\hat{\bm z}^0_{t}(\tilde{\bm f}^0_{t}-\bm H^0 \bm f_t)^\top     \\
\frac{1}{R_0}\sum(\frac{t}{T}-1)\hat{\bm z}^0_{t}(\tilde{\bm f}^0_{t}-\bm H^0 \bm f_t)^\top 
\end{bmatrix}
(\bm H^{0 \top})^{-1}\bm\alpha(1)
\\
&-
\begin{bmatrix}
\frac{1}{R_0}\sum(\frac{t}{T}-1)\hat{\bm z}^0_{t}(\tilde{\bm f}^0_{t}-\bm H^0 \bm f_t)^\top      \\
\frac{1}{R_0}\sum(\frac{t}{T}-1)^2\hat{\bm z}^0_{t}(\tilde{\bm f}^0_{t}-\bm H^0 \bm f_t)^\top 
\end{bmatrix}
(\bm H^{0\top})^{-1}\bm\alpha^{\prime}(1)
\\
&+
\begin{bmatrix}
\frac{1}{R_0}\sum(\frac{t}{T}-1)^2\hat{\bm z}^0_{t}\bm z_{t}^\top     \\
\frac{1}{R_0}\sum(\frac{t}{T}-1)^3\hat{\bm z}^0_{t} \bm z_{t}^\top 
\end{bmatrix}
\frac{1}{2}\bm\delta^{\prime\prime}(d)
+
\begin{bmatrix}
\frac{1}{R_0}\sum\hat{\bm z}^0_{t}\varepsilon_{t+h}     \\
\frac{1}{R_0}\sum(\frac{t}{T}-1)\hat{\bm z}^0_{t}\varepsilon_{t+h} 
\end{bmatrix}
.
\end{align*}
Replacing this in \Cref{vector} and expanding the last two terms, we get 
\begin{align}
\begin{bmatrix}
 \ddot{\bm\delta}(1)-\bm\delta^0(1) \\
  \ddot{\bm\delta}^{\prime}(1) -\bm\delta^{0\prime}(1)
\end{bmatrix}
&=
-\hat{\bm S}^{-1}\begin{bmatrix}
\frac{1}{R_0}\sum\hat{\bm z}^0_{t}(\tilde{\bm f}^0_{t}-\bm H^0 \bm f_t)^\top     \\
\frac{1}{R_0}\sum(\frac{t}{T}-1)\hat{\bm z}^0_{t}(\tilde{\bm f}^0_{t}-\bm H^0 \bm f_t)^\top 
\end{bmatrix}
(\bm H^{0\top})^{-1}\bm\alpha(1)\nonumber
\\
&-
\hat{\bm S}^{-1}\begin{bmatrix}
\frac{1}{R_0}\sum(\frac{t}{T}-1)\hat{\bm z}^0_{t}(\tilde{\bm f}^0_{t}-\bm H^0 \bm f_t)^\top      \\
\frac{1}{R_0}\sum(\frac{t}{T}-1)^2\hat{\bm z}^0_{t}(\tilde{\bm f}^0_{t}-\bm H^0 \bm f_t)^\top 
\end{bmatrix}
(\bm H^{0\top})^{-1}\bm\alpha^{\prime}(1)\nonumber
\\
&+
\hat{\bm S}^{-1}\begin{bmatrix}
\frac{1}{R_0}\sum(\frac{t}{T}-1)^2(\hat{\bm z}^0_{t}-\bm\Phi^0\bm z_t)\bm z_{t}^\top     \nonumber\\
\frac{1}{R_0}\sum(\frac{t}{T}-1)^3(\hat{\bm z}^0_{t}-\bm\Phi^0\bm z_t) \bm z_{t}^\top 
\end{bmatrix}
\frac{1}{2}\bm\delta^{\prime\prime}(d)\\
&+
\hat{\bm S}^{-1}\begin{bmatrix}
\bm\Phi^0\frac{1}{R_0}\sum(\frac{t}{T}-1)^2\bm z_{t}\bm z_{t}^\top     \\
\bm\Phi^0\frac{1}{R_0}\sum(\frac{t}{T}-1)^3\bm z_{t} \bm z_{t}^\top 
\end{bmatrix}
\frac{1}{2}\bm\delta^{\prime\prime}(d)\nonumber
\\
&
+
\hat{\bm S}^{-1}
\begin{bmatrix}
\frac{1}{R_0}\sum(\hat{\bm z}^0_{t}-\bm \Phi^0\bm z_t)\varepsilon_{t+h}     \\
\frac{1}{R_0}\sum(\frac{t}{T}-1)(\hat{\bm z}^0_{t}-\bm \Phi^0\bm z_t)\varepsilon_{t+h} 
\end{bmatrix}
+
\hat{\bm S}^{-1}
\begin{bmatrix}
\bm \Phi^0\frac{1}{R_0}\sum\bm z_{t}\varepsilon_{t+h}     \\
\bm \Phi^0\frac{1}{R_0}\sum(\frac{t}{T}-1)\bm z_{t}\varepsilon_{t+h} 
\end{bmatrix}.
\label{delta_expansion}
\end{align}

To find the order in probability of $ \ddot{\bm\delta}(1)-\bm\delta^0(1) $, let us first determine the order of the elements of $\hat{\bm S}^{-1}$.  To do so, we  can decompose $\hat{\bm S}$ as $\hat{\bm S}_1+\hat{\bm S}_2+\hat{\bm S}_3+\hat{\bm S}_4$, where 
\begin{align*}
\hat{\bm S}_1 =&\begin{bmatrix}
\bm\Phi^0\frac{1}{R_0}\sum\mathrm{E}(\bm z_{t}\bm z_{t}^{\top})\bm\Phi^{0\top}  & \bm\Phi^0 \frac{1}{R_0}\sum(\frac{t}{T}-1)\mathrm{E}(\bm z_{t}\bm z_{t}^{\top})\bm\Phi^{0\top}
 \\
\bm\Phi^0\frac{1}{R_0}\sum(\frac{t}{T}-1)\mathrm{E}(\bm z_{t}\bm z_{t}^{\top})\bm\Phi^{0\top}& \bm\Phi^0\frac{1}{R_0}\sum(\frac{t}{T}-1)^2\mathrm{E}(\bm z_{t}\bm z_{t}^{\top})\bm\Phi^{0\top}
\end{bmatrix}\\
\hat{\bm S}_2=&
\begin{bmatrix}
\bm\Phi^0\frac{1}{R_0}\sum(\bm z_{t}\bm z_{t}^{\top}-\mathrm{E}(\bm z_{t}\bm z_{t}^{0\top}))\bm\Phi^{0\top}  & \bm\Phi^0 \frac{1}{R_0}\sum(\frac{t}{T}-1)(\bm z_{t}\bm z_{t}^{\top}-\mathrm{E}(\bm z_{t}\bm z_{t}^{\top}))\bm\Phi^{0\top}
 \\
\bm\Phi^0\frac{1}{R_0}\sum(\frac{t}{T}-1)(\bm z_{t}\bm z_{t}^{\top}-\mathrm{E}(\bm z_{t}\bm z_{t}^{\top}))\bm\Phi^{0\top} & \bm\Phi^0\frac{1}{R_0}\sum(\frac{t}{T}-1)^2(\bm z_{t}\bm z_{t}^{\top}-\mathrm{E}(\bm z_{t}\bm z_{t}^{0\top}))\bm\Phi^{0\top}
\end{bmatrix}\\
\hat{\bm S}_3= &
\begin{bmatrix}
\frac{1}{R_0}\sum(\hat{\bm z}_{t}^0-\bm\Phi^0 z_t)\hat{\bm z}_{t}^{0\top}  &  \frac{1}{R_0}\sum(\frac{t}{T}-1)(\hat{\bm z}_{t}^0-\bm\Phi^0 z_t)\hat{\bm z}_{t}^{0\top}
 \\
\frac{1}{R_0}\sum(\frac{t}{T}-1)(\hat{\bm z}_{t}^0-\bm\Phi^0 z_t)\hat{\bm z}_{t}^{0\top}  & \frac{1}{R_0}\sum(\frac{t}{T}-1)^2(\hat{\bm z}_{t}^0-\bm\Phi^0 z_t)\hat{\bm z}_{t}^{0\top}
\end{bmatrix}
\\
\hat{\bm S}_4=&
\begin{bmatrix}
\bm\Phi^0\frac{1}{R_0}\sum\bm z_{t}(\hat{\bm z}_{t}^{0}-\bm \Phi^0\bm z_t)^\top &  \bm\Phi^0\frac{1}{R_0}\sum(\frac{t}{T}-1)\bm z_{t}(\hat{\bm z}_{t}^{0}-\bm \Phi^0\bm z_t)^\top
 \\
\bm\Phi^0\frac{1}{R_0}\sum(\frac{t}{T}-1)\bm z_{t}(\hat{\bm z}_{t}^{0}-\bm \Phi^0\bm z_t)^\top  & \bm\Phi^0\frac{1}{R_0}\sum(\frac{t}{T}-1)^2\bm z_{t}(\hat{\bm z}_{t}^{0}-\bm \Phi^0\bm z_t)^\top
\end{bmatrix}
\end{align*}
\Cref{Lemma1} (3) implies that the second term is 
$\begin{bmatrix}
O_P\big(\frac{1}{\sqrt R_0}\big) & O_P\big(\frac{1}{\sqrt R_0}\frac{R_0}{T}\big)\\
 O_P\big(\frac{1}{\sqrt R_0}\frac{R_0}{T}\big)  & O_P\big(\frac{1}{\sqrt R_0}\frac{R_0^2}{T^2}\big)
\end{bmatrix}.$
\Cref{Lemma2} (6), (9) and (12) imply the third and the fourth terms are
$\begin{bmatrix}
O_P\big(\phi_{NR_0T}^{-2}\big) & O_P\big(\phi_{NR_0T}^{-2}\frac{R_0}{T}\big)\\
 O_P\big(\phi_{NR_0T}^{-2}\frac{R_0}{T}\big)  & O_P\big(\phi_{NR_0T}^{-2}\frac{R_0^2}{T^2}\big)
\end{bmatrix}$. 
Therefore, the order of $\hat{\bm S}$ is determined by  $\hat{\bm S}_1$, because its order is 
$\begin{bmatrix}
O_P(1) & O_P\big(\frac{R_0}{T}\big)\\
 O_P\big(\frac{R_0}{T}\big)  & O_P\big(\frac{R_0^2}{T^2}\big)
\end{bmatrix}$ from \Cref{Lemma1} (1). Using a block matrix inversion formula, we get
\begin{align*}
\hat{\bm S}_1^{-1}=
\begin{bmatrix}
\bm A^{-1}+\bm A^{-1}\bm B (\bm C-\bm B\bm A^{-1}\bm B)^{-1}\bm B \bm A^{-1} &  -\bm A^{-1}\bm B (\bm C-\bm B\bm A^{-1}\bm B)^{-1}\\
-(\bm C-\bm B\bm A^{-1}\bm B)^{-1}\bm B \bm A^{-1} & (\bm C-\bm B\bm A^{-1}\bm B)^{-1} 
\end{bmatrix},
\end{align*}
where 
$ \bm A= \bm\Phi^0\frac{1}{R_0}\sum\mathrm{E}(\bm z_{t}\bm z_{t}^{\top})\bm\Phi^{0\top},$
$ \bm B= \bm\Phi^0\frac{1}{R_0}\sum(\frac{t}{T}-1)\mathrm{E}(\bm z_{t}\bm z_{t}^{\top})\bm\Phi^{0\top}$
and 
$\bm C= \bm\Phi^0\frac{1}{R_0}\sum(\frac{t}{T}-1)^2\mathrm{E}(\bm z_{t}\bm z_{t}^{\top})\bm\Phi^{0\top}.$
From \Cref{Ass4} (c), 
we have $\bm A^{-1} =O(1)$. In addition, $\bm B=O\big(\frac{R_0}{T}\big)$. 
Because $\bm\Phi^{0}$ has an invertible limit, and the limit of the middle term in 
$\frac{T^2}{R_0^2}(\bm C-\bm B\bm A^{-1}\bm B)=\bm\Phi^{0}
\big(\frac{T^2}{R_0^3}\sum(\frac{t}{T}-1)^2\mathrm{E}(\bm z_{t}\bm z_{t}^{\top})
-\frac{T}{R_0^2}\sum(\frac{t}{T}-1)\mathrm{E}(\bm z_{t}\bm z_{t}^{\top})(\frac{1}{R_0}\sum\mathrm{E}(\bm z_{t}\bm z_{t}^{\top}))^{-1}\frac{T}{R_0^2}\sum(\frac{t}{T}-1)\mathrm{E}(\bm z_{t}\bm z_{t}^{\top})\big)\bm\Phi^{0\top}$
is positive definite, 
we deduce that $(\frac{T^2}{R_0^2}(\bm C-\bm B\bm A^{-1}\bm B))^{-1}=O(1)$ and $(\bm C-\bm B\bm A^{-1}\bm B)^{-1}=\frac{T^2}{R_0^2}(\frac{T^2}{R_0^2}(\bm C-\bm B\bm A^{-1}\bm B))^{-1}=O\big(\frac{T^2}{R_0^2}\big)$. Thus,
$$\hat{\bm S}^{-1}=
\begin{bmatrix}
O(1) & O\big(\frac{R_0}{T}\big)O\big(\frac{T^2}{R_0^2}\big)\\
O\big(\frac{R_0}{T}\big)O\big(\frac{T^2}{R_0^2}\big) & O\big(\frac{T^2}{R_0^2}\big)
\end{bmatrix}
=
\begin{bmatrix}
	O(1) & O\big(\frac{T}{R_0}\big)\\
	O\big(\frac{T}{R_0}\big) & O\big(\frac{T^2}{R_0^2}\big) 
\end{bmatrix}.
$$
From \Cref{Lemma2} (6), (9) and (12), we have  $\frac{1}{R_0}\sum\hat{\bm z}^0_{t}(\tilde{\bm f}^0_{t}-\bm H^0 \bm f_t)^\top=O_P(\phi_{NR_0T}^{-2})$, 
$\frac{1}{R_0}\sum(\frac{t}{T}-1)\hat{\bm z}^0_{t}(\tilde{\bm f}^0_{t}-\bm H^0 \bm f_t)^\top=O_P\big(\phi_{NR_0T}^{-2}\frac{R_0}{T}\big)$, and
$\frac{1}{R_0}\sum(\frac{t}{T}-1)^2\hat{\bm z}^0_{t}(\tilde{\bm f}^0_{t}-\bm H^0 \bm f_t)^\top=O_P\big(\phi_{NR_0T}^{-2}\frac{R_0^2}{T^2}\big)$. 
These results allow us to obtain that
the first and second terms of
\Cref{delta_expansion} are $\begin{bmatrix}O_P\big(\phi_{NR_0T}^{-2}\big)\\
O_P\big(\phi_{NR_0T}^{-2}\frac{T}{R_0}\big)\end{bmatrix}$
and $\begin{bmatrix}O_P\big(\phi_{NR_0T}^{-2}\frac{R_0}{T}\big)\\
O_P\big(\phi_{NR_0T}^{-2}\big)	\end{bmatrix}$, respectively.
In addition, \Cref{Lemma2} (14) shows that $\frac{1}{R_0}\sum(\frac{t}{T}-1)^3(\hat{\bm z}^0_{t}-\bm\Phi^0\bm z_t) \bm z_{t}^\top=O_P\big(\phi_{NR_0T}^{-2}\frac{R_0^3}{T^3}\big)$. 
Thus, the third term of \Cref{delta_expansion} is $\begin{bmatrix} O_P\big(\phi_{NR_0T}^{-2}\frac{R_0^2}{T^2}\big)\\
	O_P\big(\phi_{NR_0T}^{-2}\frac{R_0}{T}\big)
\end{bmatrix}$.
Furthermore, \Cref{Lemma1} (1) and (3) imply 
$\frac{1}{R_0}\sum(\frac{t}{T}-1)^2\bm z_t \bm z_{t}^\top=O_P\big(\frac{R_0^2}{T^2}\big) $ and  $\frac{1}{R_0}\sum(\frac{t}{T}-1)^3\bm z_t \bm z_{t}^\top=O_P\big(\frac{R_0^3}{T^3}\big)$.
Therefore, the fourth term of \Cref{delta_expansion} is $\begin{bmatrix} O_P\big(\frac{R_0^2}{T^2}\big)\\
O_P\big(\frac{R_0}{T}\big)	\end{bmatrix}$.
Similarly, we employ  \Cref{Lemma2} (7) and (10) and \Cref{Lemma1} (1) to have the fifth and the last terms of \Cref{delta_expansion} are $\begin{bmatrix} O_P\big(\phi_{NR_0T}^{-2}\big)\\
O_P\big(\phi_{NR_0T}^{-2}\frac{T}{R_0}\big)\end{bmatrix}$ and $\begin{bmatrix} O_P\big(\frac{1}{\sqrt{R_0}}\big)\\
O_P\big(\frac{1}{\sqrt{R_0}}\frac{T}{R_0}\big)
\end{bmatrix}$, respectively. 
Therefore, 
$\ddot{\bm\delta}(1)-\bm\delta^0(1)=O_P\big(\frac{1}{\sqrt{R_0}}\big)+O_P\big(\frac{1}{ N}\big)
+O_P\big(\frac{R_0}{T}\big)=O_P\big(\frac{1}{\sqrt{R_0}}\big)+O_P\big(\frac{R_0}{T}\big)$, 
using $\sqrt{R_0}/N\rightarrow c.$

\subsubsection*{Proof of \Cref{Thm23} in \Cref{Thm2}}

In this proof, we show that $\ddot{\bm f}_T-\bm H^0\bm f_T=O_P\big(\frac{1}{\sqrt N}\big)+O_P\big(\frac{R_0}{T}\big)$.	
Recall that 
$$\bm H^0=(\tilde{\bm V}^{0})^{-1}\frac{1}{R_0}\sum\tilde{\bm f}_t^0\bm f_t^\top\big(\frac{1}{N}\sum_{i=1}^N\bm\lambda_i(\frac{t}{T})\bm\lambda_i^\top(1)\big)$$ and
$\ddot{\bm f}_T=(\tilde{\bm V}^{0})^{-1}\frac{1}{R_0}\sum\tilde{\bm f}_t^0\tilde{\bm f}_t^{0\top}\big(\frac{1}{N}\sum_{i=1}^N(\ddot{\bm\lambda}_i(1)+\frac{(t-T)}{T}\ddot{\bm\lambda}_i^\prime(1))\ddot{\bm\lambda}_i^\top(1)\big)\tilde{\bm f}_T^0.$
Therefore, we have
$\ddot{\bm f}_T-\bm H^0\bm f_T=\sum_{j=1}^{12}\bm K_j$,
where 
we have
\begin{align*}
&\bm K_1=(\tilde{\bm V}^{0})^{-1} \frac{1}{R_0}\sum\tilde{\bm f}_t\tilde{\bm f}_t^\top\big(\frac{1}{N}\sum_{i=1}^N(\ddot{\bm\lambda}_i(1)-\bm\lambda_i^0(1))(\ddot{\bm\lambda}_i(1)-\bm\lambda_i^0(1))^\top\big)\tilde{\bm f}_T^0\\
&\bm K_2=(\tilde{\bm V}^{0})^{-1} \frac{1}{R_0}\sum\tilde{\bm f}_t\tilde{\bm f}_t^\top\big(\frac{1}{N}\sum_{i=1}^N\bm\lambda_i^0(1)(\ddot{\bm\lambda}_i(1)-\bm\lambda_i^0(1))^\top\big)\tilde{\bm f}_T^0\\
&\bm K_3=(\tilde{\bm V}^{0})^{-1} \frac{1}{R_0}\sum\tilde{\bm f}_t\tilde{\bm f}_t^\top\big(\frac{1}{N}\sum_{i=1}^N(\ddot{\bm\lambda}_i(1)-\bm\lambda_i^0(1))\bm\lambda_i^{0\top}(1)\big)\tilde{\bm f}_T^0\\
\end{align*}

\begin{align*}
&\bm K_4=(\tilde{\bm V}^{0})^{-1} \frac{1}{R_0}\sum\tilde{\bm f}_t^0\tilde{\bm f}_t^{0\top}\frac{(t-T)}{T}\big(\frac{1}{N}\sum_{i=1}^N(\ddot{\bm\lambda}_i^\prime(1)-\bm\lambda_i^{0\prime}(1))(\ddot{\bm\lambda}_i(1)-\bm\lambda_i^0(1))^\top\big)\tilde{\bm f}_T^0\\
&\bm K_5=(\tilde{\bm V}^{0})^{-1}\frac{1}{R_0} \sum\tilde{\bm f}_t^0\tilde{\bm f}_t^{0\top}\frac{(t-T)}{T} \big(\frac{1}{N}\sum_{i=1}^N\bm\lambda_i^{0\prime}(1)(\ddot{\bm\lambda}_i(1)-\bm\lambda_i^0(1))^\top\big)\tilde{\bm f}_T^0\\
&\bm K_6=(\tilde{\bm V}^{0})^{-1}\frac{1}{R_0}\sum\tilde{\bm f}_t^0\tilde{\bm f}_t^{0\top}\frac{(t-T)}{T} \big(\frac{1}{N}\sum_{i=1}^N(\ddot{\bm\lambda}_i^\prime(1)-\bm\lambda_i^{0\prime}(1))\bm\lambda_i^{0\top}(1)\big)\tilde{\bm f}_T^0\\
&\bm K_7=-(\tilde{\bm V}^{0})^{-1}\frac{1}{R_0}\sum\tilde{\bm f}_t^0\tilde{\bm f}_t^{0\top}\big(\frac{1}{N}\sum_{i=1}^N({\bm\lambda}_i^0(1){\bm\lambda}_i^{0\top}(1)\big)(\tilde{\bm f}_T^0-\bm H^0 f_T)\\
&\bm K_8=-(\tilde{\bm V}^{0})^{-1}\frac{1}{R_0}\sum\tilde{\bm f}_t^0\tilde{\bm f}_t^{0\top}\frac{(t-T)}{T}\big(\frac{1}{N}\sum_{i=1}^N({\bm\lambda}_i^{0\prime}(1)){\bm\lambda}_i^{0\top}(1)\big)(\tilde{\bm f}_T^0-\bm H^0 f_T)\\
&\bm K_9=-(\tilde{\bm V}^{0})^{-1}\big[\frac{1}{R_0}\sum\tilde{\bm f}_t^0\big(\tilde{\bm f}_t^{0\top}-\bm H^{0}{\bm f}_t^{0}\big)^\top\big]\big(\frac{1}{N}\sum_{i=1}^N({\bm\lambda}_i^0(1){\bm\lambda}_i^{0\top}(1)\big)\bm H^0 {\bm f}_T\\
&\bm K_{10}=-(\tilde{\bm V}^{0})^{-1}\big[\frac{1}{R_0}\sum\tilde{\bm f}_t^0\big(\tilde{\bm f}_t^{0\top}-\bm H^{0}{\bm f}_t^{0}\big)^\top\frac{(t-T)}{T}\big]\big(\frac{1}{N}\sum_{i=1}^N({\bm\lambda}_i^{0\prime}(1)){\bm\lambda}_i^{0\top}(1)\big)\bm H^0 {\bm f}_T\\
&\bm K_{11}=-\frac{1}{2}(\tilde{\bm V}^{0})^{-1}\frac{1}{R_0}\sum\big(\tilde{\bm f}_t^{0\top}-\bm H^{0}{\bm f}_t^{0}\big)\bm f_t^\top\frac{(t-T)^2}{T^2}\big(\frac{1}{N}\sum_{i=1}^N\bm\lambda_i^{\prime\prime}(c_i)\bm\lambda_i^\top(1)\big)\\
%
&\bm K_{12}=-\frac{1}{2}(\tilde{\bm V}^{0})^{-1}\bm H^{0}\frac{1}{R_0}\sum\big({\bm f}_t\bm f_t^\top\frac{(t-T)^2}{T^2}\big)\big(\frac{1}{N}\sum_{i=1}^N\bm\lambda_i^{\prime\prime}(c_i)\bm\lambda_i^\top(1)\big).
\end{align*}

Thus, we need to find the order in probability of $\bm K_i, i=1,\ldots,12$. First, we note $\frac{1}{R_0}\sum\tilde{\bm f}_t^0\tilde{\bm f}_t^{0\top}=\bm I$. Therefore, $\frac{1}{R_0}\sum\tilde{\bm f}_t^0\tilde{\bm f}_t^{0\top}=O_P(1)$. We also observe that
$\tilde{\bm f}_T=O_P(1), \bm f_T=O_P(1), (\tilde{\bm V}^{0})^{-1}=O_P(1), \bm H^{0}=O_P(1)$. Furthermore,
$\frac{1}{R_0}\sum \bm f_t\bm f_t^\top=O_P(1),$ $\frac{1}{R_0}\sum \bm f_t\bm f_t^\top\frac{(t-T)}{T}=O_P\big(\frac{R_0}{T}\big),$ $\frac{1}{R_0}\sum \bm f_t\bm f_t^\top\frac{(t-T)^2}{T^2}=O_P\big(\frac{R_0^2}{T^2}\big)$.
We also recall from the results in \Cref{Lemma2} that
$\frac{1}{R_0}\sum\big(\tilde{\bm f}_t^{0\top}-\bm H^{0}{\bm f}_t^{0}\big)\bm f_t^\top=O_P\big(\phi_{NR_0T}^{-2}\big),
 \frac{1}{R_0}\sum\big(\tilde{\bm f}_t^{0\top}-\bm H^{0}{\bm f}_t^{0}\big)\bm f_t^\top\frac{t-T}{T}=O_P\big(\phi_{NR_0T}^{-2}\frac{R_0}{T}\big), $ and $
 \frac{1}{R_0}\sum\big(\tilde{\bm f}_t^{0\top}-\bm H^{0}{\bm f}_t^{0}\big)\bm f_t^\top\frac{(t-T)^2}{T^2}=O_P\big(\phi_{NR_0T}^{-2}\frac{R_0^2}{T^2}\big)$, and can easily deduce  that $\frac{1}{R_0}\sum\tilde{\bm f}_t^0\tilde{\bm f}_t^{0\top}\frac{(t-T)}{T}=O_P\big(\frac{R_0}{T}\big)$.
In addition \Cref{Ass1} (b) and \Cref{Ass5} (b) implies $\frac{1}{N}\sum_{i=1}^N({\bm\lambda}_i(1)){\bm\lambda}_i^{\top}(1)=O(1)$, $\frac{1}{N}\sum_{i=1}^N({\bm\lambda}_i^{\prime}(1)){\bm\lambda}_i^{\top}(1)=O(1)$ and
$\frac{1}{N}\sum_{i=1}^N({\bm\lambda}_i^{\prime\prime}(c_i)){\bm\lambda}_i^{\top}(1)=O(1)$. Because, $(\bm H^0)^{-1}=O_P(1)$, we have $\frac{1}{N}\sum_{i=1}^N({\bm\lambda}_i^0(1)){\bm\lambda}_i^{0\top}(1)=O_P(1)$ and $\frac{1}{N}\sum_{i=1}^N({\bm\lambda}_i^{0\prime}(1)){\bm\lambda}_i^{0\top}(1)=O_P(1)$.

Moreover,  we have from \Cref{Lemma3} (13) that 
$\frac{1}{N}\sum_{i=1}^N(\ddot{\bm\lambda}_i(1)-\bm\lambda_i^0(1))(\ddot{\bm\lambda}_i(1)-\bm\lambda_i^0(1))^\top=O_P\big(\phi_{NR_0T}^{-2}\big),$
$\frac{1}{N}\sum_{i=1}^N(\ddot{\bm\lambda}_i(1)-\bm\lambda_i^0(1))(\bm\lambda_i^0(1))^\top=
 O_P\big(\phi_{NR_0T}^{-2}\big)  ,$
$\frac{1}{N}\sum_{i=1}^N(\ddot{\bm\lambda}_i^\prime(1)-\bm\lambda_i^{0\prime}(1))(\ddot{\bm\lambda}_i(1)-\bm\lambda_i^0(1))^\top=
O_P\big(\frac{T}{R_0}\phi_{NR_0T}^{-2}\big)  $,
and
$\frac{1}{N}\sum_{i=1}^N(\ddot{\bm\lambda}_i^\prime(1)-\bm\lambda_i^{0\prime}(1))(\bm\lambda_i^0(1))^\top=  O_P(\frac{T}{R_0}\phi_{NR_0T}^{-2})$.
We also have
$\frac{1}{N}\sum_{i=1}^N(\ddot{\bm\lambda}_i^\prime(1)-\bm\lambda_i^{0\prime}(1))(\bm\lambda_i^{0\prime}(1))^\top=  O_P\big(\frac{T}{R_0}\phi_{NR_0T}^{-2}\big)$, using identical arguments as for $\frac{1}{N}\sum_{i=1}^N(\ddot{\bm\lambda}_i^\prime(1)-\bm\lambda_i^{0\prime}(1))(\bm\lambda_i^{0}(1))^\top$.

Combining the previous results,
we have 
\begin{align*}
\bm K_1&=O_P\big(\phi_{NR_0T}^{-2}\big),\\
\bm K_2&=O_P\big(\phi_{NR_0T}^{-2}\big)  ,\\
\bm K_3&=O_P\big(\phi_{NR_0T}^{-2}\big) ,\\
\bm K_4 & =O_P\big(\phi_{NR_0T}^{-2}\big),\\
\bm K_5&=O_P\big(\frac{R_0}{T}\phi_{NR_0T}^{-2}\big) ,\\
\bm K_6&=O_P\big(\phi_{NR_0T}^{-2}\big) ,\\
\bm K_7&=O_P\big(\frac{1}{\sqrt N}\big),\\
%
\bm K_8&=O_P\big(\frac{R_0}{T}\frac{1}{\sqrt N}\big),\\
\bm K_9&=O_P\big(\phi_{NR_0T}^{-2}\big),\\
\bm K_{10}&=O_P\big(\frac{R_0}{T}\phi_{NR_0T}^{-2}\big),\\
\bm K_{11}&=O_P\big(\frac{R_0^2}{T^2}\phi_{NR_0T}^{-2}\big),\\
\bm K_{12}&=O_P\big(\frac{R_0^2}{T^2}\big).
\end{align*}
Consequently, 
$\ddot{\bm f}_T-\bm H_0\bm f_T=\sum_{j=1}^{12}\bm K_j=
O_P\big(\frac{1}{\sqrt N}\big)+O_P\big(\frac{1}{ R_0}\big)+O_P\big(\frac{R_0}{T}\big)=O_P\big(\frac{1}{\sqrt N}\big)+O_P\big(\frac{R_0}{T}\big).$

\subsubsection*{Proof of \Cref{Thm24} in \Cref{Thm2}}

Rewrite $\ddot{\bm z}_T^\top\ddot{\bm \delta}(1)-\bm z_T^\top\bm\delta(1)$
as $\ddot{\bm z}_T^\top(\ddot{\bm \delta}(1)-\bm\delta^0(1))
+(\ddot{\bm z}_T-\bm\Phi_0\bm z_T)^\top\bm\delta^0(1),
$
where
$\ddot{\bm z}_T^\top=O_P(1)$,
$\ddot{\bm \delta}(1)-\bm\delta^0(1)=O_P\big(\frac{1}{\sqrt R_0}+\frac{R_0}{T}\big)$,
$\ddot{\bm z}_T-\bm\Phi_0\bm z_T=O_P\big(\frac{1}{\sqrt N}+\frac{R_0}{T}\big)$,
$\bm\delta^0(1)=O_P(1)$
gives
$\ddot{\bm z}_T^\top\ddot{\bm \delta}(1)-\bm z_T^\top\bm\delta(1)
=O_P\big({\frac{1}{\sqrt R_0}+\frac{1}{\sqrt N}+\frac{R_0}{T}}\big)=O_P(\delta_{NR_0T}^{-1})$.

\subsection*{Proof of \Cref{Thm25} in \Cref{Thm2}}
Using the triangle inequality, we note that
\begin{align*}
    (C_{NT}^0(\hat R))^{1/2}&=\vert \hat{\bm z}_{T,\hat R}^\top\hat{\bm\delta}_{\hat R}(1)-\bm z_T^\top\bm\delta(1)\vert
    \leq \vert \hat{\bm z}_{T,\hat R}^\top\hat{\bm\delta}_{\hat R}(1)-\ddot{\bm z}_T^\top\ddot{\bm\delta}(1)\vert+\vert \ddot{\bm z}_{T}^\top\ddot{\bm\delta}(1)-\bm z_T^\top\bm\delta(1)\vert\\
    &=(C_{NT}(\hat R))^{1/2}+\vert \ddot{\bm z}_{T}^\top\ddot{\bm\delta}(1)-\bm z_T^\top\bm\delta(1)\vert.
\end{align*}
Because, $\hat R=\arg\min_{R\in \Theta_R} C_{NT}(R)$, we have $ C_{NT}(\hat R)\leq  C_{NT}(\hat R^0)$, and therefore
\begin{align*}
    (C_{NT}^0(\hat R))^{1/2}\leq (C_{NT}(\hat R^0))^{1/2}+\vert \ddot{\bm z}_{T}^\top\ddot{\bm\delta}(1)-\bm z_T^\top\bm\delta(1)\vert.
\end{align*}
Using the triangle inequality again, we get
\begin{align*}
    (C_{NT}(\hat R^0))^{1/2}&=\vert \hat{\bm z}_{T,\hat R^0}^\top\hat{\bm\delta}_{\hat R^0}(1)-\ddot{\bm z}_T^\top\ddot{\bm\delta}(1)\vert =\vert (\hat{\bm z}_{T,\hat R}^\top\hat{\bm\delta}_{\hat R^0}(1)-\bm z_T^\top\bm\delta(1))- (\ddot{\bm z}_{T}^\top\ddot{\bm\delta}(1)-\bm z_T^\top\bm\delta(1))\vert\\
    &\leq \vert \hat{\bm z}_{T,\hat R^0}^\top\hat{\bm\delta}_{\hat R^0}(1)-\bm z_T^\top\bm\delta(1)\vert+\vert \ddot{\bm z}_{T}^\top\ddot{\bm\delta}(1)-\bm z_T^\top\bm\delta(1)\vert\\
    &=(C_{NT}^0(\hat R^0))^{1/2}+\vert \ddot{\bm z}_{T}^\top\ddot{\bm\delta}(1)-\bm z_T^\top\bm\delta(1)\vert.
\end{align*}
From this result and the previous one, we get
\begin{align*}
    (C_{NT}^0(\hat R))^{1/2}\leq (C_{NT}^0(\hat R^0))^{1/2}+2\vert \ddot{\bm z}_{T}^\top\ddot{\bm\delta}(1)-\bm z_T^\top\bm\delta(1)\vert.
\end{align*}
Since $(a+b)^2\leq 2a^2+2b^2$, we deduce that
\begin{align*}
    C_{NT}^0(\hat R)\leq 2C_{NT}^0(\hat R^0)+8(\ddot{\bm z}_{T}^\top\ddot{\bm\delta}(1)-\bm z_T^\top\bm\delta(1))^2.
\end{align*}
Consequently, we have that
\begin{align*}
   \frac{C_{NT}^0(\hat R)}{ \inf_{R\Theta_R}C_{NT}^0(R)+\delta_{NR_0T}^{-2}}\leq 2\frac{C_{NT}^0(\hat R^0)}{ \inf_{R\Theta_R}C_{NT}^0(R)+\delta_{NR_0T}^{-2}}+8\frac{(\ddot{\bm z}_{T}^\top\ddot{\bm\delta}(1)-\bm z_T^\top\bm\delta(1))^2}{ \inf_{R\Theta_R}C_{NT}^0(R)+\delta_{NR_0T}^{-2}}.
\end{align*}
Because $\frac{C_{NT}^0(\hat R^0)}{ \inf_{R\Theta_R}C_{NT}^0(R)+\delta_{NR_0T}^{-2}}\leq 1$
and
\begin{align*}
\frac{C_{NT}^0(\hat R^0)}{ \inf_{R\Theta_R}C_{NT}^0(R)+\delta_{NR_0T}^{-2}}&=\frac{\delta_{NR_0T}^{-2}}{ \inf_{R\Theta_R}C_{NT}^0(R)+\delta_{NR_0T}^{-2}}\delta_{NR_0T}^{2}(\ddot{\bm z}_{T}^\top\ddot{\bm\delta}(1)-\bm z_T^\top\bm\delta(1))^2\\
&\leq \delta_{NR_0T}^{2}(\ddot{\bm z}_{T}^\top\ddot{\bm\delta}(1)-\bm z_T^\top\bm\delta(1))^2
=O_P(1),
\end{align*}
from \Cref{Thm2} (\Cref{Thm24}), we obtain that $\frac{C_{NT}^0(\hat R)}{ \inf_{R\Theta_R}C_{NT}^0(R)+\delta_{NR_0T}^{-2}}=O_P(1)$. As a result, \Cref{Thm2} (\Cref{Thm25}) holds. 
Note that this proof only uses the definitions of $C_{NT}^0(R)$ and $C_{NT}(R)$, the fact that $\hat R$ minimizes $C_{NT}(R)$,  over $\Theta_R$, and \Cref{Thm2} (\Cref{Thm24}).

\subsection*{Proof of \Cref{Thm31} in \Cref{Thm3}}

Because $C_{NT}^0(\hat R^0)\leq C_{NT}^0(\hat R)$ and $C_{NT}(\hat R)\leq C_{NT}(\hat R^0)$, we have $    0\leq C_{NT}^0(\hat R)-\inf_{R\in \Theta_R} C_{NT}^0(R)$, which equals 
\begin{align*}
&(C_{NT}^0(\hat R)-C_{NT}(\hat R))+(C_{NT}(\hat R)-C_{NT}(\hat R^0))+(C_{NT}(\hat R^0)-C_{NT}^0(\hat R^0)),\\
    &\leq \vert C_{NT}^0(\hat R)-C_{NT}^0(\hat R) \vert + \vert C_{NT}^0(\hat R^0)-C_{NT}^0(\hat R^0)\vert 
\end{align*}
where we remove the negative middle term before applying the triangle inequality.
Consequently, we have
 \begin{align*}
 0\leq C_{NT}^0(\hat R)-\inf_{R\in \Theta_R} C_{NT}^0(R) \leq \vert C_{NT}^0(\hat R)-C_{NT}(\hat R) \vert + \vert C_{NT}^0(\hat R^0)-C_{NT}(\hat R^0)\vert.
\end{align*}
Because  $\sup_{R \in \Theta_R} \big(m_{NRT}^2\left\vert \left(C_{NT}( R)-C_{NT}^0( R)\right)\right\vert\big) =o_P(1)$ from \Cref{Lemma4} (16), we deduce that $C_{NT}^0(\hat R)-\inf_{R\in \Theta_R} C_{NT}^0(R)=o_P(m_{N\hat RT}^{-2}+m_{N\hat R^0T}^{-2})$.


\subsection*{Proof of \Cref{Thm32} in \Cref{Thm3}}

To show that $\frac{C_{NT}^0(\hat R)}{C^0_{NT(\hat R^0)}}\overset{P}{\longrightarrow}1$, we note as in \citet{IJR2017} that if 
 $\epsilon >0$,
$$P\left(\frac{\left(C_{NT}^0(\hat R)-C_{NT}^0(\hat R^0)\right)-\left(C_{NT}(\hat R)-C_{NT}(\hat R^0)\right)}{C_{NT}^0(\hat R)+C_{NT}^0(\hat R^0)} \leq \epsilon \right) \longrightarrow 1,$$ then $P\left(1 \leq \frac{C_{NT}^0(\hat R)}{C_{NT}^0(\hat R^0)} \leq \frac{1+\epsilon}{1-\epsilon} \right) \longrightarrow 1$, for $\epsilon\in (0,1)$, using the fact that   $C_{NT}(\hat R)\leq C_{NT}(\hat R^0)$ and 
$C_{NT}^0(\hat R^0)\leq C_{NT}^0(\hat R)$, where $\hat R=\arg\min_{R\in \Theta_R} C_{NT}(R)$ and $\hat R^0=\arg\min_{R\in \Theta_R} C_{NT}^0(R)$. 
Thus, to show that $ \frac{C_{NT}^0(\hat R)}{C_{NT}^0(\hat R^0)}  \longrightarrow 1$, it is enough to establish that $$\sup_{R, R^0\in \Theta_R} \left\vert \frac{\left(C_{NT}^0( R)-C_{NT}^0( R^0)\right)-\left(C_{NT}( R)-C_{NT}( R^0)\right)}{C_{NT}^0( R)+C_{NT}^0( R^0)}\right\vert \overset{P}{\longrightarrow} 0.$$ Given that   $\sup_{R, R^0\in \Theta_R} \left\vert \frac{\left(C_{NT}^0( R)-C_{NT}( R)\right)-\left(C^0_{NT}( R^0)-C_{NT}( R^0)\right)}{C_{NT}^0( R)+C^0_{NT}( R^0)}\right\vert \leq  \sup_{R \in \Theta_R} \left\vert \frac{\left(C_{NT}^0( R)-C_{NT}( R)\right)}{C_{NT}^0( R)}\right\vert + \sup_{R^0\in \Theta_R} \left\vert \frac{\left(C_{NT}^0( R^0)-C_{NT}( R^0)\right)}{C_{NT}^0( R^0)}\right\vert $, it is enough to prove 
$$\sup_{R \in \Theta_R} \left\vert \frac{\left(C_{NT}^0( R)-C_{NT}( R)\right)}{C_{NT}^0( R)}\right\vert=o_P(1).$$ 
Because, 
$\sup_{R \in \Theta_R} \left\vert \frac{\left(C_{NT}^0( R)-C_{NT}( R)\right)}{C_{NT}^0( R)}\right\vert\leq \frac{\sup_{R\in \Theta_R}  \big(\min(R^{1-2\theta},\frac{T^2}{R^2},N)\left\vert C_{NT}(R)-C_{NT}^0(R)\right\vert\big)}{\inf_{R\in \Theta_R}\big(\min(R^{1-2\theta},\frac{T^2}{R^2},N)C_{NT}^0(R)\big)},$
the result follows from the fact that $\inf_{R\in \Theta_R}\big(\min(R^{1-2\theta},\frac{T^2}{R^2},N)C_{NT}^0(R)\big)>0$ and 
 \Cref{Lemma4} (16).

\subsection*{Proof of \Cref{Thm33} in \Cref{Thm3}}
The objective in this section is to show that $\frac{C_{NT}(\hat R)}{\inf_{R\in \Theta_R}C^0_{NT}(R)}\overset{P}{\longrightarrow}1.$ 
To complete the proof, we need to show that 
$\frac{C_{NT}^0(\hat R)-C_{NT}(\hat R)}{C^0_{NT(\hat R^0)}}\overset{P}{\longrightarrow}0$, because $\frac{C_{NT}^0(\hat R)}{C^0_{NT(\hat R^0)}}\overset{P}{\longrightarrow}1$ from  \Cref{Thm3} (\Cref{Thm32} ).
To do so, we note that
$\frac{C_{NT}(\hat R)-C_{NT}^0(\hat R)}{C^0_{NT(\hat R^0)}}=\frac{C_{NT}(\hat R)-C_{NT}^0(\hat R)}{C^0_{NT}(\hat R)}\times \frac{C^0_{NT}(\hat R)}{C^0_{NT}(\hat R^0)}.$
Because $\frac{C^0_{NT}(\hat R)}{C^0_{NT}(\hat R^0)}\overset{P}{\longrightarrow} 1$, it is sufficient to show 
that
$\vert\frac{C_{NT}(\hat R)-C_{NT}^0(\hat R)}{C^0_{NT}(\hat R)}\vert\leq \sup_R\vert \frac{C_{NT}( R)-C_{NT}^0( R)}{C^0_{NT}(R)}\vert=o_P(1)$, which follows from the proof of  \Cref{Thm3} (\Cref{Thm32}). This completes the proof.
 
\subsection*{Proof of \Cref{Lemma1}}
\Cref{Lemma1} (1)-(9) correspond to results from Lemmas 3, 4, 5, 7, 8, 9, 10, 11, and 12, in \citet{IJR2017}. Therefore, their proofs are omitted.

\subsection*{Proof of \Cref{Lemma2}}
\subsubsection*{Proof of \Cref{Lemma2} (1)} 
Here we show that $\frac{1}{R^2}\sum_{s=T-R+1}^{T}\sum_{t=T-R+1}^{T} \Vert  \bm \Lambda^\top(s/T)\bm \Lambda(t/T)/N-\bm \Lambda^\top(s/T)\bm \Lambda(1)/N
\Vert^k=O\big(\frac{R^{k}}{T^{k}}\big).$
Recall that $\lambda_i(t/T)=\lambda_i(1)+\lambda_i^\prime(1)(t/T-1)+\frac{1}{2}\lambda_i^{\prime\prime}(c_i)(t/T-1)^2$ and  $\lambda_i(s/T)=\lambda_i(1)+\lambda_i^\prime(1)(s/T-1)+\frac{1}{2}\lambda_i^{\prime\prime}(c_i)(s/T-1)^2$. Therefore, 
$$ 
\Vert\frac{1}{N}\bm \Lambda^\top(s/T)\bm \Lambda(t/T)-\frac{1}{N}\bm \Lambda^\top(s/T)\bm \Lambda(1)\Vert = \Vert\frac{1}{N}\sum_{i=1}^{N}\big[\bm\lambda_i\big(t/T\big)-\bm\lambda_i(1)\big]\bm\lambda_i^{\top}(s/T)\Vert$$ and we can write $\frac{1}{N}\sum_{i=1}^{N}\big[\bm\lambda_i\big(t/T\big)-\bm\lambda_i(1)\big]\bm\lambda_i^{\top}(s/T)$ as
\begin{align*}
	&\big(\frac{t-T}{T}\big) \frac{1}{N}\sum_{i=1}^{N}\bm\lambda_i(1)\bm\lambda_i^{\prime\top}(1)+\frac{1}{2} \big(\frac{t-T}{T}\big)^2\frac{1}{N}\sum_{i=1}^{N}\bm\lambda_i(1)\bm\lambda_i^{\prime\prime\top}(c_i) \\
	&
	+ \big(\frac{t-T}{T}\big)\big(\frac{s-T}{T}\big)\frac{1}{N}\sum_{i=1}^{N}\bm\lambda_i^\prime(1)\bm\lambda_i^{\prime\top}(1)
	+\frac{1}{2} \big(\frac{t-T}{T}\big)^2\big(\frac{s-T}{T}\big)\frac{1}{N}\sum_{i=1}^{N}\bm\lambda_i^\prime(1)\bm\lambda_i^{\prime\prime\top}(c_i) 
	\\
	&+ \frac{1}{2}\big(\frac{t-T}{T}\big)\big(\frac{s-T}{T}\big)^2\frac{1}{N}\sum_{i=1}^{N}\bm\lambda_i^{\prime\prime}(c_i)\bm\lambda_i^{\prime\top}(1)+\frac{1}{4} \big(\frac{t-T}{T}\big)^2\big(\frac{s-T}{T}\big)^2\frac{1}{N}\sum_{i=1}^{N}\bm\lambda_i^{\prime\prime}(c_i)\bm\lambda_i^{\prime\prime\top}(c_i).
\end{align*}

Thus, $\frac{1}{R^2}\sum_{s=T-R+1}^{T}\sum_{t=T-R+1}^{T} \Vert  \bm \Lambda^\top(s/T)\bm \Lambda(t/T)/N-\bm \Lambda^\top(s/T)\bm \Lambda(1)/N
	\Vert^k$ is bounded by 
	\begin{align*}
		&\frac{1}{R}\sum_{t=T-R+1}^{T}\big(\frac{T-t}{T}\big)^k\big\Vert \frac{1}{N}\sum_{i=1}^{N}\bm\lambda_i(1)\bm\lambda_i^{\prime\top}(1)\big\Vert^k
		+ \frac{1}{2^k}\frac{1}{R}\sum_{t=T-R+1}^{T}  \big(\frac{T-t}{T}\big)^{2k}\big\Vert\frac{1}{N}\sum_{i=1}^{N}\bm\lambda_i(1)\bm\lambda_i^{\prime\prime\top}(c_i)\big\Vert^k\\
		& +\frac{1}{R^2}\sum_{t=T-R+1}^{T} \big(\frac{T-t}{T}\big)^k\sum_{s=T-R+1}^{T}\big(\frac{T-s}{T}\big)^k\big\Vert\frac{1}{N}\sum_{i=1}^{N}\bm\lambda_i^\prime(1)\bm\lambda_i^{\prime\top}(1)\big\Vert^{k}\\
		&+\frac{1}{R^2}\sum_{t=T-R+1}^{T} \big(\frac{T-t}{T}\big)^{2k}\sum_{s=T-R+1}^{T} \big(\frac{T-s}{T}\big)^k\big\Vert\frac{1}{N}\sum_{i=1}^{N}\bm\lambda_i^\prime(1)\bm\lambda_i^{\prime\prime\top}(c_i)\big\Vert^k \\
		& +\frac{1}{4^{2k}}\frac{1}{R^2}\sum_{t=T-R+1}^{T} \big(\frac{T-t}{T}\big)^{2k}\sum_{s=T-R+1}^{T}\big(\frac{T-s}{T}\big)^{2k}\big\Vert\frac{1}{N}\sum_{i=1}^{N}\bm\lambda_i^{\prime\prime}(c_i)\bm\lambda_i^{\prime\prime\top}(c_i)\big\Vert^k.
	\end{align*}
	
	Furthermore, we have 
	\begin{align*}
		\sum_{t=T-R+1}^{T}\big(\frac{T-t}{T}\big)^k &= \sum_{j=0}^{R-1}\big(\frac{j}{T}\big)^k=O\big(\frac{R^{k+1}}{T^{k}}\big),\\
		\sum_{t=T-R+1}^{T}\big(\frac{T-t}{T}\big)^{2k} &= \sum_{j=0}^{R-1}\big(\frac{j}{T}\big)^{2k}=O\big(\frac{R^{2k+1}}{T^{2k}}\big),
	\end{align*}
	where we use \citet[Theorem 2.27]{Davidson1994}, that $\sum_{j=1}^{R}j^k=O(R^{k+1})$, if $k\geq 1$. Note from \Cref{Ass1} (b) and \Cref{Ass5} (b) that 
	$\big\Vert \frac{1}{N}\sum_{i=1}^{N}\bm\lambda_i(1)\bm\lambda_i^{\prime\top}(1)\big\Vert,$
	$\big\Vert\frac{1}{N}\sum_{i=1}^{N}\bm\lambda_i(1)\bm\lambda_i^{\prime\prime\top}(c_i)\big\Vert,$
	$\big\Vert\frac{1}{N}\sum_{i=1}^{N}\bm\lambda_i^\prime(1)\bm\lambda_i^{\prime\prime\top}(c_i)\big\Vert,$
	$\big\Vert\frac{1}{N}\sum_{i=1}^{N}\bm\lambda_i^\prime(1)\bm\lambda_i^{\prime\top}(1)\big\Vert,$
	and 
	$\Vert\frac{1}{N}\sum_{i=1}^{N}\bm\lambda_i^{\prime\prime}(c_i)\bm\lambda_i^{\prime\prime\top}(c_i)\big\Vert$ 
	are bounded.
	Therefore, we have 
	\begin{align*}
		& \frac{1}{R^2}\sum_{s=T-R+1}^{T}\sum_{t=T-R+1}^{T} \Vert  \bm \Lambda^\top(s/T)\bm \Lambda(t/T)/N-\bm \Lambda^\top(s/T)\bm \Lambda(1)/N
		\Vert^k =O\big( \frac{R^k}{T^k} + \frac{R^{2k}}{T^{2k}}+ \frac{R^{3k}}{T^{3k}} + \frac{R^{4k}}{T^{4k}}  \big),
	\end{align*}
	which is $O\big( \frac{R^k}{T^k}\big).$ 

\subsubsection*{Proof of \Cref{Lemma2} (2)}
From the identity in \Cref{identity}, we deduce $\frac{1}{R}\sum_{t=T-R+1}^{T}\Vert\tilde{\bm f}_{t,R}-\bm H_R\bm f_t\Vert^2\leq 5\Vert\tilde{\bm V}_{R}^{-1}\Vert^2 \times \frac{1}{R}\sum_{t=T-R+1}^{T}( \Vert \bm A_{1t}\Vert^2+\Vert\bm A_{2t}\Vert^2+\Vert\bm A_{3t}\Vert^2+\Vert\bm A_{4t}\Vert^2+\Vert\bm A_{5t}\Vert^2).$ Next, we find the order in probability for $\frac{1}{R}\sum_{t=T-R+1}^{T}\Vert\bm A_{jt}\Vert^2, j=1,\ldots,5$. 

First, we note that $\Vert\bm A_{1t}\Vert\leq \big(\frac{1}{R}\sum_{s=T-R+1}^{T}\Vert\tilde{\bm f}_{s,R}\Vert^2\big)^{1/2}\big(\frac{1}{R}\sum_{s=T-R+1}^{T}\Vert\bm \nu_{st}\Vert^2\big)^{1/2}$. Therefore,
$\frac{1}{R}\sum_{t=T-R+1}^{T}\Vert\bm A_{1t}\Vert^2\leq \big(\frac{1}{R}\sum_{s=T-R+1}^{T}\Vert\tilde{\bm f}_{s,R}\Vert^2\big)\big(\frac{1}{R^2}\sum_{s=T-R+1}^{T}\sum_{t=T-R+1}^{T}\Vert\bm \nu_{st}\Vert^2\big).$ Because, $\frac{1}{R}\sum_{s=T-R+1}^{T}\Vert\tilde{\bm f}_{s,R}\Vert^2=\mathrm{Trace}\big(\frac{1}{R}\sum_{s=T-R+1}^{T}\tilde{\bm f}_{s,R}\tilde{\bm f}_{s,R}^\top\big)=\mathrm{Trace}(\bm I_r)=r,$ we deduce that  $\frac{1}{R}\sum_{s=T-R+1}^{T}\Vert\tilde{\bm f}_{s,R}\Vert^2=O_P(1).$ In addition, $\frac{1}{R^2}\sum_{s=T-R+1}^{T}\sum_{t=T-R+1}^{T}\Vert\bm \nu_{st}\Vert^2 $ is smaller than
\begin{align*}
  \big( \frac{1}{R}\sum_{s=T-R+1}^{T} \Vert \bm f_s \Vert^4 \big)^{\frac{1}{2}} \bigg( \frac{1}{R}\sum_{s=T-R+1}^{T}\big(\frac{1}{R}\sum_{t=T-R+1}^{T} \Vert  (\frac{1}{N}\bm \Lambda^\top(s/T)\bm \Lambda(t/T)-\frac{1}{N}\bm \Lambda^\top(s/T)\bm \Lambda(1))\bm f_t
 \Vert^2\big)^2 \bigg)^{\frac{1}{2}},
\end{align*}
which in turn is bounded by
\begin{align*}
  &\big( \frac{1}{R}\sum_{s=T-R+1}^{T} \Vert \bm f_s \Vert^4 \big)^{\frac{1}{2}}  \bigg( \frac{1}{R^2}\sum_{s=T-R+1}^{T}\sum_{t=T-R+1}^{T} \Vert  \bm \Lambda^\top(s/T)\bm \Lambda(t/T)/N-\bm \Lambda^\top(s/T)\bm \Lambda(1)/N
 \Vert^4 \bigg)^{1/2}\\
 &=O_P\big(\frac{R^2}{T^2}\big),
\end{align*}
because \Cref{Ass1} (a) implies $ \frac{1}{R}\sum_{s=T-R+1}^{T} \Vert \bm f_s \Vert^4=O_P(1)$  and \Cref{Lemma2} (1) implies that $\Vert\Lambda^\top(s/T)\bm \Lambda(t/T)/N-\bm \Lambda^\top(s/T)\bm \Lambda(1)/N
\Vert^4=O\big(\frac{R^4}{T^4}\big).$

Second, 
$\frac{1}{R}\sum_{t=T-R+1}^{T}\Vert\bm A_{2t}\Vert^2\leq \big(\frac{1}{R}\sum_{s=T-R+1}^{T}\Vert\tilde{\bm f}_{s,R}\Vert^2\big)\big(\frac{1}{R^2}\sum_{t=T-R+1}^{T}\sum_{s=T-R+1}^{T}\Vert\bm \eta_{st}\Vert^2\big)$, where $\frac{1}{R}\sum_{s=T-R+1}^{T}\Vert\tilde{\bm f}_{s,R}\Vert^2=O_P(1)$.  Furthermore, 
\begin{align*}
\frac{1}{R^2}\sum_{s=T-R+1}^{T}\sum_{t=T-R+1}^{T}\Vert\bm \eta_{st}\Vert^2
&=\frac{1}{R^2}\sum_{s=T-R+1}^{T}\sum_{t=T-R+1}^{T}\big\Vert \bm f_s^\top\frac{1}{N}\sum_{i=1}^{N}\bm \lambda_i\big(\frac{s}{T}\big)\bm e_{it} \big\Vert^2\\
&\leq 3\frac{1}{R^2}\sum_{s=T-R+1}^{T}\sum_{t=T-R+1}^{T}\big\Vert \bm f_s^\top\frac{1}{N}\sum_{i=1}^{N}\bm \lambda_i(1)\bm e_{it} \big\Vert^2\\
&+ 3\frac{1}{R^2}\sum_{s=T-R+1}^{T}\sum_{t=T-R+1}^{T}\big\Vert \bm f_s^\top\big(\frac{s-T}{T}\big)\frac{1}{N}\sum_{i=1}^{N}\bm \lambda_i^{\prime}(1)\bm e_{it} \big\Vert^2\\
&+3\frac{1}{R^2}\sum_{s=T-R+1}^{T}\sum_{t=T-R+1}^{T}\big\Vert \bm f_s^\top\big(\frac{s-T}{T}\big)^2\frac{1}{N}\sum_{i=1}^{N}\bm \lambda_i^{\prime\prime}(c_i)\bm e_{it} \big\Vert^2.
\end{align*}
In addition, $\frac{1}{R^2}\sum_{s=T-R+1}^{T}\sum_{t=T-R+1}^{T}\Vert\bm \eta_{st}\Vert^2$ is bounded by
\begin{align}
& \frac{3}{N}\big(\frac{1}{R}\sum_{s=T-R+1}^{T}\big\Vert \bm f_s\big\Vert^2\big)\big( \frac{1}{R}\sum_{t=T-R+1}^{T}\big\Vert\frac{1}{\sqrt N}\sum_{i=1}^{N}\bm \lambda_i(1)\bm e_{it} \big\Vert^2\big)\nonumber\\
&+ \frac{3}{N}\big(\frac{1}{R}\sum_{s=T-R+1}^{T}\big\Vert \bm f_s\big(\frac{s-T}{T}\big)\big\Vert^2\big)\big(\frac{1}{R}\sum_{t=T-R+1}^{T}\big\Vert \frac{1}{\sqrt N}\sum_{i=1}^{N}\bm \lambda_i^{\prime}(1)\bm e_{it} \big\Vert^2\big)\nonumber\\
&+\frac{3}{N}\big(\frac{1}{R}\sum_{s=T-R+1}^{T}\big\Vert \bm f_s\big(\frac{s-T}{T}\big)^2\big\Vert^2\big)\big(\frac{1}{R}\sum_{t=T-R+1}^{T}\big\Vert \frac{1}{\sqrt N}\sum_{i=1}^{N}\bm \lambda_i^{\prime\prime}(c_i)\bm e_{it} \big\Vert^2\big).
\label{eta_eta}
\end{align}
For the right side, the first term is $O_P(1/N)$. Indeed, $\frac{1}{R}\sum_{s=T-R+1}^{T}\big\Vert \bm f_s\big\Vert^2=O_P(1)$, from \Cref{Ass1} (a). Also, $ \frac{1}{R}\sum_{t=T-R+1}^{T}\big\Vert\frac{1}{\sqrt N}\sum_{i=1}^{N}\bm \lambda_i(1)\bm e_{it} \big\Vert^2=O_P(1)$, because 
\begin{align*}
\mathrm{E}\big( \frac{1}{R}\sum_{t=T-R+1}^{T}\big\Vert\frac{1}{\sqrt N}\sum_{i=1}^{N}\bm \lambda_i(1)\bm e_{it} \big\Vert^2\big)&=\mathrm{E}\big( \frac{1}{RN}\sum_{t=T-R+1}^{T}\sum_{i=1}^{N}\sum_{j=1}^{N}\bm \lambda_i^\top(1)\bm\lambda_j(1)\bm e_{it}\bm e_{jt} \big)\\
& \leq \frac{1}{RN}\sum_{t=T-R+1}^{T}\sum_{i=1}^{N}\sum_{j=1}^{N}\vert\bm \lambda_i^\top(1)\bm\lambda_j(1)\vert \times \vert\mathrm{E}\big(\bm e_{it}\bm e_{jt} \big)\vert\\
&\leq C \frac{1}{RN}\sum_{t=T-R+1}^{T}\sum_{i=1}^{N}\sum_{j=1}^{N} \vert\mathrm{E}\big(\bm e_{it}\bm e_{jt} \big)\vert =O_P(1),
\end{align*}
where  we use \Cref{Ass1} (b) to bound $\sup_{T-R+1\leq s\leq T}|\bm\lambda_i^\top(s/T)\bm\lambda_j(s/T)|$ for any $i, j, s,$ and \Cref{Ass1} (e) to bound $\frac{1}{NR} \sum_{t=T-R+1}^T \sum_{i=1}^N \sum_{j=1}^N|\mathrm{E}\big(e_{it}e_{jt}\big)|$.
For the second term and the third of the right side of \Cref{eta_eta} are $o_P(1/N)$ because $$\frac{1}{R}\sum_{s=T-R+1}^{T}\big\Vert \bm f_s\big(\frac{s-T}{T}\big)\big\Vert^2\leq \big(\frac{1}{R}\sum_{s=T-R+1}^{T}\big\Vert \bm f_s\big\Vert^4\big)^{\frac{1}{2}}\big(\frac{1}{R}\sum_{s=T-R+1}^{T}\big(\frac{s-T}{T}\big)^4\big)^{\frac{1}{2}}=O_P(1)O_P\big(\frac{R^2}{T^2}\big),$$
$$\frac{1}{R}\sum_{s=T-R+1}^{T}\big\Vert \bm f_s\big(\frac{s-T}{T}\big)^8\big\Vert^2\leq \big(\frac{1}{R}\sum_{s=T-R+1}^{T}\big\Vert \bm f_s\big\Vert^4\big)^{\frac{1}{2}}\big(\frac{1}{R}\sum_{s=T-R+1}^{T}\big(\frac{s-T}{T}\big)^8\big)^{\frac{1}{2}}=O_P(1)O_P\big(\frac{R^4}{T^4}\big),$$
 $ \frac{1}{R}\sum_{t=T-R+1}^{T}\big\Vert\frac{1}{\sqrt N}\sum_{i=1}^{N}\bm \lambda^{\prime}_i(1)\bm e_{it} \big\Vert^2=O_P(1)$ and
 $ \frac{1}{R}\sum_{t=T-R+1}^{T}\big\Vert\frac{1}{\sqrt N}\sum_{i=1}^{N}\bm \lambda^{\prime\prime}_i(c_i)\bm e_{it} \big\Vert^2=O_P(1)$
 using similar arguments as for the first term. Hence, $\frac{1}{R^2}\sum_{t=T-R+1}^{T}\sum_{s=T-R+1}^{T}\Vert\bm \eta_{st}\Vert^2=O_P(1/N)$ and $\frac{1}{R}\sum_{t=T-R+1}^{T}\Vert\bm A_{2t}\Vert^2=O_P\big(\frac{1}{ N}\big)$.
 
Third, $\frac{1}{R}\sum_{t=T-R+1}^{T}\Vert\bm A_{3t}\Vert^2\leq \big(\frac{1}{R}\sum_{s=T-R+1}^{T}\Vert\tilde{\bm f}_{s,R}\Vert^2\big)\big(\frac{1}{R^2}\sum_{t=T-R+1}^{T}\sum_{s=T-R+1}^{T}\Vert\bm \xi_{st}\Vert^2\big)$, where we noted again that $\frac{1}{R}\sum_{s=T-R+1}^{T}\Vert\tilde{\bm f}_{s,R}\Vert^2=O_P(1)$.  In addition, we similarly have that
$$\frac{1}{R^2}\sum_{t=T-R+1}^{T}\sum_{s=T-R+1}^{T}\Vert\bm \xi_{st}\Vert^2
=\frac{1}{R^2}\sum_{s=T-R+1}^{T}\sum_{t=T-R+1}^{T}\big\Vert \bm f_t^\top\frac{1}{N}\sum_{i=1}^{N}\bm \lambda_i\big(\frac{t}{T}\big)\bm e_{is} \big\Vert^2=O_P\big(\frac{1}{ N}\big).$$

We omit the proof that $\frac{1}{R}\sum_{t=T-R+1}^{T}\Vert\bm A_{4t}\Vert^2=O_P\big(\frac{1}{ R}\big)$ and $\frac{1}{R}\sum_{t=T-R+1}^{T}\Vert\bm A_{5t}\Vert^2=O_P\big(\frac{1}{ N}\big)$, as they follow identical steps as in the proof of \citet[Theorem 1]{BN2002}.  
 Combining the previous results, the proof that $\frac{1}{R}\sum_{t=T-R+1}^{T}\Vert\tilde{\bm f}_{t,R}-\bm H_R\bm f_t\Vert^2=O_P\big(\frac{1}{ N}+\frac{1}{R}+\frac{R^2}{T^2}\big)=O_P\big((\min(\sqrt R,\sqrt N,\frac{T}{R}))^{-2}\big)=O_P\big(\delta_{NRT}^{-2}\big)$ is  completed.

\subsubsection*{Proof of \Cref{Lemma2} (3)-(4)} 
 We can easily show  
$\frac{1}{R}\sum_{t=T-R+1}^{T}\Vert\tilde{\bm f}_{t,R}-\bm H_R\bm f_t\Vert^2\frac{T-t}{T}=O_P\big(\delta_{NRT}^{-2}\frac{R}{T}\big)$ and
$\frac{1}{R}\sum_{t=T-R+1}^{T}\Vert\tilde{\bm f}_{t,R}-\bm H_R\bm f_t\Vert^2\frac{(t-T)^2}{T^2}=O_P\big(\delta_{NRT}^{-2}\frac{R^2}{T^2}\big)$, using the same steps as in the proof of  \Cref{Lemma2} (2). Therefore, the proofs are omitted.

\subsubsection*{Proof of \Cref{Lemma2} (5)} 

To find the order in probability of $\frac{1}{R}\sum_{t=T-R+1}^{T-h}(\tilde{\bm f}_{t,R}-\bm H_R\bm f_t)\bm z_t^\top$, we use the identity
\begin{align*}
\frac{1}{R}\sum_{t=T-R+1}^{T-h}(\tilde{\bm f}_{t,R}-\bm H_R\bm f_t)\bm z^\top_t&=\tilde{\bm V}_{R}^{-1}\big(  \frac{1}{R}\sum_{t=T-R+1}^{T-h}\bm A_{1t}\bm{z}_t^\top+ \frac{1}{R}\sum_{t=T-R+1}^{T-h}\bm A_{2t}\bm{z}_t^\top \\
&+ \frac{1}{R}\sum_{t=T-R+1}^{T-h}\bm A_{3t}\bm{z}_t^\top+ \frac{1}{R}\sum_{t=T-R+1}^{T-h}\bm A_{4t}\bm{z}_t^\top+ \frac{1}{R}\sum_{t=T-R+1}^{T-h}\bm A_{5t}\bm{z}_t^\top\big),
\end{align*}
where we need to obtain the order in probability of $\frac{1}{R}\sum_{t=T-R+1}^{T-h}\bm A_{it}\bm{z}_t^\top, i=1,2,3,4,5.$

We note that $\frac{1}{R}\sum_{t=T-R+1}^{T-h}\bm A_{1t}\bm{z}_t^\top$ equals
\begin{align*}
	\frac{1}{R^2}\sum_{t=T-R+1}^{T-h}\sum_{s=T-R+1}^{T} \tilde{\bm f}_s\nu_{st}\bm{z}_t^\top
	=\frac{1}{R^2}\sum_{t=T-R+1}^{T-h}\sum_{s=T-R+1}^{T} \tilde{\bm f}_s\big(\bm f_s^\top\bm\Lambda^\top(s/T) [\bm\Lambda(t/T)-\bm\Lambda(1)]\bm f_{t}/N\big)\bm{z}_t^\top.
\end{align*}
It can be written as
\begin{align}
	&\frac{1}{R}\sum_{s=T-R+1}^{T} (\tilde{\bm f}_s-\bm H_R\bm f_s)\frac{1}{NR}\sum_{t=T-R+1}^{T-h}\big(\bm f_s^\top\bm\Lambda^\top(s/T) [\bm\Lambda(t/T)-\bm\Lambda(1)]\bm f_{t}\big)\bm{z}_t^\top	\label{fz1}\\
	&+	\bm H_R\frac{1}{R}\sum_{s=T-R+1}^{T} (\bm f_s\bm f_s^\top)\frac{1}{NR}\sum_{t=T-R+1}^{T-h}\bm \Lambda^\top(s/T) [\bm\Lambda(t/T)-\bm\Lambda(1)]\bm f_{t}\bm{z}_t^\top.
	\label{fz2}
\end{align}
The norm of \Cref{fz1}
is bounded by 
$$\big(\frac{1}{R}\sum_{s=T-R+1}^{T} \Vert\tilde{\bm f}_s-\bm H_R\bm f_s\Vert^2\big)^{\frac{1}{2}}\big(\frac{1}{R}\sum_{s=T-R+1}^{T}\big\Vert\frac{1}{NR}\sum_{t=T-R+1}^{T-h}\bm f_s^\top\bm\Lambda^\top(s/T) [\bm\Lambda(t/T)-\bm\Lambda(1)]\bm f_{t}\bm{z}_t^\top\big\Vert^2\big)^{\frac{1}{2}},$$
which is $O_P\big(\frac{R}{T}\delta_{NRT}^{-1}\big)$, as $\frac{1}{R}\sum_{s=T-R+1}^{T} \Vert\tilde{\bm f}_s-\bm H_R\bm f_s\Vert^2=O_P(\delta_{NRT}^{-2})$, from \Cref{Lemma2} (2) and 
$$\frac{1}{R}\sum_{s=T-R+1}^{T}\big\Vert\frac{1}{NR}\sum_{t=T-R+1}^{T-h}\bm f_s^\top\bm\Lambda^\top(s/T) [\bm\Lambda(t/T)-\bm\Lambda(1)]\bm f_{t}\bm{z}_t^\top\Vert^2=O_P\big(\frac{R^2}{T^2}\big).$$ 
To see the latter, we observe that the term can be decomposed as $\sum_{j=1}^{6} B_{1j}$, where
\begin{align*}
	 B_{11}=&\frac{1}{R}\sum_{s=T-R+1}^{T}\big\Vert\bm f_s^\top \frac{1}{N}\sum_{i=1}^{N}\bm\lambda_i(1)\bm\lambda_i^{\prime\top}(1) \frac{1}{R}\sum_{t=T-R+1}^{T-h}\big(\frac{t-T}{T}\big)\bm f_{t}\bm{z}_t^\top\big\Vert^2=O_P\big(\frac{R^2}{T^2}\big),\\
	 B_{12}= &	\frac{1}{R}\sum_{s=T-R+1}^{T}\big\Vert\bm f_s^\top \frac{1}{N}\sum_{i=1}^{N}\bm\lambda_i(1)\bm\lambda_i^{\prime\prime\top}(c_i) \frac{1}{R}\sum_{t=T-R+1}^{T-h}\big(\frac{t-T}{T}\big)^2\bm f_{t}\bm{z}_t^\top\big\Vert^2=O_P\big(\frac{R^4}{T^4}\big),\\
	 B_{13}= &\frac{1}{R}\sum_{s=T-R+1}^{T}\big\Vert\big(\frac{s-T}{T}\big)\bm f_s^\top \frac{1}{N}\sum_{i=1}^{N}\bm\lambda_i^\prime(1)\bm\lambda_i^{\prime\top}(1) \frac{1}{R}\sum_{t=T-R+1}^{T-h}\big(\frac{t-T}{T}\big)\bm f_{t}\bm{z}_t^\top\big\Vert^2=O_P\big(\frac{R^4}{T^4}\big),\\
	 B_{14}= &\frac{1}{R}\sum_{s=T-R+1}^{T}\big\Vert\big(\frac{s-T}{T}\big)\bm f_s^\top \frac{1}{N}\sum_{i=1}^{N}\bm\lambda_i^\prime(1)\bm\lambda_i^{\prime\prime\top}(c_i) \frac{1}{R}\sum_{t=T-R+1}^{T-h}\big(\frac{t-T}{T}\big)^2\bm f_{t}\bm{z}_t^\top\big\Vert^2=O_P\big(\frac{R^6}{T^6}\big),\\
	 B_{15}= &\frac{1}{R}\sum_{s=T-R+1}^{T}\big\Vert\big(\frac{s-T}{T}\big)^2\bm f_s^\top \frac{1}{N}\sum_{i=1}^{N}\bm\lambda_i^{\prime\prime}(c_i)\bm\lambda_i^{\prime\top}(1) \frac{1}{R}\sum_{t=T-R+1}^{T-h}\big(\frac{t-T}{T}\big)\bm f_{t}\bm{z}_t^\top\big\Vert^2=O_P\big(\frac{R^5}{T^5}\big),\\
 B_{16}= &\frac{1}{R}\sum_{s=T-R+1}^{T}\big\Vert\big(\frac{s-T}{T}\big)^2\bm f_s^\top \frac{1}{N}\sum_{i=1}^{N}\bm\lambda_i^{\prime\prime}(c_i)\bm\lambda_i^{\prime\prime\top}(c_i) \frac{1}{R}\sum_{t=T-R+1}^{T-h}\big(\frac{t-T}{T}\big)^2\bm f_{t}\bm{z}_t^\top\big\Vert^2=O_P\big(\frac{R^8}{T^8}\big).
\end{align*}
Thus, \Cref{fz1} is $O_P(\frac{R}{T}\delta_{NRT}^{-1})$. For $B_{11}$, note that 
$$ B_{11}\leq \frac{1}{R}\sum_{s=T-R+1}^{T}\Vert\bm f_s\Vert^2 \frac{1}{N}\sum_{i=1}^{N}\Vert\bm\lambda_i(1)\Vert^2\Vert\bm\lambda_i^{\prime}(1)\Vert^2 \frac{1}{R}\sum_{t=T-R+1}^{T-h}\big(\frac{T-t}{T}\big)^2\Vert\bm f_{t}\Vert^2 \Vert\bm{z}_t\Vert^2,$$
where $\frac{1}{R}\sum_{s=T-R+1}^{T}\Vert\bm f_s\Vert^2=O_P(1)$,  $\frac{1}{N}\sum_{i=1}^{N}\Vert\bm\lambda_i(1)\Vert^2\Vert\bm\lambda_i^{\prime}(1)\Vert^2=O(1)$, and $$\frac{1}{R}\sum_{t=T-R+1}^{T-h}\big(\frac{T-t}{T}\big)^2\mathrm{E}(\Vert\bm f_{t}\Vert^2 \Vert\bm{z}_t\Vert^2)\leq C\frac{1}{R}\sum_{t=T-R+1}^{T-h}\big(\frac{T-t}{T}\big)^2=O\big(\frac{R^2}{T^2}\big),$$
using \Cref{Ass4} (c) to bound $\mathrm{E}(\Vert \bm f_t\Vert^2\Vert \bm z_t\Vert^2)\leq \mathrm{E}(\Vert \bm z_t\Vert^4)$. For $B_{13}$, note that 
$$ B_{13}\leq \frac{1}{R}\sum_{s=T-R+1}^{T}\big(\frac{T-s}{T}\big)^2\Vert\bm f_s\Vert^2 \frac{1}{N}\sum_{i=1}^{N}\Vert\bm\lambda_i(1)\Vert^2\Vert\bm\lambda_i^{\prime}(1)\Vert^2 \frac{1}{R}\sum_{t=T-R+1}^{T-h}\big(\frac{T-t}{T}\big)^2\Vert\bm f_{t}\Vert^2 \Vert\bm{z}_t\Vert^2,$$
where we use \Cref{Ass1} (a) and \Cref{Ass4} (c)  to get $\frac{1}{R}\sum_{s=T-R+1}^{T}\big(\frac{T-s}{T}\big)^2\mathrm{E}(\Vert\bm f_s\Vert^2)=O\big(\frac{R^2}{T^2}\big)$,  $\frac{1}{N}\sum_{i=1}^{N}\Vert\bm\lambda_i(1)\Vert^2\Vert\bm\lambda_i^{\prime}(1)\Vert^2=O(1)$, and $\frac{1}{R}\sum_{t=T-R+1}^{T-h}\big(\frac{T-t}{T}\big)^2\mathrm{E}(\Vert\bm f_{t}\Vert^2 \Vert\bm{z}_t\Vert^2)=O\big(\frac{R^2}{T^2}\big).$
The orders for $B_{12},B_{14},B_{15}$ and $B_{16}$ use similar arguments as for $B_{11}$ and $B_{13}$. 

To find the order of \Cref{fz2}, we note that it can be written as $\sum_{j=1}^{6}\bm B_{2j}$, where
\begin{align*}
\bm	B_{21}=& \bm H_R\frac{1}{R}\sum_{s=T-R+1}^{T} \bm f_s\bm f_s^\top\big(\frac{1}{N}\sum_{i=1}^{N}\bm\lambda_i(1)\bm\lambda_i^{\prime\top}(1)\big)\frac{1}{R}\sum_{t=T-R+1}^{T-h}\big(\frac{t-T}{T}\big)\bm f_{t}\bm{z}_t^\top=O_P\big(\frac{R}{T}\big),\\
\bm	B_{22}=& \bm H_R\frac{1}{R}\sum_{s=T-R+1}^{T} \bm f_s\bm f_s^\top\big(\frac{1}{N}\sum_{i=1}^{N}\bm\lambda_i(1)\bm\lambda_i^{\prime\prime\top}(c_i)\big)\frac{1}{R}\sum_{t=T-R+1}^{T-h}\big(\frac{t-T}{T}\big)^2\bm f_{t}\bm{z}_t^\top=O_P\big(\frac{R^2}{T^2}\big),\\
\bm	B_{23}=& \bm H_R\frac{1}{R}\sum_{s=T-R+1}^{T}\big(\frac{s-T}{T}\big) \bm f_s\bm f_s^\top\big(\frac{1}{N}\sum_{i=1}^{N}\bm\lambda_i^\prime(1)\bm\lambda_i^{\prime\top}(1)\big)\frac{1}{R}\sum_{t=T-R+1}^{T-h}\big(\frac{t-T}{T}\big)\bm f_{t}\bm{z}_t^\top=O_P\big(\frac{R^2}{T^2}\big),\\
\bm	B_{24}=&  \bm H_R\frac{1}{R}\sum_{s=T-R+1}^{T} \big(\frac{s-T}{T}\big)\bm f_s\bm f_s^\top\big(\frac{1}{N}\sum_{i=1}^{N}\bm\lambda_i^\prime(1)\bm\lambda_i^{\prime\prime\top}(c_i)\big)\frac{1}{R}\sum_{t=T-R+1}^{T-h}\big(\frac{t-T}{T}\big)^2\bm f_{t}\bm{z}_t^\top=O_P\big(\frac{R^3}{T^3}\big),\\
\bm B_{25}=& \bm H_R\frac{1}{R}\sum_{s=T-R+1}^{T}\big(\frac{s-T}{T}\big)^2 \bm f_s\bm f_s^\top\big(\frac{1}{N}\sum_{i=1}^{N}\bm\lambda_i^{\prime\prime}(c_i)\bm\lambda_i^{\prime\top}(1)\big)\frac{1}{R}\sum_{t=T-R+1}^{T-h}\big(\frac{t-T}{T}\big)\bm f_{t}\bm{z}_t^\top=O_P\big(\frac{R^3}{T^3}\big),\\
\bm B_{26}=&  \bm H_R\frac{1}{R}\sum_{s=T-R+1}^{T} \big(\frac{s-T}{T}\big)^2\bm f_s\bm f_s^\top\big(\frac{1}{N}\sum_{i=1}^{N}\bm\lambda_i^{\prime\prime}(c_i)\bm\lambda_i^{\prime\prime\top}(c_i)\big)\frac{1}{R}\sum_{t=T-R+1}^{T-h}\big(\frac{t-T}{T}\big)^2\bm f_{t}\bm{z}_t^\top=O_P\big(\frac{R^4}{T^4}\big).
\end{align*}
The orders are obtained using the same techniques as for $B_{11}$ and $B_{13}$ on $\Vert \bm B_{2j}\Vert, j=1,\ldots,6.$
Therefore, \Cref{fz2} is $O_P\big(\frac{R}{T}\big)$. 
As, \Cref{fz1} is $O_P\big(\delta_{NRT}^{-1}\frac{R}{T}\big)$, we deduce that $$\frac{1}{R}\sum_{t=T-R+1}^{T-h}\bm A_{1t}\bm z_t^\top=O_P\big(\delta_{NRT}^{-1}\frac{R}{T}
+\frac{R}{T}\big)
=O_P\big(\frac{R}{T}\big).$$

For $\frac{1}{R}\sum_{t=T-R+1}^{T-h}\bm A_{2t}\bm{z}_t^\top=	\frac{1}{R^2}\sum_{t=T-R+1}^{T-h}\sum_{s=T-R+1}^{T} \tilde{\bm f}_s\eta_{st}\bm{z}_t^\top$, we use the decomposition
\begin{align}
	&\frac{1}{NR^2}\sum_{s=T-R+1}^{T}\sum_{t=T-R+1}^{T-h} (\tilde{\bm f}_s-\bm H_{R}\bm f_s)\big(\bm f_s^\top\bm \Lambda^\top\big(\frac{s}{T}\big)\bm e_{t}\big)\bm{z}_t^\top\label{A2z1}\\
	&+\bm H_{R}\frac{1}{NR^2}\sum_{s=T-R+1}^{T}\sum_{t=T-R+1}^{T-h}(\bm f_s\bm f_s^\top)\bm \Lambda^\top\big(\frac{s}{T}\big)\bm e_{t}\bm{z}_t^\top\label{A2z2},
\end{align}
where we show that the first term \eqref{A2z1} is $O_P\big(\delta_{NRT}^{-1}\frac{1}{\sqrt N}\big)$ and the second term \eqref{A2z2} is $O_P\big(\frac{1}{\sqrt{NR}}\big)$. Thus, $\frac{1}{R}\sum_{t=T-R+1}^{T-h}\bm A_{2t}\bm{z}_t^\top=O_P\big(\delta_{NRT}^{-1}\frac{1}{\sqrt N}\big)+O_P\big(\frac{1}{\sqrt{NR}}\big)$.

To find the order in probability of \eqref{A2z1}, we use the Cauchy-Schwarz inequality to show its norm is bounded by 
\begin{align}
	 \big(\frac{1}{R}\sum_{s=T-R+1}^{T}\Vert\tilde{\bm f}_{s,R}-\bm H_R\bm f_s\Vert^2\big)^{1/2}\big(\frac{1}{R}\sum_{s=T-R+1}^{T}\Vert \frac{1}{NR}\sum_{t=T-R+1}^{T-h}\big(\bm f_s^\top\bm \Lambda^\top\big(\frac{s}{T}\big)\bm e_{t}\big)\bm z_t\Vert^2\big)^{1/2},
\end{align}
where $\frac{1}{R}\sum_{t=T-R+1}^{T}\Vert\tilde{\bm f}_{t,R}-\bm H_R\bm f_t\Vert^2=O_P(\delta_{NRT}^{-2})$ from  \Cref{Lemma2} (2) and 
$$\frac{1}{R}\sum_{s=T-R+1}^{T}\big\Vert\frac{1}{NR}\sum_{t=T-R+1}^{T-h}\big(\bm f_s^\top\bm \Lambda^\top\big(\frac{s}{T}\big)\bm e_{t}\big)\bm z_t\big\Vert^2=O_P(\frac{1}{N}),$$ 
as shown next. 
Indeed, we have $\frac{1}{R}\sum_{s=T-R+1}^{T}\big\Vert\big(\frac{1}{R}\sum_{t=T-R+1}^{T-h}\frac{1}{N}\sum_{i=1}^{N}\bm f_s^\top\bm \lambda_i\big(\frac{s}{T}\big)\bm e_{it}\big)\bm z_t\big\Vert^2$ is lower than
$\big(\frac{1}{R^2}\sum_{s=T-R+1}^{T}\sum_{t=T-R+1}^{T-h}\big\Vert\frac{1}{N}\sum_{i=1}^{N}\bm f_s^\top\bm \lambda_i\big(\frac{s}{T}\big)\bm e_{it}\big\Vert^2\big)\big(
\frac{1}{R}\sum_{t=T-R+1}^{T-h}\big\Vert \bm z_t\big\Vert^2\big)=O_P\big(\frac{1}{N}\big)$,
given that \Cref{Ass2} (c) implies $\frac{1}{R^2}\sum_{s=T-R+1}^{T}\sum_{t=T-R+1}^{T-h}\mathrm E\big(\big\Vert\frac{1}{\sqrt N}\sum_{i=1}^{N}\bm f_s^\top\bm \lambda_i\big(\frac{s}{T}\big)\bm e_{it}\big\Vert^2\big)=O(1)$ and \Cref{Ass4} (c) leads to $\frac{1}{R}\sum_{t=T-R+1}^{T-h}\big\Vert \bm z_t\big\Vert^2=O_P(1)$. Consequently, 
$$\frac{1}{NR^2}\sum_{t=T-R+1}^{T-h}\sum_{s=T-R+1}^{T}(\tilde{\bm f}_{s,R}-\bm H_R\bm f_s)\big(\bm f_s^\top\bm \Lambda^\top\big(\frac{s}{T}\big)\bm e_{t}\big)\bm z^\top_t=O_P\big(\delta_{NRT}^{-1}\frac{1}{\sqrt N}\big).$$

To find the order in probability of \eqref{A2z2},  we rely on the 
bound
\begin{align*}
    &\bigg\Vert\bm H_{R}\frac{1}{NR^2}\sum_{s=T-R+1}^{T}\sum_{t=T-R+1}^{T-h}(\bm f_s\bm f_s^\top)\bm \Lambda^\top\big(\frac{s}{T}\big)\bm e_{t}\bm{z}_t^\top\bigg\Vert
    \\&
    \leq \Vert\bm H_R \Vert
    \bigg( \frac{1}{R}\sum_{s=1}^T\Vert \bm f_s \Vert^2 \bigg)^{\frac{1}{2}}
    \frac{1}{\sqrt{NR}}\bigg(  \frac{1}{R}\sum_{s=1}^T\big\Vert \frac{1}{\sqrt{NR}}\sum_{t=T-R+1}^{T-h}\bm \Lambda^{\top}\big(\frac{s}{T}\big)\bm e_t\bm z_t^\top \big\Vert^2
    \bigg)^{\frac{1}{2}}\\
    &=O_P(1)O_P\big(\frac{1}{\sqrt{NR}}\big)=O_P\big(\frac{1}{\sqrt{NR}}\big), 
\end{align*}
using \Cref{Ass1} (a)
and \Cref{Ass2} (d). Consequently,
 $$\frac{1}{R}\sum_{t=T-R+1}^{T-h}\bm A_{2t}\bm{z}_t^\top=O_P\big(\delta_{NRT}^{-1}\frac{1}{\sqrt N}\big)+\big(\frac{1}{\sqrt{NR}}\big).$$
Using similar arguments, we also have
 $$\frac{1}{R}\sum_{t=T-R+1}^{T-h}\bm A_{3t}\bm{z}_t^\top=O_P\big(\delta_{NRT}^{-1}\frac{1}{\sqrt N}\big)+\big(\frac{1}{\sqrt{NR}}\big).$$
 
The proof that $\frac{1}{R}\sum_{t=T-R+1}^{T-h}\bm A_{4t}\bm{z}_t^\top=O_P\big(\delta_{NRT}^{-1}\frac{1}{\sqrt R}\big)$ and $\frac{1}{R}\sum_{t=T-R+1}^{T-h}\bm A_{5t}\bm{z}_t^\top=O_P\big(\delta_{NRT}^{-1}\frac{1}{\sqrt N}\big)$ follow the same strategy as in the proof of \citet[Lemma A1 (ii)]{BN2006}. Combining the results for $\frac{1}{R}\sum_{t=T-R+1}^{T-h}\bm A_{jt}\bm{z}_t^\top, j=1,\ldots,5,$ we conclude that 
$\frac{1}{R}\sum_{t=T-R+1}^{T-h}(\tilde{\bm f}_{t,R}-\bm H_R\bm f_t)\bm z^\top_t=O_P\big(\frac{R}{T}+\delta_{NRT}^{-1}\frac{1}{\sqrt N}+\frac{1}{\sqrt{NR}}+\delta_{NRT}^{-1}\frac{1}{\sqrt R} \big)
=O_P\big(\phi_{NRT}^{-2}\big)$.

\subsubsection*{Proof of \Cref{Lemma2} (6)} 
 Note that  
 $$\frac{1}{R}\sum_{t=T-R+1}^{T}(\tilde{\bm f}_{t,R}-\bm H_R\bm f_t)\hat{\bm z}^\top_t=\frac{1}{R}\sum_{t=T-R+1}^{T}(\tilde{\bm f}_{t,R}-\bm H_R\bm f_t)(\hat{\bm z}_t-\bm z_{Rt})^\top+\frac{1}{R}\sum_{t=T-R+1}^{T}(\tilde{\bm f}_{t,R}-\bm H_R\bm f_t)\bm z_t^\top\bm \Phi_R^\top,$$
where $\bm z_{Rt}=\bm \Phi_R\bm z_t$. Furthermore,
$\big\Vert\frac{1}{R}\sum_{t=T-R+1}^{T}(\tilde{\bm f}_{t,R}-\bm H_R\bm f_t)(\hat{\bm z}_t-\bm z_{Rt})^\top\big\Vert \leq \frac{1}{R}\sum_{t=T-R+1}^{T}\big\Vert\tilde{\bm f}_{t,R}-\bm H_R\bm f_t\big\Vert^2=O_P(\delta_{NRT}^{-2}),$ from \Cref{Lemma2} (2) and 
$\frac{1}{R}\sum_{t=T-R+1}^{T}(\tilde{\bm f}_{t,R}-\bm H_R\bm f_t)\bm z_t^\top\bm \Phi_R^\top=O_P(\phi_{NRT}^{-2}) ,$ from \Cref{Lemma2} (5).
Therefore, $\frac{1}{R}\sum_{t=T-R+1}^{T}(\tilde{\bm f}_{t,R}-\bm H_R\bm f_t)\hat{\bm z}^\top_t=O_P(\phi_{NRT}^{-2})$.

\subsubsection*{Proof of \Cref{Lemma2} (7)} 
This proof is similar to the proof of \Cref{Lemma2} (5), where $\bm z_t$ is replaced with $\varepsilon_{t+h}$. Therefore, it is omitted.

\subsubsection*{Proof of \Cref{Lemma2} (8)} 
To find the order in probability of 
 $\frac{1}{R}\sum_{t=T-R+1}^{T-h}( \tilde{\bm f}_{t,R}-\bm H_{R}\bm f_t)\big(\frac{t-T}{T}\big)\bm z_{t}^\top=O_P(\phi_{NRT}^{-2}\frac{R}{T}),$
we rely on the following expansion
\begin{align*}
&\tilde{\bm V}_{R}^{-1}\big(  \frac{1}{R}\sum_{t=T-R+1}^{T-h}\bm A_{1t}\big(\frac{t-T}{T}\big)\bm{z}_t^\top+ \frac{1}{R}\sum_{t=T-R+1}^{T-h}\bm A_{2t}\big(\frac{t-T}{T}\big)\bm{z}_t^\top \\
&
+ \frac{1}{R}\sum_{t=T-R+1}^{T-h}\bm A_{3t}\big(\frac{t-T}{T}\big)\bm{z}_t^\top+ \frac{1}{R}\sum_{t=T-R+1}^{T-h}\big(\frac{t-T}{T}\big)\bm A_{4t}\bm{z}_t^\top+ \frac{1}{R}\sum_{t=T-R+1}^{T-h}\bm A_{5t}\big(\frac{t-T}{T}\big)\bm{z}_t^\top\big),
\end{align*}
where we need to obtain the order in probability of $\frac{1}{R}\sum_{t=T-R+1}^{T-h}\bm A_{jt}\big(\frac{t-T}{T}\big)\bm{z}_t^\top, j=1,2,3,4,5.$ Following the same steps as for the proofs of $\frac{1}{R}\sum_{t=T-R+1}^{T-h}\bm A_{1t}\bm{z}_t^\top$, we get 
$$\frac{1}{R}\sum_{t=T-R+1}^{T-h}\bm A_{1t}\big(\frac{t-T}{T}\big)\bm{z}_t^\top=O_P\big(\frac{R^2}{T^2}\big).$$

 For $\frac{1}{R}\sum_{t=T-R+1}^{T-h}\bm A_{2t}\big(\frac{t-T}{T}\big)\bm{z}_t^\top=	\frac{1}{R^2}\sum_{t=T-R+1}^{T-h}\sum_{s=T-R+1}^{T} \tilde{\bm f}_s\eta_{st}\bm{z}_t^\top$, we note it equals
 \begin{align}
 	&\frac{1}{NR^2}\sum_{s=T-R+1}^{T}\sum_{t=T-R+1}^{T-h} (\tilde{\bm f}_s-\bm H_{R}\bm f_s)\big(\bm f_s^\top\bm \Lambda^\top\big(\frac{s}{T}\big)\bm e_{t}\big)\big(\frac{t-T}{T}\big)\bm{z}_t^\top\label{A2z11}\\
 	&+\bm H_{R}\frac{1}{NR^2}\sum_{s=T-R+1}^{T}\sum_{t=T-R+1}^{T-h}(\bm f_s\bm f_s^\top)\bm \Lambda^\top\big(\frac{s}{T}\big)\bm e_{t}\big(\frac{t-T}{T}\big)\bm{z}_t^\top\label{A2z21},
 \end{align}
 where we show that the first term \eqref{A2z11} is $O_P\big(\delta_{NRT}^{-1}\frac{1}{\sqrt N}\frac{R}{T}\big)$ and the second term \eqref{A2z21} is $\big(\frac{1}{\sqrt{NR}}\frac{R}{T}\big)$. Thus, $\frac{1}{R}\sum_{t=T-R+1}^{T-h}\bm A_{2t}\bm{z}_t^\top\frac{t-T}{T}=O_P\big(\delta_{NRT}^{-1}\frac{1}{\sqrt N}\frac{R}{T}\big)+O_P\big(\frac{1}{\sqrt{NR}}\frac{R}{T}\big)$.

For the order in probability of \eqref{A2z11}, we use the Cauchy-Schwarz inequality to show its norm is bounded by 
 \begin{align}
 	\big(\frac{1}{R}\sum_{s=T-R+1}^{T}\Vert\tilde{\bm f}_{s,R}-\bm H_R\bm f_s\Vert^2\big)^{1/2}\big(\frac{1}{R}\sum_{s=T-R+1}^{T}\big\Vert \frac{1}{NR}\sum_{t=T-R+1}^{T-h}\big(\bm f_s^\top\bm \Lambda^\top\big(\frac{s}{T}\big)\bm e_{t}\big)\frac{t-T}{T}\bm z_t\big\Vert^2\big)^{1/2},
 \end{align}
 where $\frac{1}{R}\sum_{s=T-R+1}^{T}\Vert\tilde{\bm f}_{s,R}-\bm H_R\bm f_s\Vert^2=O_P(\delta_{NRT}^{-2})$ from  \Cref{Lemma2} (2) and 
 $$\frac{1}{R}\sum_{s=T-R+1}^{T}\big\Vert\frac{1}{NR}\sum_{t=T-R+1}^{T-h}\big(\bm f_s^\top\bm \Lambda^\top\big(\frac{s}{T}\big)\bm e_{t}\big)\frac{t-T}{T}\bm z_t\big\Vert^2=O_P(\frac{1}{N}\frac{R^2}{T^2}),$$ 
because 
it is lower than
 $\big(\frac{1}{R^2}\sum_{s=T-R+1}^{T}\sum_{t=T-R+1}^{T-h}\big\Vert\frac{1}{N}\sum_{i=1}^{N}\bm f_s^\top\bm \lambda_i\big(\frac{s}{T}\big)\bm e_{it}\big\Vert^2\big)
 \frac{1}{R}\sum_{t=T-R+1}^{T-h}\big\Vert \bm z_t\frac{t-T}{T}\big\Vert^2=O_P\big(\frac{1}{N}\frac{R^2}{T^2}\big)$,
 given that \Cref{Ass2} (c) 
 and \Cref{Ass4} (c). 
 Consequently, 
 $$\frac{1}{NR^2}\sum_{t=T-R+1}^{T}\sum_{s=T-R+1}^{T}(\tilde{\bm f}_{s,R}-\bm H_R\bm f_s)\big(\bm f_s^\top\bm \Lambda^\top\big(\frac{s}{T}\big)\bm e_{t}\big)\bm z^\top_t\frac{t-T}{T}=O_P\big(\delta_{NRT}^{-1}\frac{1}{\sqrt N}\frac{R}{T}\big).$$

For the order in probability of \eqref{A2z21},  we rely on the bound
\begin{align*}
    &\bigg\Vert\bm H_{R}\frac{1}{NR^2}\sum_{s=T-R+1}^{T}\sum_{t=T-R+1}^{T-h}(\bm f_s\bm f_s^\top)\bm \Lambda^\top\big(\frac{s}{T}\big)\bm e_{t}\frac{t-T}{T}\bm{z}_t^\top\bigg\Vert
    \\&
    \leq \Vert\bm H_R \Vert
    \bigg( \frac{1}{R}\sum_{s=1}^T\Vert \bm f_s \Vert^2 \bigg)^{\frac{1}{2}}
    \bigg(  \frac{1}{NR}\sum_{s=1}^T\Vert \bm \Lambda^{\top}\big(\frac{s}{T}\big)\bm e_t\frac{t-T}{T}\bm z_t^\top \Vert^2
    \bigg)^{\frac{1}{2}}\\
    &=O_P(1)O_P\big(\frac{1}{\sqrt{NR}}\frac{R}{T}\big)=O_P\big(\frac{1}{\sqrt{NR}}\frac{R}{T}\big), 
\end{align*}
 using \Cref{Ass1} (a) and \Cref{Ass2} (d). Consequently,
 $$\frac{1}{R}\sum_{t=T-R+1}^{T-h}\bm A_{2t}\frac{t-T}{T}\bm{z}_t^\top=O_P\big(\delta_{NRT}^{-1}\frac{1}{\sqrt N}\frac{R}{T}\big)+\big(\frac{1}{\sqrt{NR}}\frac{R}{T}\big).$$
 Using similar arguments, we also have
 $$\frac{1}{R}\sum_{t=T-R+1}^{T-h}\bm A_{3t}\frac{t-T}{T}\bm{z}_t^\top=O_P\big(\delta_{NRT}^{-1}\frac{1}{\sqrt N}\frac{R}{T}\big)+\big(\frac{1}{\sqrt{NR}}\frac{R}{T}\big).$$
 
 The proof that $\frac{1}{R}\sum_{t=T-R+1}^{T-h}\bm A_{4t}\frac{t-T}{T}\bm{z}_t^\top=O_P\big(\delta_{NRT}^{-1}\frac{1}{\sqrt R}\frac{R}{T}\big)$ and $\frac{1}{R}\sum_{t=T-R+1}^{T-h}\bm A_{5t}\frac{t-T}{T}\bm{z}_t^\top=O_P\big(\delta_{NRT}^{-1}\frac{1}{\sqrt N}\frac{R}{T}\big)$ is similar to the proof of \citet[Lemma A1 (ii)]{BN2006} with minor modifications because of the $\frac{t-T}{T}$ factor. Combining the results for $\frac{1}{R}\sum_{t=T-R+1}^{T-h}\bm A_{jt}\frac{t-T}{T}\bm{z}_t^\top, j=1,\ldots,5,$ we conclude that 
 $\frac{1}{R}\sum_{t=T-R+1}^{T-h}(\tilde{\bm f}_{t,R}-\bm H_R\bm f_t)\bm z^\top_t\frac{t-T}{T}=O_P\big(\phi_{NRT}^{-2}\frac{R}{T}\big)$.

\subsubsection*{Proof of \Cref{Lemma2} (9)} 
  We observe that  $\frac{1}{R}\sum_{t=T-R+1}^{T-h}(\tilde{\bm f}_{t,R}-\bm H_R\bm f_t)\big(\frac{t-T}{T}\big)\hat{\bm z}^\top_t$ equals
 $$\frac{1}{R}\sum_{t=T-R+1}^{T-h}(\tilde{\bm f}_{t,R}-\bm H_R\bm f_t)(\hat{\bm z}_t-\bm z_{Rt})^\top\big(\frac{t-T}{T}\big)+\frac{1}{R}\sum_{t=T-R+1}^{T-h}(\tilde{\bm f}_{t,R}-\bm H_R\bm f_t)\bm z_t^\top\bm \Phi_R^\top\big(\frac{t-T}{T}\big),$$
where $\bm z_{Rt}=\bm \Phi_R\bm z_t$. Furthermore,
$\big\Vert\frac{1}{R}\sum_{t=T-R+1}^{T-h}(\tilde{\bm f}_{t,R}-\bm H_R\bm f_t)(\hat{\bm z}_t-\bm z_{Rt})^\top \big(\frac{t-T}{T}\big)\big\Vert $ is bounded by $ \frac{1}{R}\sum_{t=T-R+1}^{T-h}\big\Vert\tilde{\bm f}_{t,R}-\bm H_R\bm f_t\big\Vert^2\big(\frac{T-t}{T}\big)=O_P(\delta_{NRT}^{-2}\frac{R}{T}),$ from \Cref{Lemma2} (3) and 
$\frac{1}{R}\sum_{t=T-R+1}^{T-h}(\tilde{\bm f}_{t,R}-\bm H_R\bm f_t)\bm z_t^\top\big(\frac{t-T}{T}\big)\bm \Phi_R^\top=O_P(\phi_{NRT}^{-2}\frac{R}{T}) ,$ from \Cref{Lemma2} (8).
Therefore, $\frac{1}{R}\sum_{t=T-R+1}^{T-h}(\tilde{\bm f}_{t,R}-\bm H_R\bm f_t)\big(\frac{t-T}{T}\big)\hat{\bm z}^\top_t=O_P(\phi_{NRT}^{-2}\frac{R}{T})$.

\subsubsection*{Proof of \Cref{Lemma2} (10)} 
 The proof of 
 $\frac{1}{R}\sum_{t=T-R+1}^{T-h}( \tilde{\bm f}_{t,R}-\bm H_{R}\bm f_t)\big(\frac{t-T}{T}\big)\varepsilon_{t+h}^\top=O_P(\phi_{NRT}^{-2}\frac{R}{T})$  
is similar to the proof of \Cref{Lemma2} (8), where $\bm z_t$ is replaced with $\varepsilon_{t+h}$.

\subsubsection*{Proof of \Cref{Lemma2} (11)-(16)} 

The proofs follow the same steps as for the proofs of \Cref{Lemma2} (8)-(10). Therefore, they are omitted.

\subsubsection*{Proof of \Cref{Lemma2} (17)} 
 First, we write $\frac{1}{R}\sum_{t=T-R+1}^{T-h}\hat{\bm z}_{t,R}\hat{\bm z}_{t,R}^\top-\bm \Phi_R^{\top}\big(\frac{1}{R}\sum_{t=T-R+1}^{T-h}\bm z_{t}\bm z_{t}^\top\big)\bm \Phi_R$ as 
 \begin{align*}
 &\frac{1}{R}\sum_{t=T-R+1}^{T-h}(\hat{\bm z}_{t,R}-\bm\Phi_R\bm z_t)\bm z_{t}^\top\bm\Phi_R^\top+\bm\Phi_R \frac{1}{R}\sum_{t=T-R+1}^{T-h}\bm z_{t}(\hat{\bm z}_{t,R}-\bm\Phi_R\bm z_t)^\top\\
 &+ \frac{1}{R}\sum_{t=T-R+1}^{T-h}(\hat{\bm z}_{t,R}-\bm\Phi_R\bm z_t)(\hat{\bm z}_{t,R}-\bm\Phi_R\bm z_t)^\top,
\end{align*}
where the first two terms are $O_P(\phi_{NRT}^{-2})$ using \Cref{Lemma2} (5) and the third term is $O_P(\delta_{NRT}^{-2})$ using \Cref{Lemma2} (2). Consequently, $\frac{1}{R}\sum_{t=T-R+1}^{T-h}\hat{\bm z}_{t,R}\hat{\bm z}_{t,R}^\top-\bm \Phi_R^{\top}\big(\frac{1}{R}\sum_{t=T-R+1}^{T-h}\bm z_{t}\bm z_{t}^\top\big)\bm \Phi_R$ is $O_P(\phi_{NRT}^{-2})$.
 
 \subsubsection*{Proof of \Cref{Lemma2} (18)} 

 We recall from \Cref{Ass4} (c) that
 $\frac{1}{R}\sum_{t=T-R+1}^{T-h}%
\mathrm{E}(\bm{z}_{t}\bm{z}_{t}^{\top})\overset{P}{\longrightarrow}%
\bm{\Sigma }_{\bm{Z}}$. 
Therefore, we have $\bm \Phi_R^{\top}\big(\frac{1}{R}\sum_{t=T-R+1}^{T-h}\bm z_{t}\bm z_{t}^\top\big)\bm \Phi_R -\bm \Phi^{\top}\bm \Sigma_{\bm Z} \bm \Phi=o_P(1),$
as
$\frac{1}{R}\sum_{t=T-R+1}^{T-h}\bm{z}_{t}\bm{z}_{t}^{\top}-\frac{1}{R}\sum_{t=T-R+1}^{T-h}%
\mathrm{E}(\bm{z}_{t}\bm{z}_{t}^{\top})\overset{P}{\longrightarrow} \bm 0$, from \Cref{Lemma1} (3), where $\bm \Phi$ is the limit in probability of $\bm \Phi_R$.

 \subsubsection*{Proof of \Cref{Lemma2} (19)} 
To show $\frac{1}{R}\sum_{t=T-R+1}^{T-h}(\hat{\bm z}_{t,R}-\bm \Phi_R\bm z_t)\bm z_{t}^\top(\bm\delta(t/T)-\bm\delta(1))=O_P(\phi_{NRT}^{-2}\frac{R}{T})$, we observe that $\frac{1}{R}\sum_{t=T-R+1}^{T-h}(\hat{\bm z}_{t,R}-\bm \Phi_R\bm z_t)\bm z_{t}^\top(\bm\delta(t/T)-\bm\delta(1))$ equals
\begin{align*}
\frac{1}{R}\sum_{t=T-R+1}^{T-h}(\hat{\bm z}_{t,R}-\bm \Phi_R\bm z_t)\bm z_{t}^\top\frac{t-T}{T}\bm\delta^{\prime}(1)
+ \frac{1}{R}\sum_{t=T-R+1}^{T-h}(\hat{\bm z}_{t,R}-\bm \Phi_R\bm z_t)\bm z_{t}^\top\frac{(t-T)^2}{T^2}\frac{1}{2}\bm\delta^{\prime\prime}(d)
\end{align*}
which is $O_P(\phi_{NRT}^{-2}\frac{R}{T})+O_P(\phi_{NRT}^{-2}\frac{R^2}{T^2})=O_P(\phi_{NRT}^{-2}\frac{R}{T})$ from \Cref{Lemma2} (8) and (11). The steps for 
$\frac{1}{R}\sum_{t=T-R+1}^{T-h}\bm z_{t}(\hat{\bm z}_{t,R}-\bm \Phi_R\bm z_t)^\top(\bm\delta(t/T)-\bm\delta(1))=O_P(\phi_{NRT}^{-2}\frac{R}{T})$ are identical.

 \subsubsection*{Proof of \Cref{Lemma2} (20)} 
For $\frac{1}{R}\sum_{t=T-R+1}^{T-h}(\hat{\bm z}_{t,R}-\bm \Phi_R\bm z_t)\hat{\bm z}_{t,R}^\top(\bm\delta_R(t/T)-\bm\delta_R(1))=O_P(\phi_{NRT}^{-2}\frac{R}{T})$, we note that it is
\begin{align*}
	&\frac{1}{R}\sum_{t=T-R+1}^{T-h}(\hat{\bm z}_{t,R}-\bm \Phi_R\bm z_t)\hat{\bm z}_{t,R}^\top\frac{t-T}{T}(\bm\Phi_R^{\top})^{-1}\bm\delta^{\prime}(1)
	+ \frac{1}{R}\sum_{t=T-R+1}^{T-h}(\hat{\bm z}_{t,R}-\bm \Phi_R\bm z_t)\hat{\bm z}_{t,R}^\top\frac{(t-T)^2}{T^2}\\
    &\times(\bm\Phi_R^{\top})^{-1}\frac{1}{2}\bm\delta^{\prime\prime}(d),
\end{align*}
which equals $O_P(\phi_{NRT}^{-2}\frac{R}{T})+O_P(\phi_{NRT}^{-2}\frac{R^2}{T^2})=O_P(\phi_{NRT}^{-2}\frac{R}{T})$, from \Cref{Lemma2} (9) and (12). The proof that
$\frac{1}{R}\sum_{t=T-R+1}^{T-h}\hat{\bm z}_{t,R}(\hat{\bm z}_{t,R}-\bm \Phi_R\bm z_t)^\top(\bm\delta_R(t/T)-\bm\delta_R(1))=O_P(\phi_{NRT}^{-2}\frac{R}{T})$ follows the same steps.

\subsection*{Proof of \Cref{Lemma3}}
\subsubsection*{Proof of \Cref{Lemma3} (1)} 
The proof follows from \Cref{Ass1} (b) and \Cref{Ass5} (b).

\subsubsection*{Proof of \Cref{Lemma3} (2)} 

Let $\sum$ denote $\sum_{t=T-R_0+1}^{T}$.
To determine the order in probability of $\bm W_{4}=\frac{1}{R_0^2}\sum\sum(\tilde{\bm f}^0_{t}-\bm H^0\bm f_t)(\tilde{\bm f}^0_{s}-\bm H^0\bm f_s)^\top \frac{1}{N}\sum_{i=1}^Ne_{it}e_{is}$, we note that it equals $\bm W_{4,1}+\bm W_{4,2}$,
where
$	\bm W_{4,1}=\frac{1}{R_0^2}\sum\sum(\tilde{\bm f}^0_{t}-\bm H^0\bm f_t)(\tilde{\bm f}^0_{s}-\bm H^0\bm f_s)^\top \frac{1}{N}\sum_{i=1}^N\mathrm{E}(e_{it}e_{is})$ 
	and
$	\bm W_{4,2}=\frac{1}{R_0^2}\sum\sum(\tilde{\bm f}^0_{t}-\bm H^0\bm f_t)(\tilde{\bm f}^0_{s}-\bm H^0\bm f_s)^\top \frac{1}{N}\sum_{i=1}^N(e_{it}e_{is}-\mathrm{E}(e_{it}e_{is})).$
The triangle inequality implies 
$$\Vert\bm W_{4,1}\Vert\leq\frac{1}{R_0}\sum\Vert\tilde{\bm f}^0_{t}-\bm H^0\bm f_t\Vert^2\big( \frac{1}{R_0^2}\sum\sum\vert\frac{1}{N}\sum_{i=1}^N\mathrm{E}(e_{it}e_{is})\vert^2\big)^{1/2}=O_P(\delta_{NR_0T}^{-2}).$$ using 
$\frac{1}{R_0}\sum\Vert\tilde{\bm f}^0_{t}-\bm H^0\bm f_t\Vert^2=O_P(\delta_{NRT}^{-2})$ and $\vert\frac{1}{N}\sum_{i=1}^N\mathrm{E}(e_{it}e_{is})\vert\leq C$ from \Cref{Ass1} (e).
In addition, 
$\Vert\bm W_{4,2}\Vert\leq\frac{1}{R_0}\sum\Vert\tilde{\bm f}^0_{t}-\bm H^0\bm f_t\Vert^2\big( \frac{1}{R_0^2}\sum\sum\vert\frac{1}{N}\sum_{i=1}^N(e_{it}e_{is}-\mathrm{E}(e_{it}e_{is}))\vert^2\big)^{1/2}=O_P(\delta_{NR_0T}^{-2}\frac{1}{\sqrt N})$,
because
$$\frac{1}{R_0^2}\sum\sum\mathrm{E}\big(\vert\frac{1}{N}\sum_{i=1}^N(e_{it}e_{is}-\mathrm{E}(e_{it}e_{is}))\vert^2\big)\leq \frac{1}{N} \frac{1}{R_0^2}\sum\sum\big(\mathrm{E}\big(\vert\frac{1}{\sqrt N}\sum_{i=1}^N(e_{it}e_{is}-\mathrm{E}(e_{it}e_{is}))\vert^4\big)\big)^{1/2}$$
is $O_P\big(\frac{1}{N}\big)$ using Jensen inequality and \Cref{Ass1} (f).
Consequently,
$\bm W_{4}=O_P(\delta_{NR_0T}^{-2})$.

For $\bm W_5$, we use the decomposition $\bm W_{5,1}+\bm W_{5,2}$, where the first term 
$\bm W_{5,1}=\frac{1}{R_0^2}\sum\sum(\frac{t}{T}-1)(\tilde{\bm f}^0_{t}-\bm H^0\bm f_t)(\tilde{\bm f}^0_{s}-\bm H^0\bm f_s)^\top \frac{1}{N}\sum_{i=1}^N\mathrm{E}(e_{it}e_{is})
$
and the second term
$\bm W_{5,2}=
\frac{1}{R_0^2}\sum\sum(\frac{t}{T}-1)(\tilde{\bm f}^0_{t}-\bm H^0\bm f_t)(\tilde{\bm f}^0_{s}-\bm H^0\bm f_s)^\top \frac{1}{N}\sum_{i=1}^N(e_{it}e_{is}-\mathrm{E}(e_{it}e_{is}))$. The norm of the first term is bounded by
$(\frac{1}{R_0}\sum (\frac{t}{T}-1)^2\Vert \tilde{\bm f}^0_{t}-\bm H^0\bm f_t\Vert^2)^{1/2}(\frac{1}{R_0}\sum\Vert\tilde{\bm f}^0_{s}-\bm H^0\bm f_s\Vert^2)^{1/2}\big( \frac{1}{R_0^2}\sum\sum\vert\frac{1}{N}\sum_{i=1}^N\mathrm{E}(e_{it}e_{is})\vert^2\big)^{1/2}$, which is $O_P(\frac{R_0}{T}\delta_{NR_0T}^{-2})$, because 
$\frac{1}{R_0}\sum (\frac{t}{T}-1)^2\Vert \tilde{\bm f}^0_{t}-\bm H^0\bm f_t\Vert^2=O_P(\frac{R_0}{T}\delta_{NR_0T}^{-2})$, $\frac{1}{R_0}\sum\Vert\tilde{\bm f}^0_{s}-\bm H^0\bm f_s\Vert^2=O_P(\delta_{NR_0T}^{-2})$,
and
$\frac{1}{R_0^2}\sum\sum\vert\frac{1}{N}\sum_{i=1}^N\mathrm{E}(e_{it}e_{is})\vert^2=O(1)$. Similarly, the second term is $O_P(\frac{R}{T}\frac{1}{\sqrt N}\delta_{NR_0T}^{-2})$. Therefore, $\bm W_{5}=O_P(\frac{R_0}{T}\delta_{NR_0T}^{-2})$. Using similar steps, we also have 
$\bm W_{6}=O_P(\frac{R_0^2}{T^2}\delta_{NR_0T}^{-2})$.

\subsubsection*{Proof of \Cref{Lemma3} (3)} 

Note that 
$\Vert\bm W_{7}\Vert\leq\frac{1}{R_0}\Vert\bm H^0\Vert^2\big(\frac{1}{N}\sum_{i=1}^N \Vert\frac{1}{\sqrt{R_0}}\sum\bm f_{t}e_{it}\Vert^2\big)^{\frac{1}{2}}\big(\frac{1}{N}\sum_{i=1}^N \Vert\frac{1}{\sqrt{R_0}}\sum\bm f_{s}e_{is}\Vert^2\big)^{\frac{1}{2}}=O_P\big(\frac{1}{R_0}\big)$ 
because $\frac{1}{N}\sum_{i=1}^N \Vert\frac{1}{\sqrt{R_0}}\sum\bm f_{t}e_{it}\Vert^2=O_P(1)$ by \Cref{Ass2} (a). 
For $\bm W_8$, we have
$$\Vert\bm W_{8}\Vert\leq \frac{1}{R_0}\frac{R_0}{T}\Vert\bm H^0\Vert^2 
\big(\frac{1}{N}\sum_{i=1}^N\Vert\frac{1}{\sqrt{R_0}}\frac{T}{R_0}\sum \bm f_{t}(\frac{t}{T}-1)e_{it}\Vert^2\big)^{\frac{1}{2}}
\big(\frac{1}{N}\sum_{i=1}^N\Vert\frac{1}{\sqrt{R_0}}\frac{T}{R_0}\sum \bm f_{s}e_{is}\Vert^2\big)^{\frac{1}{2}},
$$
which is $O_P\big(\frac{R_0}{T}\frac{1}{R_0} \big)=O_P\big(\frac{1}{T} \big)$, because $\frac{1}{N}\sum_{i=1}^N \Vert\frac{T^k}{R_0^k}\frac{1}{\sqrt{R_0}}\sum\frac{(T-t)^k}{R_0^k}\bm f_{t}e_{it}\Vert^2=O_P(1), k=0,1,$
by \Cref{Ass2} (a). 
Using similar arguments, we have $\bm W_{9}=O_P(\frac{R_0^2}{T^2}\frac{1}{R_0})=O_P\big(\frac{R_0}{T^2} \big)$.

\subsubsection*{Proof of \Cref{Lemma3} (4)} 

We start by noting that $\Vert \bm W_{12}\Vert $ is bounded by
$$ \Vert\frac{1}{R_0}\sum \tilde{\bm f}^0_{t}(\tilde{\bm f}^0_{t}-\bm H^0 \bm f_t)^\top \Vert \Vert(\bm H^{0})^{-1}\Vert
( \frac{1}{R_0}\sum
\Vert\frac{1}{N}\sum_{i=1}^N\bm\lambda_i(1) e_{is}\Vert^2)^{1/2}
(\frac{1}{R_0}\sum\Vert\tilde{\bm f}^0_{s}-\bm H^0\bm f_s \Vert^2)^{1/2},$$
which is $O_P(\phi_{NR_0T}^{-2}\delta_{NR_0T}^{-1}/\sqrt{N})$, because 
$\frac{1}{R_0}\sum \tilde{\bm f}^0_{t}(\tilde{\bm f}^0_{t}-\bm H^0 \bm f_t)^\top = O_P(\phi_{NR_0T}^{-2})$, 
$\frac{1}{R_0}\sum\Vert\tilde{\bm f}^0_{s}-\bm H^0\bm f_s \Vert^2= O_P(\delta_{NR_0T}^{-2})$,
$\frac{1}{R_0}\sum
\Vert\frac{1}{\sqrt N}\sum_{i=1}^N\bm\lambda_i(1) e_{is}\Vert^2)^{1/2}=O_P(1)$  and $(\bm H^{0})^{-1}=O_P(1)$.

Furthermore, $\Vert \bm W_{13}\Vert $ is  $O_P(\phi_{NR_0T}^{-2}\delta_{NR_0T}^{-1}/\sqrt{N}\frac{R_0}{T})$. Indeed, it is bounded by $\Vert\frac{1}{R_0}\sum \tilde{\bm f}^0_{t}(\tilde{\bm f}^0_{t}-\bm H^0 \bm f_t)^\top \Vert $ times
$$ \Vert(\bm H^{0})^{-1}\Vert
( \frac{1}{R_0}\sum
\Vert\frac{1}{N}\sum_{i=1}^N\bm\lambda_i(1) e_{is}\Vert^2)^{1/2}
(\frac{1}{R_0}\sum\Vert\tilde{\bm f}^0_{s}-\bm H^0\bm f_s \Vert^2(s/T-1)^2)^{1/2},$$
where 
$\frac{1}{R_0}\sum \tilde{\bm f}^0_{t}(\tilde{\bm f}^0_{t}-\bm H^0 \bm f_t)^\top = O_P(\phi_{NR_0T}^{-2})$, 
$\frac{1}{R_0}\sum\Vert\tilde{\bm f}^0_{s}-\bm H^0\bm f_s \Vert^2(s/T-1)^2= O_P(\frac{R_0^2}{T^2}\delta_{NR_0T}^{-2})$,
$\frac{1}{R_0}\sum
\Vert\frac{1}{\sqrt N}\sum_{i=1}^N\bm\lambda_i(1) e_{is}\Vert^2)^{1/2}=O_P(1)$  and $(\bm H^{0})^{-1}=O_P(1)$.
Similarly, we have that
 $\bm W_{14}=O_P(\phi_{NR_0T}^{-2}\delta_{NR_0T}^{-1}\frac{R_0}{T\sqrt{N}})$ and  $\bm W_{15}=O_P(\phi_{NR_0T}^{-2}\delta_{NR_0T}^{-1}\frac{R_0^2}{T^2\sqrt{N}})$.

\subsubsection*{Proof of \Cref{Lemma3} (5)-(11)}  
The proofs for  \Cref{Lemma3} (5)-(11) use the same arguments as in the proofs of  \Cref{Lemma3} (2)-(4). Therefore, they are omitted.

\subsubsection*{Proof of \Cref{Lemma3} (12)} 

Note that 
$\Vert\bm W_{45}\Vert\leq \Vert\bm H^0\Vert^2\frac{1}{\sqrt{R_0}}  
\big(\frac{1}{N}\sum_{i=1}^N \Vert \frac{1}{R_0}\sum (\tilde{\bm f}^0_{t}-\bm H^0 \bm f_t) e_{it}\Vert^2\big)^{\frac{1}{2}}
\big(\frac{1}{N}\sum_{i=1}^N \Vert \frac{1}{\sqrt{R_0}}\sum \bm f_se_{is}\Vert^2\big)^{\frac{1}{2}}$.
First, we note that
$\frac{1}{N}\sum_{i=1}^N \Vert \frac{1}{R_0}\sum (\tilde{\bm f}^0_{t}-\bm H^0 \bm f_t) e_{it}\Vert^2\leq  \big( \frac{1}{R_0}\sum \Vert\tilde{\bm f}^0_{t}-\bm H^0 \bm f_t \Vert^2 \big) \frac{1}{NR_0}\sum_{i=1}^N\sum e_{it}^2
=O_P(\delta_{NR_0T}^{-2})O_P(1)=O_P(\delta_{NR_0T}^{-2})$. In addition,
$\frac{1}{N}\sum_{i=1}^N \Vert \frac{1}{\sqrt{R_0}}\sum \bm f_se_{is}\Vert^2=O_P(1)$. Therefore, 
$\bm W_{45}=O_P\big(\frac{\delta_{NR_0T}^{-1}}{\sqrt{R_0}}\big)$.
Similarly, $\bm W_{46}= 	O_P\big(\frac{\delta_{NR_0T}^{-1}}{\sqrt{R_0}}\frac{R_0}{T}\big)$, 
$\bm W_{47}= 	O_P\big(\frac{\delta_{NR_0T}^{-1}}{\sqrt{R_0}}\frac{R_0}{T}\big)$,  and $\bm W_{48}= 	O_P\big(\frac{\delta_{NR_0T}^{-1}}{\sqrt{R_0}}\frac{R_0^2}{T^2}\big)$.

\subsubsection*{Proof of \Cref{Lemma3} (13)} 

Given the OLS estimation to obtain the alternative estimate of the latent factor loadings, we observe that 
\begin{equation}
\begin{bmatrix}
 \ddot{\bm\lambda}_i(1) \\
  \ddot{\bm\lambda}_i^{\prime}(1) 
\end{bmatrix}
=
{\underbrace{\begin{bmatrix}
\frac{1}{R_0}\sum\tilde{\bm f}_{t}^0\tilde{\bm f}_{t}^{0\top}  &  \frac{1}{R_0}\sum(\frac{t}{T}-1)\tilde{\bm f}^0_{t}\tilde{\bm f}_{t}^{0\top}
 \\
\frac{1}{R_0}\sum(\frac{t}{T}-1)\tilde{\bm f}^0_{t}\tilde{\bm f}_{t}^{0\top}  & \frac{1}{R_0}\sum(\frac{t}{T}-1)^2\tilde{\bm f}^0_{t}\tilde{\bm f}_{t}^{0\top}
\end{bmatrix}}_{\hat{\bm R}}
}^{-1}
\begin{bmatrix}
\frac{1}{R_0}\sum\tilde{\bm f}^0_{t} x_{it}     \\
\frac{1}{R_0}\sum(\frac{t}{T}-1)\tilde{\bm f}^0_{t} x_{it}
\end{bmatrix}
.
\label{vector2}
\end{equation}

In addition, we have from \Cref{panel_expansion} that
\begin{align*}
x_{it}&=\bm\lambda^{0\top}(1) \tilde{\bm f}^0_{t}+\bm\lambda^{0\prime\top}(1) \big(\frac{t}{T}-1\big) \tilde{\bm f}^0_{t} -\bm\lambda^{0\top}(1)(\tilde{\bm f}^0_{t}-\bm H^0 \bm f_t)\\
&-\bm\lambda^{0\prime\top}(1)(\tilde{\bm f}^0_{t}-\bm H^0 \bm f_t)(\frac{t}{T}-1)+\frac{1}{2}\bm\lambda^{0\prime\prime\top}(c_i)\big(\frac{t}{T}-1\big)^2\bm H^0\bm f_t+e_{it}
\end{align*}
and
\begin{align*}
\begin{bmatrix}
\frac{1}{R_0}\sum\tilde{\bm f}^0_{t}x_{it}     \\
\frac{1}{R_0}\sum(\frac{t}{T}-1)\tilde{\bm f}^0_{t}x_{it}  
\end{bmatrix}
&=\hat{\bm R}
\begin{bmatrix}
\bm\lambda_i^0(1)   \\
\bm\lambda_i^{0\prime}(1)
\end{bmatrix}
-\begin{bmatrix}
\frac{1}{R_0}\sum\tilde{\bm f}^0_{t}(\tilde{\bm f}^0_{t}-\bm H^0 \bm f_t)^\top     \\
\frac{1}{R_0}\sum(\frac{t}{T}-1)\tilde{\bm f}^0_{t}(\tilde{\bm f}^0_{t}-\bm H^0 \bm f_t)^\top 
\end{bmatrix}
(\bm H^{0 \top})^{-1}\bm\lambda_i(1)
\\
&-
\begin{bmatrix}
\frac{1}{R_0}\sum(\frac{t}{T}-1)\tilde{\bm f}^0_{t}(\tilde{\bm f}^0_{t}-\bm H^0 \bm f_t)^\top      \\
\frac{1}{R_0}\sum(\frac{t}{T}-1)^2\tilde{\bm f}^0_{t}(\tilde{\bm f}^0_{t}-\bm H^0 \bm f_t)^\top 
\end{bmatrix}
(\bm H^{0\top})^{-1}\bm\lambda_i^{\prime}(1)
\\
&+
\begin{bmatrix}
\frac{1}{R_0}\sum(\frac{t}{T}-1)^2\tilde{\bm f}^0_{t}\bm f_{t}^\top     \\
\frac{1}{R_0}\sum(\frac{t}{T}-1)^3\tilde{\bm f}^0_{t} \bm f_{t}^\top 
\end{bmatrix}
\frac{1}{2}\bm\lambda_i^{\prime\prime}(c_i)
+
\begin{bmatrix}
\frac{1}{R_0}\sum\tilde{\bm f}^0_{t}e_{it}     \\
\frac{1}{R_0}\sum(\frac{t}{T}-1)\tilde{\bm f}^0_{t}e_{it} 
\end{bmatrix}
.
\end{align*}
Replacing the latter in \Cref{vector2} and after some expansions, we obtain 
\begin{align}
\begin{bmatrix}
 \ddot{\bm\lambda}_i(1)-\bm\lambda_i^0(1) \\
  \ddot{\bm\lambda}_i^{\prime}(1) -\bm\lambda_i^{0\prime}(1)
\end{bmatrix}
=&
-\hat{\bm R}^{-1}\begin{bmatrix}
\frac{1}{R_0}\sum\tilde{\bm f}^0_{t}(\tilde{\bm f}^0_{t}-\bm H^0 \bm f_t)^\top     \\
\frac{1}{R_0}\sum(\frac{t}{T}-1)\tilde{\bm f}^0_{t}(\tilde{\bm f}^0_{t}-\bm H^0 \bm f_t)^\top 
\end{bmatrix}
(\bm H^{0\top})^{-1}\bm\lambda_i(1)\nonumber
\\
&-
\hat{\bm R}^{-1}\begin{bmatrix}
\frac{1}{R_0}\sum(\frac{t}{T}-1)\tilde{\bm f}^0_{t}(\tilde{\bm f}^0_{t}-\bm H^0 \bm f_t)^\top      \\
\frac{1}{R_0}\sum(\frac{t}{T}-1)^2\tilde{\bm f}^0_{t}(\tilde{\bm f}^0_{t}-\bm H^0 \bm f_t)^\top 
\end{bmatrix}
(\bm H^{0\top})^{-1}\bm\lambda_i^{\prime}(1)\nonumber
\\
&+
\hat{\bm R}^{-1}\begin{bmatrix}
\frac{1}{R_0}\sum(\frac{t}{T}-1)^2(\tilde{\bm f}^0_{t}-\bm H^0\bm f_t)\bm f_{t}^\top     \nonumber\\
\frac{1}{R_0}\sum(\frac{t}{T}-1)^3(\tilde{\bm f}^0_{t}-\bm H^0\bm f_t) \bm f_{t}^\top 
\end{bmatrix}
\frac{1}{2}\bm\lambda_i^{\prime\prime}(c_i)\\
&+
\hat{\bm R}^{-1}\begin{bmatrix}
\bm H^0\frac{1}{R_0}\sum(\frac{t}{T}-1)^2\bm f_{t}\bm f_{t}^\top     \\
\bm H^0\frac{1}{R_0}\sum(\frac{t}{T}-1)^3\bm f_{t} \bm f_{t}^\top 
\end{bmatrix}
\frac{1}{2}\bm\lambda_i^{\prime\prime}(c_i)\nonumber
\\
&
+
\hat{\bm R}^{-1}
\begin{bmatrix}
\frac{1}{R_0}\sum(\tilde{\bm f}^0_{t}-\bm H^0\bm f_t) e_{it}     \\
\frac{1}{R_0}\sum(\frac{t}{T}-1)(\tilde{\bm f}^0_{t}-\bm H^0\bm f_t) e_{it} 
\end{bmatrix}\nonumber
\\
&+
\hat{\bm R}^{-1}
\begin{bmatrix}
\bm H^0\frac{1}{R_0}\sum\bm f_{t}e_{it}     \\
\bm H^0\frac{1}{R_0}\sum(\frac{t}{T}-1)\bm f_{t}e_{it} 
\end{bmatrix},
\label{lambda_expansion}
\end{align}
where we decompose $\hat{\bm R}$ as $\hat{\bm R}_1+\hat{\bm R}_2+\hat{\bm R}_3+\hat{\bm R}_4$, where 
\begin{align*}
\hat{\bm R}_1 =&\begin{bmatrix}
\bm H^0\frac{1}{R_0}\sum\mathrm{E}(\bm f_{t}\bm f_{t}^{\top})\bm H^{0\top}  & \bm H^0 \frac{1}{R_0}\sum(\frac{t}{T}-1)\mathrm{E}(\bm f_{t}\bm f_{t}^{\top})\bm H^{0\top}
 \\
\bm H^0\frac{1}{R_0}\sum(\frac{t}{T}-1)\mathrm{E}(\bm f_{t}\bm f_{t}^{\top})\bm H^{0\top}& \bm H^0\frac{1}{R_0}\sum(\frac{t}{T}-1)^2\mathrm{E}(\bm f_{t}\bm f_{t}^{\top})\bm H^{0\top}
\end{bmatrix}\\
\hat{\bm R}_2=&
\begin{bmatrix}
\bm H^0\frac{1}{R_0}\sum(\bm f_{t}\bm f_{t}^{\top}-\mathrm{E}(\bm f_{t}\bm f_{t}^{\top}))\bm H^{0\top}  & \bm H^0 \frac{1}{R_0}\sum(\frac{t}{T}-1)(\bm f_{t}\bm f_{t}^{\top}-\mathrm{E}(\bm f_{t}\bm f_{t}^{\top}))\bm H^{0\top}
 \\
\bm H^0\frac{1}{R_0}\sum(\frac{t}{T}-1)(\bm f_{t}\bm f_{t}^{\top}-\mathrm{E}(\bm f_{t}\bm f_{t}^{\top}))\bm H^{0\top} & \bm H^0\frac{1}{R_0}\sum(\frac{t}{T}-1)^2(\bm f_{t}\bm f_{t}^{\top}-\mathrm{E}(\bm f_{t}\bm f_{t}^{\top}))\bm H^{0\top}
\end{bmatrix}\\
\hat{\bm R}_3= &
\begin{bmatrix}
\frac{1}{R_0}\sum(\tilde{\bm f}_{t}^0-\bm H^0 f_t)\tilde{\bm f}_{t}^{0\top}  &  \frac{1}{R_0}\sum(\frac{t}{T}-1)(\tilde{\bm f}_{t}^0-\bm H^0 f_t)\tilde{\bm f}_{t}^{0\top}
 \\
\frac{1}{R_0}\sum(\frac{t}{T}-1)(\tilde{\bm f}_{t}^0-\bm H^0 f_t)\tilde{\bm f}_{t}^{0\top}  & \frac{1}{R_0}\sum(\frac{t}{T}-1)^2(\tilde{\bm f}_{t}^0-\bm H^0 f_t)\tilde{\bm f}_{t}^{0\top}
\end{bmatrix}
\\
\hat{\bm R}_4=&
\begin{bmatrix}
\bm H^0\frac{1}{R_0}\sum\bm f_{t}(\tilde{\bm f}_{t}^{0}-\bm H^0\bm f_t)^\top &  \bm H^0\frac{1}{R_0}\sum(\frac{t}{T}-1)\bm f_{t}(\tilde{\bm f}_{t}^{0}-\bm H^0\bm f_t)^\top
 \\
\bm H^0\frac{1}{R_0}\sum(\frac{t}{T}-1)\bm f_{t}(\tilde{\bm f}_{t}^{0}-\bm H^0\bm f_t)^\top  & \bm H^0\frac{1}{R_0}\sum(\frac{t}{T}-1)^2\bm f_{t}(\tilde{\bm f}_{t}^{0}-\bm H^0\bm f_t)^\top
\end{bmatrix}.
\end{align*}
Note that $\hat{\bm R}_1,\hat{\bm R}_2,\hat{\bm R}_3,\hat{\bm R}_4,$ resemble $\hat{\bm S}_1,\hat{\bm S}_2, \hat{\bm S}_3,\hat{\bm S}_4$, in the proof of \Cref{Thm2} (\ref{Thm21}). 
Using the same steps as in that proof,  it follows that 
$$\hat{\bm R}^{-1}=
\begin{bmatrix}
	O(1) & O\big(\frac{T}{R_0}\big)\\
	O\big(\frac{T}{R_0}\big) & O\big(\frac{T^2}{R_0^2}\big) 
\end{bmatrix}.
$$

For the order of $ \frac{1}{N}\sum_{i=1}^N(\ddot{\bm\lambda}_i(1)-\bm\lambda^0(1))(\ddot{\bm\lambda}_i(1)-\bm\lambda_i^0(1))^\top $ and 
 $ \frac{1}{N}\sum_{i=1}^N(\ddot{\bm\lambda}_i^{\prime}(1)-\bm\lambda^{0\prime}(1))(\ddot{\bm\lambda}_i(1)-\bm\lambda_i^0(1))^\top $, we use \Cref{lambda_expansion} to get
$$ \frac{1}{N}\sum_{i=1}^N
\begin{bmatrix}
 \ddot{\bm\lambda}_i(1)-\bm\lambda_i^0(1) \\
  \ddot{\bm\lambda}_i^{\prime}(1) -\bm\lambda_i^{0\prime}(1)
\end{bmatrix}
\begin{bmatrix}
 \ddot{\bm\lambda}_i(1)-\bm\lambda_i^0(1) \\
  \ddot{\bm\lambda}_i^{\prime}(1) -\bm\lambda_i^{0\prime}(1)
\end{bmatrix}^\top
=\sum_{j=1}^{6} \bm U_j
+
\sum_{j=7}^{21} (\bm U_j+\bm U_j^\top)
$$
with
\begin{align*}
\bm U_1 & = \hat{\bm R}^{-1}\begin{bmatrix}
\frac{1}{R_0}\sum\tilde{\bm f}^0_{t}(\tilde{\bm f}^0_{t}-\bm H^0 \bm f_t)^\top     \\
\frac{1}{R_0}\sum(\frac{t}{T}-1)\tilde{\bm f}^0_{t}(\tilde{\bm f}^0_{t}-\bm H^0 \bm f_t)^\top 
\end{bmatrix}
\bm W_{1}
\begin{bmatrix}
\frac{1}{R_0}\sum\tilde{\bm f}^0_{t}(\tilde{\bm f}^0_{t}-\bm H^0 \bm f_t)^\top     \\
\frac{1}{R_0}\sum(\frac{t}{T}-1)\tilde{\bm f}^0_{t}(\tilde{\bm f}^0_{t}-\bm H^0 \bm f_t)^\top 
\end{bmatrix}
^\top
(\hat{\bm R}^{-1})^{\top}
\\
\bm U_2 & =\hat{\bm R}^{-1}\begin{bmatrix}
\frac{1}{R_0}\sum(\frac{t}{T}-1)\tilde{\bm f}^0_{t}(\tilde{\bm f}^0_{t}-\bm H^0 \bm f_t)^\top     \\
\frac{1}{R_0}\sum(\frac{t}{T}-1)^2\tilde{\bm f}^0_{t}(\tilde{\bm f}^0_{t}-\bm H^0 \bm f_t)^\top 
\end{bmatrix}
\bm W_{2}
\begin{bmatrix}
\frac{1}{R_0}\sum(\frac{t}{T}-1)\tilde{\bm f}^0_{t}(\tilde{\bm f}^0_{t}-\bm H^0 \bm f_t)^\top     \\
\frac{1}{R_0}\sum(\frac{t}{T}-1)^2\tilde{\bm f}^0_{t}(\tilde{\bm f}^0_{t}-\bm H^0 \bm f_t)^\top 
\end{bmatrix}
^\top
(\hat{\bm R}^{-1})^{\top}
 \\
\bm U_3 & = 
\hat{\bm R}^{-1}\begin{bmatrix}
\frac{1}{R_0}\sum(\frac{t}{T}-1)^2(\tilde{\bm f}^0_{t}-\bm H^0\bm f_t)\bm f_{t}^\top    \\
\frac{1}{R_0}\sum(\frac{t}{T}-1)^3(\tilde{\bm f}^0_{t}-\bm H^0\bm f_t) \bm f_{t}^\top 
\end{bmatrix}
\bm W_{3}
\begin{bmatrix}
\frac{1}{R_0}\sum(\frac{t}{T}-1)^2(\tilde{\bm f}^0_{t}-\bm H^0\bm f_t)\bm f_{t}^\top    \\
\frac{1}{R_0}\sum(\frac{t}{T}-1)^3(\tilde{\bm f}^0_{t}-\bm H^0\bm f_t) \bm f_{t}^\top 
\end{bmatrix}
^\top
(\hat{\bm R}^{-1})^{\top}
 \\
\bm U_4 & = 
\hat{\bm R}^{-1}\begin{bmatrix}
\bm H^0\frac{1}{R_0}\sum(\frac{t}{T}-1)^2\bm f_{t}\bm f_{t}^\top     \\
\bm H^0\frac{1}{R_0}\sum(\frac{t}{T}-1)^3\bm f_{t} \bm f_{t}^\top 
\end{bmatrix}
\bm W_3
\begin{bmatrix}
\bm H^0\frac{1}{R_0}\sum(\frac{t}{T}-1)^2\bm f_{t}\bm f_{t}^\top     \\
\bm H^0\frac{1}{R_0}\sum(\frac{t}{T}-1)^3\bm f_{t} \bm f_{t}^\top 
\end{bmatrix}
^\top 
(\hat{\bm R}^{-1})^{\top}\\
\bm U_5 & = 
\hat{\bm R}^{-1}
\begin{bmatrix}
\bm W_4  & \bm W_5    \\
\bm W_5  & \bm W_6    \\
\end{bmatrix}
(\hat{\bm R}^{-1})^{\top}\\
\bm U_6 & = 
\hat{\bm R}^{-1}
\begin{bmatrix}
\bm W_7  & \bm W_8    \\
\bm W_8  & \bm W_9    \\
\end{bmatrix}
(\hat{\bm R}^{-1})^{\top}
\\
\bm U_7 & =\hat{\bm R}^{-1}
\begin{bmatrix}
\frac{1}{R_0}\sum\tilde{\bm f}^0_{t}(\tilde{\bm f}^0_{t}-\bm H^0 \bm f_t)^\top     \\
\frac{1}{R_0}\sum(\frac{t}{T}-1)\tilde{\bm f}^0_{t}(\tilde{\bm f}^0_{t}-\bm H^0 \bm f_t)^\top 
\end{bmatrix}
\bm W_{10}
\begin{bmatrix}
\frac{1}{R_0}\sum(\frac{t}{T}-1)\tilde{\bm f}^0_{t}(\tilde{\bm f}^0_{t}-\bm H^0 \bm f_t)^\top     \\
\frac{1}{R_0}\sum(\frac{t}{T}-1)^2\tilde{\bm f}^0_{t}(\tilde{\bm f}^0_{t}-\bm H^0 \bm f_t)^\top 
\end{bmatrix}
^\top
(\hat{\bm R}^{-1})^{\top}
 \\
\bm U_8 & =- 
\hat{\bm R}^{-1}
\begin{bmatrix}
\frac{1}{R_0}\sum\tilde{\bm f}^0_{t}(\tilde{\bm f}^0_{t}-\bm H^0 \bm f_t)^\top     \\
\frac{1}{R_0}\sum(\frac{t}{T}-1)\tilde{\bm f}^0_{t}(\tilde{\bm f}^0_{t}-\bm H^0 \bm f_t)^\top 
\end{bmatrix}
\bm W_{11}
\begin{bmatrix}
\frac{1}{R_0}\sum(\frac{t}{T}-1)^2(\hat{\bm f}^0_{t}-\bm H^0\bm f_t)\bm f_{t}^\top    \\
\frac{1}{R_0}\sum(\frac{t}{T}-1)^3(\hat{\bm f}^0_{t}-\bm H^0\bm f_t) \bm f_{t}^\top 
\end{bmatrix}
^\top
(\hat{\bm R}^{-1})^{\top}
 \\
\bm U_9 & = -
\hat{\bm R}^{-1}
\begin{bmatrix}
\frac{1}{R_0}\sum\tilde{\bm f}^0_{t}(\tilde{\bm f}^0_{t}-\bm H^0 \bm f_t)^\top     \\
\frac{1}{R_0}\sum(\frac{t}{T}-1)\tilde{\bm f}^0_{t}(\tilde{\bm f}^0_{t}-\bm H^0 \bm f_t)^\top 
\end{bmatrix}
\bm W_{11}
\begin{bmatrix}
\bm H^0\frac{1}{R_0}\sum(\frac{t}{T}-1)^2\bm f_{t}\bm f_{t}^\top     \\
\bm H^0\frac{1}{R_0}\sum(\frac{t}{T}-1)^3\bm f_{t} \bm f_{t}^\top 
\end{bmatrix}
^\top 
(\hat{\bm R}^{-1})^{\top}
\\
\end{align*}

\begin{align*}
\bm U_{10} & = -
\hat{\bm R}^{-1}
\begin{bmatrix}
\bm W_{12}  & \bm W_{13}    \\
\bm W_{14}  & \bm W_{15}   \\
\end{bmatrix}
(\hat{\bm R}^{-1})^{\top}\\
\bm U_{11} & = -
\hat{\bm R}^{-1}
\begin{bmatrix}
\bm W_{16}  & \bm W_{17}    \\
\bm W_{18}  & \bm W_{19}   \\
\end{bmatrix}
(\hat{\bm R}^{-1})^{\top}
 \\
\bm U_{12} & =- 
\hat{\bm R}^{-1}
\begin{bmatrix}
\frac{1}{R_0}\sum\tilde{\bm f}^0_{t}(\frac{t}{T}-1)(\tilde{\bm f}^0_{t}-\bm H^0 \bm f_t)^\top     \\
\frac{1}{R_0}\sum(\frac{t}{T}-1)^2\tilde{\bm f}^0_{t}(\tilde{\bm f}^0_{t}-\bm H^0 \bm f_t)^\top 
\end{bmatrix}
\bm W_{20}
\begin{bmatrix}
\frac{1}{R_0}\sum(\frac{t}{T}-1)^2(\tilde{\bm f}^0_{t}-\bm H^0\bm f_t)\bm f_{t}^\top    \\
\frac{1}{R_0}\sum(\frac{t}{T}-1)^3(\tilde{\bm f}^0_{t}-\bm H^0\bm f_t) \bm f_{t}^\top 
\end{bmatrix}
^\top
(\hat{\bm R}^{-1})^{\top}
 \\
\bm U_{13} & = -
\hat{\bm R}^{-1}
\begin{bmatrix}
\frac{1}{R_0}\sum(\frac{t}{T}-1)\tilde{\bm f}^0_{t}(\tilde{\bm f}^0_{t}-\bm H^0 \bm f_t)^\top     \\
\frac{1}{R_0}\sum(\frac{t}{T}-1)^2\tilde{\bm f}^0_{t}(\tilde{\bm f}^0_{t}-\bm H^0 \bm f_t)^\top 
\end{bmatrix}
\bm W_{20}
\begin{bmatrix}
\bm H^0\frac{1}{R_0}\sum(\frac{t}{T}-1)^2\bm f_{t}\bm f_{t}^\top     \\
\bm H^0\frac{1}{R_0}\sum(\frac{t}{T}-1)^3\bm f_{t} \bm f_{t}^\top 
\end{bmatrix}
^\top 
(\hat{\bm R}^{-1})^{\top}
\\
%
\bm U_{14} & = -
\hat{\bm R}^{-1}
\begin{bmatrix}
\bm W_{21}  & \bm W_{22}    \\
\bm W_{23}  & \bm W_{24}   \\
\end{bmatrix}
(\hat{\bm R}^{-1})^{\top}\\
\bm U_{15} & = -
\hat{\bm R}^{-1}
\begin{bmatrix}
\bm W_{25}  & \bm W_{26}    \\
\bm W_{27}  & \bm W_{28}   \\
\end{bmatrix}
(\hat{\bm R}^{-1})^{\top}
\\
\bm U_{16} & = 
\hat{\bm R}^{-1}
\begin{bmatrix}
\frac{1}{R_0}\sum(\frac{t}{T}-1)^2(\tilde{\bm f}^0_{t}-\bm H^0 \bm f_t){\bm f}_{t}^{\top}     \\
\frac{1}{R_0}\sum(\frac{t}{T}-1)^3(\tilde{\bm f}^0_{t}-\bm H^0 \bm f_t){\bm f}_{t}^{\top} 
\end{bmatrix}
\bm W_{3}
\begin{bmatrix}
\bm H^0\frac{1}{R_0}\sum(\frac{t}{T}-1)^2\bm f_{t}\bm f_{t}^\top     \\
\bm H^0\frac{1}{R_0}\sum(\frac{t}{T}-1)^3\bm f_{t} \bm f_{t}^\top 
\end{bmatrix}
^\top 
(\hat{\bm R}^{-1})^{\top}
\\
\bm U_{17} & = 
\hat{\bm R}^{-1}
\begin{bmatrix}
\bm W_{29}  & \bm W_{30}    \\
\bm W_{31}  & \bm W_{32}   \\
\end{bmatrix}
(\hat{\bm R}^{-1})^{\top}\\
\bm U_{18} & = 
\hat{\bm R}^{-1}
\begin{bmatrix}
\bm W_{33}  & \bm W_{34}    \\
\bm W_{35}  & \bm W_{36}   \\
\end{bmatrix}
(\hat{\bm R}^{-1})^{\top}
\\
\bm U_{19} & = 
\hat{\bm R}^{-1}
\begin{bmatrix}
\bm W_{37}  & \bm W_{38}    \\
\bm W_{39}  & \bm W_{40}   \\
\end{bmatrix}
(\hat{\bm R}^{-1})^{\top}\\
\bm U_{20} & = 
\hat{\bm R}^{-1}
\begin{bmatrix}
\bm W_{41}  & \bm W_{42}    \\
\bm W_{43}  & \bm W_{44}   \\
\end{bmatrix}
(\hat{\bm R}^{-1})^{\top}
\\
\bm U_{21} & = 
\hat{\bm R}^{-1}
\begin{bmatrix}
\bm W_{45}  & \bm W_{46}    \\
\bm W_{47}  & \bm W_{48}   \\
\end{bmatrix}
(\hat{\bm R}^{-1})^{\top}
\end{align*}
where the definition and the order of probability of the terms
$\bm W_{j}, j=1,\ldots,48$ are given \Cref{Lemma3} (1)-(12).
Combining their orders in probability with the result for $\hat{\bm R}^{-1}$ and using similar arguments as for orders of the terms in  \Cref{delta_expansion}, we get
\begin{align*}
\bm U_{1} & = 
\begin{bmatrix}
O_P(\phi_{NR_0T}^{-4})  & O_P(\frac{T}{R_0}\phi_{NR_0T}^{-4})   \\
O_P(\frac{T}{R_0}\phi_{NR_0T}^{-4})   &  O_P(\frac{T^2}{R^2_0}\phi_{NR_0T}^{-4})   \\
\end{bmatrix},
\\
\bm U_{2} & = 
\begin{bmatrix}
O_P(\frac{R_0^2}{T^2}\phi_{NR_0T}^{-4})  & O_P(\frac{R_0}{T}\phi_{NR_0T}^{-4}) \\
O_P(\frac{R_0}{T}\phi_{NR_0T}^{-4})  &  O_P(\phi_{NR_0T}^{-4})  \\
\end{bmatrix},
\\
\bm U_{3} & = 
\begin{bmatrix}
O_P(\frac{R_0^4}{T^4}\phi_{NR_0T}^{-4})  & O_P(\frac{R_0^3}{T^3}\phi_{NR_0T}^{-4})  \\
O_P(\frac{R_0^3}{T^3}\phi_{NR_0T}^{-4})  &  O_P(\frac{R_0^2}{T^2}\phi_{NR_0T}^{-4})  \\
\end{bmatrix},\\
\bm U_{4} & = 
\begin{bmatrix}
O_P(\frac{R_0^4}{T^4})  & O_P(\frac{R_0^3}{T^3})   \\
O_P(\frac{R_0^3}{T^3})   &  O_P(\frac{R_0^2}{T^2})   \\
\end{bmatrix},
\\
\bm U_{5} & = 
\begin{bmatrix}
O_P(\delta_{NR_0T}^{-2})  & O_P(\frac{T}{R_0}\delta_{NR_0T}^{-2})   \\
O_P(\frac{T}{R_0}\delta_{NR_0T}^{-2})   &  O_P(\frac{T^2}{R_0^2}\delta_{NR_0T}^{-2})   \\
\end{bmatrix},
\\
\bm U_{6} & = 
\begin{bmatrix}
O_P(\frac{1}{R_0}) & O_P(\frac{T}{R_0^2})   \\
 O_P(\frac{T}{R_0^2} )  &   O_P(\frac{T^2}{R_0^3})  \\
\end{bmatrix},
\\
\bm U_{7} & = 
\begin{bmatrix}
O_P(\frac{R_0}{T}\phi_{NR_0T}^{-4})  & O_P(\phi_{NR_0T}^{-4})   \\
O_P(\phi_{NR_0T}^{-4})   &  O_P(\frac{T}{R_0}\phi_{NR_0T}^{-4})   \\
\end{bmatrix},
\\
\bm U_{8} & = 
\begin{bmatrix}
O_P(\frac{R_0^2}{T^2}\phi_{NR_0T}^{-4})  & O_P(\frac{R_0}{T}\phi_{NR_0T}^{-4})   \\
O_P(\frac{R_0}{T}\phi_{NR_0T}^{-4})   &  O_P(\phi_{NR_0T}^{-4})   \\
\end{bmatrix},
\\
\bm U_{9} & = 
\begin{bmatrix}
O_P(\frac{R_0^2}{T^2}\phi_{NR_0T}^{-2})  & O_P(\frac{R_0}{T}\phi_{NR_0T}^{-2})   \\
O_P(\frac{R_0}{T}\phi_{NR_0T}^{-2})   &  O_P(\phi_{NR_0T}^{-2})   \\
\end{bmatrix},
\\
\bm U_{10} & = 
\begin{bmatrix}
O_P(\frac{\phi_{NR_0T}^{-2}\delta_{NR_0T}^{-1}}{\sqrt N})  & O_P(\frac{T}{R_0}\frac{\phi_{NR_0T}^{-2}\delta_{NR_0T}^{-1}}{\sqrt N})   \\
O_P(\frac{T}{R_0}\frac{\phi_{NR_0T}^{-2}\delta_{NR_0T}^{-1}}{\sqrt N})   &  O_P(\frac{T^2}{R_0^2}\frac{\phi_{NR_0T}^{-2}\delta_{NR_0T}^{-1}}{\sqrt N})   \\
\end{bmatrix},
\\
\bm U_{11} & = 
\begin{bmatrix}
O_P(\frac{\phi_{NR_0T}^{-2}}{\sqrt N})  & O_P(\frac{T}{R_0}\frac{\phi_{NR_0T}^{-2}}{\sqrt N})   \\
O_P(\frac{T}{R_0}\frac{\phi_{NR_0T}^{-2}}{\sqrt N})   &  O_P(\frac{T^2}{R_0^2}\frac{\phi_{NR_0T}^{-2}}{\sqrt N})   \\
\end{bmatrix},
\\
\bm U_{12} & = 
\begin{bmatrix}
O_P(\frac{R_0^3}{T^3}\phi_{NR_0T}^{-4})  & O_P(\frac{R_0^2}{T^2}\phi_{NR_0T}^{-4})   \\
O_P(\frac{R_0^2}{T^2}\phi_{NR_0T}^{-4})   &  O_P(\frac{R_0}{T}\phi_{NR_0T}^{-4})   \\
\end{bmatrix},
\\
\end{align*}

\begin{align*}
\bm U_{13} & = 
\begin{bmatrix}
O_P(\frac{R_0^3}{T^3}\phi_{NR_0T}^{-2})  & O_P(\frac{R_0^2}{T^2}\phi_{NR_0T}^{-2})   \\
O_P(\frac{R_0^2}{T^2}\phi_{NR_0T}^{-2})   &  O_P(\frac{R_0}{T}\phi_{NR_0T}^{-2})   \\
\end{bmatrix},
\\
\bm U_{14} & = 
\begin{bmatrix}
O_P(\frac{R_0}{T}\frac{\phi_{NR_0T}^{-2}\delta_{NR_0T}^{-1}}{\sqrt N})  & O_P(\frac{\phi_{NR_0T}^{-2}\delta_{NR_0T}^{-1}}{\sqrt N})   \\
O_P(\frac{\phi_{NR_0T}^{-2}\delta_{NR_0T}^{-1}}{\sqrt N})   &  O_P(\frac{T}{R_0}\frac{\phi_{NR_0T}^{-2}\delta_{NR_0T}^{-1}}{\sqrt N})   \\
\end{bmatrix},
\\
\bm U_{15} & = 
\begin{bmatrix}
O_P(\frac{R_0}{T}\frac{\phi_{NR_0T}^{-2}}{\sqrt N})  & O_P(\frac{\phi_{NR_0T}^{-2}}{\sqrt N})   \\
O_P(\frac{\phi_{NR_0T}^{-2}}{\sqrt N})   &  O_P(\frac{T}{R_0}\frac{\phi_{NR_0T}^{-2}}{\sqrt N})   \\
\end{bmatrix},
\\
\bm U_{16} & = 
\begin{bmatrix}
O_P(\frac{R_0^4}{T^4}\phi_{NR_0T}^{-2})  & O_P(\frac{R_0^3}{T^3}\phi_{NR_0T}^{-2})   \\
O_P(\frac{R_0^3}{T^3}\phi_{NR_0T}^{-2})   &  O_P(\frac{R_0^2}{T^2}\phi_{NR_0T}^{-2})   \\
\end{bmatrix},
\\
\bm U_{17} & = 
\begin{bmatrix}
O_P(\frac{R_0^2}{T^2}\frac{\phi_{NR_0T}^{-2}\delta_{NR_0T}^{-1}}{\sqrt N})  & O_P(\frac{R_0}{T}\frac{\phi_{NR_0T}^{-2}\delta_{NR_0T}^{-1}}{\sqrt N})   \\
O_P(\frac{R_0}{T}\frac{\phi_{NR_0T}^{-2}\delta_{NR_0T}^{-1}}{\sqrt N})   &  O_P(\frac{\phi_{NR_0T}^{-2}\delta_{NR_0T}^{-1}}{\sqrt N})   \\
\end{bmatrix},\\
%
\bm U_{18} & = 
\begin{bmatrix}
O_P(\frac{R_0^2}{T^2}\frac{\phi_{NR_0T}^{-2}}{\sqrt N})  & O_P(\frac{R_0}{T}\frac{\phi_{NR_0T}^{-2}}{\sqrt N})   \\
O_P(\frac{R_0}{T}\frac{\phi_{NR_0T}^{-2}}{\sqrt N})   &  O_P(\frac{\phi_{NR_0T}^{-2}}{\sqrt N})   \\
\end{bmatrix},
\\
\bm U_{19} & = 
\begin{bmatrix}
O_P(\frac{R_0^2}{T^2}\frac{\delta_{NR_0T}^{-1}}{\sqrt N})  & O_P(\frac{R_0}{T}\frac{\delta_{NR_0T}^{-1}}{\sqrt N})   \\
O_P(\frac{R_0}{T}\frac{\delta_{NR_0T}^{-1}}{\sqrt N})   &  O_P(\frac{\delta_{NR_0T}^{-1}}{\sqrt N})   \\
\end{bmatrix},
\\
\bm U_{20} & = 
\begin{bmatrix}
O_P(\frac{R_0^2}{T^2}\frac{1}{\sqrt N})  & O_P(\frac{R_0}{T}\frac{1}{\sqrt N})   \\
O_P(\frac{R_0}{T}\frac{1}{\sqrt N})   &  O_P(\frac{1}{\sqrt N})   \\
\end{bmatrix},
\\
\bm U_{21} & = 
\begin{bmatrix}
O_P(\frac{\delta_{NR_0T}^{-1}}{\sqrt{R_0}})  & O_P(\frac{T}{R_0}\frac{\delta_{NR_0T}^{-1}}{\sqrt{R_0}})   \\
O_P(\frac{T}{R_0}\frac{\delta_{NR_0T}^{-1}}{\sqrt{R_0}})   &  O_P(\frac{T^2}{R_0^2}\frac{\delta_{NR_0T}^{-1}}{\sqrt{R_0}})   \\
\end{bmatrix}.
\end{align*} 
Consequently,
\begin{align*}
\frac{1}{N}\sum_{i=1}^N
\begin{bmatrix}
 \ddot{\bm\lambda}_i(1)-\bm\lambda_i^0(1) \\
  \ddot{\bm\lambda}_i^{\prime}(1) -\bm\lambda_i^{0\prime}(1)
\end{bmatrix}
\begin{bmatrix}
 \ddot{\bm\lambda}_i(1)-\bm\lambda_i^0(1) \\
  \ddot{\bm\lambda}_i^{\prime}(1) -\bm\lambda_i^{0\prime}(1)
\end{bmatrix}^\top
&=\sum_{j=1}^{6} \bm U_j+\sum_{j=7}^{21} (\bm U_j+\bm U_j^\top)\\
&=
\begin{bmatrix}
O_P(\phi_{NR_0T}^{-2})  & O_P(\frac{T}{R_0}\phi_{NR_0T}^{-2})   \\
O_P(\frac{T}{R_0}\phi_{NR_0T}^{-2})   &  O_P(\frac{T^2}{R_0^2}\phi_{NR_0T}^{-2})   \\
\end{bmatrix}.
\end{align*}

For the order of $ \frac{1}{N}\sum_{i=1}^N(\ddot{\bm\lambda}_i(1)-\bm\lambda^0(1))(\bm\lambda_i^0(1))^\top $ and 
 $ \frac{1}{N}\sum_{i=1}^N(\ddot{\bm\lambda}_i^{\prime}(1)-\bm\lambda^{0\prime}(1))(\bm\lambda_i^0(1))^\top $, we use 
 similar arguments to get
$$ \frac{1}{N}\sum_{i=1}^N
\begin{bmatrix}
 (\ddot{\bm\lambda}_i(1)-\bm\lambda_i^0(1))(\bm\lambda_i^0(1))^\top  \\
 ( \ddot{\bm\lambda}_i^{\prime}(1) -\bm\lambda_i^{0\prime}(1))(\bm\lambda_i^0(1))^\top 
\end{bmatrix}
= 
\sum_{j=1}^{6} \bm V_j
=\begin{bmatrix}
O_P(\phi_{NR_0T}^{-2} )   \\
 O_P(\frac{T}{R_0}\phi_{NR_0T}^{-2})   \\
\end{bmatrix}.
$$
with
\begin{align*}
\bm V_1 & = \hat{\bm R}^{-1}\begin{bmatrix}
\frac{1}{R_0}\sum\tilde{\bm f}^0_{t}(\tilde{\bm f}^0_{t}-\bm H^0 \bm f_t)^\top     \\
\frac{1}{R_0}\sum(\frac{t}{T}-1)\tilde{\bm f}^0_{t}(\tilde{\bm f}^0_{t}-\bm H^0 \bm f_t)^\top 
\end{bmatrix}
\bm W_{1}
=
\begin{bmatrix}
O_P(\phi_{NR_0T}^{-2})   \\
 O_P(\frac{T}{R_0}\phi_{NR_0T}^{-2})   \\
\end{bmatrix},
\\
\bm V_2 & =\hat{\bm R}^{-1}\begin{bmatrix}
\frac{1}{R_0}\sum(\frac{t}{T}-1)\tilde{\bm f}^0_{t}(\tilde{\bm f}^0_{t}-\bm H^0 \bm f_t)^\top     \\
\frac{1}{R_0}\sum(\frac{t}{T}-1)^2\tilde{\bm f}^0_{t}(\tilde{\bm f}^0_{t}-\bm H^0 \bm f_t)^\top 
\end{bmatrix}
\bm W_{10}^\top
=
\begin{bmatrix}
O_P(\frac{R_0}{T}\phi_{NR_0T}^{-2})   \\
 O_P(\phi_{NR_0T}^{-2})   \\
\end{bmatrix},
 \\
\bm V_3 & = 
\hat{\bm R}^{-1}\begin{bmatrix}
\frac{1}{R_0}\sum(\frac{t}{T}-1)^2(\hat{\bm f}^0_{t}-\bm H^0\bm f_t)\bm f_{t}^\top    \\
\frac{1}{R_0}\sum(\frac{t}{T}-1)^3(\hat{\bm f}^0_{t}-\bm H^0\bm f_t) \bm f_{t}^\top 
\end{bmatrix}
\bm W_{11}^\top
=
\begin{bmatrix}
O_P(\frac{R_0^2}{T^2}\phi_{NR_0T}^{-2})   \\
 O_P(\frac{R_0}{T}\phi_{NR_0T}^{-2})   \\
\end{bmatrix},
 \\
\bm V_4 & = 
\hat{\bm R}^{-1}\begin{bmatrix}
\bm H^0\frac{1}{R_0}\sum(\frac{t}{T}-1)^2\bm f_{t}\bm f_{t}^\top     \\
\bm H^0\frac{1}{R_0}\sum(\frac{t}{T}-1)^3\bm f_{t} \bm f_{t}^\top 
\end{bmatrix}
\bm W_{11}^\top
=
\begin{bmatrix}
O_P(\frac{R_0^2}{T^2})   \\
 O_P(\frac{R_0}{T})   \\
\end{bmatrix},
 \\
\bm V_5 & = 
\hat{\bm R}^{-1}
\begin{bmatrix}
 \frac{1}{R_0}\sum(\tilde{\bm f}^0_{t}-\bm H^0\bm f_t) \frac{1}{N}\sum_{i=1}^Ne_{it}\bm\lambda_i^{\top}(1)(\bm H^{0})^{-1} \\
\frac{1}{R_0}\sum(\frac{t}{T}-1)(\tilde{\bm f}^0_{t}-\bm H^0\bm f_t) \frac{1}{N}\sum_{i=1}^Ne_{it}\bm\lambda_i^{\top}(1)(\bm H^{0})^{-1}  
\end{bmatrix}
=
\begin{bmatrix}
O_P(\frac{\delta_{NR_0T}^{-1}}{\sqrt N})   \\
 O_P(\frac{T}{R_0}\frac{\delta_{NR_0T}^{-1}}{\sqrt N})   \\
\end{bmatrix},
\\
\bm V_6 & = 
\hat{\bm R}^{-1}
\begin{bmatrix}
\bm H^0\frac{1}{R_0}\sum\bm f_{t} \frac{1}{N}\sum_{i=1}^N e_{it} \bm\lambda_i^{\top}(1)(\bm H^{0})^{-1}  \\
\bm H^0\frac{1}{R_0}\sum(\frac{t}{T}-1)\bm f_{t}  \frac{1}{N}\sum_{i=1}^N e_{it}\bm\lambda_i^{\top}(1)(\bm H^{0})^{-1}
\end{bmatrix}
=
\begin{bmatrix}
O_P(\frac{1}{\sqrt{NR_0}})  \\
 O_P(\frac{T}{R_0}\frac{1}{\sqrt{ NR_0}})  \\
\end{bmatrix},
\end{align*}
where the last equality uses \Cref{Ass2} (d).
Similarly,
$$ \frac{1}{N}\sum_{i=1}^N
\begin{bmatrix}
 (\ddot{\bm\lambda}_i(1)-\bm\lambda_i^0(1))(\bm\lambda_i^{0\prime}(1))^\top  \\
 ( \ddot{\bm\lambda}_i^{\prime}(1) -\bm\lambda_i^{0\prime}(1))(\bm\lambda_i^{0\prime}(1))^\top 
\end{bmatrix}
=\begin{bmatrix}
O_P(\phi_{NR_0T}^{-2} )   \\
 O_P(\frac{T}{R_0}\phi_{NR_0T}^{-2})   \\
\end{bmatrix}.
$$

 \subsection*{Proof of \Cref{Lemma4}}
 
  \subsubsection*{Proof of \Cref{Lemma4} (1)} 
 To prove that
 $\sup_{R} \big(\frac{1}{R}\sum_{s=T-R+1}^{T}  \Vert \bm f_s\Vert^k\big)=O_P(1)$, $k=1,2,3,4$, we first observe that 
 $$\sup_R\big(\frac{1}{R}\sum_{t=T-R+1}^{T}\Vert \bm f_t\Vert^k\big)\leq \sup_R\big(\frac{1}{R}\sum_{t=T-R+1}^{T}\mathrm E(\Vert \bm{f}_t\Vert^k)\big)+\sup_R\big(\big\vert\frac{1}{R}\sum_{t=T-R+1}^{T}(\Vert \bm{f}_t\Vert^k-\mathrm E (\Vert \bm{f}_t\Vert)^k)\big\vert\big).$$
 Because, $\mathrm E(\Vert\bm f_t\Vert^k)\leq C<\infty,$ from \Cref{Ass1} (a), we have $\sup_R\big(\frac{1}{R}\sum_{t=T-R+1}^{T}\mathrm E(\Vert \bm{f}_t\Vert^k)\big)=O(1)$.

 We follow similar steps as in \citet{IJR2017}, and note also that for any $\theta\in (0,1/2)$, 
 \begin{equation}
 \sup_R\big(\big\vert\frac{1}{R}\sum_{t=T-R+1}^{T}(\Vert \bm{f}_t\Vert^k-\mathrm E (\Vert \bm{f}_t\Vert)^k)\big\vert\big)\leq \sup_R\big(R^{\frac{1}{2}-\theta}\big\vert\frac{1}{R}\sum_{t=T-R+1}^{T}(\Vert \bm{f}_t\Vert^k-\mathrm E (\Vert \bm{f}_t\Vert)^k)\big\vert\big),
 \end{equation}
because $1\leq R$ and $1 \leq R^{\frac{1}{2}-\theta}.$ In addition, we have 
 \begin{equation*}
 P(\sup_R(R^{-\frac{1}{2}-\theta}\big\vert\sum_{t=T-R+1}^{T}(\Vert \bm{f}_t\Vert^k-\mathrm E (\Vert \bm{f}_t\Vert)^k)\big\vert)> \epsilon ) \leq  \sum_{\theta\in \Theta_R}P(R^{-\frac{1}{2}-\theta}\big\vert\sum_{t=T-R+1}^{T}(\Vert \bm{f}_t\Vert^k-\mathrm E (\Vert \bm{f}_t\Vert)^k)\big\vert> \epsilon ) .
 \end{equation*}
 which is lower than $|\Theta_R| \sup_R P((R^{-\frac{1}{2}-\theta}\big\vert\sum_{t=T-R+1}^{T}(\Vert \bm{f}_t\Vert^k-\mathrm E (\Vert \bm{f}_t\Vert)^k)\big\vert)> \epsilon )$.
Applying Chebyshev's inequality, it is bounded by
 $ |\Theta_R| \sup_R\Vert\mathrm{Vec}(\mathrm{Var}(R^{-\frac{1}{2}-\theta}\sum_{t=T-R+1}^{T}(\Vert \bm{f}_t\Vert^k-\mathrm E (\Vert \bm{f}_t\Vert)^k)))\Vert /\epsilon^2  .
$ 
Using $\Vert\mathrm{Vec}(\mathrm{Var}(R^{-\frac{1}{2}-\theta}\sum_{t=T-R+1}^{T}(\Vert \bm{f}_t\Vert^k-\mathrm E (\Vert \bm{f}_t\Vert)^k)))\Vert\leq C,$ from \Cref{Ass4} (b), we deduce that 
 \begin{equation*}
 P(\sup_R(R^{-\frac{1}{2}-\theta}\big\vert\sum_{t=T-R+1}^{T}(\Vert \bm{f}_t\Vert^k-\mathrm E (\Vert \bm{f}_t\Vert)^k)\big\vert)> \epsilon ) \leq  |\Theta_R| \sup_R \frac{C}{R^{2\theta}\epsilon^2}=|\Theta_R| \frac{C}{\underline{R}^{2\theta}\epsilon^2}=O(1),
 \end{equation*}
given \Cref{Ass7} (b). Thus, we have the result.

\subsubsection*{Proof of \Cref{Lemma4} (2)} 
 Now we show that $\sup_R\frac{T^4}{R^6}\sum_{s=T-R+1}^{T}\sum_{t=T-R+1}^{T} \Vert  \bm \Lambda^\top(s/T)\bm \Lambda(t/T)/N-\bm \Lambda^\top(s/T)\bm \Lambda(1)/N
\Vert^4=O(1).$
Using the decomposition in the proof of \Cref{Lemma2} (1), it is bounded by 
	\begin{align*}
		&\sup_R\frac{T^4}{R^5}\sum_{t=T-R+1}^{T}\big(\frac{T-t}{T}\big)^4\big\Vert \frac{1}{N}\sum_{i=1}^{N}\bm\lambda_i(1)\bm\lambda_i^{\prime\top}(1)\big\Vert^4\\
		&+\sup_R \frac{1}{4}\frac{T^4}{R^5}\sum_{t=T-R+1}^{T}  \big(\frac{T-t}{T}\big)^{8}\big\Vert\frac{1}{N}\sum_{i=1}^{N}\bm\lambda_i(1)\bm\lambda_i^{\prime\prime\top}(c_i)\big\Vert^4\\
		& +\sup_R\frac{T^4}{R^6}\sum_{t=T-R+1}^{T} \big(\frac{T-t}{T}\big)^4\sum_{s=T-R+1}^{T}\big(\frac{T-s}{T}\big)^4\big\Vert\frac{1}{N}\sum_{i=1}^{N}\bm\lambda_i^\prime(1)\bm\lambda_i^{\prime\top}(1)\big\Vert^{4}\\
		&+\sup_R\frac{T^4}{R^6}\sum_{t=T-R+1}^{T} \big(\frac{T-t}{T}\big)^{8}\sum_{s=T-R+1}^{T} \big(\frac{T-s}{T}\big)^4\big\Vert\frac{1}{N}\sum_{i=1}^{N}\bm\lambda_i^\prime(1)\bm\lambda_i^{\prime\prime\top}(c_i)\big\Vert^4 \\
		& +\sup_R\frac{1}{4^{8}}\frac{T^4}{R^6}\sum_{t=T-R+1}^{T} \big(\frac{T-t}{T}\big)^{8}\sum_{s=T-R+1}^{T}\big(\frac{T-s}{T}\big)^{8}\big\Vert\frac{1}{N}\sum_{i=1}^{N}\bm\lambda_i^{\prime\prime}(c_i)\bm\lambda_i^{\prime\prime\top}(c_i)\big\Vert^4.
	\end{align*}
We use the fact that
		$\sum_{t=T-R+1}^{T}\big(\frac{T-t}{T}\big)^k=\frac{R^{k+1}}{T^{k}},
        $
	and 
	$\big\Vert \frac{1}{N}\sum_{i=1}^{N}\bm\lambda_i(1)\bm\lambda_i^{\prime\top}(1)\big\Vert,$
	$\big\Vert\frac{1}{N}\sum_{i=1}^{N}\bm\lambda_i(1)\bm\lambda_i^{\prime\prime\top}(c_i)\big\Vert,$
	$\big\Vert\frac{1}{N}\sum_{i=1}^{N}\bm\lambda_i^\prime(1)\bm\lambda_i^{\prime\prime\top}(c_i)\big\Vert,$
	$\big\Vert\frac{1}{N}\sum_{i=1}^{N}\bm\lambda_i^\prime(1)\bm\lambda_i^{\prime\top}(1)\big\Vert,$
	and 
	$\Vert\frac{1}{N}\sum_{i=1}^{N}\bm\lambda_i^{\prime\prime}(c_i)\bm\lambda_i^{\prime\prime\top}(c_i)\big\Vert$ 
	are bounded, to have
	$\sup_R\frac{T^4}{R^6}\sum_{s=T-R+1}^{T}\sum_{t=T-R+1}^{T} \Vert  \bm \Lambda^\top(s/T)\bm \Lambda(t/T)/N-\bm \Lambda^\top(s/T)\bm \Lambda(1)/N
		\Vert^k$ equals $O\big(\sup_R( \frac{T^4}{R^4}\frac{R^4}{T^4} + \frac{T^4}{R^4}\frac{R^{8}}{T^{8}}+ \frac{T^4}{R^4}\frac{R^{12}}{T^{12}} + \frac{T^4}{R^4}\frac{R^{16}}{T^{16}} ) \big)$,
	which is $O(1).$

\subsubsection*{Proof of \Cref{Lemma4} (3)} 
 The identity \eqref{identity} implies $\sup_{R}(\delta_{NRT}^{2}\frac{1}{R}\sum_{t=T-R+1}^{T}\Vert\tilde{\bm f}_{t,R}-\bm H_{R}\bm f_t\Vert^2))$ is lower than $5\sup_{R}\Vert\tilde{\bm V}_{R}^{-1}\Vert^2 \times \sum_{j=1}^{5}\sup_{R}(\delta_{NRT}^{2} \frac{1}{R}\sum_{t=T-R+1}^{T}( \Vert \bm A_{jt}\Vert^2)).$ Thus, we only have to show $\sup_{R}(\delta_{NRT}^{2}\frac{1}{R}\sum_{t=T-R+1}^{T}\Vert\bm A_{jt}\Vert^2)=O_P(1), j=1,\ldots,5$.

We start by noting that 
$$\sup_R\big(\delta_{NRT}^{2} \frac{1}{R}\sum_{t=T-R+1}^{T}\Vert\bm A_{1t}\Vert^2\big)\leq \sup_R\big(\frac{1}{R}\sum_{s=T-R+1}^{T}\Vert\tilde{\bm f}_{s,R}\Vert^2\big)\sup_R\big(\frac{\delta_{NRT}^{2}}{R^2}\sum_{s=T-R+1}^{T}\sum_{t=T-R+1}^{T}\Vert\bm \nu_{st}\Vert^2\big).$$ We observe that $\sup_R\big(\frac{1}{R}\sum_{s=T-R+1}^{T}\Vert\tilde{\bm f}_{s,R}\Vert^2\big)
\sup_R\mathrm{Trace}\big(\frac{1}{R}\sum_{s=T-R+1}^{T}\tilde{\bm f}_{s,R}\tilde{\bm f}_{s,R}^\top\big)=\mathrm{Trace}(\bm I_r)=r=O_P(1)$.
Also, $\sup_R\frac{\delta_{NRT}^{2}}{R^2}\sum_{s=T-R+1}^{T}\sum_{t=T-R+1}^{T}\Vert\bm \nu_{st}\Vert^2 $ is less than $ (\sup_R\frac{1}{R}\sum_{s=T-R+1}^{T} \Vert \bm f_s \Vert^4 )^{\frac{1}{2}}$ times
  $\bigg(\sup_R \frac{1}{R}\sum_{s=T-R+1}^{T}\big(\frac{\delta_{NRT}^{2}}{R}\sum_{t=T-R+1}^{T} \Vert  (\frac{1}{N}\bm \Lambda^\top(s/T)\bm \Lambda(t/T)-\frac{1}{N}\bm \Lambda^\top(s/T)\bm \Lambda(1)/N)\bm f_t
 \Vert^2\big)^2 \bigg)^{\frac{1}{2}}.$
 Using the Cauchy-Schwarz inequality, the bound becomes
 $$ \big(\sup_R\frac{1}{R}\sum_{s=T-R+1}^{T} \Vert \bm f_s \Vert^4 \big)
 \bigg(\sup_R \frac{\delta_{NRT}^{2}}{R^2}\sum_{s=T-R+1}^{T}\sum_{t=T-R+1}^{T} \Vert  \frac{1}{N}\bm \Lambda^\top(s/T)\bm \Lambda(t/T)-\frac{1}{N}\bm \Lambda^\top(s/T)\bm \Lambda(1)
 \Vert^4 \bigg)^{\frac{1}{4}}.$$
Since \Cref{Lemma4} (1) and (2) imply these two terms are $O_P(1)$, we deduce that
$$\sup_R\big(\frac{\delta_{NRT}^{2}}{R^2}\sum_{s=T-R+1}^{T}\sum_{t=T-R+1}^{T}\Vert\bm \nu_{st}\Vert^2\big)=O_P(1),$$ and
$\sup_R\big(\delta_{NRT}^{2} \frac{1}{R}\sum_{t=T-R+1}^{T}\Vert\bm A_{1t}\Vert^2\big)=O_P(1)$.

Second, we have 
$$\sup_{R}(\delta_{NRT}^{2}\frac{1}{R}\sum_{t=T-R+1}^{T}\Vert\bm A_{2t}\Vert^2)\leq \sup_{R}\big(\frac{1}{R}\sum_{s=T-R+1}^{T}\Vert\tilde{\bm f}_{s,R}\Vert^2\big)\sup_{R}\big(\frac{\delta_{NRT}^{2}}{R^2}\sum_{s=T-R+1}^{T}\sum_{t=T-R+1}^{T}\Vert\bm \eta_{st}\Vert^2\big),$$
where $\sup_{R}\frac{1}{R}\sum_{s=T-R+1}^{T}\Vert\tilde{\bm f}_{s,R}\Vert^2=O_P(1).$ Also, 
 $\sup_{R}\frac{\delta_{NRT}^{2}}{R^2}\sum_{s=T-R+1}^{T}\sum_{t=T-R+1}^{T}\Vert\bm \eta_{st}\Vert^2$ is bounded by
$\sup_{R} \big(\frac{1}{R}\sum_{s=T-R+1}^{T}  \Vert \bm f_s\Vert^2\big)\sup_{R} \big(\frac{\delta_{NRT}^{2}}{N}\sup_{T-R+1\leq s\leq T}\frac{1}{R}\sum_{t=T-R+1}^{T}\Vert\frac{1}{\sqrt N} \bm \Lambda^\top(s/T)\bm e_{t} \Vert^2\big)$, where
$\sup_{R} \big(\frac{1}{R}\sum_{s=T-R+1}^{T}  \Vert \bm f_s\Vert^2\big)=O_P(1)$  from  \Cref{Lemma4} (1). In addition,
the second term is bounded by $\sup_{R} \big(\frac{\delta_{NRT}^{2}}{N}\sup_{T-R+1\leq s,t\leq T}\Vert \frac{1}{\sqrt N} \bm \Lambda^\top(s/T)\bm e_{t} \Vert^2\big)=O_P(1)$ from  \Cref{Ass6} (a). Consequently, $\sup_{R}\big(\frac{\delta_{NRT}^{2}}{R^2}\sum_{s=T-R+1}^{T}\sum_{t=T-R+1}^{T}\Vert\bm \eta_{st}\Vert^2\big)=O_P(1)$, and we deduce that $\sup_{R}(\delta_{NRT}^{2}\frac{1}{R}\sum_{t=T-R+1}^{T}\Vert\bm A_{2t}\Vert^2)=O_P(1)$.
 
Third, $\sup_{R}(\delta_{NRT}^{2}\frac{1}{R}\sum_{t=T-R+1}^{T}\Vert\bm A_{3t}\Vert^2)$ is lower than 
$$\sup_{R} \big(\frac{1}{R}\sum_{s=T-R+1}^{T}\Vert\tilde{\bm f}_{s,R}\Vert^2\big)\sup_{R}\big(\frac{\delta_{NRT}^{2}}{R^2}\sum_{s=T-R+1}^{T}\sum_{t=T-R+1}^{T}\Vert\bm \xi_{st}\Vert^2\big),$$ 
where we noted again that $\sup_{R}\frac{1}{R}\sum_{s=T-R+1}^{T}\Vert\tilde{\bm f}_{s,R}\Vert^2=O_P(1)$.  In addition,  we observe that $\sup_{R}\big(\frac{\delta_{NRT}^{2}}{R^2}\sum_{t=T-R+1}^{T}\sum_{s=T-R+1}^{T}\Vert\bm \xi_{st}\Vert^2\big)$ is bounded by
$  \sup_{R} \big( \frac{1}{R}\sum_{t=T-R+1}^{T}\Vert \bm f_t\Vert^2\big)=O_P(1)$
times $\sup_{R} \big( \frac{\delta_{NRT}^{2}}{N} \sup_{T-R+1\leq t\leq T}\frac{1}{R}\sum_{s=T-R+1}^{T}\Vert \bm e_s^\top\frac{1}{\sqrt N} \bm \Lambda(t/T)\Vert^2\big)=O_P(1),$ where the probability orders come from \Cref{Lemma4} (1) and \Cref{Ass6} (a). Therefore, we have that $\sup_{R}\big(\frac{\delta_{NRT}^{2}}{R^2}\sum_{t=T-R+1}^{T}\sum_{s=T-R+1}^{T}\Vert\bm \xi_{st}\Vert^2\big)=O_P(1)$ and $\sup_{R}(\delta_{NRT}^{2}\frac{1}{R}\sum_{t=T-R+1}^{T}\Vert\bm A_{3t}\Vert^2)=O_P(1)$ follows.

 Fourth, $\sup_{R}(\delta_{NRT}^{2}\frac{1}{R}\sum_{t=T-R+1}^{T}\Vert\bm A_{4t}\Vert^2)$ is lower than 
$$\sup_{R} \big(\frac{1}{R}\sum_{s=T-R+1}^{T}\Vert\tilde{\bm f}_{s,R}\Vert^2\big)\sup_{R}\big(\frac{\delta_{NRT}^{2}}{R^2}\sum_{s=T-R+1}^{T}\sum_{t=T-R+1}^{T}\Vert\bm \gamma_{st}\Vert^2\big),$$ 
where $\sup_{R}\frac{1}{R}\sum_{s=T-R+1}^{T}\Vert\tilde{\bm f}_{s,R}\Vert^2=O_P(1)$. Also, $\sup_{R}\big(\frac{\delta_{NRT}^{2}}{R^2}\sum_{s=T-R+1}^{T}\sum_{t=T-R+1}^{T}\Vert\bm \gamma_{st}\Vert^2\big)\leq \sup_{R}\big(\frac{1}{R}\sum_{s=T-R+1}^{T}\sum_{t=T-R+1}^{T}\vert\mathrm E(\bm e_s^\top\bm e_t/N)\vert^2\big) =O(1)$ using the same steps as in \citet[Proof of Lemma 1 (i)]{BN2002}. Therefore,  $\sup_{R}(\delta_{NRT}^{2}\frac{1}{R}\sum_{t=T-R+1}^{T}\Vert\bm A_{4t}\Vert^2)=O_P(1).$  

Fifth, we noted from the proof of \citet[Theorem 1]{BN2002} that 
$$\frac{1}{R}\sum_{t=T-R+1}^{T}\Vert\bm A_{5t}\Vert\leq \big(\frac{1}{R}\sum_{s=T-R+1}^{T}\Vert\tilde{\bm f}_{s,R}\Vert^2\big)\big( \frac{1}{R^4}\sum_{s=T-R+1}^{T}\sum_{u=T-R+1}^{T}\big(\sum_{t=T-R+1}^{T}\zeta_{st}\zeta_{ut}\big)^2\big)^{1/2}.$$ 
In consequence, 
$\sup_{R} \frac{\delta_{NRT}^{2}}{R}\sum_{t=T-R+1}^{T}\Vert\bm A_{5t}\Vert^2$ 
is lower than
$ \sup_{R}\big(\frac{1}{R}\sum_{s=T-R+1}^{T}\Vert\tilde{\bm f}_{s,R}\Vert^2\big)$
times
$\big( \sup_{R}\big(\frac{\delta_{NRT}^{2}}{R^4}\sum_{s=T-R+1}^{T}\sum_{u=T-R+1}^{T}\big(\sum_{t=T-R+1}^{T}\zeta_{st}\zeta_{ut}\big)^2\big)^{1/2}.$
In addition, we note that
$$ \sup_{R}\big(\frac{\delta_{NRT}^{2}}{R^4}\sum_{s=T-R+1}^{T}\sum_{u=T-R+1}^{T}\big(\sum_{t=T-R+1}^{T}\zeta_{st}\zeta_{ut}\big)^2\leq \sup_{R}\big(\frac{\delta_{NRT}^{2}}{N^2}\frac{R^{4}}{R^4}(\sup_{s,t}\vert N^{-1/2}(\bm e_t^\top\bm e_s-\mathrm{E}(\bm e_t^\top\bm e_s)) \vert)^4\big),$$ 
is $O_P(1)$ because $\sup_R\sup_{s,t}\vert N^{-1/2}(\bm e_t^\top\bm e_s-\mathrm{E}(\bm e_t^\top\bm e_s)) \vert=O_P(1)$ from \Cref{Ass6} (b), we obtain 
$$ \sup_{R}\frac{\delta_{NRT}^{2}}{R}\sum_{t=T-R+1}^{T}\Vert\bm A_{5t}\Vert^2=O_P(1).$$
Combining the previous results completes the proof.

\subsubsection*{Proof of \Cref{Lemma4} (4)} 

We show that if $\sup_R\big( q_{NRT}\Vert \mathrm{Vec}(\bm Q_R)-\mathrm{Vec}(\bm Q_{R,0})\Vert\big)=O_P(1),$
then $\sup_R\big( q_{NRT}\Vert \mathrm{Vec}(\bm Q_R^{-1})-\mathrm{Vec}(\bm Q_{R,0}^{-1})\Vert\big)=O_P(1),$ as $N,T\rightarrow\infty$, with $q_{NRT}\rightarrow \infty $. 

To prove this, some arguments are similar to those in \citet[Proof of Lemma 6]{IJR2017}. Define a continuously differentiable function $g(\cdot)$, such that $g(\mathrm{Vec}(\bm Q_{R}))=\mathrm{Vec}(\bm Q_{R}^{-1})$. Therefore, there is an open neighborhood, $N(\mathrm{Vec}(\bm Q_{R,0}))$ of $\mathrm{Vec}(\bm Q_{R,0})$ such that 
$\frac{\partial g(\bm q)}{\partial \bm q}$ is bounded for any $\bm q\in N(\mathrm{Vec}(\bm Q_{R,0}))$. From a first-order Taylor expansion,
$\mathrm{Vec}(\bm Q_R^{-1})-\mathrm{Vec}(\bm Q_{R,0}^{-1})=g(\mathrm{Vec}(\bm Q_{R}))-g(\mathrm{Vec}(\bm Q_{R,0}))=\big(\frac{\partial g(\bm q_0)}{\partial \bm q}\big)^\prime(\mathrm{Vec}(\bm Q_{R})-\mathrm{Vec}(\bm Q_{R,0}))$, where $\Vert \bm q_0 -\mathrm{Vec}(\bm Q_{R,0})\Vert \leq\Vert \mathrm{Vec}(\bm Q_R)-\mathrm{Vec}(\bm Q_{R,0})\Vert$. Consequently,
\begin{equation}
\sup_R (q_{NRT}\Vert \mathrm{Vec}(\bm Q_R^{-1})-\mathrm{Vec}(\bm Q_{R,0}^{-1})\Vert) \leq \sup_R\big\Vert \frac{\partial g(\bm q_0)}{\partial \bm q}\big\Vert \sup_R (q_{NRT}\Vert \mathrm{Vec}(\bm Q_R)-\mathrm{Vec}(\bm Q_{R,0}) \Vert).
\label{taylor}
\end{equation}
By assumption,  $\sup_R\big( q_{NRT}\Vert \mathrm{Vec}(\bm Q_R)-\mathrm{Vec}(\bm Q_{R,0})\Vert\big)=O_P(1).$ As a result,
$\mathrm{Vec}(\bm Q_R)\overset{P}{\longrightarrow}\mathrm{Vec}(\bm Q_{R,0})$, for any $R$. Therefore, there exists $R$ sufficiently large such that $\bm q_0\in N(\mathrm{Vec}(\bm Q_{R,0}))$ with probability greater than $1-\vartheta$, for some $\vartheta>0$, and $\sup_R\big\Vert \frac{\partial g(\bm q_0)}{\partial \bm q}\big\Vert \leq C$. 
 Therefore, \Cref{taylor} implies $\sup_R (q_{NRT}\Vert \mathrm{Vec}(\bm Q_R^{-1})-\mathrm{Vec}(\bm Q_{R,0}^{-1})\Vert)=O_P(1)$.

\subsubsection*{Proof of \Cref{Lemma4} (5)}

Now, we show that $\sup_{R}\big(\min(\sqrt{N},\sqrt R\delta_{NRT},R^{7/8-\mu})((\bm H_{R}^{-1})^{\top}\bm\alpha(1))^\top(\tilde{\bm f}_{T,R}-\bm H_R\bm f_T)\big)$ is lower than $ \Vert\bm\alpha(1)\Vert \sup_R\Vert \bm H_{R}^{-1}\Vert \sup_R\Vert \min(\sqrt{N},\sqrt R\delta_{NRT},R^{7/8-\mu})(\tilde{\bm f}_{T,R}-\bm H_R\bm f_T)\Vert$ 
is $O_P(1)$, for some $\mu \in (0,1/8)$. Note that $\Vert\bm\alpha(1)\Vert=O(1)$ and $\sup_R\Vert \bm H_{R}^{-1}\Vert=O_P(1)$. We also recall from \Cref{identity} that
$\tilde{\bm f}_{T,R}-\bm H_R\bm f_T=\tilde{\bm V}_{R}^{-1}(\bm A_{1T}+\bm A_{2T}+\bm A_{3T}+\bm A_{4T}+\bm A_{5T}).$ Using $\sup_R\Vert \tilde{\bm V}_{R}^{-1}\Vert =O_P(1)$ and $\bm A_{1T}=0$, it is enough to show that $\sup_R \Vert\sqrt N\bm A_{jT}\Vert, j=2,3,5,$ and  $\sup_R \Vert\min(\sqrt R\delta_{NRT},R^{7/8-\mu})\bm A_{4T}\Vert$ are $O_P(1)$.

First, we use
\begin{equation}
\sup_R (\Vert \sqrt N\bm A_{2T}\Vert)\leq \sup_R\bigg\Vert\frac{\sqrt{N}}{R}\sum_{s=T-R+1}^{T} (\tilde{\bm f}_s-\bm H_R \bm f_s)\eta_{sT}\bigg\Vert+\sup_R\bigg\Vert\bm H_R\frac{\sqrt{N}}{R}\sum_{s=T-R+1}^{T}  \bm f_s\eta_{sT}\bigg\Vert.
\label{supA2}
\end{equation}
We employ the fact that
$\sup_R\big\Vert\frac{\sqrt{N}}{R}\sum_{s=T-R+1}^{T} (\tilde{\bm f}_s-\bm H_R\bm f_s)\eta_{sT}\big\Vert $
is bounded by
$$ \big( \sup_R\frac{1}{R}\sum_{s=T-R+1}^{T} \Vert\tilde{\bm f}_s-\bm H_R \bm f_s\Vert^2\big)^{1/2} \big( \sup_R\frac{N}{R}\sum_{s=T-R+1}^{T} \Vert\eta_{sT}\Vert^2\big)^{1/2} 
	=O_P(1)O_P(1)=O_P(1).$$
Indeed, $\sup_R\frac{1}{R}\sum_{s=T-R+1}^{T} \Vert\tilde{\bm f}_s-\bm H_R \bm f_s\Vert^2\leq \sup_R\frac{\delta_{NRT}^2}{R}\sum_{s=T-R+1}^{T} \Vert\tilde{\bm f}_s-\bm H_R \bm f_s\Vert^2=O_P(1)$ by \Cref{Lemma4} (3). Because $\sup_R\sup_{s,t}\big\Vert \frac{1}{\sqrt N}\sum_{i=1}^{N}\bm \lambda_i\big(\frac{s}{T}\big)\bm e_{it} \big\Vert^4=O_P(1)$, we also have that $\sup_R\frac{N}{R}\sum_{s=T-R+1}^{T} \Vert\eta_{sT}\Vert^2$ is bounded by
$$ \big(\sup_R\frac{1}{R}\sum_{s=T-R+1}^{T}\Vert\bm f_s\Vert^4\big)^{1/2}\big(\sup_R\frac{1}{R}\sum_{s=T-R+1}^{T}\big\Vert \frac{1}{\sqrt N}\sum_{i=1}^{N}\bm \lambda_i\big(\frac{s}{T}\big)\bm e_{iT} \big\Vert^4\big)^{1/2}=O_P(1),$$
using \Cref{Lemma4} (1) and \Cref{Ass6} (a).

To show that $\sup_R\Vert\sqrt N \bm A_{3T}\Vert=O_P(1),$ we decompose $\sqrt N \bm A_{3T}$ as
$$\frac{\sqrt N}{R}\sum_{s=T-R+1}^{T}\tilde{\bm f}_s\xi_{sT}=\frac{1}{R\sqrt N}\sum_{s=T-R+1}^{T} (\tilde{\bm f}_s-\bm H_R \bm f_s)\bm e_s^\top\bm \Lambda(1)\bm f_T+\bm H_R\frac{1}{R\sqrt N}\sum_{s=T-R+1}^{T}  \bm f_s \bm e_s^\top\bm \Lambda(1)\bm f_T.$$
Next, we use 
$\sup_R\Vert\frac{1}{R\sqrt N}\sum_{s=T-R+1}^{T} (\tilde{\bm f}_s-\bm H_R \bm f_s)\bm e_s^\top\bm \Lambda(1)\bm f_T\Vert\leq 
\big(\sup_R\frac{1}{R}\sum_{s=T-R+1}^{T} \Vert\tilde{\bm f}_s-\bm H_R \bm f_s\Vert^2\big)^{1/2}\big(\sup_R \frac{1}{R}\sum_{s=T-R+1}^{T}\Vert \frac{1}{\sqrt N}\bm e_s^\top\bm \Lambda(1)\bm f_T\Vert^2\big)^{1/2}=O_P(1)$, where we use the fact that
$\sup_R\frac{1}{R}\sum_{s=T-R+1}^{T} \Vert\tilde{\bm f}_s-\bm H_R \bm f_s\Vert^2 =O_P(1)$
and $\sup_R \frac{1}{R}\sum_{s=T-R+1}^{T}\Vert \frac{1}{\sqrt N}\bm e_s^\top\bm \Lambda(1)\bm f_T\Vert^2$ is bounded by 
$ \Vert\bm f_T\Vert^2\sup_R \sup_{s,t}\frac{1}{R}\sum_{t=T-R+1}^{T}\Vert \frac{1}{\sqrt N} \bm e_s^\top \bm \Lambda(\frac{t}{T})\Vert^2=O_P(1),$
using \Cref{Ass6} (a). Also, we note that $\sup_R\Vert\bm H_R\frac{1}{R\sqrt N}\sum_{s=T-R+1}^{T}  \bm f_s \bm e_s^\top\bm \Lambda(1)\bm f_T\Vert $ is bounded by
$$\sup_R\Vert\bm H_R\Vert \big(\sup_R\frac{1}{R}\sum_{s=T-R+1}^{T} \Vert \bm f_s\Vert^2\big)^{1/2}\big(\sup_R\frac{1}{R}\sum_{s=T-R+1}^{T}  \Vert \frac{1}{\sqrt N}\bm e_s^\top\bm \Lambda(1)\bm f_T\Vert^2\big)^{1/2}=O_P(1),$$
because each of the three components is $O_P(1)$. Thus, $\sup_R\Vert\sqrt N \bm A_{3T}\Vert=O_P(1).$ 

Furthermore, $\min(\sqrt R\delta_{NRT},R^{7/8-\mu})\bm A_{4T}=\frac{\min(\sqrt R\delta_{NRT},R^{7/8-\mu})}{R}\sum_{s=T-R+1}^{T} \tilde{\bm f}_s\gamma_N(s,T) $ is
$$\frac{\min(\sqrt R\delta_{NRT},R^{7/8-\mu})}{R}\sum_{s=T-R+1}^{T} (\tilde{\bm f}_s-\bm H_R\bm f_s)\mathrm E(\frac{1}{N}\bm e_s^\top\bm e_T)$$
$$+\bm H_R\frac{\min(\sqrt R\delta_{NRT},R^{7/8-\mu})}{R}\sum_{s=T-R+1}^{T} \bm f_s\mathrm E(\frac{1}{N}\bm e_s^\top\bm e_T).$$
The supremum of the norm of the first term 
is below
$\big( \sup_R\frac{\delta_{NRT}^2}{R}\sum_{s=T-R+1}^{T} \Vert\tilde{\bm f}_s-\bm H_R\bm f_s\Vert^2\big)^{1/2}\times\big(\sup_R\sum_{s=T-R+1}^{T}\vert\mathrm E(\frac{1}{N}\bm e_s^\top\bm e_T)\vert^2\big)^{1/2}=O_P(1)$, as $\sup_R\sum_{s=T-R+1}^{T}\vert\mathrm E(\frac{1}{N}\bm e_s^\top\bm e_t)\vert^2\leq C< \infty$. The supremum of the norm of the second term $\sup_R\Vert \bm H_R\frac{R^{7/8-\mu}}{R}\sum_{s=T-R+1}^{T} \bm f_s\mathrm E(\frac{1}{N}\bm e_s^\top\bm e_T)\Vert$ is bounded by 
$\sup_R\Vert \bm H_R\Vert\big(\sup_R\sup_s(R^{-1/8-\mu}\Vert\bm f_s\Vert)\big)\times\big(\sup_R\sum_{s=T-R+1}^{T}\vert\mathrm E(\frac{1}{N}\bm e_s^\top\bm e_T)\vert\big)=O_P(1),$
because $\mathrm{P}\big( \sup_R\sup_s(R^{-1/8-\mu}\Vert\bm f_s\Vert>\epsilon\big)\leq \vert\Theta_R\vert\sup_{R} \frac{\sum_s\mathrm E(\Vert \bm f_s\Vert^8)}{R^{1+8\mu}\epsilon^8}\leq \frac{\vert\Theta_R\vert C}{\underline R^{8\mu}\epsilon^8}=O(1)$,  using \Cref{Ass1} (a), \Cref{Ass7} (b) for some $\mu\in (0,\frac{1}{8})$,
and from \Cref{Ass1} (g).
Consequently, $\sup_R\Vert\min(\sqrt R\delta_{NRT},R^{7/8-\mu})\bm A_{4t}\Vert=O_P(1)$.
Finally, we write $\sqrt N\bm A_{5t}=\frac{\sqrt N}{R}\sum_{s=T-R+1}^{T} \tilde{\bm f}_s\zeta_{st}$ as
$$\frac{\sqrt N}{R}\sum_{s=T-R+1}^{T} (\tilde{\bm f}_s-\bm H_R\bm f_s)(\bm e_s^\top\bm e_t/N-\gamma_{st})+\bm H_R\frac{\sqrt N}{R}\sum_{s=T-R+1}^{T} \bm f_s(\bm e_s^\top\bm e_t/N-\gamma_{st}).$$
Here, the supremum of the norm of the first term is below $\big( \sup_R\frac{1}{R}\sum_{s=T-R+1}^{T} \Vert\tilde{\bm f}_s-\bm H_R\bm f_s\Vert^2\big)^{1/2}\times\big(\sup_R\frac{1}{R}\sum_{s=T-R+1}^{T}\vert N^{-1/2}(\bm e_s^\top\bm e_T-\mathrm E(\bm e_s^\top\bm e_T))\vert^2\big)^{1/2}=O_P(1)$, because we have from \Cref{Ass6} (b) that $\sup_R\sup_{s,t}\vert N^{-1/2}(\bm e_s^\top\bm e_T-\mathrm E(\bm e_s^\top\bm e_T))\vert=O_P(1)$. The supremum of the norm of the second term 
is bounded by $$\sup_R\Vert \bm H_R\Vert\big( \sup_R\frac{1}{R}\sum_{s=T-R+1}^{T} \Vert\bm f_s\Vert^2\big)^{1/2}\times\big(\sup_R\frac{1}{R}\sum_{s=T-R+1}^{T}\vert N^{-1/2}(\bm e_s^\top\bm e_T-\mathrm E(\bm e_s^\top\bm e_T))\vert^2\big)^{1/2}=O_P(1).$$ Therefore, $\sup_R\Vert\sqrt N\bm A_{5t}\Vert=O_P(1)$.
In conclusion, 
$\sup_R\Vert \min(\sqrt{N},\sqrt R\delta_{NRT},R^{7/8-\mu})(\tilde{\bm f}_{T,R}-\bm H_R\bm f_T)\Vert=O_P(1)$, and we have the result.

\subsubsection*{Proof of \Cref{Lemma4} (6)} 
For $\sup_{R\in \Theta_R} \big\Vert\frac{\min(\phi_{NRT}^{2},R^{\frac{3}{4}-\nu})}{R}\sum_{t=T-R+1}^{T-h}( \tilde{\bm f}_{t,R}-\bm H_{R}\bm f_t)\bm z_{t}^\top\big\Vert=O_P(1),$ for $\nu \in (0,1/4)$, 
we use $$\frac{1}{R}\sum_{t=T-R+1}^{T-h}( \tilde{\bm f}_{t,R}-\bm H_{R}\bm f_t)\bm z_{t}^\top=\tilde{\bm V}^{-1}_{R}\frac{1}{R}\sum_{t=T-R+1}^{T-h}(\bm A_{1t}\bm z_t^\top+\bm A_{2t}\bm z_t^\top+\bm A_{3t}\bm z_t^\top+\bm A_{4t}\bm z_t^\top+\bm A_{5t}\bm z_t^\top),$$
and we complete the proofs by showing that $\sup_{R\in \Theta_R} \big\Vert\frac{\phi_{NRT}^{2}}{R}\sum_{t=T-R+1}^{T-h}\bm A_{jt}\bm z_t^\top\big\Vert=O_P(1), j=1,2,3,5,$
and $\sup_{R\in \Theta_R} \big\Vert\frac{\min(\phi_{NRT}^{2},R^{\frac{3}{4}-\nu})}{R}\sum_{t=T-R+1}^{T-h}\bm A_{4t}\bm z_t^\top\big\Vert=O_P(1).
$


We observe that $\frac{\phi_{NRT}^2}{R}\sum_{t=T-R+1}^{T-h}\bm A_{1t}\bm{z}_t^\top$ equals
\begin{align}
	&\frac{\phi_{NRT}^2}{R}\sum_{s=T-R+1}^{T} (\tilde{\bm f}_s-\bm H_R\bm f_s)\frac{1}{NR}\sum_{t=T-R+1}^{T-h}\big(\bm f_s^\top\bm \Lambda^\top(s/T) [\Lambda(t/T)-\Lambda(1)]\bm f_{t}\big)\bm{z}_t^\top	\label{fz1sup}\\
	&+	\bm H_R\frac{\phi_{NRT}^2}{R}\sum_{s=T-R+1}^{T} (\bm f_s\bm f_s^\top)\frac{1}{NR}\sum_{t=T-R+1}^{T-h}\bm \Lambda^\top(s/T) [\Lambda(t/T)-\Lambda(1)]\bm f_{t}\bm{z}_t^\top.
	\label{fz2sup}
\end{align}
The supremum of the norm of \Cref{fz1sup}
is bounded by $\big(\frac{\phi_{NRT}^2}{R}\sum_{s=T-R+1}^{T} \Vert\tilde{\bm f}_s-\bm H_R\bm f_s\Vert^2\big)^{\frac{1}{2}}$ times
$$\big(\frac{\phi_{NRT}^2}{R}\sum_{s=T-R+1}^{T}\big\Vert\frac{1}{NR}\sum_{t=T-R+1}^{T-h}\bm f_s^\top\bm \Lambda^\top(s/T) [\Lambda(t/T)-\Lambda(1)]\bm f_{t}\bm{z}_t^\top\big\Vert^2\big)^{\frac{1}{2}},$$
which is $O_P(1)$, as $\sup_R\frac{1}{R}\sum_{s=T-R+1}^{T}\Vert\tilde{\bm f}_s-\bm H_R\bm f_s\Vert^2=O_P(1)$ from \Cref{Lemma4} (3) and 
$$\sup_R\frac{\phi_{NRT}^2}{R}\sum_{s=T-R+1}^{T}\big\Vert\frac{1}{NR}\sum_{t=T-R+1}^{T-h}\bm f_s^\top\bm \Lambda^\top(s/T) [\Lambda(t/T)-\Lambda(1)]\bm f_{t}\bm{z}_t^\top\Vert=O_P(1),$$
using a similar decomposition as $\bm B_1$ in the proof of \Cref{Lemma2} (5).
To show that the supremum of the norm of \Cref{fz2sup} is $O_P(1)$, we also follow similar steps as for $\bm B_2$ in the proof of \Cref{Lemma2} (5).
Thus, \Cref{fz2sup} is $O_P(1)$. Because, \Cref{fz1sup} is $O_P(1)$, we deduce that $\sup_R\Vert\frac{\phi^{2}_{NRT}}{R}\sum_{t=T-R+1}^{T-h}\bm A_{1t}\bm z_t^\top\Vert=O_P(1).$

For $\sup_R\Vert \frac{\phi_{NRT}^{2}}{R}\sum_{t=T-R+1}^{T-h}\bm A_{2t}\bm{z}_t^\top\Vert =	\sup_R\Vert\frac{\phi_{NRT}^{2}}{R^2}\sum_{t=T-R+1}^{T-h}\sum_{s=T-R+1}^{T} \tilde{\bm f}_s\eta_{st}\bm{z}_t^\top\Vert$, we use the decomposition of $\frac{\phi_{NRT}^{2}}{R}\sum_{t=T-R+1}^{T-h}\bm A_{2t}\bm{z}_t^\top $ as
\begin{align}
	&\frac{\phi_{NRT}^{2}}{NR^2}\sum_{s=T-R+1}^{T}\sum_{t=T-R+1}^{T-h} (\tilde{\bm f}_s-\bm H_{R}\bm f_s)\big(\bm f_s^\top\bm \Lambda^\top\big(\frac{s}{T}\big)\bm e_{t}\big)\bm{z}_t^\top\label{A2z1sup}\\
	&+\bm H_{R}\frac{\phi_{NRT}^{2}}{NR^2}\sum_{s=T-R+1}^{T}\sum_{t=T-R+1}^{T-h}(\bm f_s\bm f_s^\top)\bm \Lambda^\top\big(\frac{s}{T}\big)\bm e_{t}\bm{z}_t^\top\label{A2z2sup},
\end{align}
where the supremum of their norms is $O_P(1)$ using
similar steps as for \eqref{A2z1} and \eqref{A2z2}. Thus, $\sup_R\Vert\frac{\phi_{NRT}^{2}}{R}\sum_{t=T-R+1}^{T-h}\bm A_{2t}\bm{z}_t^\top\Vert=O_P(1)$.
Similarly, 
 $\sup_R\Vert\frac{\phi_{NRT}^{2}}{R}\sum_{t=T-R+1}^{T-h}\bm A_{3t}\bm{z}_t^\top\Vert=O_P(1).$
 
 To show that
 $\sup_R\Vert\frac{\min(\phi_{NRT}^{2},R^{\frac{3}{4}-\nu})}{R}\sum_{t=T-R+1}^{T-h}\bm A_{4t}\bm{z}_t^\top\Vert=O_P(1)$, we decompose $\frac{\min(\phi_{NRT}^{2},R^{\frac{3}{4}-\nu})}{R}\times\sum_{t=T-R+1}^{T-h}\bm A_{4t}\bm{z}_t^\top$ as
 \begin{align*}
&
\frac{\min(\phi_{NRT}^{2},R^{\frac{3}{4}-\nu})}{R^2}\sum_{s=T-R+1}^{T-h}\sum_{t=T-R+1}^{T-h}(\tilde{\bm f}_{s,R}-\bm H_R\bm f_s)(\mathrm{E}(\bm e_s^\top\bm e_t/N))\bm{z}_t^\top\\
&+\bm H_R\frac{\min(\phi_{NRT}^{2},R^{\frac{3}{4}-\nu})}{R^2}\sum_{s=T-R+1}^{T}\sum_{t=T-R+1}^{T-h}\bm f_s(\mathrm{E}(\bm e_s^\top\bm e_t/N))\bm{z}_t^\top.
\end{align*}
For the first component, we observe that its norm is bounded by
$$ \big(\frac{\phi_{NRT}^{2}}{R}\sum_{s=T-R+1}^T \Vert\tilde{\bm f}_{t,R}-\bm H_R\bm f_s \Vert^2 \big)^{\frac{1}{2}}\big(\frac{\phi_{NRT}^{2}}{R^2}\sum_{s=T-R+1}^T\sum_{t=T-R+1}^{T-h}\vert\mathrm{E}(\bm e_s^\top\bm e_t/N)\vert^2\big)^{\frac{1}{2}}\big(\frac{1}{R}\sum_{t=T-R+1}^{T-h} \Vert\bm z_{t}\Vert^2\big)^{\frac{1}{2}},$$
which is lower than
$ \big(\frac{\phi_{NRT}^{2}}{R}\sum_{s=T-R+1}^T \Vert\tilde{\bm f}_{t,R}-\bm H_R\bm f_s \Vert^2 \big)^{\frac{1}{2}}\big(\frac{1}{R}\sum_{s=T-R+1}^T\sum_{t=T-R+1}^{T-h}\vert\mathrm{E}(\bm e_s^\top\bm e_t/N)\vert\big)^{\frac{1}{2}}\times \big(\frac{1}{R}\sum_{t=T-R+1}^{T-h} \Vert\bm z_{t}\Vert^2\big)^{\frac{1}{2}}$. Thus, $\sup_R\Vert\frac{\phi_{NRT}^{2}}{R^2}\sum_{s=T-R+1}^{T-h}\sum_{t=T-R+1}^{T-h}(\tilde{\bm f}_{s,R}-\bm H_R\bm f_s)(\mathrm{E}(\bm e_s^\top\bm e_t/N))\bm{z}_t^\top\Vert=O_P(1)O_P(1)O_P(1)=O_P(1),$
using \Cref{Lemma4} (3), \citet[Lemma 1 (i)]{BN2002} and \Cref{Ass1} (e). The proof that $\sup_R\frac{1}{R}\sum_{t=T-R+1}^{T-h} \Vert\bm z_{t}\Vert^2=O_P(1)$ follows similar steps at the proof of \Cref{Lemma4} (1).
For the second component, the sup of its norm is bounded by
$\sup_R\Vert\bm H_R\Vert(\sup_R R^{-1/4-\nu}\sup_{s,t}\Vert \bm f_s \Vert \Vert \bm z_t \Vert )\sup_R (
\frac{1}{R}\sum_{s=T-R+1}^{T}\sum_{t=T-R+1}^{T-h}(\vert\mathrm{E}(\bm e_s^\top\bm e_t/N)))\vert.$ 
Because $\mathrm{P}\big( \sup_R\sup_{s,t}(R^{-1/4-\nu}\Vert\bm f_s\Vert\Vert\bm z_t\Vert>\epsilon\big)\leq \vert\Theta_R\vert\sup_{R} \frac{\sum_{s,t}\mathrm (E(\Vert \bm f_s\Vert^8))^{1/2}(E(\Vert \bm z_t\Vert^8))^{1/2}}{R^{2+4\nu}\epsilon^4}\leq \frac{\vert\Theta_R\vert C}{\underline R^{4\nu}\epsilon^4}=O(1)$,  using \Cref{Ass4} (c), \Cref{Ass7} (b) for some $\nu\in (0,\frac{1}{4})$,
and from \Cref{Ass1} (e), we have 
$\sup_R R^{-1/4-\nu}\sup_{s,t}\Vert \bm f_s \Vert \Vert \bm z_t \Vert =O_P(1)$. Also, we have from \Cref{Ass1} (e) that $\sup_R (
\frac{1}{R}\sum_{s=T-R+1}^{T}\sum_{t=T-R+1}^{T-h}(\vert\mathrm{E}(\bm e_s^\top\bm e_t/N)\vert))\leq C$.
As a result, we have that
$$\sup_R\Vert\bm H_R\frac{\min(\phi_{NRT}^{2},R^{\frac{3}{4}-\nu})}{R^2}\sum_{s=T-R+1}^{T}\sum_{t=T-R+1}^{T-h}\bm f_s(\mathrm{E}(\bm e_s^\top\bm e_t/N))\bm{z}_t^\top\Vert=O_P(1).$$
Thus,  $\sup_R\Vert\frac{\min(\phi_{NRT}^{2},R^{\frac{3}{4}-\nu})}{R}\sum_{t=T-R+1}^{T-h}\bm A_{4t}\bm{z}_t^\top\Vert=O_P(1)$.

Furthermore, we have 
\begin{align*}
\frac{\phi_{NRT}^{2}}{R}\sum_{t=T-R+1}^{T-h}\bm A_{5t}\bm{z}_t^\top
&=
\frac{\phi_{NRT}^{2}}{R^2}\sum_{s=T-R+1}^{T-h}\sum_{t=T-R+1}^{T-h}(\tilde{\bm f}_{s,R}-\bm H_R\bm f_s)(\bm e_s^\top\bm e_t/N-\gamma_{st})\bm{z}_t^\top\\
&+\bm H_R\frac{\phi_{NRT}^{2}}{R^2}\sum_{s=T-R+1}^{T}\sum_{t=T-R+1}^{T-h}\bm f_s(\bm e_s^\top\bm e_t/N-\gamma_{st})\bm{z}_t^\top.
\end{align*}
We note the first term is lower than $\big(\frac{\phi_{NRT}^{2}}{R}\sum_{s=T-R+1}^T \Vert\tilde{\bm f}_{t,R}-\bm H_R\bm f_s \Vert^2 \big)^{1/2}$ times
$$ \big(\frac{\phi_{NRT}^{2}}{R^2}\sum_{s=T-R+1}^T\sum_{t=T-R+1}^{T-h}\Vert\bm e_s^\top\bm e_t/N-\gamma_{st}\Vert^2\big)^{1/2}\big(\frac{1}{R}\sum_{t=T-R+1}^{T-h} \Vert\bm z_{t}\Vert^2\big)^{1/2}.$$
Thus, its sup is bounded by
 $\big(\sup_R\frac{\delta_{NRT}^{2}}{R}\sum_{s=T-R+1}^T \Vert\tilde{\bm f}_{t,R}-\bm H_R\bm f_s \Vert^2 \big)^{1/2}\big(\sup_R\sup_{s,t}\vert\frac{1}{\sqrt N}(\bm e_s^\top\bm e_t-\mathrm{E}(\bm e_s^\top\bm e_t)\vert^2\big)^{1/2}\big(\sup_R\frac{1}{R}\sum_{t=T-R+1}^{T-h} \Vert\bm z_{t}\Vert^2\big)^{1/2}=O_P(1)O_P(1)O_P(1)=O_P(1)$.
For the second term, we note that
$\sup_R\Vert\bm H_R\frac{\phi_{NRT}^{2}}{R^2}\sum_{s=T-R+1}^{T}\sum_{t=T-R+1}^{T-h}\bm f_s(\bm e_s^\top\bm e_t/N-\gamma_{st})\bm{z}_t^\top\Vert$ is lower than $
\sup_R\Vert\bm H_R\Vert (\sup_R\frac{1}{R}\sum_{s=T-R+1}^{T} \Vert\bm f_{s}\Vert^2)^{\frac{1}{2}}
(\sup_R\frac{1}{R}\sum_{s=T-R+1}^{T}\Vert\frac{1}{\sqrt{RN}}\sum_{t=T-R+1}^{T-h}(\bm e_s^\top\bm e_t-\mathrm{E}(\bm e_s^\top\bm e_t))\bm z_t\Vert^2)^{\frac{1}{2}},$ which is $
O_P(1),$
using \Cref{Ass6} (c).
Hence, $\sup_R\Vert\frac{\phi_{NRT}^{2}}{R}\sum_{t=T-R+1}^{T-h}\bm A_{5t}\bm{z}_t^\top\Vert=O_P(1)$.
Therefore, we conclude 
$\sup_R\Vert\frac{\min(\phi_{NRT}^{2},R^{\frac{3}{4}-\nu})}{R}\sum_{t=T-R+1}^{T-h}(\tilde{\bm f}_{t,R}-\bm H_R\bm f_t)\bm z^\top_t\Vert=O_P(1)$.

\subsubsection*{Proof of \Cref{Lemma4} (7)} 
For $\sup_{R\in \Theta_R} \big\Vert\min(\phi_{NRT}^{2},R^{\frac{3}{4}-\nu})\frac{1}{R}\sum_{t=T-R+1}^{T-h}( \tilde{\bm f}_{t,R}-\bm H_{R}\bm f_t)\hat{\bm z}_{t}^\prime\big\Vert=O_P(1),$
we use the fact that $\frac{\min(\phi_{NRT}^{2},R^{\frac{3}{4}-\nu})}{R}\sum_{t=T-R+1}^{T-h}( \tilde{\bm f}_{t,R}-\bm H_{R}\bm f_t)\hat{\bm z}_{t}^\prime$ equals
$\frac{\min(\phi_{NRT}^{2},R^{\frac{3}{4}-\nu})}{R}\sum_{t=T-R+1}^{T-h}( \tilde{\bm f}_{t,R}-\bm H_{R}\bm f_t)\bm z_{t}^\prime \Phi_{R}^\prime+\frac{\min(\phi_{NRT}^{2},R^{\frac{3}{4}-\nu})}{R}\sum_{t=T-R+1}^{T-h}( \tilde{\bm f}_{t,R}-\bm H_{R}\bm f_t)(\hat{\bm z}_{t}-\bm \Phi_{R} \bm z_t)^\prime.$ In particular, we use the fact that 
$\sup_{R} \big\Vert\frac{\min(\phi_{NRT}^{2},R^{\frac{3}{4}-\nu})}{R}\sum_{t=T-R+1}^{T-h}( \tilde{\bm f}_{t,R}-\bm H_{R}\bm f_t)\hat{\bm z}_{t}^\prime\big\Vert\leq\sup_{R} \big\Vert\frac{\min(\phi_{NRT}^{2},R^{\frac{3}{4}-\nu})}{R}\times\sum_{t=T-R+1}^{T-h}( \tilde{\bm f}_{t,R}-\bm H_{R}\bm f_t)\bm z_{t}^\prime \big\Vert\sup_R\Vert \bm\Phi_R\Vert + \sup_{R} \frac{\delta_{NRT}^{2}}{R}\sum_{t=T-R+1}^{T-h}\big\Vert \tilde{\bm f}_{t,R}-\bm H_{R}\bm f_t\big\Vert^2=O_P(1).$

\subsubsection*{Proof of \Cref{Lemma4} (8)} 
The proof of $\sup_{R\in \Theta_R} \big\Vert\min(\phi_{NRT}^{2},R^{\frac{3}{4}-\nu})\frac{1}{R}\sum_{t=T-R+1}^{T-h}( \tilde{\bm f}_{t,R}-\bm H_{R}\bm f_t)\varepsilon_{t+h}\big\Vert=O_P(1)$ follows the same steps as in the proof of \Cref{Lemma4} (6) after replacing $\bm z_t$ with $\varepsilon_{t+h}$.
 
\subsubsection*{Proof of \Cref{Lemma4} (9)} 
 To show that $\sup_{R}\big\Vert \min(R^{\frac{1}{2}-\theta},\phi_{NRT}^2) \frac{1}{R}\sum_{t=T-R+1}^{T-h}\hat{\bm z}_{t,R}\varepsilon_{t+h}\big\Vert=O_P(1)$ for some $\theta \in (0,1/2),$ we observe that $\sup_{R}\big\Vert \min(R^{\frac{1}{2}-\theta},\phi_{NRT}^2) \frac{1}{R}\sum_{t=T-R+1}^{T-h}\hat{\bm z}_{t,R}\varepsilon_{t+h}\big\Vert$ is bounded above by
 $$ \sup_{R}\big\Vert R^{-\frac{1}{2}-\theta} \sum_{t=T-R+1}^{T-h}\bm\Phi_{R}\bm z_t\varepsilon_{t+h}\big\Vert+\sup_{R}\big\Vert  \frac{\min(\phi_{NRT}^{2},R^{\frac{3}{4}-\nu})}{R}\sum_{t=T-R+1}^{T-h}(\hat{\bm z}_{t,R}-\bm\Phi_{R} \bm z_t)\varepsilon_{t+h} \big\Vert,$$ for some $\nu\in (0,1/4)$.
 Note that 
 $$\sup_{R}\big\Vert R^{-\frac{1}{2}-\theta} \bm\Phi_{R}\sum_{t=T-R+1}^{T-h}\bm z_t\varepsilon_{t+h}\big\Vert\leq    \sup_{R}\big\Vert\bm\Phi_{R}\big\Vert \times \sup_{R}\big\Vert R^{-\frac{1}{2}-\theta} \sum_{t=T-R+1}^{T-h}\bm z_t\varepsilon_{t+h}\big\Vert.$$
Because, $\sup_{R}\big\Vert\bm\Phi_{R}\big\Vert=O_P(1)$ and $\sup_{R}\big\Vert R^{-\frac{1}{2}-\theta} \sum_{t=T-R+1}^{T-h}\bm z_t\varepsilon_{t+h}\big\Vert=O_P(1)$ from \Cref{Lemma1} (2), we deduce that $ \sup_{R}\big\Vert R^{-\frac{1}{2}-\theta} \bm\Phi_{R}\frac{1}{R}\sum_{t=T-R+1}^{T-h}\bm z_t\varepsilon_{t+h}\big\Vert=O_P(1).$ In addition, we note that 
$ \sup_{R}\big\Vert  \frac{\min(\phi_{NRT}^{2},R^{\frac{3}{4}-\nu})}{R}\sum_{t=T-R+1}^{T-h}(\hat{\bm z}_{t,R}-\bm\Phi_{R} \bm z_t)\varepsilon_{t+h} \big\Vert=O_P(1)$ from \Cref{Lemma4} (8), and conclude that $\min(R^{\frac{1}{2}-\theta},\phi_{NRT}^2) \big\Vert\frac{1}{R}\sum_{t=T-R+1}^{T-h}\hat{\bm z}_{t,R}\varepsilon_{t+h}\big\Vert=O_P(1).$

 \subsubsection*{Proof of \Cref{Lemma4} (10)} 
 To show that $\sup_{R}\big\Vert \frac{1}{R}\sum_{t=T-R+1}^{T-h}\hat{\bm z}_{t,R}\hat{\bm z}_{t,R}^\top\big\Vert=O_P(1)$, we observe that is bounded by
 $$\sup_{R}\big\Vert\frac{1}{R}\sum_{t=T-R+1}^{T-h}\hat{\bm z}_{t,R}\hat{\bm z}_{t,R}^\top
-
\bm \Phi_R^{\top}\frac{1}{R}\sum_{t=T-R+1}^{T-h}\mathrm{E}(\bm z_{t}\bm z_{t}^\top)\bm \Phi_R \big\Vert+\sup_R \big\Vert 
 \bm \Phi_R^{\top}\frac{1}{R}\sum_{t=T-R+1}^{T-h}\mathrm{E}(\bm z_{t}\bm z_{t}^\top)\bm \Phi_R \big\Vert,$$
and it is enough to show that each term is $O_P(1)$.

 We note that 
$\frac{1}{R}\sum_{t=T-R+1}^{T-h}\hat{\bm z}_{t,R}\hat{\bm z}_{t,R}^\top
-
\bm \Phi_R^{\top}\big(\frac{1}{R}\sum_{t=T-R+1}^{T-h}\mathrm{E}(\bm z_{t}\bm z_{t}^\top)\big)\bm \Phi_R$
can be written as
$$\big(\frac{1}{R}\sum_{t=T-R+1}^{T-h}\hat{\bm z}_{t,R}\hat{\bm z}_{t,R}^\top
-\bm \Phi_R^{\top}\big(\frac{1}{R}\sum_{t=T-R+1}^{T-h}\bm z_{t}\bm z_{t}^\top\big)\bm \Phi_R
\big)
+\bm \Phi_R^{\top}\big(\frac{1}{R}\sum_{t=T-R+1}^{T-h}(\bm z_{t}\bm z_{t}^\top-\mathrm{E}(\bm z_{t}\bm z_{t}^\top))\big)\bm \Phi_R.$$
Thus, the supremum of the norm is bounded by
$\sup_R\min(\phi_{NRT}^2,R^{\frac{1}{2}-\theta})\Vert\frac{1}{R}\sum_{t=T-R+1}^{T-h}\hat{\bm z}_{t,R}\hat{\bm z}_{t,R}^\top
-
\bm \Phi_R^{\top}\big(\frac{1}{R}\sum_{t=T-R+1}^{T-h}\mathrm{E}(\bm z_{t}\bm z_{t}^\top)\big)\bm \Phi_R\Vert$, which is below 
$\sup_R\Vert \min(\phi_{NRT}^{2},R^{\frac{3}{4}-\nu})\times 
\big(\frac{1}{R}\sum_{t=T-R+1}^{T-h}\hat{\bm z}_{t,R}\hat{\bm z}_{t,R}^\top
-\bm \Phi_R^{\top} \big(\frac{1}{R}\sum_{t=T-R+1}^{T-h}\bm z_{t}\bm z_{t}^\top\big)\bm \Phi_R
\big)\Vert+\sup_R\Vert\bm \Phi_R^{\top}\big(\frac{R^{\frac{1}{2}-\theta}}{R}\sum_{t=T-R+1}^{T-h}(\bm z_{t}\bm z_{t}^\top-\mathrm{E}(\bm z_{t}\bm z_{t}^\top))\big)\bm \Phi_R\Vert$.
Using the decomposition in the proof of \Cref{Lemma2} (17), 
$\sup_R\big\Vert\min(\phi_{NRT}^{2},R^{\frac{3}{4}-\nu})\big(\frac{1}{R}\sum_{t=T-R+1}^{T-h}\hat{\bm z}_{t,R}\hat{\bm z}_{t,R}^\top-\bm \Phi_R^{\top}\big(\frac{1}{R}\sum_{t=T-R+1}^{T-h}\bm z_{t}\bm z_{t}^\top\big)\bm \Phi_R\big)\big\Vert$ 
is lower than
 $2\sup_R\big\Vert\frac{\min(\phi_{NRT}^{2},R^{\frac{3}{4}-\nu})}{R}\sum_{t=T-R+1}^{T-h}(\hat{\bm z}_{t,R}-\bm\Phi_R\bm z_t)\bm z_{t}^\top\big\Vert\times \sup_R\big\Vert\bm\Phi_R\big\Vert+ \sup_R\frac{\phi_{NRT}^2}{R}\sum_{t=T-R+1}^{T-h}\big\Vert\hat{\bm z}_{t}-\bm\Phi_R\bm z_t\big\Vert^2 \sup_R\Vert\bm\Phi_R\Vert=O_P(1),
$
where we use \Cref{Lemma4} (6) and (3), to have 
$$\sup_R\big\Vert\frac{\min(\phi_{NRT}^{2},R^{\frac{3}{4}-\nu})}{R}\sum_{t=T-R+1}^{T-h}(\hat{\bm z}_{t,R}-\bm\Phi_R\bm z_t)\bm z_{t}^\top\big\Vert =O_P(1) $$ and $$\sup_R\frac{\phi_{NRT}^2}{R}\sum_{t=T-R+1}^{T-h}\big\Vert\hat{\bm z}_{t,R}-\bm\Phi_R\bm z_t\big\Vert^2 \leq \sup_R\frac{\delta_{NRT}^2}{R}\sum_{t=T-R+1}^{T-h}\big\Vert\tilde{\bm f}_{t,R}-\bm H_R\bm f_t\big\Vert^2 =O_P(1).$$ Therefore,
 $$\sup_R\big\Vert\min(\phi_{NRT}^2,R^{\frac{1}{2}-\theta})\big(\frac{1}{R}\sum_{t=T-R+1}^{T-h}\hat{\bm z}_{t,R}\hat{\bm z}_{t,R}^\top-\bm \Phi_R^{\top}\big(\frac{1}{R}\sum_{t=T-R+1}^{T-h}\bm z_{t}\bm z_{t}^\top\big)\bm \Phi_R\big)\big\Vert=O_P(1).$$ In addition,
 $\sup_R\Vert\bm \Phi_R^{\top}\big(\frac{R^{\frac{1}{2}-\theta}}{R}\sum_{t=T-R+1}^{T-h}(\bm z_{t}\bm z_{t}^\top-\mathrm{E}(\bm z_{t}\bm z_{t}^\top))\big)\bm \Phi_R\Vert
 \leq 
 \sup_R\Vert\frac{R^{\frac{1}{2}-\theta}}{R}\sum_{t=T-R+1}^{T-h}(\bm z_{t}\bm z_{t}^\top-\mathrm{E}(\bm z_{t}\bm z_{t}^\top))\sup_R\Vert\bm \Phi_R\Vert^2
 =O_P(1)O_P(1)=O_P(1),$ where we use \Cref{Lemma1} (4). Thus, 
\begin{equation}
\sup_R\Vert\min(\phi_{NRT}^2,R^{\frac{1}{2}-\theta})\bigg(\frac{1}{R}\sum_{t=T-R+1}^{T-h}\hat{\bm z}_{t,R}\hat{\bm z}_{t,R}^\top
-
\bm \Phi_R^{\top}\frac{1}{R}\sum_{t=T-R+1}^{T-h}\mathrm{E}(\bm z_{t}\bm z_{t}^\top)\bm \Phi_R\bigg)\Vert=O_P(1).
\label{zznew}
\end{equation}

To complete the proof, we note that 
$$\sup_R\big\Vert\bm\Phi_R^{\top}\big(\frac{1}{R}\sum_{t=T-R+1}^{T-h} \mathrm{E}(\bm z_{t}\bm z_{t}^\top)\big)\bm\Phi_R \big\Vert
\leq \sup_R\big\Vert\bm \Phi_R \big\Vert^2 \sup_R\frac{1}{R}\sum_{t=T-R+1}^{T-h}\big\Vert\mathrm{E}(\bm z_{t}\bm z_{t}^\top)\big\Vert=O_P(1),$$
using the fact that is uniformly bounded in \Cref{Ass4} (d).

\subsubsection*{Proof of \Cref{Lemma4} (11)} 
 To get $\sup_{R}\big\Vert \big(\frac{1}{R}\sum_{t=T-R+1}^{T-h}\hat{\bm z}_{t,R}\hat{\bm z}_{t,R}^\top\big)^{-1}\big\Vert=O_P(1)$,
we note that
$\sup_{R}\big\Vert \big(\frac{1}{R}\sum_{t=T-R+1}^{T-h}\hat{\bm z}_{t,R}\hat{\bm z}_{t,R}^\top\big)^{-1}\big\Vert
$
is bounded by
\begin{align*}
& \sup_{R}\big\Vert\big(\frac{1}{R}\sum_{t=T-R+1}^{T-h}\hat{\bm z}_{t,R}\hat{\bm z}_{t,R}^\prime\big)^{-1}-\big(\frac{1}{R}\sum_{t=T-R+1}^{T-h}\bm\Phi_R\mathrm{E}(\bm z_{t}\bm z_{t}^\prime)\bm\Phi_R^\prime\big)^{-1}\big\Vert +
\sup_{R}\big\Vert\big(\frac{1}{R}\sum_{t=T-R+1}^{T-h}\bm\Phi_R\mathrm{E}(\bm z_{t}\bm z_{t}^\prime)\bm\Phi_R^\prime\big)^{-1}\big\Vert.
\end{align*}
Combining \Cref{zznew} and \Cref{Lemma4} (4), the first term is $O_P(1)$.
Because 
$$\sup_{R}\big\Vert\big(\frac{1}{R}\sum_{t=T-R+1}^{T-h}\bm\Phi_R\mathrm{E}(\bm z_{t}\bm z_{t}^\prime)\bm\Phi_R^\prime\big)^{-1}\big\Vert\leq \sup_{R}\big\Vert\big(\frac{1}{R}\sum_{t=T-R+1}^{T-h}\mathrm{E}(\bm z_{t}\bm z_{t}^\prime)\big)^{-1}\big\Vert\sup_R\Vert \bm\Phi_R^{-1}\Vert^2=O_P(1),$$
we have the results.

 \subsubsection*{Proof of \Cref{Lemma4} (12)}

To show $\sup_R\bigg\Vert \frac{T}{R^2}\sum_{t=T-R+1}^{T-h}\hat{\bm z}_{t,R}\hat{\bm z}_{t,R}^\top(\bm\delta_R(t/T)-\bm\delta_R(1))\bigg\Vert =O_P(1)$, we use the fact that
\begin{align*}
&\sup_R\bigg\Vert \frac{T}{R^2}\sum_{t=T-R+1}^{T-h}\hat{\bm z}_{t,R}\hat{\bm z}_{t,R}^\top(\bm\delta_R(t/T)-\bm\delta_R(1))\bigg\Vert\\ 
&\leq \sup_R\bigg\Vert \frac{T}{R^2}\sum_{t=T-R+1}^{T-h}\hat{\bm z}_{t,R}\hat{\bm z}_{t,R}^\top\frac{t-T}{T}\bigg\Vert\Vert\bm\delta^{0\prime}(1)\Vert\sup_R\Vert\bm\Phi_R^{-1}\Vert\\
&+ \sup_R\bigg\Vert \frac{T}{R^2}\sum_{t=T-R+1}^{T-h}\hat{\bm z}_{t,R}\hat{\bm z}_{t,R}^\top\frac{(t-T)^2}{T^2}\bigg\Vert\Vert\bm\delta^{0\prime\prime}(d)\Vert\sup_R\Vert\bm\Phi_R^{-1}\Vert,
\end{align*}
where similar steps as for the previous results lead to $\sup_R\bigg\Vert \frac{T}{R^2}\sum_{t=T-R+1}^{T-h}\hat{\bm z}_{t,R}\hat{\bm z}_{t,R}^\top\frac{t-T}{T}\bigg\Vert=O_P(1)$ and
$\sup_R\bigg\Vert \frac{T}{R^2}\sum_{t=T-R+1}^{T-h}\hat{\bm z}_{t,R}\hat{\bm z}_{t,R}^\top\frac{(t-T)^2}{T^2}\bigg\Vert=O_P(1)$.
Thus, we have the results.

 \subsubsection*{Proof of \Cref{Lemma4} (13)}

We note that  $\sup_R\bigg\Vert
\frac{\min(\phi_{NRT}^{2},R^{\frac{3}{4}-\nu})}{R}\sum_{t=T-R+1}^{T-h}\hat{\bm z}_{t,R}(\tilde{\bm f}_{t,R}-\bm H_R  \bm f_t)^\top(\bm H_R^{\top} )^{-1}\bm\alpha(t/T)
\bigg\Vert$ is bounded by
\begin{align*}
&     \sup_R\bigg\Vert
\frac{\min(\phi_{NRT}^{2},R^{\frac{3}{4}-\nu})}{R}\sum_{t=T-R+1}^{T-h}\hat{\bm z}_{t,R}(\tilde{\bm f}_{t,R}-\bm H_R  \bm f_t)^\top
\bigg\Vert\sup_R\Vert (\bm H_R^{\top} )^{-1} \Vert \Vert \bm\alpha(1)\Vert \\
&+
\sup_R\bigg\Vert
\frac{\min(\phi_{NRT}^{2},R^{\frac{3}{4}-\nu})}{R}\sum_{t=T-R+1}^{T-h}\hat{\bm z}_{t,R}(\tilde{\bm f}_{t,R}-\bm H_R  \bm f_t)^\top \frac{t-T}{T}
\bigg\Vert \sup_R\Vert (\bm H_R^{\top} )^{-1} \Vert \Vert \bm\alpha^\prime(1)\Vert
\\
&+
 \sup_R\bigg\Vert
\frac{\min(\phi_{NRT}^{2},R^{\frac{3}{4}-\nu})}{R}\sum_{t=T-R+1}^{T-h}\hat{\bm z}_{t,R}(\tilde{\bm f}_{t,R}-\bm H_R  \bm f_t)^\top \frac{(t-T)^2}{T^2}
\bigg\Vert \sup_R\Vert (\bm H_R^{\top} )^{-1} \Vert \Vert \bm\alpha^{\prime\prime}(d)\Vert.
\end{align*}
Using similar steps as in previous results, they are bounded in probability. We conclude 
 $\sup_R\bigg\Vert
\frac{\min(\phi_{NRT}^{2},R^{\frac{3}{4}-\nu})}{R}\sum_{t=T-R+1}^{T-h}\hat{\bm z}_{t,R}(\tilde{\bm f}_{t,R}-\bm H_R  \bm f_t)^\top(\bm H_R^{\top} )^{-1}\bm\alpha(t/T)
\bigg\Vert=O_p(1)$.

\subsubsection*{Proof of \Cref{Lemma4} (14)} 
Here, our goal is to show that $\sup_{R}\big(\min\{R^{1/2-\theta},\frac{T}{R},N\}\hat{\bm z}_{T,R}^\top(\hat{\bm\delta}_R(1)-\bm\delta_R(1))\big)=O_P(1)$.
Recall that $\hat{\bm\delta}_R(1)-\bm\delta_R(1)=\bm C_1+\bm C_2-\bm C_3.$  Therefore, $\sup_{R}\Vert \min(R^{\frac{1}{2}-\theta},\frac{T}{R}, N)(\hat{\bm\delta}_R(1)-\bm\delta_R(1))\Vert\leq\sup_{R}\Vert \min(R^{\frac{1}{2}-\theta},\phi_{NRT}^2)\bm C_1\Vert+\sup_{R} \Vert \frac{T}{R}\bm C_2\Vert+\sup_{R}\Vert \min(\phi_{NRT}^{2},R^{\frac{3}{4}-\nu})\bm C_3\Vert.$ 
For the first term, 
$\sup_{R}\Vert \min(R^{\frac{1}{2}-\theta},\phi_{NRT}^2) \bm C_1\Vert $ is lower than 
$$ \sup_{R}\bigg\Vert \bigg(\frac{1}{R}\sum_{t=T-R+1}^{T-h}\hat{\bm z}_{t,R}\hat{\bm z}_{t,R}^\top\bigg)^{-1}\bigg\Vert \sup_{R}\bigg\Vert \min(R^{\frac{1}{2}-\theta}, \phi_{NRT}^2) \frac{1}{R}\sum_{t=T-R+1}^{T-h}\hat{\bm z}_{t,R}\varepsilon_{t+h}\bigg\Vert,$$
where we know that from \Cref{Lemma4} (11) and (9) that $\sup_{R}\big\Vert \big(\frac{1}{R}\sum_{t=T-R+1}^{T-h}\hat{\bm z}_{t,R}\hat{\bm z}_{t,R}^\top\big)^{-1}\big\Vert=O_P(1)$ and $\sup_{R}\big\Vert \min(R^{\frac{1}{2}-\theta},\phi_{NRT}^2) \frac{1}{R}\sum_{t=T-R+1}^{T-h}\hat{\bm z}_{t,R}\varepsilon_{t+h}\big\Vert=O_P(1)$. We also observe that $\sup_{R} \Vert \frac{T}{R}\bm C_2\Vert $ is bounded by
$$ 
\bigg(\sup_R \bigg\Vert\bigg(\frac{1}{R}\sum_{t=T-R+1}^{T-h}\hat{\bm z}_{t,R}\hat{\bm z}_{t,R}^\top\bigg)^{-1}\bigg\Vert\bigg)\bigg(
\sup_R \bigg\Vert \frac{T}{R^2}\sum_{t=T-R+1}^{T-h}\hat{\bm z}_{t,R}\hat{\bm z}_{t,R}^\top(\bm\delta_R(t/T)-\bm\delta_R(1))\bigg\Vert \bigg)=O_P(1),$$
where we use the fact that the last term is $O_P(1)$, using \Cref{Lemma4} (12). Thus, $\sup_{R} \Vert \frac{T}{R}\bm C_2\Vert=O_P(1)$. Also, 
$\sup_R\bigg\Vert
\frac{\min(\phi_{NRT}^{2},R^{\frac{3}{4}-\nu})}{R}\sum_{t=T-R+1}^{T-h}\hat{\bm z}_{t,R}(\tilde{\bm f}_{t,R}-\bm H_R  \bm f_t)^\top(\bm H_R^{\top} )^{-1}\bm\alpha(t/T)
\bigg\Vert=O_p(1)$. Thus, $\sup_R\Vert
\min(\phi_{NRT}^{2},R^{\frac{3}{4}-\nu})\bm C_3\Vert=O_P(1)$. Consequently, $\sup_{R}\Vert \min(R^{\frac{1}{2}-\theta},\frac{T}{R}, N)(\hat{\bm\delta}_R(1)-\bm\delta_R(1))\Vert=O_P(1)$.

\subsubsection*{Proof of \Cref{Lemma4} (15)} 
Here, we show that $\sup_R[\min(R^{\frac{1}{2}-\theta},\frac{T}{R},\sqrt{N})(\hat{\bm z}_{T,R}^\top\hat{\bm{\delta}}_R(1)-\bm z_T^\top\bm\delta(1))]=O_P(1)$. 
We consider the decomposition
$\hat{\bm z}_{T,R}^\top\hat{\bm{\delta}}_R(1)-\bm z_T^\top\bm\delta(1)=
\hat{\bm z}_{T,R}^\top(\hat{\bm{\delta}}_R(1)-\bm\delta_R(1))+(\hat{\bm z}_{T,R}-\bm \Phi_R\bm z_T)^\top\bm\delta_R(1),$
and deduce $\sup_R[\min(R^{\frac{1}{2}-\theta},\frac{T}{R},\sqrt N)(\hat{\bm z}_{T,R}^\top\hat{\bm{\delta}}_R(1)-\bm z_T^\top\bm\delta(1))]\leq
\sup_R\Vert \min(R^{\frac{1}{2}-\theta},\frac{T}{R},  N)\hat{\bm z}_{T,R}^\top(\hat{\bm{\delta}}_R(1)-\bm\delta_R(1))\Vert
+
\sup_R \Vert(\min(\sqrt{N},\sqrt R\delta_{NRT},R^{7/8-\mu}) (\hat{\bm z}_{T,R}-\bm \Phi_R\bm z_T)^\top\bm\delta_R(1)\Vert
$,
where we observe that
$\sup_R\Vert \min(R^{\frac{1}{2}-\theta},\frac{T}{R},N)\hat{\bm z}_{T,R}^\top(\hat{\bm{\delta}}_R(1)-\bm\delta_R(1))\Vert=O_P(1)$
and 
$\sup_R \Vert \min(\sqrt{N},\sqrt R\delta_{NRT},R^{7/8-\mu})\times(\hat{\bm z}_{T,R}-\bm \Phi_R\bm z_T)^\top\bm\delta_R(1)\Vert=O_P(1)$, from \Cref{Lemma4} (14) and (5). Hence, the result follows.

\subsubsection*{Proof of \Cref{Lemma4} (16)} 
Recall that $C_{NT}^0(R) =(\hat{\bm z}_{T,R}^\top\hat{\bm\delta}_R(1)-\bm z_{T}^\top\bm\delta(1))^2$. Therefore, \Cref{Lemma4} (15) implies that we have 
$\sup_R(\min(R^{1-2\theta},\frac{T^2}{R^2},N) C_{NT}^0(R))=O_P(1).$
Furthermore, we observe that $C_{NT}(R)-C_{NT}^0(R)= \big[(\hat{\bm z}_{T,R}^\top\hat{\bm{\delta}}_R(1)-\bm z_T^\top\bm\delta(1)) -(\ddot{\bm z}_T^\top\ddot{\bm \delta}(1))-\bm z_T^\top\bm\delta(1))\big]^2- C_{NT}^0(R)$ and can be written as
$(\ddot{\bm z}_T^\top\ddot{\bm \delta}(1))-\bm z_T^\top\bm\delta(1))^2-2(\hat{\bm z}_{T,R}^\top\hat{\bm{\delta}}_R(1)-\bm z_T^\top\bm\delta(1))(\ddot{\bm z}_T^\top\ddot{\bm \delta}(1)-\bm z_T^\top\bm\delta(1))$. Consequently,
$\sup_{R\in \Theta_R}  \big(\min(R^{1-2\theta},\frac{T^2}{R^2},N)\left\vert C_{NT}(R)-C_{NT}^0(R)\right\vert\big)$ is lower than
\begin{align*}
 &  \sup_R \bigg\vert 
 \frac{\min(R^{1-2\theta},\frac{T^2}{R^2},N)}{ \delta_{NR_0T}^2  }
 \times
  [ \delta_{NR_0T} (\ddot{\bm z}_T^\top\ddot{\bm \delta}(1))-\bm z_T^\top\bm\delta(1))]^2 
 \bigg\vert
 +
  \sup_R\bigg\vert 
 \frac{\min(R^{1-2\theta},\frac{T^2}{R^2},N)}{\delta_{NR_0T} \min(R^{\frac{1}{2}-\theta},\frac{T}{R},\sqrt N)  }
 \\
 &\times 
 2 [ \min(R^{\frac{1}{2}-\theta},\frac{T}{R},\sqrt N)(\hat{\bm z}_{T,R}^\top\hat{\bm{\delta}}_R(1)-\bm z_T^\top\bm\delta(1))][ \delta_{NR_0T}(\ddot{\bm z}_T^\top\ddot{\bm \delta}(1))-\bm z_T^\top\bm\delta(1))] 
 \bigg\vert.
\end{align*}
Given that $ \delta_{NR_0T}(\ddot{\bm z}_T^\top\ddot{\bm \delta}(1))-\bm z_T^\top\bm\delta(1))=O_P(1) $ from \Cref{Thm2} (\ref{Thm24}) and the fact that $  \sup_R\min(R^{\frac{1}{2}-\theta},\frac{T}{R},\sqrt N)(\hat{\bm z}_{T,R}^\top\hat{\bm{\delta}}_R(1)-\bm z_T^\top\bm\delta(1))=O_P(1)$ from \Cref{Lemma4} (15), 
we deduce that $\sup_{R\in \Theta_R}  \big(\min(R^{1-2\theta},\frac{T^2}{R^2},N)\left\vert C_{NT}(R)-C_{NT}^0(R)\right\vert\big)$ is
\begin{align*}
&O_P\left(
\sup_R \bigg\vert 
 \frac{\min(R^{1-2\theta},\frac{T^2}{R^2},N)}{ \delta_{NR_0T}^2  }\bigg| 
+
  \sup_R\bigg\vert 
 \frac{\min(R^{1-2\theta},\frac{T^2}{R^2},N)}{\delta_{NR_0T} \min(R^{\frac{1}{2}-\theta},\frac{T}{R},\sqrt N)  }
 \bigg\vert\right)\\
 &=O_P\left(
\sup_R \bigg\vert 
 \frac{\min(R^{1-2\theta},\frac{T^2}{R^2},N)}{\min(R_0,\frac{T^2}{R^2_0},N)  }\bigg\vert\right)
 +O_P\left(
\sup_R \bigg\vert 
 \frac{\min(R^{1/2-\theta},\frac{T}{R},\sqrt N)}{\min(\sqrt{R_0},\frac{T}{R_0},\sqrt N)  }\bigg\vert\right).
 \end{align*}
 Thus,  $\sup_R \bigg( 
 \frac{\min(R^{\frac{1}{2}-\theta},\frac{T}{R},\sqrt N)}{\min(\sqrt{R_0},\frac{T}{R_0},\sqrt N)  }\bigg)=o(1)$ for some $\theta\in (0,1/2)$,  is sufficient. 
To have the last equality, suppose first that $\min(\sqrt{R_0}, \sqrt N)\leq \frac{T}{R_0}$. Then
$\frac{\min(R^{\frac{1}{2}-\theta},\frac{T}{R},N)}{\min(\sqrt{R_0},\frac{T}{R_0},N)  }=\frac{\min(R^{\frac{1}{2}-\theta},\frac{T}{R},\sqrt N)}{\min(\sqrt{R_0},\sqrt N)  }\leq \frac{R^{\frac{1}{2}-\theta}}{\min(\sqrt{R_0},\sqrt N)  }=o(1),$ as long as $ R^{1-2\theta}=o(\min (R_0,N))$, for some $\theta \in (0, 1/2)$. A sufficient condition to have the result is $R \ll \min(R_0,N)$. Suppose now that $\min(\sqrt{R_0}, \sqrt N)\geq \frac{T}{R_0}$. Then
$\frac{\min(R^{\frac{1}{2}-\theta},\frac{T}{R},\sqrt N)}{\min(\sqrt{R_0},\frac{T}{R_0},\sqrt N)  }=\frac{\min(R^{\frac{1}{2}-\theta},\frac{T}{R},\sqrt N)}{\frac{T}{R_0}}\leq \frac{\sqrt N}{\frac{T}{R_0} }=\frac{\sqrt NR_0}{T}=o(1)$ from \Cref{Ass7} (a) is enough.

\clearpage
\section*{Appendix B:  More Simulation  Results}
\label{AppendixB}
\renewcommand{\theequation}{B.\arabic{equation}} \renewcommand{\thesection}{B} \setcounter{equation}{0}
\renewcommand{\thelemma}{B.\arabic{lemma}} \renewcommand{\thesection}{B} \setcounter{lemma}{0}
\renewcommand{\thetable}{B.\arabic{table}} \renewcommand{\thesection}{B} \setcounter{table}{0}

\begin{table}[h!]
\begin{center}
\caption{Relative RMSFE when $N=100$  and $\rho_f=0.5$ for the latent factor}
\label{tab4}
\begin{tabular}{lcccccc}
\hline\hline
$T=100$ & DGP 7 & DGP 8 & DGP 9 & DGP 10 & DGP 11 & DGP 12 \\ \hline
Method 1 & 1.02 & 1.01 & 0.50 & 0.50 & 0.67 & 0.65 \\ 
Method 2 & 1.01 & 1.01 & 0.46 & 0.47 & 0.58 & 0.57 \\ 
Method 3 & 0.99 & 0.97 & 0.43 & 0.44 & 0.58 & 0.57 \\ 
\hline
$T=200$ & DGP 7 & DGP 8 & DGP 9 & DGP 10 & DGP 11 & DGP 12 \\ \hline
Method 1 & 1.01 & 1.01 & 0.50 & 0.51 & 0.61 & 0.62 \\ 
Method 2 & 1.01 & 1.01 & 0.46 & 0.47 & 0.54 & 0.55 \\ 
Method 3 & 0.99 & 0.98 & 0.45 & 0.45 & 0.53 & 0.55 \\ 

\hline
$T=300$ & DGP 7 & DGP 8 & DGP 9 & DGP 10 & DGP 11 & DGP 12 \\ \hline
Method 1 & 1.01 & 1.00 & 0.48 & 0.50 & 0.62 & 0.62 \\ 
Method 2 & 1.01 & 1.00 & 0.45 & 0.47 & 0.54 & 0.54 \\ 
Method 3 & 0.99 & 0.98 & 0.44 & 0.46 & 0.54 & 0.54 \\ 
\hline
$T=400$ & DGP 7 & DGP 8 & DGP 9 & DGP 10 & DGP 11 & DGP 12 \\ \hline
Method 1 & 1.01 & 1.01 & 0.51 & 0.49 & 0.58 & 0.60 \\ 
Method 2 & 1.01 & 1.01 & 0.46 & 0.46 & 0.52 & 0.52 \\ 
Method 3 & 0.99 & 0.98 & 0.45 & 0.45 & 0.52 & 0.52 \\ 
\hline
$T=500$ & DGP 7 & DGP 8 & DGP 9 & DGP 10 & DGP 11 & DGP 12 \\\hline
Method 1 & 1.00 & 1.01 & 0.49 & 0.48 & 0.61 & 0.60 \\ 
Method 2 & 1.00 & 1.01 & 0.46 & 0.45 & 0.52 & 0.52 \\ 
Method 3 & 0.99 & 0.98 & 0.44 & 0.44 & 0.52 & 0.52 \\

\hline
$T=600$ & DGP 7 & DGP 8 & DGP 9 & DGP 10 & DGP 11 & DGP 12 \\ \hline
Method 1 & 1.00 & 1.01 & 0.49 & 0.48 & 0.60 & 0.61 \\ 
Method 2 & 1.00 & 1.00 & 0.46 & 0.45 & 0.51 & 0.51 \\ 
Method 3 & 0.99 & 0.98 & 0.45 & 0.43 & 0.51 & 0.51 \\
\hline\hline
\end{tabular}\\
\textit{Note:} 
See note for \Cref{tab1}. 
\end{center}
\end{table}

\begin{table}[h!]
\begin{center}
\caption{Relative RMSFE when $N=200$ and $\rho_f=0.5$ for the latent factor}
\label{tab5}
\begin{tabular}{lcccccc}
\hline\hline
$T=100$ & DGP 7 & DGP 8 & DGP 9 & DGP 10 & DGP 11 & DGP 12 \\ \hline
Method 1 & 1.01 & 1.02 & 0.49 & 0.51 & 0.64 & 0.64 \\ 
Method 2 & 1.02 & 1.02 & 0.46 & 0.47 & 0.57 & 0.57 \\ 
Method 3 & 0.98 & 0.98 & 0.43 & 0.44 & 0.57 & 0.57 \\ 
\hline
$T=200$ & DGP 7 & DGP 8 & DGP 9 & DGP 10 & DGP 11 & DGP 12 \\ \hline
Method 1 & 1.00 & 1.01 & 0.49 & 0.48 & 0.62 & 0.62 \\ 
Method 2 & 1.01 & 1.02 & 0.46 & 0.45 & 0.54 & 0.55 \\ 
Method 3 & 0.99 & 0.99 & 0.45 & 0.43 & 0.54 & 0.54 \\

\hline
$T=300$ & DGP 7 & DGP 8 & DGP 9 & DGP 10 & DGP 11 & DGP 12 \\ \hline
Method 1 & 1.01 & 1.00 & 0.49 & 0.50 & 0.61 & 0.63 \\ 
Method 2 & 1.01 & 1.00 & 0.46 & 0.46 & 0.54 & 0.54 \\ 
Method 3 & 0.99 & 0.99 & 0.45 & 0.45 & 0.53 & 0.54 \\
\hline
$T=400$ & DGP 7 & DGP 8 & DGP 9 & DGP 10 & DGP 11 & DGP 12 \\ \hline
Method 1 & 1.01 & 1.01 & 0.49 & 0.47 & 0.61 & 0.60 \\ 
Method 2 & 1.00 & 1.01 & 0.45 & 0.44 & 0.53 & 0.52 \\ 
Method 3 & 0.99 & 0.99 & 0.43 & 0.43 & 0.52 & 0.52 \\
\hline
$T=500$ & DGP 7 & DGP 8 & DGP 9 & DGP 10 & DGP 11 & DGP 12 \\\hline
Method 1 & 1.01 & 1.01 & 0.50 & 0.49 & 0.62 & 0.61 \\ 
Method 2 & 1.01 & 1.01 & 0.46 & 0.46 & 0.52 & 0.52 \\ 
Method 3 & 0.99 & 0.99 & 0.45 & 0.45 & 0.52 & 0.52 \\ 

\hline
$T=600$ & DGP 7 & DGP 8 & DGP 9 & DGP 10 & DGP 11 & DGP 12 \\ \hline
Method 1 & 1.00 & 1.00 & 0.48 & 0.50 & 0.60 & 0.59 \\ 
Method 2 & 1.00 & 1.00 & 0.45 & 0.46 & 0.51 & 0.51 \\ 
Method 3 & 1.00 & 0.99 & 0.44 & 0.45 & 0.51 & 0.51 \\ 
\hline\hline
\end{tabular}
\newline
\textit{Note:} See note for \Cref{tab1}.
\end{center}
\end{table}

\begin{table}[h!]
\begin{center}
\caption{Relative RMSFE when $N=300$ and $\rho_f=0.5$ for the latent factor}
\label{tab6}
\begin{tabular}{lcccccc}
\hline\hline
$T=100$ & DGP 7 & DGP 8 & DGP 9 & DGP 10 & DGP 11 & DGP 12 \\ \hline
Method 1 & 1.02 & 1.01 & 0.49 & 0.49 & 0.64 & 0.65 \\ 
Method 2 & 1.03 & 1.01 & 0.45 & 0.45 & 0.58 & 0.57 \\ 
Method 3 & 0.99 & 0.99 & 0.42 & 0.43 & 0.57 & 0.56 \\ 
\hline
$T=200$ & DGP 7 & DGP 8 & DGP 9 & DGP 10 & DGP 11 & DGP 12 \\ \hline
Method 1 & 1.02 & 1.01 & 0.49 & 0.47 & 0.62 & 0.63 \\ 
Method 2 & 1.02 & 1.01 & 0.46 & 0.46 & 0.54 & 0.54 \\ 
Method 3 & 1.00 & 0.99 & 0.45 & 0.44 & 0.54 & 0.53 \\

\hline
$T=300$ & DGP 7 & DGP 8 & DGP 9 & DGP 10 & DGP 11 & DGP 12 \\ \hline
Method 1 & 1.00 & 1.01 & 0.49 & 0.50 & 0.61 & 0.61 \\ 
Method 2 & 1.01 & 1.01 & 0.45 & 0.46 & 0.54 & 0.54 \\ 
Method 3 & 0.99 & 0.99 & 0.44 & 0.44 & 0.54 & 0.54 \\
\hline
$T=400$ & DGP 7 & DGP 8 & DGP 9 & DGP 10 & DGP 11 & DGP 12 \\ \hline
Method 1 & 1.00 & 1.01 & 0.48 & 0.50 & 0.61 & 0.62 \\ 
Method 2 & 1.01 & 1.01 & 0.45 & 0.47 & 0.52 & 0.53 \\ 
Method 3 & 0.99 & 0.99 & 0.45 & 0.46 & 0.52 & 0.53 \\
\hline
$T=500$ & DGP 7 & DGP 8 & DGP 9 & DGP 10 & DGP 11 & DGP 12 \\\hline
Method 1 & 1.01 & 1.00 & 0.49 & 0.50 & 0.59 & 0.60 \\ 
Method 2 & 1.00 & 1.01 & 0.46 & 0.46 & 0.51 & 0.51 \\ 
Method 3 & 1.00 & 0.99 & 0.45 & 0.45 & 0.51 & 0.51 \\ 

\hline
$T=600$ & DGP 7 & DGP 8 & DGP 9 & DGP 10 & DGP 11 & DGP 12 \\ \hline
Method 1 & 1.01 & 1.00 & 0.48 & 0.51 & 0.60 & 0.59 \\ 
Method 2 & 1.00 & 1.00 & 0.44 & 0.47 & 0.50 & 0.52 \\ 
Method 3 & 0.99 & 0.99 & 0.44 & 0.46 & 0.50 & 0.52 \\
\hline\hline
\end{tabular}
\newline
\textit{Note:} See note for \Cref{tab1}.
\end{center}
\end{table}

\begin{table}[h!]
\begin{center}
\caption{Relative RMSFE when $N=100$ and there is an observed  regressor}
\label{tab7}
\begin{tabular}{lcccccc}
\hline\hline
$T=100$ & DGP 13 & DGP 14 & DGP 15 & DGP 16 & DGP 17 & DGP 18\\ \hline
Method 1 & 1.03 & 1.02 & 0.70 & 0.70 & 0.76 & 0.76 \\ 
Method 2 & 1.05 & 1.03 & 0.56 & 0.55 & 0.62 & 0.60 \\ 
Method 3 & 0.98 & 0.96 & 0.50 & 0.49 & 0.59 & 0.59 \\ 
\hline
$T=200$ & DGP 13 & DGP 14 & DGP 15 & DGP 16 & DGP 17 & DGP 18 \\ \hline
Method 1 & 1.02 & 1.02 & 0.58 & 0.56 & 0.73 & 0.76 \\ 
Method 2 & 1.03 & 1.03 & 0.49 & 0.49 & 0.58 & 0.60 \\ 
Method 3 & 0.99 & 0.97 & 0.46 & 0.45 & 0.57 & 0.60 \\  

\hline
$T=300$ & DGP 13 & DGP 14 & DGP 15 & DGP 16 & DGP 17 & DGP 18 \\ \hline
Method 1 & 1.01 & 1.01 & 0.57 & 0.57 & 0.78 & 0.84 \\ 
Method 2 & 1.02 & 1.02 & 0.44 & 0.47 & 0.58 & 0.62 \\ 
Method 3 & 0.99 & 0.98 & 0.42 & 0.44 & 0.57 & 0.61 \\  
\hline
$T=400$ & DGP 13 & DGP 14 & DGP 15 & DGP 16 & DGP 17 & DGP 18\\ \hline
Method 1 & 1.00 & 1.00 & 0.56 & 0.58 & 0.79 & 0.85 \\ 
Method 2 & 1.01 & 1.01 & 0.45 & 0.47 & 0.58 & 0.61 \\ 
Method 3 & 0.98 & 0.98 & 0.43 & 0.45 & 0.58 & 0.61 \\
\hline
$T=500$ & DGP 13 & DGP 14 & DGP 15 & DGP 16 & DGP 17 & DGP 18 \\\hline
Method 1 & 1.01 & 1.00 & 0.54 & 0.59 & 0.83 & 0.78 \\ 
Method 2 & 1.01 & 1.00 & 0.47 & 0.48 & 0.58 & 0.59 \\ 
Method 3 & 0.99 & 0.98 & 0.46 & 0.46 & 0.58 & 0.58 \\ 

\hline
$T=600$ & DGP 13 & DGP 14 & DGP 15 & DGP 16 & DGP 17 & DGP 18 \\ \hline
Method 1 & 1.00 & 1.01 & 0.55 & 0.52 & 0.82 & 0.91 \\ 
Method 2 & 1.01 & 1.01 & 0.46 & 0.46 & 0.59 & 0.63 \\ 
Method 3 & 0.98 & 0.98 & 0.44 & 0.44 & 0.59 & 0.62 \\
\hline\hline
\end{tabular}\\
\textit{Note:} 
See note for \Cref{tab1}. 
\end{center}
\end{table}

\begin{table}[h!]
\begin{center}
\caption{Relative RMSFE when $N=200$  and there is an observed  regressor}
\label{tab8}
\begin{tabular}{lcccccc}
\hline\hline
$T=100$ & DGP 13 & DGP 14 & DGP 15 & DGP 16 & DGP 17 & DGP 18 \\ \hline
Method 1 & 1.02 & 1.03 & 0.66 & 0.70 & 0.78 & 0.81 \\ 
Method 2 & 1.04 & 1.05 & 0.53 & 0.54 & 0.63 & 0.64 \\ 
Method 3 & 0.98 & 0.97 & 0.48 & 0.48 & 0.61 & 0.62 \\ 
\hline
$T=200$ & DGP 13 & DGP 14 & DGP 15 & DGP 16 & DGP 17 & DGP 18 \\ \hline
Method 1 & 1.00 & 1.01 & 0.60 & 0.59 & 0.81 & 0.85 \\ 
Method 2 & 1.01 & 1.02 & 0.51 & 0.48 & 0.62 & 0.65 \\ 
Method 3 & 0.99 & 0.97 & 0.48 & 0.46 & 0.62 & 0.64 \\ 

\hline
$T=300$ & DGP 13 & DGP 14 & DGP 15 & DGP 16 & DGP 17 & DGP 18 \\ \hline
Method 1 & 1.00 & 1.01 & 0.57 & 0.59 & 0.86 & 0.90 \\ 
Method 2 & 1.01 & 1.01 & 0.48 & 0.47 & 0.64 & 0.67 \\ 
Method 3 & 0.99 & 0.98 & 0.45 & 0.45 & 0.63 & 0.66 \\ 
\hline
$T=400$ & DGP 13 & DGP 14 & DGP 15 & DGP 16 & DGP 17 & DGP 18 \\ \hline
Method 1 & 1.01 & 1.01 & 0.56 & 0.57 & 0.89 & 0.95 \\ 
Method 2 & 1.02 & 1.01 & 0.47 & 0.45 & 0.66 & 0.69 \\ 
Method 3 & 0.99 & 0.99 & 0.46 & 0.44 & 0.66 & 0.69 \\ 
\hline
$T=500$ & DGP 13 & DGP 14 & DGP 15 & DGP 16 & DGP 17 & DGP 18 \\\hline
Method 1 & 1.00 & 1.00 & 0.52 & 0.57 & 0.91 & 0.95 \\ 
Method 2 & 1.01 & 1.00 & 0.44 & 0.46 & 0.65 & 0.67 \\ 
Method 3 & 0.99 & 0.98 & 0.43 & 0.44 & 0.65 & 0.66 \\ 

\hline
$T=600$ & DGP 13 & DGP 14 & DGP 15 & DGP 16 & DGP 17 & DGP 18 \\ \hline
Method 1 & 1.00 & 1.01 & 0.55 & 0.56 & 0.91 & 0.96 \\ 
Method 2 & 1.01 & 1.01 & 0.46 & 0.45 & 0.65 & 0.67 \\ 
Method 3 & 0.99 & 0.99 & 0.45 & 0.44 & 0.65 & 0.67 \\ 
\hline\hline
\end{tabular}
\newline
\textit{Note:} See note for \Cref{tab1}.
\end{center}
\end{table}

\begin{table}[h!]
\begin{center}
\caption{Relative RMSFE when $N=300$  and there is an observed  regressor}
\label{tab9}
\begin{tabular}{lcccccc}
\hline\hline
$T=100$ & DGP 13 & DGP 14 & DGP 15 & DGP 16 & DGP 17 & DGP 18 \\ \hline
Method 1 & 1.01 & 1.03 & 0.68 & 0.72 & 0.81 & 0.83 \\ 
Method 2 & 1.02 & 1.05 & 0.52 & 0.55 & 0.65 & 0.66 \\ 
Method 3 & 0.98 & 0.99 & 0.46 & 0.49 & 0.63 & 0.64 \\ 
\hline
$T=200$ & DGP 13 & DGP 14 & DGP 15 & DGP 16 & DGP 17 & DGP 18 \\ \hline
Method 1 & 1.01 & 1.01 & 0.59 & 0.58 & 0.86 & 0.85 \\ 
Method 2 & 1.02 & 1.02 & 0.52 & 0.50 & 0.66 & 0.67 \\ 
Method 3 & 0.99 & 0.98 & 0.48 & 0.46 & 0.65 & 0.66 \\ 

\hline
$T=300$ & DGP 13 & DGP 14 & DGP 15 & DGP 16 & DGP 17 & DGP 18 \\ \hline
Method 1 & 1.01 & 1.01 & 0.57 & 0.57 & 0.92 & 0.94 \\ 
Method 2 & 1.01 & 1.02 & 0.47 & 0.49 & 0.67 & 0.69 \\ 
Method 3 & 0.99 & 0.98 & 0.45 & 0.47 & 0.67 & 0.69 \\  
\hline
$T=400$ & DGP 13 & DGP 14 & DGP 15 & DGP 16 & DGP 17 & DGP 18 \\ \hline
Method 1 & 1.00 & 1.01 & 0.58 & 0.59 & 0.95 & 0.94 \\ 
Method 2 & 1.01 & 1.01 & 0.48 & 0.46 & 0.69 & 0.67 \\ 
Method 3 & 0.99 & 0.99 & 0.46 & 0.45 & 0.69 & 0.67 \\  
\hline
$T=500$ & DGP 13 & DGP 14 & DGP 15 & DGP 16 & DGP 17 & DGP 18 \\\hline
Method 1 & 1.00 & 1.00 & 0.55 & 0.57 & 0.98 & 0.93 \\ 
Method 2 & 1.00 & 1.00 & 0.47 & 0.47 & 0.67 & 0.67 \\ 
Method 3 & 0.99 & 1.00 & 0.46 & 0.46 & 0.67 & 0.66 \\ 

\hline
$T=600$ & DGP 13 & DGP 14 & DGP 15 & DGP 16 & DGP 17 & DGP 18 \\ \hline
Method 1 & 1.00 & 1.01 & 0.57 & 0.54 & 0.97 & 0.98 \\ 
Method 2 & 1.01 & 1.01 & 0.46 & 0.46 & 0.65 & 0.68 \\ 
Method 3 & 0.99 & 0.99 & 0.45 & 0.45 & 0.65 & 0.68 \\  
\hline\hline
\end{tabular}
\newline
\textit{Note:} See note for \Cref{tab1}.
\end{center}
\end{table}
\begin{table}[h!]
\begin{center}
\caption{Average window sizes when $N=100$ and $\rho_f=0.9$ for the latent factor}
\label{tab1b}
\begin{tabular}{lcccccc}
\hline\hline
$T=100$ & DGP 1 & DGP 2 & DGP 3 & DGP 4 & DGP 5 & DGP 6 \\ \hline
Method 1 & 63 & 63 & 54 & 54 & 57 & 56 \\ 
Method 2 & 59 & 57 & 50 & 50 & 46 & 46 \\ 
Method 3 & 68 & 68 & 49 & 50 & 46 & 46 \\ 

\hline
$T=200$ & DGP 1 & DGP 2 & DGP 3 & DGP 4 & DGP 5 & DGP 6 \\ \hline
Method 1 & 113 & 111 & 91 & 93 & 95 & 93 \\ 
Method 2 & 105 & 105 & 94 & 95 & 76 & 78 \\ 
Method 3 & 123 & 125 & 94 & 94 & 75 & 77 \\  

\hline
$T=300$ & DGP 1 & DGP 2 & DGP 3 & DGP 4 & DGP 5 & DGP 6 \\ \hline
Method 1 & 168 & 167 & 138 & 137 & 138 & 137 \\ 
Method 2 & 152 & 154 & 139 & 135 & 104 & 105 \\ 
Method 3 & 184 & 180 & 136 & 139 & 105 & 104 \\ 

\hline
$T=400$ & DGP 1 & DGP 2 & DGP 3 & DGP 4 & DGP 5 & DGP 6 \\ \hline
Method 1 & 223 & 220 & 177 & 176 & 178 & 177 \\ 
Method 2 & 194 & 198 & 180 & 180 & 131 & 131 \\ 
Method 3 & 233 & 237 & 180 & 181 & 129 & 131 \\ 

\hline
$T=500$ & DGP 1 & DGP 2 & DGP 3 & DGP 4 & DGP 5 & DGP 6 \\ \hline
Method 1 & 276 & 269 & 220 & 218 & 214 & 210 \\ 
Method 2 & 236 & 229 & 221 & 221 & 155 & 154 \\ 
Method 3 & 281 & 280 & 224 & 223 & 154 & 154 \\

\hline
$T=600$ & DGP 1 & DGP 2 & DGP 3 & DGP 4 & DGP 5 & DGP 6 \\ \hline
Method 1 & 325 & 314 & 254 & 256 & 245 & 246 \\ 
Method 2 & 282 & 282 & 259 & 263 & 178 & 180 \\ 
Method 3 & 337 & 332 & 262 & 271 & 178 & 178 \\

\hline\hline
\end{tabular}
\end{center}
\textit{Note:} This table reports the average window sizes out of $M=1000$ simulations by minimizing $C_{NT}(R)$ (Method 2), $C_{NT}^0(R)$ (Method 3), and the feasible criterion, where we treat estimated factors as observed (Method 1). The numbers are rounded to the closest integer.
\end{table}

\begin{table}[h!]
\begin{center}
\caption{Average window sizes when $N=200$ and $\rho_f=0.9$ for the latent factor}
\label{tab2b}
\begin{tabular}{lcccccc}
\hline\hline
$T=100$ & DGP 1 & DGP 2 & DGP 3 & DGP 4 & DGP 5 & DGP 6 \\ \hline
Method 1 & 64 & 63 & 54 & 54 & 57 & 55 \\ 
Method 2 & 58 & 58 & 50 & 51 & 46 & 46 \\ 
Method 3 & 69 & 68 & 49 & 49 & 46 & 46 \\ 
\hline
$T=200$ & DGP 1 & DGP 2 & DGP 3 & DGP 4 & DGP 5 & DGP 6 \\ \hline
Method 1 & 113 & 114 & 92 & 94 & 96 & 94 \\ 
Method 2 & 106 & 105 & 94 & 94 & 77 & 77 \\ 
Method 3 & 129 & 127 & 94 & 93 & 76 & 76 \\ 

\hline
$T=300$ & DGP 1 & DGP 2 & DGP 3 & DGP 4 & DGP 5 & DGP 6 \\ \hline
Method 1 & 172 & 167 & 140 & 137 & 138 & 134 \\ 
Method 2 & 149 & 153 & 139 & 138 & 104 & 106 \\ 
Method 3 & 186 & 182 & 138 & 137 & 104 & 107 \\  
\hline
$T=400$ & DGP 1 & DGP 2 & DGP 3 & DGP 4 & DGP 5 & DGP 6 \\ \hline
Method 1 & 227 & 217 & 175 & 175 & 178 & 176 \\ 
Method 2 & 197 & 193 & 178 & 179 & 132 & 130 \\ 
Method 3 & 240 & 234 & 178 & 178 & 132 & 130 \\
\hline
$T=500$ & DGP 1 & DGP 2 & DGP 3 & DGP 4 & DGP 5 & DGP 6 \\ \hline
Method 1 & 274 & 275 & 218 & 214 & 214 & 218 \\ 
Method 2 & 235 & 244 & 219 & 217 & 155 & 154 \\ 
Method 3 & 298 & 288 & 221 & 222 & 155 & 154 \\  

\hline
$T=600$ & DGP 1 & DGP 2 & DGP 3 & DGP 4 & DGP 5 & DGP 6 \\ \hline
Method 1 & 325 & 314 & 255 & 255 & 254 & 249 \\ 
Method 2 & 280 & 278 & 257 & 261 & 180 & 180 \\ 
Method 3 & 347 & 331 & 261 & 269 & 178 & 179 \\ 
\hline\hline
\end{tabular}
\newline
\textit{Note:} See note for \Cref{tab1b}.
\end{center}
\end{table}

\begin{table}[h!]
\begin{center}
\caption{Average window sizes when $N=300$ and $\rho_f=0.9$ for the latent factor}
\label{tab3b}
\begin{tabular}{lcccccc}
\hline\hline
$T=100$ & DGP 1 & DGP 2 & DGP 3 & DGP 4 & DGP 5 & DGP 6 \\ \hline
Method 1 & 64 & 63 & 54 & 54 & 56 & 56 \\ 
Method 2 & 59 & 59 & 50 & 50 & 46 & 46 \\ 
Method 3 & 69 & 69 & 49 & 49 & 45 & 45 \\ 

\hline
$T=200$ & DGP 1 & DGP 2 & DGP 3 & DGP 4 & DGP 5 & DGP 6 \\ \hline
Method 1 & 115 & 115 & 93 & 93 & 96 & 93 \\ 
Method 2 & 107 & 107 & 94 & 94 & 77 & 77 \\ 
Method 3 & 127 & 127 & 93 & 93 & 76 & 77 \\ 

\hline
$T=300$ & DGP 1 & DGP 2 & DGP 3 & DGP 4 & DGP 5 & DGP 6 \\ \hline
Method 1 & 167 & 169 & 137 & 138 & 139 & 136 \\ 
Method 2 & 151 & 150 & 137 & 137 & 106 & 105 \\ 
Method 3 & 186 & 182 & 136 & 138 & 105 & 105 \\ 

\hline
$T=400$ & DGP 1 & DGP 2 & DGP 3 & DGP 4 & DGP 5 & DGP 6 \\ \hline
Method 1 & 225 & 222 & 176 & 176 & 179 & 179 \\ 
Method 2 & 191 & 194 & 180 & 176 & 132 & 131 \\ 
Method 3 & 241 & 244 & 176 & 178 & 132 & 131 \\ 

\hline
$T=500$ & DGP 1 & DGP 2 & DGP 3 & DGP 4 & DGP 5 & DGP 6 \\ \hline
Method 1 & 273 & 272 & 213 & 216 & 217 & 214 \\ 
Method 2 & 241 & 234 & 219 & 216 & 159 & 155 \\ 
Method 3 & 295 & 281 & 219 & 222 & 156 & 155 \\ 

\hline
$T=600$ & DGP 1 & DGP 2 & DGP 3 & DGP 4 & DGP 5 & DGP 6 \\ \hline
Method 1 & 330 & 323 & 254 & 254 & 242 & 250 \\ 
Method 2 & 277 & 279 & 262 & 261 & 177 & 177 \\ 
Method 3 & 350 & 340 & 257 & 263 & 176 & 176 \\
\hline\hline
\end{tabular}
\newline
\textit{Note:} See note for \Cref{tab1b}.
\end{center}
\end{table}

\begin{table}[h!]
\begin{center}
\caption{Average window sizes when  $N=100$  and $\rho_f=0.5$ for the latent factor}
\label{tab4b}
\begin{tabular}{lcccccc}
\hline\hline
$T=100$ & DGP 7 & DGP 8 & DGP 9 & DGP 10 & DGP 11 & DGP 12 \\ \hline
Method 1 & 60 & 60 & 52 & 52 & 51 & 52 \\ 
Method 2 & 59 & 58 & 50 & 50 & 42 & 43 \\ 
Method 3 & 69 & 68 & 48 & 49 & 42 & 42 \\ 
\hline
$T=200$ & DGP 7 & DGP 8 & DGP 9 & DGP 10 & DGP 11 & DGP 12 \\ \hline
Method 1 & 113 & 110 & 96 & 97 & 89 & 86 \\ 
Method 2 & 108 & 107 & 93 & 93 & 72 & 72 \\ 
Method 3 & 127 & 124 & 92 & 93 & 72 & 72 \\

\hline
$T=300$ & DGP 7 & DGP 8 & DGP 9 & DGP 10 & DGP 11 & DGP 12 \\ \hline
Method 1 & 158 & 157 & 138 & 138 & 125 & 127 \\ 
Method 2 & 153 & 151 & 135 & 136 & 100 & 100 \\ 
Method 3 & 178 & 179 & 134 & 137 & 99 & 99 \\ 
\hline
$T=400$ & DGP 7 & DGP 8 & DGP 9 & DGP 10 & DGP 11 & DGP 12 \\ \hline
Method 1 & 206 & 202 & 183 & 181 & 158 & 163 \\ 
Method 2 & 195 & 194 & 178 & 176 & 124 & 124 \\ 
Method 3 & 239 & 236 & 176 & 180 & 124 & 125 \\ 
\hline
$T=500$ & DGP 7 & DGP 8 & DGP 9 & DGP 10 & DGP 11 & DGP 12 \\\hline
Method 1 & 245 & 254 & 224 & 227 & 194 & 191 \\ 
Method 2 & 238 & 242 & 218 & 218 & 148 & 147 \\ 
Method 3 & 294 & 280 & 220 & 221 & 147 & 147 \\

\hline
$T=600$ & DGP 7 & DGP 8 & DGP 9 & DGP 10 & DGP 11 & DGP 12 \\ \hline
Method 1 & 298 & 285 & 268 & 265 & 233 & 233 \\ 
Method 2 & 280 & 279 & 257 & 259 & 170 & 171 \\ 
Method 3 & 331 & 327 & 261 & 264 & 170 & 171 \\ 
\hline\hline
\end{tabular}\\
\textit{Note:} 
See note for \Cref{tab1b}. 
\end{center}
\end{table}

\begin{table}[h!]
\begin{center}
\caption{Average window sizes when  $N=200$ and $\rho_f=0.5$ for the latent factor}
\label{tab5b}
\begin{tabular}{lcccccc}
\hline\hline
$T=100$ & DGP 7 & DGP 8 & DGP 9 & DGP 10 & DGP 11 & DGP 12 \\ \hline
Method 1 & 60 & 58 & 51 & 52 & 51 & 51 \\ 
Method 2 & 58 & 58 & 49 & 50 & 43 & 42 \\ 
Method 3 & 69 & 68 & 48 & 48 & 42 & 42 \\ 
\hline
$T=200$ & DGP 7 & DGP 8 & DGP 9 & DGP 10 & DGP 11 & DGP 12 \\ \hline
Method 1 & 112 & 108 & 97 & 96 & 89 & 90 \\ 
Method 2 & 108 & 104 & 94 & 93 & 72 & 72 \\ 
Method 3 & 129 & 127 & 93 & 93 & 71 & 71 \\ 

\hline
$T=300$ & DGP 7 & DGP 8 & DGP 9 & DGP 10 & DGP 11 & DGP 12 \\ \hline
Method 1 & 157 & 155 & 139 & 141 & 126 & 128 \\ 
Method 2 & 151 & 152 & 135 & 135 & 100 & 99 \\ 
Method 3 & 186 & 178 & 134 & 136 & 99 & 99 \\ 
\hline
$T=400$ & DGP 7 & DGP 8 & DGP 9 & DGP 10 & DGP 11 & DGP 12 \\ \hline
Method 1 & 209 & 206 & 181 & 178 & 158 & 162 \\ 
Method 2 & 198 & 194 & 177 & 174 & 124 & 125 \\ 
Method 3 & 240 & 238 & 176 & 178 & 123 & 124 \\ 
\hline
$T=500$ & DGP 7 & DGP 8 & DGP 9 & DGP 10 & DGP 11 & DGP 12 \\\hline
Method 1 & 255 & 248 & 226 & 223 & 198 & 194 \\ 
Method 2 & 240 & 242 & 217 & 218 & 148 & 148 \\ 
Method 3 & 290 & 286 & 218 & 221 & 147 & 147 \\ 

\hline
$T=600$ & DGP 7 & DGP 8 & DGP 9 & DGP 10 & DGP 11 & DGP 12 \\ \hline
Method 1 & 293 & 285 & 268 & 271 & 227 & 228 \\ 
Method 2 & 281 & 274 & 264 & 261 & 171 & 171 \\ 
Method 3 & 350 & 339 & 259 & 264 & 170 & 171 \\
\hline\hline
\end{tabular}
\newline
\textit{Note:} See note for \Cref{tab1b}.
\end{center}
\end{table}

\begin{table}[h!]
\begin{center}
\caption{Average window sizes when $N=300$ and $\rho_f=0.5$ for the latent factor}
\label{tab6b}
\begin{tabular}{lcccccc}
\hline\hline
$T=100$ & DGP 7 & DGP 8 & DGP 9 & DGP 10 & DGP 11 & DGP 12 \\ \hline
Method 1 & 61 & 61 & 51 & 51 & 51 & 51 \\ 
Method 2 & 57 & 59 & 49 & 50 & 42 & 43 \\ 
Method 3 & 69 & 68 & 48 & 48 & 42 & 42 \\
\hline
$T=200$ & DGP 7 & DGP 8 & DGP 9 & DGP 10 & DGP 11 & DGP 12 \\ \hline
Method 1 & 110 & 109 & 96 & 94 & 91 & 90 \\ 
Method 2 & 105 & 105 & 93 & 93 & 72 & 72 \\ 
Method 3 & 132 & 129 & 92 & 92 & 71 & 71 \\  

\hline
$T=300$ & DGP 7 & DGP 8 & DGP 9 & DGP 10 & DGP 11 & DGP 12 \\ \hline
Method 1 & 156 & 159 & 141 & 139 & 124 & 123 \\ 
Method 2 & 150 & 153 & 136 & 134 & 99 & 99 \\ 
Method 3 & 188 & 182 & 135 & 135 & 99 & 99 \\ 
\hline
$T=400$ & DGP 7 & DGP 8 & DGP 9 & DGP 10 & DGP 11 & DGP 12 \\ \hline
Method 1 & 203 & 203 & 180 & 184 & 160 & 162 \\ 
Method 2 & 194 & 193 & 178 & 179 & 124 & 124 \\ 
Method 3 & 240 & 243 & 176 & 178 & 124 & 124 \\ 
\hline
$T=500$ & DGP 7 & DGP 8 & DGP 9 & DGP 10 & DGP 11 & DGP 12 \\\hline
Method 1 & 253 & 248 & 226 & 222 & 193 & 199 \\ 
Method 2 & 244 & 239 & 220 & 218 & 148 & 148 \\ 
Method 3 & 302 & 286 & 218 & 219 & 147 & 147 \\ 

\hline
$T=600$ & DGP 7 & DGP 8 & DGP 9 & DGP 10 & DGP 11 & DGP 12 \\ \hline
Method 1 & 296 & 288 & 268 & 271 & 232 & 224 \\ 
Method 2 & 287 & 279 & 260 & 259 & 171 & 171 \\ 
Method 3 & 352 & 336 & 261 & 262 & 171 & 171 \\ 
\hline\hline
\end{tabular}
\newline
\textit{Note:} See note for \Cref{tab1b}.
\end{center}
\end{table}

\begin{table}[h!]
\begin{center}
\caption{Average window sizes when  $N=100$ and there is an observed  regressor}
\label{tab7b}
\begin{tabular}{lcccccc}
\hline\hline
$T=100$ & DGP 13 & DGP 14 & DGP 15 & DGP 16 & DGP 17 & DGP 18\\ \hline
Method 1 & 63 & 63 & 59 & 59 & 65 & 66 \\ 
Method 2 & 58 & 59 & 56 & 54 & 55 & 53 \\ 
Method 3 & 68 & 68 & 53 & 54 & 54 & 54 \\ 
\hline
$T=200$ & DGP 13 & DGP 14 & DGP 15 & DGP 16 & DGP 17 & DGP 18 \\ \hline
Method 1 & 112 & 113 & 97 & 98 & 113 & 115 \\ 
Method 2 & 106 & 106 & 102 & 104 & 96 & 98 \\ 
Method 3 & 129 & 125 & 100 & 102 & 96 & 98 \\  

\hline
$T=300$ & DGP 13 & DGP 14 & DGP 15 & DGP 16 & DGP 17 & DGP 18 \\ \hline
Method 1 & 167 & 170 & 145 & 145 & 171 & 175 \\ 
Method 2 & 151 & 155 & 148 & 150 & 136 & 132 \\ 
Method 3 & 181 & 182 & 148 & 151 & 137 & 131 \\ 
\hline
$T=400$ & DGP 13 & DGP 14 & DGP 15 & DGP 16 & DGP 17 & DGP 18\\ \hline
Method 1 & 222 & 223 & 189 & 187 & 225 & 230 \\ 
Method 2 & 194 & 195 & 194 & 192 & 168 & 169 \\ 
Method 3 & 237 & 229 & 193 & 195 & 165 & 169 \\ 
\hline
$T=500$ & DGP 13 & DGP 14 & DGP 15 & DGP 16 & DGP 17 & DGP 18 \\\hline
Method 1 & 269 & 265 & 228 & 233 & 279 & 272 \\ 
Method 2 & 235 & 241 & 241 & 239 & 207 & 205 \\ 
Method 3 & 289 & 288 & 239 & 240 & 205 & 209 \\ 

\hline
$T=600$ & DGP 13 & DGP 14 & DGP 15 & DGP 16 & DGP 17 & DGP 18 \\ \hline
Method 1 & 321 & 322 & 263 & 270 & 322 & 329 \\ 
Method 2 & 276 & 273 & 280 & 285 & 236 & 243 \\ 
Method 3 & 336 & 337 & 281 & 285 & 234 & 243 \\  
\hline\hline
\end{tabular}\\
\textit{Note:} 
See note for \Cref{tab1b}. 
\end{center}
\end{table}

\begin{table}[h!]
\begin{center}
\caption{Average window sizes when  $N=200$  and there is an observed regressor}
\label{tab8b}
\begin{tabular}{lcccccc}
\hline\hline
$T=100$ & DGP 13 & DGP 14 & DGP 15 & DGP 16 & DGP 17 & DGP 18 \\ \hline
Method 1 & 63 & 62 & 59 & 58 & 65 & 67 \\ 
Method 2 & 58 & 58 & 56 & 54 & 55 & 54 \\ 
Method 3 & 68 & 68 & 53 & 54 & 54 & 54 \\ 
\hline
$T=200$ & DGP 13 & DGP 14 & DGP 15 & DGP 16 & DGP 17 & DGP 18 \\ \hline
Method 1 & 114 & 111 & 98 & 99 & 116 & 115 \\ 
Method 2 & 107 & 106 & 104 & 102 & 94 & 98 \\ 
Method 3 & 128 & 126 & 100 & 101 & 93 & 98 \\ 

\hline
$T=300$ & DGP 13 & DGP 14 & DGP 15 & DGP 16 & DGP 17 & DGP 18 \\ \hline
Method 1 & 167 & 169 & 143 & 147 & 173 & 175 \\ 
Method 2 & 149 & 152 & 150 & 151 & 133 & 135 \\ 
Method 3 & 186 & 183 & 145 & 148 & 132 & 133 \\ 
\hline
$T=400$ & DGP 13 & DGP 14 & DGP 15 & DGP 16 & DGP 17 & DGP 18 \\ \hline
Method 1 & 220 & 226 & 190 & 187 & 224 & 227 \\ 
Method 2 & 195 & 197 & 194 & 191 & 172 & 168 \\ 
Method 3 & 243 & 231 & 191 & 191 & 170 & 167 \\ 
\hline
$T=500$ & DGP 13 & DGP 14 & DGP 15 & DGP 16 & DGP 17 & DGP 18 \\\hline
Method 1 & 270 & 268 & 221 & 231 & 280 & 284 \\ 
Method 2 & 237 & 237 & 241 & 236 & 207 & 204 \\ 
Method 3 & 299 & 284 & 234 & 236 & 206 & 201 \\ 

\hline
$T=600$ & DGP 13 & DGP 14 & DGP 15 & DGP 16 & DGP 17 & DGP 18 \\ \hline
Method 1 & 316 & 315 & 267 & 273 & 334 & 332 \\ 
Method 2 & 280 & 276 & 281 & 281 & 241 & 239 \\ 
Method 3 & 344 & 340 & 282 & 279 & 240 & 242 \\ 
\hline\hline
\end{tabular}
\newline
\textit{Note:} See note for \Cref{tab1b}.
\end{center}
\end{table}

\begin{table}[h!]
\begin{center}
\caption{Average window sizes when  $N=300$  and there is an observed regressor}
\label{tab9b}
\begin{tabular}{lcccccc}
\hline\hline
$T=100$ & DGP 13 & DGP 14 & DGP 15 & DGP 16 & DGP 17 & DGP 18 \\ \hline
Method 1 & 63 & 61 & 58 & 58 & 66 & 66 \\ 
Method 2 & 59 & 58 & 55 & 55 & 54 & 54 \\ 
Method 3 & 70 & 69 & 53 & 53 & 54 & 54 \\ 
\hline
$T=200$ & DGP 13 & DGP 14 & DGP 15 & DGP 16 & DGP 17 & DGP 18 \\ \hline
Method 1 & 113 & 114 & 98 & 99 & 116 & 114 \\ 
Method 2 & 107 & 107 & 102 & 102 & 97 & 96 \\ 
Method 3 & 132 & 128 & 98 & 101 & 95 & 95 \\ 

\hline
$T=300$ & DGP 13 & DGP 14 & DGP 15 & DGP 16 & DGP 17 & DGP 18 \\ \hline
Method 1 & 165 & 167 & 144 & 143 & 174 & 176 \\ 
Method 2 & 150 & 150 & 150 & 149 & 136 & 135 \\ 
Method 3 & 188 & 185 & 145 & 144 & 135 & 134 \\ 
\hline
$T=400$ & DGP 13 & DGP 14 & DGP 15 & DGP 16 & DGP 17 & DGP 18 \\ \hline
Method 1 & 220 & 219 & 186 & 191 & 225 & 227 \\ 
Method 2 & 193 & 190 & 195 & 191 & 170 & 167 \\ 
Method 3 & 241 & 238 & 190 & 192 & 170 & 167 \\ 
\hline
$T=500$ & DGP 13 & DGP 14 & DGP 15 & DGP 16 & DGP 17 & DGP 18 \\\hline
Method 1 & 274 & 273 & 228 & 233 & 285 & 278 \\ 
Method 2 & 240 & 241 & 240 & 237 & 206 & 202 \\ 
Method 3 & 295 & 286 & 235 & 237 & 207 & 200 \\ 

\hline
$T=600$ & DGP 13 & DGP 14 & DGP 15 & DGP 16 & DGP 17 & DGP 18 \\ \hline
Method 1 & 328 & 321 & 269 & 261 & 336 & 331 \\ 
Method 2 & 273 & 276 & 277 & 275 & 235 & 238 \\ 
Method 3 & 354 & 344 & 273 & 274 & 237 & 239 \\
\hline\hline
\end{tabular}
\newline
\textit{Note:} See note for \Cref{tab1b}.
\end{center}
\end{table}

\clearpage
\bibliography{references.bib}

@article{CH2012,
 ISSN = {00129682, 14680262},
 URL = {http://www.jstor.org/stable/41493847},
 author = {Bin Chen and Yongmiao Hong},
 journal = {Econometrica},
 number = {3},
 pages = {1157--1183},
 publisher = {[Wiley, Econometric Society]},
 title = {Testing for smooth structural changes in time series models via nonparametric regression},
 urldate = {2026-10-24},
 volume = {80},
 year = {2012}
}

@article{PT2007,
title = {Selection of estimation window in the presence of breaks},
journal = {Journal of Econometrics},
volume = {137},
number = {1},
pages = {134-161},
year = {2007},
issn = {0304-4076},
doi = {https://doi.org/10.1016/j.jeconom.2006.03.010},
url = {https://www.sciencedirect.com/science/article/pii/S0304407606000418},
author = {M. Hashem Pesaran and Allan Timmermann}
}

@article{Djogbenou2021,
author = {Antoine A. Djogbenou},
title = {Model selection in factor-augmented regressions with estimated factors},
journal = {Econometric Reviews},
volume = {40},
number = {5},
pages = {470--503},
year = {2021},
publisher = {Taylor \& Francis}
}

@article{Djogbenou2024,
author = {Djogbenou, Antoine A.},
title = {Identifying oil price shocks with global, developed, and emerging latent real economy activity factors},
journal = {Journal of Applied Econometrics},
volume = {39},
number = {1},
pages = {128-149},
doi = {https://doi.org/10.1002/jae.3017},
url = {https://onlinelibrary.wiley.com/doi/abs/10.1002/jae.3017},
year = {2024}
}

@article{DS2024,
title = {Tests for group-specific heterogeneity in high-dimensional factor models},
journal = {Journal of Multivariate Analysis},
volume = {199},
pages = {105233},
year = {2024},
issn = {0047-259X},
doi = {https://doi.org/10.1016/j.jmva.2023.105233},
url = {https://www.sciencedirect.com/science/article/pii/S0047259X23000799},
author = {Djogbenou, Antoine A. and Sufana, Razvan}
}

@Book{Davidson1994,
  author={Davidson, James},
  title={{Stochastic limit theory: An introduction for econometricians}},
  publisher={Oxford University Press},
  year=1994,
  month={},
  volume={}
  }

@ARTICLE{IJR2017,
	title = {Rolling window selection for out-of-sample forecasting with time-varying parameters},
	author = {Inoue, Atsushi and Jin, Lu and Rossi, Barbara},
	year = {2017},
	journal = {Journal of Econometrics},
	volume = {196},
	number = {1},
	pages = {55-67}
}

@ARTICLE{SW2017,
	title = {On time-varying factor models: Estimation and testing},
	author = {Su, Liangjun and Wang, Xia},
	year = {2017},
	journal = {Journal of Econometrics},
	volume = {198},
	number = {1},
	pages = {84-101}
}

@Article{DD2013,
  author={Jean-Marie Dufour and Dalibor Stevanovic},
  title={{Factor-augmented VARMA models With macroeconomic applications}},
  journal={Journal of Business \& Economic Statistics},
  year=2013,
  volume={31},
  number={4},
  pages={491-506}
}

@TechReport{CM2003,
  author={Todd E. Clark and Michael W. McCracken},
  title={Forecast-based model selection in the presence of structural breaks},
  year=2002,
  institution={Federal Reserve Bank of Kansas City},
  type={Research Working Paper},
  url={https://ideas.repec.org/p/fip/fedkrw/rwp02-05.html},
  number={RWP 02-05},
}

@article{BB2005,
    author = {Bernanke, Ben S. and Boivin, Jean and Eliasz, Piotr},
    title = {{Measuring the effects of monetary policy: A factor-augmented vector autoregressive (FAVAR) approach}},
    journal = {The Quarterly Journal of Economics},
    volume = {120},
    number = {1},
    pages = {387-422},
    year = {2005}
}

@article{Rossi2005,
 ISSN = {02664666, 14694360},
 author = {Barbara Rossi},
 journal = {Econometric Theory},
 number = {5},
 pages = {962--990},
 publisher = {Cambridge University Press},
 title = {Optimal Tests for Nested Model Selection with Underlying Parameter Instability},
 volume = {21},
 year = {2005}
}

@ARTICLE{BE2011,
title = {Testing for structural breaks in dynamic factor models},
author = {Breitung, Jorg and Eickmeier, Sandra},
year = {2011},
journal = {Journal of Econometrics},
volume = {163},
number = {1},
pages = {71-84}
}

@article{CM2001,
title = "Tests of equal forecast accuracy and encompassing for nested models",
journal = "Journal of Econometrics",
volume = "105",
number = "1",
pages = "85-110",
year = "2001",
issn = "0304-4076",
doi = "https://doi.org/10.1016/S0304-4076(01)00071-9",
url = "http://www.sciencedirect.com/science/article/pii/S0304407601000719",
author = "Todd E Clark and Michael W McCracken"
}

@article{Mccracken2007,
title = {{Asymptotics for out-of-sample tests of Granger causality}},
journal = "Journal of Econometrics",
volume = "140",
number = "2",
pages = "719-752",
year = "2007",
issn = "0304-4076",
doi = "https://doi.org/10.1016/j.jeconom.2006.07.020",
url = "http://www.sciencedirect.com/science/article/pii/S0304407606001539",
author = "Michael W. McCracken"
}

@ARTICLE{HI2015,
 author={Han, Xu and Inoue, Atsushi},
  title={Tests For Parameter Instability In Dynamic Factor Models},
  journal={Econometric Theory},
  year=2015,
  volume={31},
  number={5},
  pages={1117-1152}
}

@Article{Djogbenou2020,
 author={Djogbenou, Antoine A.},
  title={Comovements in the Real Activity of Developed and Emerging Economies: A Test of Global versus Specific International Factors},
  year=2020,
  Number = {3},
Volume = {35},
pages = {344--370},
journal = {Journal of Applied Econometrics}
}

@Article{ABT2015,
  author = {Knut Are Aastveit and Hilde C. Bj{\o}rnland and Leif Anders Thorsrud},
  title = {{What drives oil prices? Emerging versus developed economies}},
  journal = {Journal of Applied Econometrics},
  year = {2015},
  volume = {30},
  pages = {1013-1028}
}

@article{DGP2015,
	Author = {Djogbenou, Antoine and Gon{\c c}alves, S{\'\i}lvia and Perron, Benoit},
	Journal = {Journal of Time Series Analysis},
	Number = {3},
	Pages = {481--502},
	Title = {Bootstrap Inference in Regressions with Estimated Factors and Serial Correlation},
	Volume = {36},
	Year = {2015}}

@article{GMP2017,
	Author = {Gon{\c c}alves, S{\'\i}lvia and McCracken, Michael W. and Perron, Benoit},
	Journal = {Journal of Econometrics},
	Number = {2},
	Pages = {231-252},
	Title = {{Tests of equal accuracy for nested models with estimated factors}},
	Volume = {198},
	Year = 2017}

@article{BN2009,
	Author = {Bai, Jushan and Ng, Serena},
	Journal = {Journal of Applied Econometrics},
	Number = {4},
	Pages = {607-629},
	Title = {Boosting diffusion indices},
	Volume = {24},
	Year = {2009}}

@article{BN2002,
	Author = {Bai, Jushan and Ng, Serena},
	Journal = {Econometrica},
	Number = {1},
	Pages = {191-221},
	Title = {Determining the Number of Factors in Approximate Factor Models},
	Volume = {70},
	Year = {2002}}

@article{BN2013,
	Author = {Bai, Jushan and Ng, Serena},
	Journal = {Journal of Econometrics},
	Number = {1},
	Pages = {18-29},
	Title = {Principal Component Estimation and Identification of Static Factor Model},
	Volume = {176},
	Year = {2013}}

@article{Bai2003,
	Author = {Bai, Jushan},
	Journal = {Econometrica},
	Number = {1},
	Pages = {135-171},
	Title = {Inferential Theory for Factor Models of Large Dimensions},
	Volume = {71},
	Year = {2003}}

@article{BN2006,
	Author = {Bai, Jushan and Ng, Serena},
	Journal = {Econometrica},
	Number = {4},
	Pages = {1133-1150},
	Title = {Confidence Intervals for Diffusion Index Forecasts and Inference for Factor-Augmented Regressions},
	Volume = {74},
	Year = {2006}}

@article{GP2014,
	Author = {Gon{\c c}alves, S{\'\i}lvia and Perron, Benoit},
	Journal = {Journal of Econometrics},
	Number = {1},
	Pages = {156-173},
	Title = {{Bootstrapping factor-augmented regression models}},
	Volume = {182},
	Year = 2014}

@techreport{sargent1977business,
  author      = {Sargent, Thomas J. and Sims, Christopher A.},
  title       = {Business Cycle Modeling Without Pretending to Have Too Much A Priori Economic Theory},
  institution = {Federal Reserve Bank of Minneapolis},
  type        = {Working Paper},
  number      = {55},
  year        = {1977}
}

@incollection{geweke1977dynamic,
  author    = {Geweke, John},
  title     = {The dynamic factor analysis of economic time series},
  booktitle = {Latent Variables in Socio-Economic Models},
  editor    = {Aigner, Dennis J. and Goldberger, Arthur S.},
  pages     = {365--383},
  year      = {1977},
  publisher = {North-Holland},
  address   = {Amsterdam}
}

@article{stock2002forecasting,
  title={Forecasting Using Principal Components from a Large Number of Predictors},
  author={Stock, James H and Watson, Mark W},
  journal={Journal of the American Statistical Association},
  volume={97},
  number={460},
  pages={1167--1179},
  year={2002},
  publisher={Taylor \& Francis}
}

\end{document}